%% file: Q_correl_sub_arXiv.tex
\documentclass[10pt]{article}
\usepackage{amssymb,amsthm,latexsym,authblk,amsmath,pstricks}
\usepackage{graphicx}
\usepackage{color} 
\usepackage{psfrag}
\usepackage{subfig}
\usepackage{epsfig}
\usepackage{epstopdf} 
\newcommand{\re}{\mathrm{Re}\, }

\newcommand{\I}{{\rm{i}}}
\newcommand{\D}{{\rm{d}}}
\newcommand{\E}{e}
\newcommand{\tr}{\operatorname{tr}}
\newcommand{\trace}{\operatornamewithlimits{\tr}}

\newcommand{\dss}{\displaystyle}
\newcommand{\nn}{\nonumber}

\newcommand{\be}{\begin{equation}}
\newcommand{\ee}{\end{equation}}
\newcommand{\bea}{\begin{eqnarray}}
\newcommand{\eea}{\end{eqnarray}}
\newcommand{\range}{\operatorname{ran}}
\newcommand{\Span}{\operatorname{span}}

\newcommand{\rank}{\operatorname{rank}}
\newcommand{\ket}[1]{| #1 \rangle}
\newcommand{\bra}[1]{\langle #1 |}
\newcommand{\braket}[2]{\langle #1 | #2 \rangle}
\newcommand{\ketbra}[2]{| #1 \rangle \langle #2 |}

\def\real{{\mathbb{R}}}

\def\complex{{\mathbb{C}}}

\def\proba{{\rm I\kern -.18em P}}
\def\condexpectation{{\mathbb{E}}}
\newcommand{\Proof}{\noindent {\it Proof. }}
\newcommand{\Proofof}[1]{\noindent {\it Proof of #1. }}
\newcommand{\finpro}{\hfill $\Box$}

\newcommand{\ie}{i.e.\;}
\newcommand{\ifif}{if and only if\;\,}
\newcommand{\RHS}{right-hand side\;\,}
\newcommand{\LHS}{left-hand side\;\,}
\newcommand{\onehalf}{\frac{1}{2}}
\newcommand{\out}{{\rm out}}
\newcommand{\inp}{{\rm in}}
\newcommand{\lsm}{{\rm lsm}}
\newcommand{\mmax}{{\rm max}}
\newcommand{\mmin}{{\rm min}}
\newcommand{\ONB}{orthonormal basis\;\,}
\newcommand{\ONBs}{orthonormal bases\;\,}
\newcommand{\QO}{quantum operation\;\,}
\newcommand{\QOs}{quantum operations\;\,}

\newcommand{\meas}{measurement\;\,}
\newcommand{\clas}{{\rm clas}}
\newcommand{\sep}{{\rm sep}}
\newcommand{\Aclass}{{{\sf A}\rm{-cl}}}

\newcommand{\opt}{{\rm{opt}}}
\newcommand{\vN}{{\rm{v.N.}}}
\newcommand{\observables}{{\cal{B}} ({\cal{H}} ) }
\newcommand{\saobservables}{ {\cal{B}} ({\cal{H}} )_{\rm s.a.} }
\newcommand{\states}{{\cal E}}
\newcommand{\EoF}{ E_{\rm EoF}}

\newtheorem{theorem}{Theorem}[subsection]
\newtheorem{definition}[theorem]{Definition}
\newtheorem{proposition}[theorem]{Proposition}
\newtheorem{lemma}[theorem]{Lemma}
\newtheorem{corollary}[theorem]{Corollary}

\newtheorem{exercice}[theorem]{Remark}

\newcommand{\av}{{\bf{a}}}
\newcommand{\bv}{{\bf{b}}}
\newcommand{\cv}{{\bf{c}}}

\newcommand{\mv}{{\bf{m}}}
\newcommand{\nv}{{\bf{n}}}

\newcommand{\pv}{{\bf{p}}}
\newcommand{\qv}{{\bf{q}}}

\newcommand{\uv}{{\bf{u}}}

\newcommand{\xv}{{\bf{x}}}
\newcommand{\yv}{{\bf{y}}}

\newcommand{\AAA}{{\sf A}}
\newcommand{\Atwice}{{\sf AA}}
\newcommand{\AB}{{\sf AB}}

\newcommand{\AAE}{{\sf AE}}
\newcommand{\AP}{{\sf AP}}

\newcommand{\ABC}{{\sf ABC}}
\newcommand{\ABE}{{\sf ABE}}

\newcommand{\ARP}{{\sf ARP}}
\newcommand{\ABCE}{{\sf ABCE}}
\newcommand{\BC}{{\sf BC}}

\newcommand{\BB}{{\sf B}}
\newcommand{\Btwice}{{\sf BB}}

\newcommand{\CC}{{\sf C}}

\newcommand{\EE}{{\sf E}}

\newcommand{\PP}{{\sf P}}

\newcommand{\RR}{{\sf R}}

\newcommand{\RP}{{\sf RP}}
\newcommand{\SSS}{{\sf S}}

\newcommand{\SE}{{\sf SE}}
\newcommand{\SP}{{\sf SP}}

\newcommand{\Aa}{{\cal A}}
\newcommand{\Bb}{{\cal B}}
\newcommand{\Cc}{{\cal C}}

\newcommand{\Ee}{{\cal E}}
\newcommand{\Ff}{{\cal F}}

\newcommand{\Hh}{{\cal H}}

\newcommand{\Kk}{{\cal K}}
\newcommand{\Ll}{{\cal L}}
\newcommand{\Mm}{{\cal M}}

\newcommand{\Oo}{{\cal O}}

\newcommand{\Rr}{{\cal R}}
\newcommand{\Ss}{{\cal S}}
\newcommand{\Ttt}{{\cal T}}
\newcommand{\Uu}{{\cal U}}
\newcommand{\Vv}{{\cal V}}

\newcommand{\Xx}{{\cal X}}

\begin{document}

\title{Quantum correlations and distinguishability of quantum states}
\author{Dominique Spehner\\
{\it Univ. Grenoble Alpes and CNRS, Institut Fourier, F-38000 Grenoble, France}
\\
\& \\ 
{\it CNRS and univ. Grenoble Alpes, LPMMC, F-38000 Grenoble, France}}

\date{\today}


\maketitle

\begin{abstract}
A survey of various concepts in quantum information is given, with a main emphasis on the distinguishability of quantum states 
and quantum correlations. Covered topics include generalized and least square measurements, state discrimination, 
quantum relative entropies, the Bures distance on the set of quantum states, the quantum Fisher information, the quantum Chernoff bound, bipartite entanglement, the
quantum discord, and geometrical measures of quantum correlations.
The article is intended both for physicists interested not
only by collections of results but also by the mathematical methods justifying them, and
for mathematicians looking for an up-to-date introductory course on these subjects, which are mainly developed in the physics 
literature.
\end{abstract}

\tableofcontents

\newpage
\section{Introduction}

The fundamental role played by the theory of information in physics has been demonstrated in the last century along with the development of 
statistical physics~\cite{Balian_book}.
More recently,  it has been recognized that  information 
is also at the heart of quantum physics, leading to the emergence of a new field called quantum information.
In few words, quantum information theory is concerned with the use of quantum systems to accomplish information-processing tasks which  
are either not feasible classically or are done classically much less efficiently~\cite{Nielsen}.
These tasks can be related to a computational problem or to communication, for instance, sending encrypted information in a secure way.
Computational tasks are performed on a quantum computer made of qubits. Such qubits are two-level quantum systems in arbitrary superpositions of 
$\ket{0}$ and $\ket{1}$ instead of being either in state $0$ or $1$ as with classical bits. 
A quantum algorithm is a unitary quantum evolution on
a set of qubits followed by a  measurement, the   outcomes of which should provide the solution of the problem. 
For example, the celebrated Shor algorithm factorizes an integer with $N$ digits into prime numbers in a time
${\cal O} (N^2 \ln N \ln ( \ln N))$~\cite{Shor97}, instead of the exponential time required by all known classical algorithms. 
Quantum computers with a few qubits have been implemented in physics laboratories. There is still a lot of debate about whether we will 
be able in the future 
to manipulate coherently many qubits and address them locally during a sufficiently long computational time, and which quantum 
systems are the most promising~\cite{Nielsen,Bouwmeester}.  

The fact that quantum algorithms and communication protocols can  outperform their classical analogs is usually attributed  to
quantum correlations. Such  correlations  in composite quantum systems are  at the origin of 
the violation of the Bell inequalities, which has been confirmed experimentally~\cite{Peres}. These quantum correlations
are quite different in nature from classical correlations in stochastic processes. For a long time they have been identified with entanglement. 
However, there is now
increasing evidence that other types of quantum correlations  in mixed states, which may be present even in unentangled states
and are captured notably by the quantum discord~\cite{Ollivier01,Henderson01}, might
be of relevance in  certain quantum algorithms and communication protocols~\cite{Datta08,Lanyon08,Passante11,Madhok11,Cavalcanti11,Gu2012,Dakic12}.

In this survey article, we review the basic properties of the entanglement measures and quantum discord and present a geometrical
description of these notions based on the Bures distance on the set of quantum states.
In this approach, the quantum discord turns out to be related to the problem of discriminating non-orthogonal quantum states.
Two central questions guide the discussion in this article and can be formulated as follows.
How well can one distinguish unknown quantum states pertaining to a given ensemble   by performing a measurement on a system?
If this system consists of  several particles, does the amount of information one gets from 
measurements on a single particle tell us something about the way the particles are correlated? 
Quantum measurements and entropies obviously come into the game in these two questions. They constitute the subjects of 
 Secs.~\ref{sec-measurements},~\ref{sec-tranpose_op_and_lsm}, and \ref{sec-entropies}. Some answers to the first question are given in 
Secs.~\ref{sec_qsd} and~\ref{sec-QHT_and_parameter_est}, devoted respectively to
state discrimination  and to related topics called hypothesis testing and parameter estimation.
The Bures distance and Uhlmann fidelity are introduced   in Sec.~\ref{sec-Bures_distance}. A detailed account of their properties is given there.
The remaining sections (Secs.~\ref{sec-entanglement}, \ref{sec-discord}, and \ref{sec-geometric_measures}) address the  
problem of quantifying quantum correlations and provide answers to the second question.
It is neither our purpose to discuss thoroughly the (huge amount of) 
quantum correlation measures found in the literature nor to  study how these correlations could explain the quantum efficiencies. 
Well-documented surveys on quantum entanglement already exist, see e.g.~\cite{Horedecki_review,Toth_review}, as well as on the quantum discord and related
measures~\cite{Modi_review}.
The precise role of entanglement  as a resource in quantum computing and quantum communication  is still not completely understood,
in spite of recent progresses (such as the proof that, in order to offer an exponential speedup over classical algorithms, a
quantum algorithm using pure states must produce  entanglement which is not 
restricted to blocks of qubits of fixed size as the system size increases~\cite{Jozsa03}).
The role played  by the discord as a quantum resource is, in turn, still poorly understood and constitutes   
a challenging issue (see~\cite{Modi_review}).  

We concentrate in  our exposition on the mathematical and fundamental aspects of the theory. In particular, we will not investigate  here the
physical implementations and  the system-dependent irreversible dynamical processes destroying (or sometimes producing) quantum correlations.
We present the detailed proofs of some selected fundamental results, instead of relating all important achievements 
obtained so far.
Most of these results have been published in 
physics journals, and are sometimes explained in the original papers without full mathematical rigor in their derivation. 
Others have been published in mathematical journals 
with full proofs, which are nevertheless given here for completeness. We try to emphasize how the results are connected between themselves
 and to stress the similarities in the arguments used to derive them. This sometimes leads to new proofs.

Quantum information is a rapidly growing  field of research
and the amount of articles and surveys devoted to it is already considerable.
Researchers who got  interested by this subject recently (including the author) may fear to have difficulties to form  a clear opinion
about the most pertinent  open questions.
Significant contributions have been made by physicists, mathematicians, and computer scientists, who constitute a broad community
with different viewpoints. 
We hope that this article may be useful to mathematicians, by providing examples of interesting problems  and explaining the mathematical tools 
used to tackle them. 
It may also be of help to physicists wishing to get acquainted with such tools, which could be useful to derive new results. 
The paper is written as an introductory course. Certain statements appear as remarks, which play the role of exercises, with
the main arguments to justify them. We encourage the reader to complete these proofs by himself. 
This work is intended to be complementary to  other surveys containing
collections of results without explicit derivations and to more introductory monographs like~\cite{Nielsen}, which do not include 
the most recent advances. 

The following comments on the structure of the article might be helpful.
The contents of Sec.~\ref{sec_qsd}, Sec.~\ref{sec-Bures_distance}, and Secs.~\ref{sec-entanglement}-\ref{sec-discord}
are largely independent. 
On the other hand, Sec.~\ref{sec_qsd} is partly related to Sec.~\ref{sec-tranpose_op_and_lsm}, and 
Sec.~\ref{sec-QHT_and_parameter_est} makes use of the results of  Sec.~\ref{sec_qsd} and  Sec.~\ref{sec-Bures_metric}.
The material of Secs.~\ref{sec-von_Neumann_wentropy} and \ref{sec-relative_entropy} is relevant for Secs.~\ref{sec-entanglement} and \ref{sec-discord}.
Section~\ref{sec-geometric_measures} needs more or less the knowledge of all previous sections. 
The main definitions and theorems presented in Secs.~\ref{sec-states} and~\ref{sec-measurements} are used in the whole article.
Two appendices contain textbook issues about operator convex functions and some less standard trace inequalities.

Before closing this introduction, let us warn the reader that
we will be exclusively concerned  by quantum systems with {\it finite-dimensional Hilbert spaces}. This  is motivated for two reasons. Firstly, this is the 
case of most systems in quantum information theory. Secondly, in this way  one avoids the
technical complications of infinite-dimensional spaces and concentrates oneself on the main ideas and concepts.
Some of these concepts have been originally worked out in the general setting of $C^\ast$-algebras, but we shall present here
simpler proofs applying  to the finite-dimensional case only.

\section{Quantum states} \label{sec-states}

In this section we review the basic definitions of pure and mixed states,  entangled states, and 
the pure state decompositions and purifications of mixed states. 
Before that, we introduce in Sec.~\ref{sec-states_and_observables} some notation and define a few  mathematical objects from the theory of operator algebras,
which will be used repeatedly in this article. 
In Sec.~\ref{eq-Schmist_dec} we discuss an extremely useful result from linear algebra, namely,  the Schmidt decomposition.

In all what follows, capital letters $\AAA$, $\BB$, etc., refer to quantum systems,
$\Hh_\AAA$, $\Hh_\BB$, etc., denote their Hilbert spaces, and
$n_\AAA= \dim \Hh_\AAA$, $n_\BB =\dim \Hh_\BB$, etc., the
dimensions of these spaces. These dimensions are always assumed to be finite.
A bipartite system  $\AB$ formed by putting together the systems $\AAA$ and $\BB$ has Hilbert space
given by the tensor product $\Hh_\AB = \Hh_\AAA \otimes \Hh_\BB$. For instance, if $\AAA$ and $\BB$ are two qubits with Hilbert spaces 
$\Hh_\AAA \simeq \Hh_\BB \simeq \complex^2$, the space of these two qubits is $\Hh_\AB =\complex^2 \otimes \complex^2 \simeq \complex^4$. 
Similarly, $\Hh_{\AAA_1 \ldots \AAA_k} = \Hh_{\AAA_1}  \otimes \cdots \otimes \Hh_{\AAA_k}$ is the Hilbert space of the multipartite system
formed by putting together the systems $\AAA_1, \ldots ,\AAA_k$. The tensor
product vectors $\ket{\psi_\AAA} \otimes \ket{\phi_\BB} \in \Hh_\AB$ will be denoted either by
$\ket{\psi_\AAA \otimes \phi_\BB}$ or, more often\footnote{
As common in the physics literature we do not write the tensor product symbol $\otimes$ explicitly.
},
by $\ket{\psi_\AAA} \ket{\phi_\BB}$.

\subsection{Quantum states and observables} \label{sec-states_and_observables}

A {\it state} of a quantum system with Hilbert space $\Hh$ is given by a density matrix $\rho$, that is, a non-negative operator on $\Hh$ with 
unit trace $\tr \rho = 1$. We write $\states ( \Hh )$ the convex cone formed by all states on $\Hh$.
States will  always be denoted by the letters $\rho$, $\sigma$, or $\tau$, with subscripts referring to the corresponding system
if necessary. The  extreme points of the cone
$\states ( \Hh) $ are the {\it pure states} $\rho_\psi = \ketbra{\psi}{\psi}$, with $\ket{\psi} \in \Hh$, $\| \psi\|=1$
(here $\ketbra{\psi}{\psi}$ designates the rank-one orthogonal projector onto $\complex \ket{\psi}$). 
The pure states can be identified with elements of the projective space $P \Hh$, that is, the set
of equivalence classes  of normalized vectors in $\Hh$ modulo a phase factor. The vectors 
$\E^{\I \theta} \ket{\psi} \in \Hh$ with $0 \leq \theta < 2 \pi$ are called the representatives of $\rho_{\psi} \in P \Hh$. We will abusively write $\ket{\psi}$ 
instead of $\rho_\psi$, except when this may be a source of confusion.  
If $\rho$ is a state of a bipartite system $\AB$ with Hilbert space $\Hh_{\AB} = \Hh_\AAA \otimes \Hh_\BB$, 
the {\it reduced states} of  $A$ and $B$ are defined by partial tracing $\rho$ 
over the other subsystem. They  are denoted by $\rho_A = \tr_{B} (\rho) \in \states ( \Hh_\AAA)$ and $\rho_B=\tr_A(\rho) \in \states ( \Hh_\BB)$.
These reduced states correspond to the marginals of a joint probability in classical probability theory.

The $C^\ast$-algebra  of bounded linear operators from 
$\Hh $ to $\Hh' $ is denoted by $\Bb (\Hh , \Hh ')$, and we write $\Bb ( \Hh ) = \Bb (\Hh , \Hh )$. 
In our finite-dimensional setting,  $\Bb (\Hh , \Hh ')$ is the algebra of all $n' \times n$ finite complex matrices, with
 $\dim \Hh = n$ and $\dim \Hh'=n'$. The Hilbert-Schmidt scalar product on $\Bb (\Hh , \Hh ')$ is defined by
\begin{equation} \label{eq-Hilbert_Schmidt_product}
\langle X \, , \, Y \rangle = \tr ( X^\ast Y )\;,
\end{equation}
where $X^\ast$ denotes the adjoint operator of $X$. The associated norm is $\| X \|_2 = [\tr ( X^\ast X )]^\onehalf$.
The set of states $\states ( \Hh) $ can be endowed with the distances\footnote{
We shall see in Sec.~\ref{sec-Bures_distance} that there are other  more natural distances on 
$\states (\Hh)$ from a quantum information point of view.
}
%
\begin{equation} \label{eq-Lp_norm}
d_p ( \rho, \sigma) = \| \rho-\sigma\|_p = \bigl[ \tr ( |\rho-\sigma|^p ) \bigr]^{\frac{1}{p}}
\end{equation}
with $p \geq 1$. Here $|X|$ denotes the non-negative operator $|X|= \sqrt{X^\ast X }$. 
 When $p \rightarrow \infty$, $\| X \|_p$ converges to the operator norm $\| X \|_\infty = \| X \|$ of $X$, that is, the maximal eigenvalue of 
$| X |$.  The H\"older inequality reads
\begin{equation} \label{eq-main_property_L_P_norm}
\| X  \|_p = \max_{Y, \| Y \|_q = 1} | \tr ( X Y ) |
\end{equation}
with  $p>1$ and $q = p/(p-1)$. This  still holds for $p=1$ and $q=\infty$, as 
can be shown by using the Cauchy-Schwarz inequality for the scalar product (\ref{eq-Hilbert_Schmidt_product}). 
In that case the maximum is achieved \ifif  $Y U | X|^\onehalf = e^{\I \theta} | X|^\onehalf$ with $\theta \in [0,2\pi)$ and 
$U$ a unitary such that $X = U | X |$ (polar decomposition). 

A self-adjoint operator $O \in \observables$ is called an  {\it  observable}. 
The real vector space of all observables on $\Hh$ is denoted by $\observables_{\rm s.a.}$. 
If $\AB$ is a bipartite system, one says that $O \in \Bb (
\Hh_{\AB})_{\rm s.a}$ is a {\it local observable} if either $O = A \otimes 1$ or $O = 1 \otimes B$, with 
$A \in \Bb ( \Hh_\AAA)_{\rm s.a.}$ and $B \in \Bb (\Hh_\BB)_{\rm s.a.}$.
Here and in the following, $1$ stands for the identity operator on
$\Hh_\AAA$, $\Hh_\BB$, or another space.

A  linear map\footnote{
Operators acting  on the vector space  of observables $\observables_{\rm s.a.}$ 
or on the whole algebra $\Bb ( \Hh)$ are always denoted by calligraphic letters.
}
$\Mm : \Bb ( \Hh ) \rightarrow \Bb ( \Hh ' )$ is positive if it transforms a non-negative operator into a non-negative operator.
It  is completely positive (CP) if the map
\begin{equation}
\Mm \otimes 1 : X \in \Bb ( \Hh   \otimes \complex^m)\; \mapsto \; \sum_{k,l=1}^m \Mm ( X_{kl} ) \otimes \ketbra{k}{l} \in \Bb ( \Hh ' \otimes \complex^m)
\end{equation}
is positive for any integer $m \geq 1$.

Given two orthonormal bases $\{ \ket{i} \}_{i=1}^{n_\AAA}$ of $\Hh_\AAA$ and $\{ \ket{j} \}_{j=1}^{n_\BB}$ of $\Hh_\BB$, one can identify 
any operator $O : \Hh_\BB \rightarrow \Hh_\AAA$ with a vector $\ket{\widetilde{\Psi}_O} \in \Hh_\AAA \otimes \Hh_\BB$ thanks to the bijection
\begin{equation} \label{eq-ismoemtry_operators_vectors}
O \mapsto \ket{\widetilde{\Psi}_O} = \sum_{i,j} \bra{i} O \ket{j} \ket{i} \ket{j}
\;.
\end{equation}
This bijection is an isomorphism between the Hilbert spaces $\Bb ( \Hh_\BB, \Hh_\AAA)$ 
(endowed with the scalar product (\ref{eq-Hilbert_Schmidt_product})) and $\Hh_\AB$.
Similarly, one can represent the linear map $\Mm : \Bb ( \Hh_\BB) \rightarrow \Bb ( \Hh_\AAA)$
by an operator $O_\Mm$ acting on $\Hh_\Btwice = \Hh_\BB \otimes \Hh_\BB$ with values in 
$\Hh_\Atwice = \Hh_\AAA \otimes \Hh_\AAA$. The matrix elements of this operator in the
product bases  $\{ \ket{k} \ket{l} \}_{k,l=1}^{n_\BB}$  of  $\Hh_\Btwice$ and $\{ \ket{i} \ket{j} \}_{i,j=1}^{n_\AAA}$ of $\Hh_\Atwice$ are given by 
$
(O_\Mm)_{ij,kl} = \bra{i} \Mm ( \ketbra{k}{l} ) \ket{j}
$.
This representation is an $\ast$-isomorphism between the $C^\ast$-algebras $\Bb ( \Bb ( \Hh_\BB), \Bb(\Hh_\AAA))$ and 
$\Bb ( \Hh_\Btwice , \Hh_\Atwice )$. 
The so-called reshuffling operation~\cite{Bengtsson} associates to $O_\Mm$ the operator  $O_\Mm^\Rr \in \Bb ( \Hh_\AB )$ 
with matrix elements $(O^\Rr_\Mm)_{ik,jl} = (O_\Mm)_{ij,kl}$, which satisfies
\begin{equation} \label{eq-reshuffling_op}
\langle A \otimes B \, ,\,    O^\Rr_\Mm \rangle = \bra{\widetilde{\Psi}_A} O_\Mm J \ket{\widetilde{\Psi}_B}
= \langle A ,  \Mm ( \overline{B}) \rangle
\end{equation}
for any $A \in \Bb ( \Hh_\AAA)$ and $B \in \Bb ( \Hh_\BB)$. 
Here $J$ denotes the anti-unitary operator  on $\Hh_\Btwice$ defined by 
$\bra{k} \bra{l} J \ket{\Psi} = \overline{ \bra{k} \braket{l}{\Psi}}$
(complex conjugation in the canonical basis) and 
$\overline{B} = \sum_{k,l} \overline{\bra{k} B \ket{l}} \ketbra{k}{l}$ is the operator associated to 
$J \ket{\widetilde{\Psi}_B}$ via the isomorphism (\ref{eq-ismoemtry_operators_vectors}).  
With these definitions,  $\Mm : \Bb ( \Hh_\BB) \rightarrow \Bb ( \Hh_\AAA)$ is CP \ifif  $O^\Rr_\Mm \geq 0$, that is, 
$O^\Rr_\Mm$ has non-negative eigenvalues\footnote{
Actually, $O^\Rr_\Mm \geq 0$ is equivalent to $O^\Rr_\Mm = A^\ast A$ for some $A \in \Bb ( \Hh_\AB )$, that is, to 
$(O^\Rr_\Mm)_{ik,jl} = \bra{i} \Mm ( \ketbra{k}{l} ) \ket{j} 
= \sum_{p,q} \overline{A_{pq,ik}} A_{pq,jl}$ for all
$i,j = 1, \ldots, n_\AAA$ and $k,l=1,\ldots , n_\BB$. Setting $A_{pq} = \sum_{i,k} \overline{A_{pq,ik}} \ketbra{i}{k}$, 
it follows that     $O^\Rr_\Mm \geq 0$ \ifif $\Mm ( X) = \sum_{pq} A_{pq} X A_{pq}^\ast$ for all $X \in \Bb ( \Hh_\BB)$, which is equivalent to 
$\Mm$ being CP by the Kraus representation theorem (Theorem~\ref{theo-Kraus} below).
}.

The left and right
multiplications $\Ll_X$ and  $\Rr_X$  by $X \in \observables$ are the operators from $\Bb ( \Hh)$ into itself defined by\footnote{
In the $C^\ast$-algebra setting, the map $X \mapsto \Ll_X$ is the
Gelfand-Neumark-Segal representation of the $C^\ast$-algebra~\cite{Bratteli}.  
}
%
\begin{equation} \label{eq-left_and_right_multilication}
\Ll_X ( Y ) = X Y 
\quad , \quad 
\Rr_X ( Y ) = Y X
\quad , \forall\; Y \in \observables\;.
\end{equation}
They are represented on $\Bb ( \Hh \otimes \Hh)$  by local operators $X
\otimes 1$ and $1 \otimes X^T$, respectively, where $T$ stands for the transposition in the basis $\{ \ket{i}\}$.
Given two states $\rho$ and $\sigma \in \states (\Hh )$ with $\rho>0$, the Araki relative modular operator $\Delta_{\sigma|\rho}$ is defined by~\cite{Araki82}
\begin{equation} \label{eq-def_modular_op}
\Delta_{\sigma|\rho} ( Y ) = \sigma Y \rho^{-1} = \Ll_{\sigma} \circ \Rr_{\rho^{-1}} (Y)
\quad , \forall\; Y \in \observables\;.
\end{equation} 
It is a self-adjoint non-negative operator on the Hilbert space $\Bb ( \Hh)$ (for the scalar product
 (\ref{eq-Hilbert_Schmidt_product})).

\subsection{The Schmidt decomposition} \label{eq-Schmist_dec}

The following standard result is very useful in quantum information theory.   

\vspace{1mm}

\begin{theorem} {\rm{ (Schmidt decomposition)}} \label{theo_Schmidt_dec}
Any pure state 
$\ket{\Psi} \in \Hh_\AAA \otimes \Hh_\BB$ of a bipartite system admits a decomposition
\be \label{eq-Schmidt_decomposition}
\ket{\Psi} = \sum_{i=1}^{n} \sqrt{\mu_i} \ket{\alpha_i} \ket{\beta_i}
\ee
where $n =\min\{ n_\AAA , n_\BB \} $,  $\mu_i \geq 0$, and  $\{ \ket{\alpha_i} \}_{i=1}^{n}$ (respectively $\{ \ket{\beta_i} \}_{i=1}^{n}$) is
an orthonormal family of $\Hh_\AAA$ (respectively of $\Hh_\BB$).
The $\mu_i$ and $ \ket{\alpha_i} $ (respectively $\ket{\beta_i}$) are the eigenvalues and
 eigenvectors of the reduced state
$\rho_\AAA = \trace_\BB ( \ketbra{\Psi}{\Psi} ) $  (respectively $\rho_\BB = \tr_\AAA ( \ketbra{\Psi}{\Psi} )$). 
Thus, if the eigenvalues $\mu_i$ are non-degenerate then the decomposition  (\ref{eq-Schmidt_decomposition}) is unique.
\end{theorem}

The non-negative numbers $\mu_i$ are called the {\it Schmidt coefficients} of $\ket{\Psi}$. They satisfy $\sum_i \mu_i=\| \Psi\|^2 = 1$.

\vspace{2mm}

\proof 
Let $\{ \ket{i} \}_{i=1}^{n_\AAA}$ and $\{ \ket{j} \}_{j=1}^{n_\BB}$ be some fixed orthonormal bases of $\Hh_\AAA$ and $\Hh_\BB$. By using the isomorphism 
$\ket{\Psi} \mapsto O_\Psi = \sum_{i,j} \braket{i \otimes j}{\Psi}  \ketbra{i}{j}$ between  $\Hh_\AB$  
and the space of $n_\AAA \times n_\BB$ matrices (see Sec.~\ref{sec-states_and_observables}), we observe that the
decomposition (\ref{eq-Schmidt_decomposition}) corresponds to the singular value decomposition of $O_\Psi$, that is, 
$O_\Psi = U_\AAA \sum_i \sqrt{\mu_i} \ketbra{i}{i} U^\ast_\BB$ with $\mu_i$  the eigenvalues  of $O_\Psi^\ast O_\Psi$
and $U_\AAA$ and $U_\BB$  unitaries on $\Hh_\AAA$ and $\Hh_\BB$. Then
$U_\AAA \ket{i}=\ket{\alpha_i}$ and $U_\BB \ket{i} = \ket{\beta_i^\ast}$ are eigenvectors of
$O_\Psi O_\Psi^\ast$ and $O_\Psi^\ast O_\Psi$, respectively. Denoting by $J$ is the complex conjugation in the basis 
$\{ \ket{j}\}$ (see above), one has $\ket{\beta_i} = J \ket{\beta_i^\ast}$.  
\finpro

\vspace{2mm}

The Schmidt decomposition can be generalized to mixed states by considering $\rho$ as a vector in the 
Hilbert space $\Bb (\Hh_\AAA) \otimes \Bb ( \Hh_\BB)$. Any $\rho \in \states ( \Hh_\AB)$ can be written as
\begin{equation} \label{eq-mixed_state_Schmidt_dec}
\rho = \sum_{m=1}^{n^2} \sqrt{\mu_{m}} X_{m} \otimes Y_{m}\;,
\end{equation}  
where $\{ X_{m} \}_{m=1}^{n_\AAA^2}$ and $\{ Y_{m}\}_{m=1}^{n_\BB^2}$ are orthonormal bases of $\Bb ( \Hh_\AAA)$ and $\Bb ( \Hh_\BB)$ for the scalar product
(\ref{eq-Hilbert_Schmidt_product}) and $\mu_{m}$ are the eigenvalues of the $n^2_\AAA \times n^2_\AAA$ matrix $R \geq 0$ defined by
\begin{equation}
R_{ij,i'j'} = \Bigl\langle \rho\,\ketbra{i}{i'} \otimes 1 \, ,\, \ketbra{j}{j'} \otimes 1 \, \rho  \Bigr\rangle 
\end{equation}
(the $R_{ij,i'j'}$ are the matrix elements  in the orthonormal basis $ \{ \ketbra{i}{j}\}_{i,j=1}^{n_\AAA^2}$ of $\Bb ( \Hh_\AAA)$ of 
the operator playing the role of the reduced state in Theorem~\ref{theo_Schmidt_dec}).
Note that $\sum_m \mu_m= \tr ( \rho^2) \leq 1$, with equality \ifif $\rho$ is a pure state.

\vspace{1mm}

\begin{exercice}
Alternatively, the  $\mu_{m}$ are the square roots of the singular values of $\rho^\Rr \in \Bb( \Hh_\Btwice , \Hh_\Atwice)$, where $\Rr$ is
 the reshuffling operation (Sec.~\ref{sec-states_and_observables}),
and $X_m$ and $Y_m$ are given in terms of the eigenvectors $\ket{\chi_m}$ 
 and $\ket{\psi_m}$ of  $\rho^\Rr (\rho^\Rr )^\ast$ and $(\rho^\Rr )^\ast \rho^\Rr$ by
$X_m= \sum_{i,j} \braket{i \otimes j}{\chi_m} \ketbra{i}{j}$ and
$Y_m= \sum_{k,l} \overline{\braket{k \otimes l}{\psi_m}} \ketbra{k}{l}$, respectively.

\vspace{1mm}

\noindent Proof. {\rm 
Considering $\rho$ as a vector in $\Bb (\Hh_\AAA) \otimes \Bb ( \Hh_\BB)$ and introducing two  orthonormal bases
$\{ A_p \}$ of $\Bb ( \Hh_\AAA)$ and $\{ B_q \}$ of $\Bb ( \Hh_\BB)$, 
according to the proof of Theorem~\ref{theo_Schmidt_dec}, $\sqrt{\mu_m}$ are the singular values of
the $n_\AAA^2 \times n_\BB^2$ matrix $(\langle A_p \otimes B_q , \rho \rangle )_{p,q}$. Denote by
$\{ \ket{\alpha_p} \}$ and $\{ \ket{\beta_q} \}$  the
orthonormal bases of $\Hh_\Atwice$ and $\Hh_\Btwice$ associated to $\{ A_p \}$ and $\{ B_q \}$ via the isomorphism (\ref{eq-ismoemtry_operators_vectors}).
The statement follows by choosing $A_p = \ketbra{i}{j}$ and $B_q = \ketbra{k}{l}$ and using the identity
$\bra{\alpha_p} \rho^\Rr J \ket{\beta_q} = \langle A_p \otimes B_q , \rho \rangle$, see (\ref{eq-reshuffling_op}). 
\finpro
}
\end{exercice}

\subsection{Purifications and pure state decompositions of mixed states}
\label{sec-purification}

\vspace{2mm}

\begin{definition}
Let $\rho$ be an arbitrary state on $\Hh$ and $\Kk$ be another Hilbert space.  A pure state $\ket{\Psi}
\in \Hh \otimes \Kk$ such that $\rho = \tr_\Kk ( \ketbra{\Psi}{\Psi}
)$ is called a purification of $\rho$ on $\Hh \otimes \Kk$.
\end{definition}

In the language of $C^\ast$-algebras, a purification is an example of cyclic representation of a state~\cite{Bratteli}.  An example
of purification of $\rho$ on $\Hh \otimes \Hh$ is
\be \label{eq-example_purification} \ket{\Psi} = \sum_{k=1}^n
\sqrt{p_k} \ket{k} \ket{k}\;,  
\ee
where $\rho = \sum_k p_k \ketbra{k}{k}$ is a spectral decomposition of $\rho$.
If $\ket{\Psi}$ and $\ket{\Phi}$ are two purifications of $\rho$ on
the same space $\Hh \otimes \Kk$, then there exists a unitary operator
$U$ acting on $\Kk$ such that $\ket{\Phi} = 1 \otimes U
\ket{\Psi}$. In fact, one infers from  the Schmidt decomposition that
$\ket{\Psi} = \sum_k \sqrt{p_k} \ket{k} \ket{f_k}$ and $\ket{\Phi} =\sum_k \sqrt{p_k} \ket{k} \ket{g_k}$, 
where $\{ \ket{f_k}\}_{k=1}^n$ and $\{ \ket{g_k}\}_{k=1}^n$ are two orthonormal families of $\Kk$. Thus $\ket{g_k} = U \ket{f_k}$ for some unitary $U$.

We will often be interested in the sequel by families of quantum states of a system $\SSS$,  $\rho_i \in \states ( \Hh_\SSS)$, $i=1,\ldots, m$, 
to which we attach some probabilities $\eta_i \geq 0$, $\sum_i \eta_i = 1$. Following the terminology employed  by physicists in statistical physics, 
we call $\{ \rho_i, \eta_i \}_{i=1}^m$ an {\it ensemble of quantum states}  (or more simply an {\it ensemble}). 
A {\it convex decomposition} of $\rho$ is an ensemble $\{ \rho_i , \eta_i \}_{i=1}^m$ such that $\rho = \sum_i \eta_i \rho_i$.  A {\it pure state
  decomposition}  of $\rho$ is a convex decomposition in terms of finitely many pure
states $\rho_i = \ketbra{\psi_i}{\psi_i}$, \ie
\be \label{eq-convex_decomp} \rho = \sum_{i=1}^m \eta_i
\ketbra{\psi_i}{\psi_i}\;.  \ee
If the vectors $\ket{\psi_i}$ are orthogonal, then (\ref{eq-convex_decomp}) coincides with the spectral decomposition,
but we will see that there are infinitely many other ways to decompose $\rho$.  Physically,
(\ref{eq-convex_decomp}) describes a {\it state preparation}: it means that the system has been prepared  in the pure state
$\ket{\psi_i}$ with probability $\eta_i$.  The non-uniqueness of the
decomposition can be interpreted as follows. If a receiver is given two ensembles
$\{ \ket{\psi_i},\eta_i \}_{i=1}^m$ and $\{ \ket{\phi_j }, \xi_j \}_{j=1}^{p}$  corresponding to different 
state preparations of two identical systems in the same state $\rho$, then he cannot make any difference between them 
if he has no prior knowledge on the state preparation.  
Indeed, any measurement 
performed by him gives rise to the same  distribution of outcomes  for the two ensembles. In other words, 
the full information that the receiver can collect on the system via measurements is encoded in $\rho$, and {\it not} 
in the ensemble involved in the state preparation.
This very important fact  has consequences that are sometimes disconcerting to people
unfamiliar with the conceptual aspects of quantum mechanics. For instance, if a preparer gives a
maximally mixed state $\rho = 1/n$ to a receiver, the latter has no way to decide whether this state was prepared from $n$ equiprobable orthonormal
pure states (which are only known by the preparer) or if it was prepared by another procedure involving more than $n$ states. 
It is also worth mentioning that the process  transforming the ensemble $\{ \rho_i, \eta_i \}_{i=1}^m$ into the average state $\rho = \sum_i \eta_i \rho_i$,
which can be viewed as the inverse of a convex decomposition, corresponds physically to a  loss of information about the state preparation.

Given a fixed \ONB $\{ \ket{f_i}\}_{i=1}^{p}$ of $\Kk$ with $p \geq \range ( \rho)=r$, 
there is a  one-to-one correspondence between pure state decompositions of $\rho$ 
containing at most $p$ states and
purifications of $\rho$ on $\Hh \otimes \Kk$.  Actually, given the pure state decomposition (\ref{eq-convex_decomp}),
\be \label{eq-purification_and_convex_decomp} 
\ket{\Psi} = \sum_{i=1}^p \sqrt{\eta_i} \ket{\psi_i}  \ket{f_i} 
\ee
defines a purification of $\rho$ on $\Hh \otimes \Kk$ (we have set $\eta_i=0$ for $m < i \leq  p$). 
Reciprocally, let $\ket{\Psi}$ be a purification of $\rho$ on  $\Hh \otimes \Kk$. Denote as before the eigenvalues and orthonormal eigenvectors of $\rho$ by
$p_k$ and $\ket{k}$.
As  argued above, one can find a unitary $U$ on $\Kk$ such that
\begin{equation} \label{eq-time_to_go_to_bed}
\ket{\Psi} = \sum_{k=1}^r \sqrt{p_k} \ket{k} U \ket{f_k} = \sum_{i=1}^p \sum_{k=1}^r  \sqrt{p_k} \bra{f_i} U \ket{f_k}  \ket{k} \ket{f_i}
= \sum_{i=1}^p \sqrt{\eta_i} \ket{\psi_i}  \ket{f_i} 
\end{equation}
with $\sqrt{\eta_i} \ket{\psi_i}  = \sum_{k}  \sqrt{p_k} \bra{f_i} U \ket{f_k}  \ket{k}$. 
Hence $\ket{\Psi}$ has the form (\ref{eq-purification_and_convex_decomp}). Taking the partial trace over $\Kk$, one can associate to it a unique
pure state decomposition, which is given by (\ref{eq-convex_decomp}).  

Since two purifications $\ket{\Psi}$ and $\ket{\Phi}$ of the same state $\rho$ are related by a local unitary $U$ acting on the
ancilla space $\Kk$, this implies that any two pure state decompositions 
$\rho = \sum_{i=1}^m \eta_i \ketbra{\psi_i}{\psi_i}$ and $\rho= \sum_{j=1}^p \xi_j \ketbra{\phi_j}{\phi_j}$ are related by
\begin{equation} \label{eq-link_between_pure_state_dec}
 \sqrt{\xi_j} \ket{\phi_j} = \sum_{i=1}^{\max \{ m, p\}} u_{ji} \sqrt{\eta_i} \ket{\psi_i}\;,
\end{equation}
where $(u_{ji})$ is a unitary matrix with size  $\max\{ m,p\}$ (if $m < i \leq p$ we set as before $\eta_i =0$).

\subsection{Entangled and separable states} \label{sec-def_entanglement}

Let us consider a bipartite system $\AB$. 
If this system is in a tensor product state $\ket{\Psi_{\rm sep}} = \ket{\psi_\AAA} \ket{\phi_\BB}$ with 
$\ket{\psi_\AAA} \in \Hh_\AAA$ and  $\ket{\phi_\BB} \in \Hh_\BB$, then
the expectation value of the product of two local observables  $A\otimes 1$ and $1 \otimes B$ coincides with the product of the expectations values, \ie
\begin{equation} \label{eq-correlation_function}
G_{AB} ( \ket{\Psi_{\rm sep}}) =
\bra{\Psi_{\rm sep}} A \otimes B \ket{\Psi_{\rm sep}} - \bra{\Psi_{\rm sep}} A \otimes 1 \ket{\Psi_{\rm sep}} \bra{\Psi_{\rm sep}} 1 \otimes B  \ket{\Psi_{\rm sep}}
= 0\;.
\end{equation}
This means that the random outcomes of measurements of the local observables $A \otimes 1$
and $1 \otimes B$ are uncorrelated. More generally, if one thinks of $\AB$ as a pair of particles located far apart (e.g. a photon pair shared by two observers
Alice and Bob), 
this pair is in  a product state
\ifif there are no correlations between the results of arbitrary local measurements performed independently on each particle (for instance, if Alice 
sends her photon through a polarizer  and then to a  photodetector, and Bob does the same with his photon, 
the clicks of the two detectors will be uncorrelated whatever the polarizer angles). 
One says that $\ket{\Psi_{\rm sep}} = \ket{\psi_\AAA} \ket{\phi_\BB}$ is a  {\it separable state}. If 
the pure state $\ket{\Psi} \in \Hh_\AB$ is not a product state one says that it is {\it entangled}.

By applying the Schmidt decomposition, one sees that  $\ket{\Psi}$ is separable \ifif all its Schmidt coefficients  vanish except one, that is, \ifif 
its reduced states $\rho_\AAA$ and $\rho_\BB$  are pure.
In the opposite, if either $\rho_\AAA$ or $\rho_\BB$ is proportional to the identity matrix (maximally mixed state), 
we say that $\ket{\Psi}$ is {\it maximally entangled}. Such states have the form
\begin{equation}
\ket{\Psi_{\rm ent}} = \frac{1}{\sqrt{n}} \sum_{i=1}^n \ket{\alpha_i} \ket{\beta_i}\;,
\end{equation} 
where $\{ \ket{\alpha_i} \}_{i=1}^n$ and  $\{ \ket{\beta_i} \}_{i=1}^n$  are orthonormal families in $\Hh_\AAA$ and $\Hh_\BB$ and $n = \min \{ n_\AAA, n_\BB\}$.
For instance, denoting by $\ket{0}$ and $\ket{1}$ the canonical basis vectors of $\complex^2$,
the EPR (or Bell) states 
$\ket{\Phi_\pm} = ( \ket{0} \ket{0} \pm \ket{1}\ket{1})/\sqrt{2}$ and $\ket{\Psi_\pm} = ( \ket{0}\ket{1} \pm \ket{1} \ket{0})/\sqrt{2}$
are maximally entangled states of two qubits, 
and any maximally entangled two-qubit state is an EPR state, up to a local unitary transformation $U_\AAA \otimes U_\BB$.

For mixed states, entanglement is no longer equivalent to being a product state. The ``good'' definition of mixed 
state entanglement is due to Werner~\cite{Werner89}.

\vspace{2mm}

\begin{definition} \label{eq-def_entangled_state}
A mixed state $\rho$ of a bipartite system $\AB$ is separable if it admits a pure state decomposition
\begin{equation}  \label{eq-def_separable_states}
\rho = \sum_i \eta_i \ketbra{ \psi_i \otimes \phi_i}{\psi_i \otimes \phi_i}
\end{equation}
in terms of pure separable states $\ket{\psi_i \otimes \phi_i} \in \Hh_\AB$.
If such a decomposition does not exist then $\rho$ is entangled.
The set  of all separable states of $\AB$ forms a convex subset of $\states (\Hh_\AB)$, which is denoted by  $\Ss_\AB$.
\end{definition}

It follows from the Carath\'eodory theorem that the number of pure product states in the decomposition (\ref{eq-def_separable_states}) can always be chosen
to be smaller or equal to $(n_\AAA n_\BB)^2+1$.

According to this definition, a state is separable if it {\it could} have been prepared from pure product states only. 
This does not mean that it has  actually been prepared using such states. For example, if one prepares two qubits in the maximally entangled states
 $\ket{\Phi_+}$ and $\ket{\Phi_-}$ with equal probabilities, the corresponding state
\begin{equation}
\rho = \frac{1}{2} \ketbra{\Phi_+}{\Phi_+} + \frac{1}{2} \ketbra{\Phi_-}{\Phi_-}
   =  \frac{1}{2} \ketbra{0}{0} \otimes \ketbra{0}{0} + \frac{1}{2} \ketbra{1}{1} \otimes \ketbra{1}{1}
\end{equation}
is separable!  This unexpected result is inherent to the ambiguity of the state preparation discussed in the preceding subsection. This quantum ambiguity
unfortunately obliges us to look for all possible state preparations of  a given mixed state $\rho$ to decide whether $\rho$ is entangled or not.
This makes this problem highly non-trivial.

An explicit complete characterization of $\Ss_\AB$ is known for qubit-qubit and qubit-qutrit systems only, that is, for $(n_\AAA ,n_\BB)= (2,2)$, $(2,3)$, and $(3,2)$.
In such a case, the Peres-Horodecki criterion~\cite{Peres96,Horodecki96b,Horodecki97}  gives a necessary and sufficient condition for $\rho$ to be entangled.
This criterion is formulated in terms of the partial transpose. Given two orthonormal bases $\{ \ket{i} \}$  of $\Hh_\AAA$ and 
$\{ \ket{k} \}$ of $\Hh_\BB$, the partial transpose $\rho^{T_\BB}$ of $\rho$ with respect to $\BB$ has matrix elements in the product basis $\{ \ket{i}\ket{k}\}$
 given by
\begin{equation}
\bra{i} \bra{k} \rho^{T_\BB} \ket{j} \ket{l}  = \bra{i} \bra{l} \rho \ket{j} \ket{k} \;. 
\end{equation}
One defines similarly $\rho^{T_\AAA}$ and note that  $\rho^{T_\AAA}= (\rho^{T_\BB} )^T$.
It follows from Definition~\ref{eq-def_entangled_state} that if $\rho$ is separable then $\rho^{T_\AAA} \geq 0$ and $\rho^{T_\BB} \geq 0$, \ie
$\rho^{T_\AAA}$ and $\rho^{T_\BB} $ are states of $\AB$. Thus, if $\rho^{T_\AAA}$ (or, equivalently, $\rho^{T_\BB}$) has negative eigenvalues 
then $\rho$ is necessarily entangled. Since the transpose is a positive but not CP  map, such negative eigenvalues  may indeed exist.
However, if $n_\AAA n_\BB > 6$, certain  entangled states have non-negative partial transposes~\cite{Horodecki97}. 
It is remarkable that this does not happen when $n_\AAA n_\BB \leq 6$: then $\rho^{T_\AAA}\geq 0$ \ifif $\rho \in \Ss_\AB$~\cite{Horodecki96b}. 
Two remarks should be made at this point. First, states with non-negative partial transposes
cannot undergo entanglement distillation and therefore form an interesting subset of $\states (\Hh_\AB)$ on their own, which contains $\Ss_\AB$
(see~\cite{Horedecki_review} for more detail).
Second, extending the Peres criterion to all positive but not CP linear maps
$\Lambda_\BB : \Bb ( \Hh_\BB) \rightarrow \Bb ( \Hh_\AAA)$ (\ie asking that $1 \otimes \Lambda_\BB  (\rho) \geq 0$ for any such map) yields 
a necessary and sufficient condition for entanglement, valid whatever the space dimensions $n_\AAA$ and $n_\BB$~\cite{Horodecki96b}. 
Due to the lack of an explicit characterization of such maps (except for $(n_\AAA,n_\BB)= (2,2)$ or $(3,2)$)\footnote{
When  $(n_\AAA,n_\BB)= (2,2)$ or $(3,2)$, any positive map $\Lambda: \Bb ( \Hh_\BB) \rightarrow \Bb ( \Hh_\AAA)$ can be written as
$\Lambda = \Mm_1 + \Mm_2 \circ T$, where $\Mm_1$ and $\Mm_2$ are CP and $T$ is the transposition~\cite{Woronowicz76}. 
The fact that the partial transpose criterion is sufficient for entanglement follows from this characterization~\cite{Horodecki96b}.
},
this condition is unfortunately not very helpful in general.

Let us also mention another necessary but not sufficient (even for two qubits)  condition for entanglement, which relies on the Schmidt decomposition 
(\ref{eq-mixed_state_Schmidt_dec}) for  mixed states. By using the fact that $\sum_m \sqrt{\mu_m}$  defines a norm on $\states (\Hh_\AB)$, one can 
show that if $\rho \in \Ss_\AB$ then $\sum_m \sqrt{\mu_m} \leq 1$~\cite{Chen03}. 
Hence $\sum_m \sqrt{\mu_m} > 1$ implies that $\rho$ is entangled. 

Once a state has been recognized as separable, it may be of relevance to determine  its decomposition(s) into pure product states. 
This problem has been addressed  in~\cite{Wootters98,Sanpera98,Sylvain_these} for two qubits.

Definition~\ref{eq-def_entangled_state} 
can be extended straightforwardly to multipartite systems $\AAA_1 \ldots \AAA_k$. Then different kinds of entanglement
can be defined according to the chosen partition of $\{ \AAA_1 , \ldots ,\AAA_k\}$. In this article we will not  consider  multipartite entanglement, which is
a challenging subject in its own~\cite{Toth_review,Horedecki_review}.

\newpage
\section{Quantum measurements} \label{sec-measurements}

In this section we review the notions of quantum operations and generalized measurements and give the basic theorems, namely,
the Stinespring theorem, the Kraus decomposition, and the Neumark extension theorem. We start by a physical description of a
von Neumann measurement.

\subsection{Physical realization of a measurement process} \label{sec-realization_meas}

A measurement on a quantum system $\SSS$ is realized by coupling $\SSS$ with a measurement apparatus.  This apparatus consists of a
macroscopic pointer $\PP$ interacting with an environment $\EE$ playing the role of an infinite bath. 
One may think of $\PP$  as the center of mass of the needle of a meter. The environment $\EE$ then includes all the other
degrees of freedom of the macroscopic apparatus.  The coupling of the measured
system $\SSS$ with the pointer transforms the initially uncorrelated state $\ket{\psi}\otimes \ket{0}$ of the composite system $\SSS \PP$
into an entangled state,
\begin{equation} \label{eq-premeasurement}
\ket{\psi}\otimes \ket{0} \longrightarrow
\ket{\Psi_{\sf{SP}}^{\rm{ent}}} = \sum_{i,l} c_{il} \ket{\alpha_{il}}
\otimes \ket{i}\;.
\end{equation}
Our assumption that $\SSS$ and $\PP$ are initially in pure states is made to simplify the foregoing discussion and can be easily relaxed.
The states $\ket{\alpha_{il}}$ form an orthonormal basis of the system Hilbert space $\Hh_{\SSS}$ (measurement basis), which is the eigenbasis of the measured
observable $A$, \ie  $A \ket{\alpha_{il}} = a_i \ket{\alpha_{il}}$. The index $l$ labels if necessary the different orthogonal eigenstates of $A$ with the same
degenerate eigenvalue $a_i$. In ideal measurements $c_{il} = \braket{\alpha_{il}}{\psi}$.  The states $\ket{i}$ are the pointer
states of the apparatus. After a sufficiently long coupling time between $\SSS$ and $\PP$, these states are macroscopically
distinct and thus nearly orthogonal, $\braket{i}{j}
\simeq \delta_{ij}$ (hereafter $\delta_{ij}$ stands for the Kronecker symbol, equal to $1$ if $i=j$ and zero otherwise).  
The transformation (\ref{eq-premeasurement}) is a unitary transformation, \ie $\ket{\Psi_{\sf{SP}}^{\rm{ent}}}= U_{\sf{SP}} \ket{\psi} \ket{0}$ where $U_{\sf{SP}}$ is a
unitary evolution operator on $\Hh_{\SSS \PP}$. One usually calls such a transformation the {\it pre-measurement}~\cite{Giulini}. This unitary evolution 
induces quantum correlations between $\SSS$ and $\PP$, such that each eigenprojector $\Pi_i = \sum_l
\ketbra{\alpha_{il}}{\alpha_{il}}$ of $A$ is in one-to-one correspondence with a pointer state $\ket{i}$. The resulting state
(\ref{eq-premeasurement}) is a superposition of macroscopically distinct states, sometimes referred to as a ``Schr\"odinger cat state''.
The pointer states are singled out by their robustness against environment-induced decoherence. More precisely, if the pointer $\PP$
is initially in the state $\ket{i}$, its interaction with the environment $\EE$ does not entangle $\PP$ and $\EE$.  Letting $\PP$
and $\EE$ interact during a time $t$ much larger than the decoherence time, the $\SSS \PP$-entangled state $\ket{\Psi_{\SP}^{\rm{ent}}}$ is 
transformed into a statistical mixture in which all the coherences between the  pointer states $\ket{i}$ have disappeared. After tracing out 
the environment degrees of freedom, the reduced state of  $\SSS \PP$ is modified according to 
\begin{equation} \label{eq-decoherence_meas}
\ketbra{\Psi_{\sf{SP}}^{\rm{ent}}}{ \Psi_{\sf{SP}}^{\rm{ent}}}
\longrightarrow \rho_{\sf{SP}}^{\rm p.m.} = \sum_{i k l}
c_{ik}\overline{c_{il}} \ketbra{\alpha_{ik}}{\alpha_{il}} \otimes
\ketbra{i}{i} = \sum_i \Pi_i \,\rho  \,\Pi_i \otimes
\ketbra{i}{i}\;,
\end{equation}
$\rho = \ketbra{\psi}{\psi}$ being the initial system
state.  The final  $\SSS \PP$-state  has no quantum correlations but is
classically correlated: indeed, each pointer state $\ket{i}$ goes hand in hand with the system state
\begin{equation} \label{eq-conditional_state_and_proba}
\rho_{S|i} = p_i^{-1} \Pi_i \,\rho \, \Pi_i\quad , \quad p_i = \tr ( \Pi_i
\rho ) \;.
\end{equation}
Concrete models for the pointer and its coupling with the system and the environment have been investigated in~\cite{Balian01,Balian03,moi1,moi2};
in these works the aforementioned  decoherence time and the time duration of the measurement are estimated 
in the more realistic situation where the two transformations (\ref{eq-premeasurement}) and (\ref{eq-decoherence_meas}) occur simultaneously. 
The readout of the pointer (that is, the observation of the position of the needle) cannot significantly alter the macroscopic
state $\ket{i}$. It merely selects one of the measurement outcomes,
\begin{equation} \label{eq-WP_reduction}
\text{{\bf outcome} $i$:} \qquad \rho_{\SP}^{\rm p.m.} \; \longrightarrow\;
\rho_{\SP |i} = \rho_{\SSS |i} \otimes \ketbra{i}{i} \qquad
\text{(wavepacket reduction).}
\end{equation}
After the measurement yielding the outcome $i$ the measured system is in the conditional state $\rho_{S|i}$, and this outcome occurs with
probability $p_i$ (Born rule).  The transformation (\ref{eq-WP_reduction}) results from the knowledge of the
random outcome, it should not be regarded as a true dynamical process. It is actually analog to a state preparation (see
Sec.~\ref{sec-purification}).  In mathematical terms, it corresponds
to a convex decomposition of $\rho_{\SP}^{\rm p.m.}= \sum_i p_i \rho_{\SP |i}$. 

We point out that recent progresses in the understanding of quantum measurement processes via dynamical models and their interpretation 
with a statistical physics viewpoint have been made by Allahverdyan, Balian, and Nieuwenhuizen~\cite{Balian13}.

\subsection{Quantum operations} \label{sec-QO}

In the absence of readout of the measurement result, one does not know which state $\rho_{S|i}$ has been prepared and 
the system is after the measurement in the average state
\begin{equation} \label{eq-von_Neumann_measurement}
\Mm_\Pi ( \rho) = \sum_i \Pi_i \,\rho \, \Pi_i\;,
\end{equation}
where $\rho$ is the state before the measurement.

Since $\{ \Pi_i\}$ is the spectral measure of the self-adjoint operator $A$, the $\Pi_i$ form a family of projectors in $\Bb ( \Hh_\SSS)_{\rm s.a.}$ satisfying
$\Pi_i \Pi_j =\delta_{ij} \Pi_i$ and $\sum_i \Pi_i = 1$. We will refer in the sequel to such a family as an {\it orthonormal family of projectors}.
 It is easy to show that the map $\Mm_\Pi$ is  CP (as a sum of CP maps) and trace-preserving. In quantum information, such maps are called 
{\it  quantum operations}.

\vspace{2mm}

\begin{definition} \label{def-QOs}
A quantum operation $\Mm: \Bb (\Hh_\SSS) \rightarrow \Bb ( \Hh_\SSS')$ is a trace-preserving CP map from $\Bb (\Hh_\SSS)$ into $\Bb ( \Hh_\SSS')$. 
\end{definition}

\vspace{1mm}

A necessary and sufficient  condition for a linear map $\Mm : \Bb (\Hh_\SSS) \rightarrow \Bb ( \Hh_\SSS')$ to be CP
is that it satisfies $\Mm \otimes 1 ( \ketbra{\Psi_{\rm ent}}{\Psi_{\rm ent}} ) \geq 0$ for the  maximally entangled state  
$\ket{\Psi_{\rm ent}} = n_\SSS^{-1/2} \sum_k \ket{k} \ket{k}$ in $\Hh_\SSS \otimes \Hh_\SSS$, where $\{ \ket{k}\}$ is an \ONB of $\Hh_\SSS$.   
In fact, $\Mm \otimes 1 ( \ketbra{\Psi_{\rm ent}}{\Psi_{\rm ent}} )$ coincides with the operator $O_\Mm^\Rr$ defined in Sec.~\ref{sec-states_and_observables} up to a factor
$1/n_\SSS$, and it has been argued above that $\Mm$ is CP \ifif $O_\Mm^\Rr \geq 0$.  

A quantum operation is the quantum analog of a stochastic matrix $\Mm^\clas$
giving the transition probabilities $q(j|i)$ of a classical Markov
process, 
\begin{equation} \label{eq-stochastic matrix}
\pv = ( p_1,\ldots , p_n ) \; \mapsto \; \Mm^\clas \pv \quad {\rm{with}} \quad   
(\Mm^\clas \pv )_j = \sum_{i=1}^n q(j|i) \,p_i  \;\; , \;\; q(j|i) \geq 0 \;\; ,  \;\; \sum_{j=1}^n q(j|i) = 1\;.
\end{equation} 
Save for the wavepacket reduction (\ref{eq-WP_reduction}), all physical dynamical processes on quantum systems are given by quantum
operations\footnote{
In order to include the transformation (\ref{eq-WP_reduction}), many authors define a more general notion of
quantum operation by relaxing the trace-preserving condition and
replacing it by $\tr( \Mm (\rho) ) \leq 1$ for any $\rho \in \states (\Hh)$. The state transformation is then given by the non-linear map 
$\rho \mapsto \Mm(\rho)/\tr ( \Mm(\rho))$.  Theorems~\ref{theo-Stinespring} and \ref{theo-Kraus} can be
easily adapted to this more general definition. In particular, the Kraus decomposition (\ref{eq-Kraus_decomp}) 
holds, with Kraus operators $A_i$ satisfying 
$\sum_i A_i^\ast A_i \leq 1$. 
}.
Let a system $\SSS$ interact with another system $\EE$ at times $t\geq 0$. If  $\SSS$ and $\EE$ are  initially in a
product state $\rho (0) \otimes \rho_\EE (0)$ and $\SSS\EE$ can be considered as an isolated system, so that its dynamics is governed
by the Schr\"odinger equation, then the reduced state of $\SSS$ at time $t$ reads
\begin{equation} \label{eq-reduced_dynamics}
\rho (t) = \tr_{\EE} \bigl( e^{-\I t H_{\SSS \EE} } \rho (0) \otimes \rho_{\EE} (0) e^{\I t H_{\SSS \EE} } \bigr)\;.
\end{equation}
Here $H_{\SSS \EE}= H_\SSS + H_\EE + \lambda H_{\rm int}$ is the Hamiltonian of $\SSS \EE$, where
$H_\SSS$ and $H_\EE$ are the Hamiltonians of $\SSS$ and $\EE$, $H_{\rm int}$ their coupling Hamiltonian, 
and $\lambda$ the coupling constant. The time-evolved state (\ref{eq-reduced_dynamics}) is related to 
the initial state $\rho (0)$ by a quantum operation $\Mm_t$, \ie $\rho(t) = \Mm_t \rho(0)$. 
 The Stinespring theorem says that any
quantum operation $\Mm$ can be viewed as a reduced evolution of the
system coupled to an auxiliary system (ancilla).

\vspace{2mm}

\begin{theorem} {\rm{(Stinespring~\cite{Stinespring55})}}  \label{theo-Stinespring}
Let $\Mm$ be a quantum operation $\Bb ( \Hh_\SSS)\rightarrow \Bb ( \Hh_\SSS)$.
Then one can find an  ancilla Hilbert space  $\Hh_\EE$, a state
$\ket{\epsilon_0} \in\Hh_\EE$, and a unitary operator $U$ on $\Hh_{\SSS\EE}$ such that 
$\Mm ( \rho) = \tr_{\EE} ( U \rho \otimes \ketbra{\epsilon_0}{\epsilon_0} \,U^\ast )$.
\end{theorem} 

\vspace{1mm}

It is appropriate at this point to review a few well-known facts from the theory of CP maps on $C^\ast$-algebras.  
The {\it adjoint} $\Mm^\ast$ with respect to the trace of $\Mm : \Bb (\Hh_\SSS ) \rightarrow \Bb ( \Hh_\SSS ')$ is the map
$\Mm^\ast : \Bb (\Hh_\SSS ') \rightarrow \Bb ( \Hh_\SSS )$
defined by $\tr [ A \Mm(\rho)]= \tr[ \Mm^\ast (A) \rho]$ for any $A \in \Bb ( \Hh_\SSS ')$ and $\rho \in \Bb ( \Hh_\SSS)$.
If $\Mm$ is a \QO then $\Mm^\ast$ is also a CP map and is unity-preserving, 
$\Mm^\ast ( 1) = 1$.  According to Stinespring's theorem, one has
\begin{equation}
\Mm^\ast ( X) = \bra{\epsilon_0} U^\ast X \otimes 1 U \ket{\epsilon_0}
\end{equation}
for any $X \in \Bb ( \Hh)$.  
It follows that $\Mm^\ast$ satisfies the Kadyson-Schwarz inequality
\begin{equation} \label{eq-Kadison-Schwarz}
| \Mm^\ast ( X) |^2 \leq \Mm^\ast ( | X|^2 )\;.
\end{equation}

\vspace{1mm}

\begin{theorem} {\rm{(Kraus~\cite{Kraus70})}} \label{theo-Kraus}
A linear map $\Mm$ from $\Bb (\Hh_\SSS) $ into itself is a \QO \ifif it admits the representation
\begin{equation} \label{eq-Kraus_decomp}
\Mm (\rho) = \sum_i A_i \rho A_i^\ast\;,
\end{equation}
where $\{ A_i \}$ is a countable family of operators on $\Hh_\SSS$
satisfying $\sum_i A_i^\ast A_i = 1$.
\end{theorem}

\vspace{1mm}

For infinite dimensional Hilbert spaces and in the more general C$^\ast$-algebra
setting, the Kraus decomposition holds under the additional assumption that $\Mm$ is normal, that is, ultra-weakly continuous. One usually
deduces it from Stinespring's theorem.  In our finite-dimensional setting, however, a simple direct proof of Theorem~\ref{theo-Kraus}
exists (see Remark~\ref{exo-proof_Kraus} below). One can then obtain the Stinespring theorem from the Kraus decomposition as
follows. Let $\{ \ket{k}\}_{k=1}^{n_\SSS}$ be an \ONB of $\Hh_\SSS$ and $\Hh_\EE$ be a (possibly infinite-dimensional) Hilbert space with
\ONB $\{ \ket{\epsilon_i}\}$.  Define the vectors $\ket{\Psi_{k0}} = \sum_i A_i \ket{k} \ket{\epsilon_i}$.  Using
$\sum_i A_i^\ast A_i = 1$, one finds that these vectors form an orthonormal family in  $\Hh_{\SE}$, which can be completed so as to get an \ONB 
$\{ \ket{\Psi_{kl}} \}$. Then $\Mm^\ast ( X ) = \bra{\epsilon_0} U^\ast X \otimes 1 \, U \ket{\epsilon_0}$ for any 
$X \in \Bb( \Hh_\SSS)$, where the unitary $U$ on $\Hh_\SE$ is defined by $U \ket{k} \ket{\epsilon_l} = \ket{\Psi_{kl}}$ for any $k$ and
$l$.

\vspace{1mm}

\begin{exercice}  \label{exo-proof_Kraus} 
Any \QO $\Bb ( \Hh_\SSS ) \rightarrow \Bb ( \Hh_\SSS)$ with $\dim \Hh_\SSS = n_\SSS < \infty$  admits a Kraus decomposition
 (\ref{eq-Kraus_decomp}) with at most $n^2_\SSS$ operators $A_i$. Consequently, 
 one can choose the ancilla space $\Hh_\EE$ in Theorem~\ref{theo-Stinespring} of dimension $\dim  \Hh_\EE = n^2_\SSS$. 

\vspace{1mm}

\noindent Sketch  the proof
{\rm \cite{Nielsen}. To show that $\Mm$ has the form
(\ref{eq-Kraus_decomp}), consider the operator $B = \Mm \otimes 1 (  \ketbra{{\Psi}_{\rm ent}}{{\Psi}_{\rm ent}} )$
with $\ket{{\Psi}_{\rm ent}} = n_\SSS^{-1/2} \sum_k \ket{k}   \ket{k} \in \Hh_{\SSS \SSS}$ as above.  Since $\Mm$ is
CP, one has $B \geq 0$.  Let $\ket{\widetilde{\Phi}_i}$  be orthogonal eigenvectors of $B$, normalized in such a way that 
$n_\SSS B   = \sum_i \ketbra{\widetilde{\Phi}_i}{\widetilde{\Phi}_i}$.  Then define the Kraus operators $A_i$ as the operators associated to 
$\ket{\widetilde{\Phi}_i}$ by the isomorphism  (\ref{eq-ismoemtry_operators_vectors}) between $\Bb ( \Hh_\SSS)$ and $\Hh_{\SSS \SSS}$.
\finpro
}
\end{exercice}

\vspace{1mm}

It is important to realize that the Kraus decomposition is not unique. For indeed, if $\{ A_i \}_{i=1}^p$ is a family of Kraus operators for $\Mm$ and
$(u_{ji})_{i,j=1}^q$ is a unitary matrix of size $q\geq p$, then the operators
\begin{equation}  \label{eq-equivalence_Kraus_op}
B_j = \sum_{i=1}^p \overline{u}_{ji} A_i \quad , \quad j = 1 ,\ldots , q\;,
\end{equation}
define another family of Kraus operators for $\Mm$.  Conversely, two families $\{ A_i \}_{i=1}^p$ and $\{ B_j\}_{j=1}^q$ 
of Kraus operators for $\Mm$ with $p\leq q < \infty$ are related to each other by (\ref{eq-equivalence_Kraus_op}).
Actually, let $B$ and $\ket{{\Psi}_{\rm ent}}$ be defined as
in the Remark~\ref{exo-proof_Kraus} above.  Then $B= \sum_i \ketbra{\widetilde{\mu}_i}{\widetilde{\mu}_i} = \sum_j \ketbra{\widetilde{\nu}_j}{\widetilde{\nu}_j}$ with
\begin{equation}
\ket{\widetilde{\mu}_i} 
= n_\SSS^{-\onehalf } \sum_k ( A_i \ket{k}) \ket{k}
\quad , \quad 
\ket{\widetilde{\nu}_j} 
= n_\SSS^{-\onehalf} \sum_k ( B_j \ket{k}) \ket{k} \;.
\end{equation}
In view of the link (\ref{eq-link_between_pure_state_dec}) between pure state decompositions of a non-negative
operator, one has $\ket{\widetilde{\nu}_j} = \sum_{i} \overline{u}_{ji} \ket{\widetilde{\mu}_i}$ with $(u_{ji})_{i,j=1}^q$
unitary. 
This implies (\ref{eq-equivalence_Kraus_op}).

Given a purification $\ket{\Psi}$ of $\rho$ on $\Hh_\SSS \otimes \Hh_\RR$ and
a \QO $\Mm : \Bb ( \Hh_\SSS) \rightarrow \Bb(\Hh_\SSS' )$, it is natural to ask about purifications of $\Mm ( \rho)$. A slight generalization
of Theorem~\ref{theo-Stinespring} ensures that
there exist  a vector $ \ket{\epsilon_0} \in \Hh_\EE$ and a unitary $U: \Hh_\SSS \otimes \Hh_\EE \rightarrow \Hh_\SSS ' \otimes \Hh_\EE'$
such that $\Mm ( \rho) = \tr_{\EE'} ( U \rho \otimes \ketbra{\epsilon_0}{\epsilon_0} U^\ast )$. Therefore, 
\begin{equation} \label{eq-purification_Mm_rho}
\ket{\Psi_\Mm} = 1_{\RR} \otimes U \ket{\Psi} \ket{\epsilon_0} =
\sum_{k=1}^n \sum_{i=1}^p \sqrt{p_k} (A_i \ket{k}) \ket{f_k} \ket{\epsilon_i'}
\end{equation}
is a purification of $\Mm (\rho)$ on $\Hh_\SSS ' \otimes \Hh_\RR \otimes \Hh_\EE'$. 
In the second equality, $\{ \ket{k} \}$ is an orthonormal eigenbasis of $\rho$, $\{ \ket{f_k}\}$ is the orthonormal family of $\Hh_\RR$ such that
$\ket{\Psi} = \sum_k \sqrt{p_k} \ket{k} \ket{f_k}$, and
  $\{ \ket{\epsilon_i'} \}$ is an \ONB of $\Hh_\EE'$ such that $U \ket{k} \ket{\epsilon_0} = \sum_i ( A_i \ket{k} ) \ket{\epsilon_i'}$
(see the expression of $U$ in terms of the Kraus operators after Theorem~\ref{theo-Kraus}).

\subsection{Generalized measurements} \label{sec-generalized_meas}

For the \QO  $\Mm_\Pi$ defined by (\ref{eq-von_Neumann_measurement}), the orthogonal projectors $\Pi_i$ form a family of Kraus operators.  One
may wonder if more general quantum operations, given by Kraus operators $A_i$ which are not necessarily orthogonal projectors,
correspond to some kind of measurements.  The answer is yes: such operations can always be obtained by coupling the system $\sf{S}$
to an auxiliary system $\sf{E}$ (the ancilla) and subsequently performing a von Neumann measurement on $\sf{E}$.

\vspace{2mm}

\begin{theorem} \label{eq-th-Neumark} {\rm  (Neumark extension theorem)} Let  $\{ A_i\}_{i=1}^p$ be a family of operators such that
$\sum_i A_i^\ast A_i = 1$. Then there exist a space $\Hh_\EE$ with dimension $\dim \Hh_\EE = p$, a pure state $\ket{\epsilon_0} \in  \Hh_\EE$,  
an orthonormal family $\{ \pi_i^\EE \}$ of projectors in $\Bb ( \Hh_\EE)$, and a unitary operator $U$ on $\Hh_{\SSS \EE}$
such that for any density matrix $\rho \in \states (\Hh_\SSS)$,
\begin{equation} \label{eq-Neumark_extension}
A_i \rho A_i^\ast = \tr_\EE ( 1 \otimes \pi_i^\EE \,U \rho \otimes
\ketbra{\epsilon_0}{\epsilon_0} U^\ast 1 \otimes \pi_i^\EE )\;.
\end{equation}
\end{theorem}

\proof Use the same arguments as in the above proof of Stinespring's
theorem from Theorem~\ref{theo-Kraus}, and define $\pi_i^\EE
=\ketbra{\epsilon_i}{\epsilon_i}$.  \finpro

\vspace{2mm}

\begin{definition} \label{def-generalized_meas}
A generalized measurement is given by a family $\{ M_i \}$ of non-negative operators $M_i$ satisfying $\sum_i M_i = 1$ (positive operator valued
measure, abbreviated as POVM) together with a family of operators $\{ A_i\}$ such
that $M_i = A_i^\ast A_i$.  The conditional state $\rho_{S|i}$ given outcome $i$  and the probability of this outcome read
\begin{equation} \label{eq-conditional_state}
\rho_{S|i} = p_i^{-1} A_i \rho A_i^\ast \quad , \quad p_i = \tr ( M_i \rho) \;.
\end{equation}
\end{definition}

According to Theorem~\ref{eq-th-Neumark}, any generalized \meas can be realized by letting the system $\SSS$ interact with an ancilla
$\EE$ in the state $\ket{\epsilon_0}$ and subsequently performing a von Neumann \meas on $\EE$, that is, coupling $\EE$ to a macroscopic
apparatus with pointer $\PP$.  The interaction between $\SSS$ and
$\EE$ first transforms the initial state $\rho_\SSS \otimes \ketbra{\epsilon_0}{\epsilon_0}$ into 
$\rho_\SE = U \rho_\SSS \otimes \ketbra{\epsilon_0}{\epsilon_0} U^\ast$, $U$ being a unitary evolution 
operator on $\Hh_\SE$, and the subsequent von Neumann \meas leads to the wavepacket reduction for the system $\SP$ (compare with
(\ref{eq-conditional_state_and_proba}) and (\ref{eq-WP_reduction}))
\begin{equation} \label{eq-cond_state_gen_meas}
\text{ {\bf outcome} $i$:} \qquad \rho_{\SP} \rightarrow \rho_{\SP |i}
 = p_i^{-1} \tr_\EE ( 1 \otimes \pi_i^\EE \, \rho_\SE 1 \otimes \pi_i^\EE ) \otimes \ketbra{i}{i}
= p_i^{-1} A_i \rho_\SSS A_i^\ast \otimes \ketbra{i}{i} \;,
\end{equation}
where $p_i = \tr ( 1 \otimes \pi_i^\EE \rho_\SE )= \tr( M_i \rho_\SSS)$ is the probability of outcome $i$, in agreement with
(\ref{eq-conditional_state}). 

One has $A_i = U_i  M_i^{1/2}$ (polar decomposition) for some unitary operator $U_i$ depending on $i$. The conditional states $\rho_{\SSS |i}$ are thus characterized by the POVM $\{ M_i\}$ up to unitary conjugations, which
introduce a freedom in choosing the output state associated to each measurement outcome.  For instance, if 
$M_i = \ketbra{\widetilde{\mu}_i}{\widetilde{\mu}_i}$ are of rank one then $A_i = \ketbra{i}{\widetilde{\mu}_i}$ for some arbitrary normalized vector
$\ket{i}$ and the output conditional states are $\rho_{\SSS | i} = \ketbra{i}{i}$.
One usually takes the vectors $\ket{i}$ to form an orthonormal basis (which can be identified to the pointer state basis of Sec.~\ref{sec-realization_meas}), 
in such a way that the states $\rho_{\SSS | i} $ be perfectly distinguishable 
(this happens  if the $\rho_{\SSS | i} $ are orthogonal only, see Sec.~\ref{sec_qsd} below). 
One should keep in mind, however,  that the probability $p_i = \bra{\widetilde{\mu}_i} \rho \ket{\widetilde{\mu}_i}$ of outcome $i$ is
independent of the choice of $\{ \ket{i} \}$. 
 If one is interested only in functions of the post-measurement states $\rho_{\SSS|i}$ which are
invariant under unitary conjugations (as, for instance, the von Neumann entropy), then the generalized
measurement can be fully specified by the measurement operators $M_i$.
Thanks  to the Neumark extension theorem, these operators may be written as
\begin{equation} \label{eq-Neumark_theo_for_M_i}
M_i = A_i^\ast A_i = \bra{\epsilon_0} U^\ast 1 \otimes \pi_i^\EE  U \ket{\epsilon_0} \;.
\end{equation}
As stressed above, in the absence of
read-out the  state  of the system after the \meas is the average of the conditional states,
\be \label{eq-convex_decomp_meas} \Mm ( \rho) = \sum_i p_i \rho_{S|i}
= \sum_i A_i \rho A_i^\ast\;, 
\ee
in analogy with (\ref{eq-von_Neumann_measurement}). This defines a \QO $\Mm$, the Kraus
decomposition of which specifies the state preparation associated with the wavepacket reduction.

Writing the spectral decomposition of each operator $M_i$, one observes that  
\begin{equation} \label{eq-spectral_decomp_M_i}
M_i= \sum_{k=1}^{r_i} \ketbra{\widetilde{\mu}_{ik}}{\widetilde{\mu}_{ik}}
\quad , \quad 
\sum_i M_i =\sum_{i,k} \ketbra{\widetilde{\mu}_{ik}}{\widetilde{\mu}_{ik}} =1 \;,
\end{equation}
where  $r_i = \rank ( M_i)$ and  $\ket{\widetilde{\mu}_{ik}}$ are unnormalized eigenvectors with norms equal to the square roots of the 
corresponding eigenvalues. The last 
condition in (\ref{eq-spectral_decomp_M_i}) implies that  either $ \{ \ket{\widetilde{\mu}_{ik}}\}$ is an orthonormal basis, in which case
$\{ M_i\}$ is an orthonormal family of projectors (von Neumann measurement), 
or $ \{ \ket{\widetilde{\mu}_{ik}}\}$ is a non-orthogonal family containing more than $n_\SSS$
vectors, in which case at least two eigenvalues $\| \widetilde{\mu}_{ik} \|$ are strictly smaller than one and
 $\{ M_i\}$ is not a von Neumann measurement.

The set of all POVMs is a convex set. Its boundary and extremal points have been studied in~\cite{D_Ariano05}.

\begin{exercice}  \label{exo-Peres_Found_Phys} 
An alternative version of Theorem~\ref{eq-th-Neumark} states that if
 $m= \sum_i r_i $ with $r_i= \rank ( M_i)$, then there exist a space  $\Hh_\EE$ with dimension 
$m-n_\SSS + 1$, a state $\ket{\epsilon_0} \in  \Hh_\EE$, and a von Neumann \meas $\{ \Pi_i^\SE \}$ on $\Hh_{\SSS \EE}$ such that
\begin{equation}  \label{eq-Neumark_extension_bis}
M_i = \bra{\epsilon_0} \Pi_i^\SE \ket{\epsilon_0} \;.
\end{equation}
The interesting point is that the dimension of the ancilla space $\Hh_\EE$ can be smaller than $p$ in Theorem~\ref{eq-th-Neumark} (for instance 
$\dim \Hh_\EE = p-n_\SSS + 1$ for rank-one operators $M_i$).

\vspace{1mm}

\noindent Sketch of the proof {\rm ~\cite{Peres90}.  Note that $m\geq n_\SSS$ by the observation  above. Define
\begin{equation}
\ket{\zeta_{ik}} = \ket{\widetilde{\mu}_{ik}} \ket{\epsilon_0} + \sum_{l=1}^{m-n_\SSS } c_{ik,l} \ket{\phi} \ket{\epsilon_l}\;,
\end{equation}
where $\ket{\widetilde{\mu}_{ik}}$ is as in (\ref{eq-spectral_decomp_M_i}), $\ket{\phi} \in \Hh_\SSS$ is an arbitrary state, and 
$\{ \ket{\epsilon_l}\}_{l=0}^{m-n_\SSS }$ is an \ONB of $\Hh_\EE$. The coefficients $c_{ik,l}$ may be chosen such that 
$\{ \ket{\zeta_{ik}}  \}$ is an orthonormal family of $\Hh_\SE$. To establish this statement,
set $c_{ik,l}= \braket{l}{\widetilde{\mu}_{ik}}$ for $m-n_\SSS < l \leq  m$,
with $\{ \ket{l} \}_{l=m-n_\SSS +1}^m$ an \ONB of $\Hh_\SSS$, and let $\cv_l \in \complex^{m}$ be the vector with components $c_{ik,l}$. Then 
$\cv_l \cdot \cv_{l'} = \delta_{l l'}$ for any $l,l' > m-n_\SSS$, as a result of $\sum_i M_i =1$. One can choose the $(m-n_\SSS)$ other vectors
$\cv_l$ in such a way that $( \cv_1, \ldots , \cv_m )$ forms a $m\times m$ unitary matrix.  Then $\Pi_i^\SE = \sum_k \ketbra{\zeta_{ik}}{\zeta_{ik}}$ has the desired property.
\finpro 
}
\end{exercice}

\subsection{Connections between POVMs, quantum operations, and state ensembles}
\label{sec-link_meas_QO_ensemble}

To each POVM one can associate a quantum operation and vice-versa. Similarly, there is a canonical way to associate to a \QO a  state ensemble and vice-versa.
These correspondences depend on  an \ONB $\{ \ket{i}\}_{i=1}^m$  of a fictitious pointer $\PP$ with $m$-dimensional space $\Hh_\PP$.
It has been already seen above that one can associate to a POVM $\{ M_i\}_{i=1}^m$ on $\SSS$ a \QO   with Kraus operators $A_i$ such that $M_i = A_i^\ast A_i$.
This operation implements the state changes in the measurement process in the absence of readout. If we imagine that $\SSS$ is coupled to
$\PP$ and that the \meas is performed on both $\SSS$ and $\PP$, 
one may consider the  Kraus operators $A_{ik} = \ket{k} \ket{i} \bra{\widetilde{\mu}_{ik}}$ such that $M_i = \sum_k A_{ik}^\ast A_{ik}$, 
where $\{ \ket{k} \}_{k=1}^{n_\SSS}$ is an \ONB of $\Hh_\SSS$ and
$\ket{\widetilde{\mu}_{ik}}$ are the unnormalized eigenvectors of $M_i$ in (\ref{eq-spectral_decomp_M_i}). 
Provided that there is no readout of the \meas on $\SSS$, one may trace the post-\meas states over $\Hh_\SSS$. The conditional states of $\PP$
are given by $\rho_{\PP | i} = p_{ik}^{-1} \tr_\SSS ( A_{ik} \rho A_{ik}^\ast)= \ketbra{i}{i}$ with $p_{ik} = \bra{\widetilde{\mu}_{ik}} \rho \ket{\widetilde{\mu}_{ik}}$,
and the corresponding probability is $p_i = \sum_k p_{ik} = \tr ( M_i \rho)$. 
The state changes in the absence of readout are implemented by the \QO $\Mm : \Bb ( \Hh_\SSS) \rightarrow \Bb ( \Hh_\PP)$ defined by
\begin{equation} \label{eq-correspondance_POVM_QO}
\Mm ( \rho ) = \sum_i \tr ( M_i \rho) \ketbra{i}{i} \;\; , \;\; \rho \in \states (\Hh_\SSS)
\quad \Leftrightarrow \quad  \Mm^\ast ( \ketbra{i}{j})= M_i \delta_{ij}\;\;,\;\;i,j=1,\ldots, m\;.
 \end{equation}
Conversely, if $\Mm$ is a \QO $\Bb ( \Hh_\SSS ) \rightarrow \Bb ( \Hh_\PP )$ then $M_i = \Mm^\ast ( \ketbra{i}{i} )$ defines a POVM $\{ M_i \}_{i=1}^m$
(actually, $M_i \geq 0$ by the positivity of $\Mm^\ast$ and $\sum_i M_i = \Mm^\ast ( 1) = 1$).
Therefore, for a given \ONB $\{ \ket{i}\}_{i=1}^m$ of  $\Hh_\PP$, there is a one-to-one correspondence between
POVMs $\{ M_i\}_{i=1}^m$ on $\Hh_\SSS$ and  \QOs $\Mm : \Bb ( \Hh_\SSS) \rightarrow \Bb ( \Hh_\PP)$ of the form (\ref{eq-correspondance_POVM_QO}). 

A similar one-to-one correspondence can be found between state  ensembles on $\Hh_\SSS$ with fixed  probabilities $\{ \eta_i\}_{i=1}^m$
 and quantum operations $\Bb ( \Hh_\PP) \rightarrow \Bb ( \Hh_\SSS)$ such that $\Mm ( \ketbra{i}{j})=0$ for $i \not= j$. 
This correspondence is given by
\begin{equation} \label{eq-falta_poco}
\rho_i = \Mm ( \ketbra{i}{i} ) \;\; ,\;\; i=1,\ldots,m\;.
\end{equation}
In fact, if $\Mm : \Bb ( \Hh_\PP) \rightarrow \Bb ( \Hh_\SSS)$ is a \QO then $\{ \rho_i , \eta_i\}_{i=1}^m$ is clearly  an ensemble on $\Hh_\SSS$.
Conversely, if  $\{ \rho_i , \eta_i\}_{i=1}^m$ is an ensemble of $m$ states, let us write the spectral decompositions 
$\rho_i = \sum_{k} p_{ik} \ketbra{\psi_{ik}}{\psi_{ik}}$.
Then the operation with Kraus operators $A_{ik} = \sqrt{p_{ik}} \ketbra{\psi_{ik}}{i}$ has the required property.

\newpage
\section{Transpose operation and least square measurement}  \label{sec-tranpose_op_and_lsm}
\subsection{Recovery operation in quantum error correction}  \label{sec-transpose_operation}

The notion of transpose operation was introduced by Ohya and Petz in their monograph~\cite{Petz}.
It plays the role of an approximate reversal of a quantum operation, in a sense that will be made more precise
below.   

\vspace{2mm}

\begin{definition} \label{def-transpose_operation}
Let $\Mm : \Bb ( \Hh ) \rightarrow \Bb ( \Hh ')$ be a \QO and $\rho \in \states (\Hh)$ be a state such that $\Mm ( \rho)>0$. 
The transpose operation of
$\Mm$ for $\rho$ is the \QO $\Rr_{\Mm,\rho} :  \Bb ( \Hh ' ) \rightarrow \Bb ( \Hh )$ with Kraus operators
$R_i = \rho^{\frac{1}{2}} A_i^\ast \Mm (\rho )^{-\frac{1}{2}}$, where
$\{ A_i \}$ is a family of Kraus operators for $\Mm$. It is
independent of the Kraus decomposition of $\Mm$. Actually, for any $\sigma \in \states (\Hh')$,
\begin{equation} \label{eq-def_widehat_Mm}
\Rr_{\Mm,\rho} (\sigma) = \rho^{\frac{1}{2}} \Mm^\ast \bigl( \Mm (\rho )^{-\frac{1}{2}} \sigma \Mm (\rho )^{-\frac{1}{2}} \bigr)
\rho^{\frac{1}{2}} \;.
\end{equation}
\end{definition}

One easily
checks that $\sum_i R_i^\ast R_i = 1$, so that $\Rr_{\Mm,\rho}$ is
indeed a quantum operation, and that $\Rr_{\Mm,\rho} \circ \Mm ( \rho ) = \rho$. Furthermore, transposing twice amounts to do
nothing, that is, the  transpose of $\Rr_{\Mm,\rho}$ for the state $\Mm (\rho)$ is equal to $\Mm$.

The operation $\Rr_{\Mm,\rho }$ appears naturally in the context of quantum error correction. The problem of quantum error correction is
to send a state $\rho$ over a noisy quantum communication channel in such a way that $\rho$ is resilient
to the effect of the noise in the channel.  The state $\rho$ is encoded via a unitary transformation into a subspace $\Hh_C$ of the
Hilbert space $\Hh$ of the quantum channel.  The noise is described by some \QO $\Mm$.

\vspace{2mm}

\begin{proposition}
Let $\Mm$ be a \QO on $\observables$ with Kraus operators $\{ A_i\}$.  Let $\Pi_C$ denote the
orthogonal projector onto a subspace $\Hh_C \subset \Hh$ and $\condexpectation_C : \rho \mapsto \Pi_C \,\rho \, \Pi_C$ be the
conditional expectation onto the space of operators supported on $\Hh_C$. There exists a
recovery \QO $\Rr$ on $\observables$ satisfying $\Rr \circ \Mm \circ
\condexpectation_C = \condexpectation_C$ \ifif the following condition holds:
\begin{equation} \label{quantum_error_corr}
\condexpectation_C ( A_i^\ast A_j ) = a_{ij} \Pi_C\;,
\end{equation}
where  $(a_{ij})$ is a self-adjoint matrix. If this condition is satisfied then  for any $\rho $ with support $\range ( \rho ) \subset \Hh_C$,
the transpose operation $\Rr_{\Mm , \rho }$ is a recovery quantum operation.
\end{proposition}

 We refer the reader to the book of Nielsen and Chuang~\cite{Nielsen} for a proof of the necessary and sufficient condition
(\ref{quantum_error_corr}). Some bibliographic information on this topic can also be found there.

\vspace{2mm}

\Proofof{the second statement} By taking advantage of the non-uniqueness of the Kraus decomposition, (\ref{quantum_error_corr}) can be transformed into
$\condexpectation_C (B_i^\ast B_j ) = p_i \delta_{ij} \Pi_C$, where the Kraus operators $B_i$ are given by
(\ref{eq-equivalence_Kraus_op}) with $( u_{ij}) (a_{ij})  (u_{ij})^\ast$ the diagonal matrix with entries $p_i$. 
Together with the polar decomposition, this implies 
$B_j \Pi_C = \sqrt{p_j} W_j$ with $W_j = V_j \Pi_C$  satisfying $W_i^\ast W_j = \delta_{ij} \Pi_C$, the $V_j$ being some unitary operators. 
Thus the subspaces $V_j \Hh_C$ are orthogonal for different $j$'s and the restriction of $\sum_j W_j W_j^\ast$ to the
subspace $\Vv = \oplus_{j} V_j \Hh_C$ equals the identity.
If  $\rho = \condexpectation_C (\rho)$ and the restriction of $\rho$ to $\Hh_C$ is invertible, then $\Mm (\rho) = \sum_j p_j W_j \rho W_j^\ast$ and 
$\Mm (\rho)^{-1/2} = \sum_j  W_j \rho^{-1/2} W_j^\ast/\sqrt{p_j}$, the last operator being defined on $\Vv$.
 A simple calculation then shows that $\Rr_{\Mm,\rho} \circ \Mm \circ \condexpectation_C = \condexpectation_C $, as stated in the Proposition.
\finpro

\subsection{Transpose operation as an approximate reverse operation}
\label{sec-approximate_reverse_operation}

Since the condition (\ref{quantum_error_corr}) is not always fulfilled, it is natural to ask whether one can find an optimal
imperfect recovery map, which would enable to recover a given ensemble $\{ \rho_i, \eta_i \}$ subject to some noise
 with a maximal fidelity.  A notion of fidelity has been introduced by Schumacher~\cite{Schumacher96}. 
Its definition is as follows (for more detail and  motivations from classical information
theory, see~\cite{Nielsen}). Given a state $\rho \in \states (\Hh_\SSS)$, consider a purification $\ket{\Psi_{\rho}}$ of $\rho$ on $\Hh_\SSS
\otimes \Hh_\RR$, where $\RR$ is a reference system with Hilbert space $\Hh_\RR \simeq \Hh_\SSS$.  For instance, $\ket{\Psi_{\rho}}$
can be given by (\ref{eq-example_purification}).  If $\rho$ is a mixed state then $\ket{\Psi_{\rho}}$ is $\SSS \RR$-entangled 
(Sec.~\ref{sec-def_entanglement}). The {\it entanglement fidelity} of $\rho$ quantifies  how well this entanglement 
is preserved when the system $\SSS$ is subject to some noise  modelized by
a \QO $\Mm$ on $\Bb( \Hh_\SSS)$. It is defined by
\begin{equation} \label{eq-def_entanglement_fidelity} 
F_{\rm e} ( \rho , \Mm ) =
\bra{\Psi_{\rho}} \Mm \otimes 1  ( \ketbra{\Psi_\rho}{\Psi_\rho} )
\ket{\Psi_\rho}\;.  
\end{equation}
Since different purifications of $\rho$ on $\Hh_{\SSS \RR}$ are related by unitaries acting on $\Hh_\RR$, the \RHS of
(\ref{eq-def_entanglement_fidelity}) does not depend on the chosen purification.  
 As a consequence of the positivity and the trace-preserving property of $\Mm$, one has 
$0 \leq F_{\rm e} ( \rho, \Mm) \leq \tr_{\SSS \RR} [ \Mm \otimes 1 ( \ketbra{\Psi_\rho}{\Psi_\rho} )] = \tr [ \Mm (\rho)] =1$. 
Plugging
(\ref{eq-example_purification}) and (\ref{eq-Kraus_decomp}) into
(\ref{eq-def_entanglement_fidelity}), a simple calculation yields
\begin{equation}  \label{eq-entanglement_fidelity} 
F_{\rm e} ( \rho , \Mm ) = \sum_j \bigl|
\tr ( A_j \rho ) \bigr|^2 \;,
\end{equation}
where $\{ A_j\}$ is a family of Kraus operators for $\Mm$. Note that the sum in the \RHS does not depend on the choice of Kraus decomposition
(this follows from (\ref{eq-equivalence_Kraus_op})), as it should be. 
For a pure state $\rho_\psi = \ketbra{\psi}{\psi}$, the entanglement fidelity reduces to the
input-output fidelity $F ( \rho_\psi, \Mm ) = \bra{\psi} \Mm ( \ketbra{\psi}{\psi} ) \ket{\psi}$.
One infers from (\ref{eq-entanglement_fidelity}) that $F_{\rm e} ( \rho , \Mm )$ is a convex function of $\rho$.

Let us now consider an ensemble of states $\{ \rho_i , \eta_i \}_{i=1}^m$. 
The corresponding average entanglement fidelity is defined by 
\begin{equation} \label{eq-average_entanglement_fidelity}
\overline{F}_{\rm e} ( \{ \rho_i,\eta_i\},  \Mm ) = \sum_i \eta_i F_{\rm e} ( \rho_i , \Mm )\;.  
\end{equation}
This fidelity belongs to the interval $[0,1]$.  

\vspace{2mm}

\begin{proposition} {\rm (Barnum and Knill~\cite{Barnum02})} \label{prop_transpose_operation}
If the states $\rho_i$ commute with $\rho = \sum_i \eta_i \rho_i$, then
\begin{equation} \label{eq_transpose_operation}
 \overline{F}_{\rm e} \bigl( \{ \rho_i,\eta_i\}, \Rr_{\Mm,\rho} \circ \Mm
\bigr) \geq \overline{F}_{\rm e} \bigl( \{ \rho_i,\eta_i\}, \Rr_\opt \circ \Mm
\bigr)^2 \;,
\end{equation}
where $\Rr_{\Mm,\rho}$ is the transpose operation of $\Mm$ for $\rho$ and $\Rr_\opt$ the  optimal recovery \QO $\Rr$ maximizing
$\overline{F}_e ( \{ \rho_i,\eta_i\}, \Rr \circ \Mm)$.
\end{proposition}

\vspace{1mm}

Hence, if the minimal fidelity error is $1 - \overline{F}_e ( \{ \rho_i,\eta_i\}, \Rr_\opt \circ \Mm) = \eta$, then the fidelity error by using
$\Rr_{\rho,\Mm}$ as the recovery operation is at most twice larger than this minimal error.

\vspace{1mm}

\proof Taking advantage of the non-uniqueness of
the Kraus  decomposition, one can choose for any fixed $i$ some families  $\{ R_j^{\opt \,(i)}\}$ and 
$\{ A_k^{(i)}\}$ of  Kraus  operators for $\Rr^\opt$ and $\Mm$
satisfying
\begin{equation} \label{eq-proof_Barnum_Knill}
\tr \bigl( R_j^{\opt \,(i)} A_k^{(i)} \rho_i \bigr)= 0 \quad , \quad j \not= k\;. 
\end{equation}
Actually,
given any families $\{ R_m^\opt \}$ for $\Rr^\opt$ and $\{ A_l \}$ for $\Mm$, the operators
$R_j^{\opt  \,(i)} = \sum_m {u}_{jm}^{(i)} R_m^\opt$ and $A_k^{(i)} = \sum_l \overline{v}_{kl}^{(i)} A_l$  have the required property if $(u_{jm}^{(i)})$
and   $(v_{kl}^{(i)})$ are the unitary matrices in  the singular decomposition of  $(\tr ( R_m^\opt A_l \rho_i ))$.
Since $\{ R_j^{\opt \,(i)} A_k^{(i)} \}$ is a Kraus family for
$\Rr^\opt \circ \Mm$, one obtains from (\ref{eq-entanglement_fidelity}), (\ref{eq-average_entanglement_fidelity}), and (\ref{eq-proof_Barnum_Knill})
\begin{equation} \label{eq-proof_Barnum_Knill1} 
\overline{F}_{\rm e} \bigl( \{ \rho_i,\eta_i\}, \Rr^\opt \circ \Mm \bigr) = \sum_{i,j} \eta_i \bigl| \tr (
R_j^{\opt (i)} A_j^{(i)} \rho_i ) \bigr|^2 \;.  
\end{equation}
We first consider the case $\rho_\Mm = \Mm ( \rho)  >0$. Without loss of generality, we may assume that 
$\range ( R_j^{\opt (i)} ) \subset \range \rho_i \subset  \range \rho$, so that the 
 operators 
\begin{equation}
X_{ij} = \eta_i^{\frac{1}{4}} \rho_\Mm^{- \frac{1}{4}} A_j^{(i)} \rho^{ \frac{1}{4}}\rho_i^{  \frac{1}{2}}  \quad , 
\quad Y_{ij} = \eta_i^{\frac{1}{4}} \rho_\Mm^{- \frac{1}{4}} B_j^{(i)} \rho^{ \frac{1}{4}}\rho_i^{ \frac{1}{2}} 
\quad  \text{ and } \quad  (B_j^{(i)} )^\ast = \rho^{- \frac{1}{2}} R_j^{\opt\,(i)} \rho_\Mm^{\frac{1}{2}}   
\end{equation}
are well-defined.
Since $[\rho_i,\rho]=0$, one finds by using twice the Cauchy-Schwarz inequality
\begin{eqnarray}
 \label{eq-proof_Barnum_Knill2} \nn 
\overline{F}_{\rm e} ( \{ \rho_i,\eta_i\}, \Rr^\opt \circ \Mm )^2 
& = & 
\biggl( \sum_{i,j} \bigl| \tr ( Y_{ij}^\ast X_{ij} ) \bigr|^2 \biggr)^2 
\leq \sum_{i,j} \bigl( \tr ( Y_{ij}^\ast Y_{ij} ) \bigr)^2 \sum_{i,j} \bigl( \tr ( X_{ij}^\ast X_{ij} ) \bigr)^2 
\\ 
& \leq & 
\sum_{i,j,k} \bigl| \tr ( Y_{ij}^\ast Y_{ik} ) \bigr|^2
\sum_{i,j,k} \bigl| \tr ( X_{ij}^\ast X_{ik} ) \bigr|^2 \;.
\end{eqnarray}
The transpose operation  $\Rr_{\rho,\Mm}$ has Kraus operators $R_j^{(i)}= \rho^\onehalf (A_j^{(i)})^\ast \rho_\Mm^{- \frac{1}{2}}$. As a result,
\begin{equation}
 \overline{F}_{\rm e} \bigl( \{ \rho_i,\eta_i\} , \Rr_{\rho,\Mm} \circ \Mm \bigr) =
\sum_{i,j,k} \eta_i \bigl| \tr ( R_j^{(i)} A_k^{(i)} \rho_i ) \bigr|^2
= \sum_{i,j,k} \bigl| \tr ( X_{ij}^\ast X_{ik}) \bigr|^2  \;.
\end{equation}
The first sum in the last member of (\ref{eq-proof_Barnum_Knill2}) is
equal to $\overline{F}_{\rm e} ( \{ \rho_i,\eta_i\},  \Rr^\opt \circ \Bb )$,
where $\Bb$ is the CP map defined by $\Bb (\sigma) = \sum_k B_k^{(i)} \sigma (B_k^{(i)})^\ast$ 
(note that $\Bb$ does not depend on $i$).  Even if $\Bb$ is not trace-preserving, with the help of  (\ref{eq-def_entanglement_fidelity}) 
this fidelity can be bounded  from above by $\tr [ \Rr^\opt \circ \Bb (\rho) ]$,  which equals unity
thanks to the identity $\Bb (\rho) = \Mm (\rho)$. This yields the inequality (\ref{eq_transpose_operation}).
If $\rho_\Mm$ is not invertible, one approximates $\Mm$ by some \QOs $\Mm_\varepsilon$ satisfying $\Mm_\varepsilon ( \rho) >0$ for $\varepsilon >0$ and 
$\Mm_\varepsilon \rightarrow \Mm$ as $\varepsilon \rightarrow 0$, and obtains the result by continuity.
 \finpro

\subsection{Least square measurement}  \label{sec-least_square_meas}

Let us consider an ensemble $\{ \rho_i, \eta_i\}_{i=1}^m$ of states of the system $\SSS$ forming a convex decomposition of 
$\rho_\out = \sum_i \eta_i \rho_i$. For any $i$, we denote by 
$\rho_i = \sum_k p_{ik} \ketbra{\psi_{ik}}{\psi_{ik}}$ the spectral  decomposition of $\rho_i$ and set
$\rho_i = A_i A_i^\ast$,
where $A_i= \sqrt{\rho_i} U_i$ is defined up to a unitary $U_i$. Introducing   as in Sec.~\ref{sec-link_meas_QO_ensemble} 
an arbitrary \ONB $\{ \ket{k}\}_{k=1}^{n_\SSS}$ of $\Hh_\SSS$ and  a fictitious pointer with $m$-dimensional space $\Hh_\PP$ and
\ONB $\{ \ket{i}\}_{i=1}^m$, one can choose 
\begin{equation} \label{eq-square_root_rho_i}
A_i = \sum_{k=1}^{n_\SSS} \sqrt{p_{ik}} \ket{\psi_{ik}} \bra{k} \bra{i} \;\in\;  \Bb (\Hh_\SP,  \Hh_\SSS) \;.
\end{equation}
We remark that $A_{i}$ is associated to a purification
of $\rho_i \otimes \ketbra{i}{i}$ on $\Hh_\SP \otimes \Hh_\SSS$ via the isometry (\ref{eq-ismoemtry_operators_vectors}) between $\Bb (\Hh_\SP,  \Hh_\SSS)$ 
and $\Hh_\SP \otimes \Hh_\SSS$, namely, $\ket{\Psi_i} = \sum_k  \sqrt{p_{ik}} \ket{\psi_{ik}}   \ket{i} \ket{k}$.
Moreover, $\ket{\Psi_\out}= \sum_i \sqrt{\eta_i} \ket{\Psi_i}$ is a purification of $\rho_\out$ on the same space.

The {\it least square measurement}\;\footnote{
This measurement bears several names: it was referred to as the
``pretty good measurement'' in~\cite{Hausladen94} and is also called ``square-root measurement'' 
by many authors.
}
associated to $\{ \rho_i, \eta_i\}_{i=1}^m$ is given by the Kraus and \meas operators
\begin{equation} \label{eq-definition_LSM}
R_i^\lsm = \sqrt{\eta_i} A_i^\ast \rho_\out^{-\onehalf} = \sum_k \sqrt{\eta_i p_{ik}}  \ket{k}  \ket{i}\bra{\psi_{ik}} \rho_\out^{-\onehalf} 
\quad ,\quad  
M_i^\lsm = \bigl|  R_i^\lsm \bigr|^2 = \eta_i  \rho_\out^{-\onehalf} \rho_i \rho_\out^{-\onehalf}
\end{equation}
for $i=1,\ldots, m$.
One indeed checks that $\sum_i M_i^\lsm = 1$, so that (\ref{eq-definition_LSM}) defines  a generalized \meas in the sense of Definition~\ref{def-generalized_meas}.
While the operators $M_i^\lsm$ and thus the outcome probabilities $q_i = \tr ( M_i^\lsm \sigma_\SSS )$ 
(here $\sigma_\SSS$ is the system state)  only depend on $\{ \rho_i, \eta_i\}$, 
the post-measurement states also depend on the choice of the basis $\{ \ket{i}\}$,  as highlighted in Sec.~\ref{sec-measurements}. 
The conditional and average post-measurement states of the pointer $\PP$ are 
\begin{eqnarray} \label{eq-QO_lsm}
\text{{\bf outcome} $i$:} & & 
\sigma_\SSS \;\mapsto\; \sigma_{\PP | i} = q_i^{-1} \tr_\SSS ( R_i^\lsm \sigma_\SSS  (R_i^\lsm)^\ast )=  \ketbra{i}{i} 
\\[1mm]
\text{{\bf no readout}:} & & \sigma_\SSS \;\mapsto\;  \sigma_\PP =  \Mm^\lsm ( \sigma_\SSS ) = \sum_{i=1}^m q_i \sigma_{\PP | i} = \sum_{i=1}^m q_i  \ketbra{i}{i} \;.
\end{eqnarray}
For a pure state ensemble $\{ \ket{\psi_i} , \eta_i\}_{i=1}^m$, the least square \meas  consists of rank-one \meas opera\-tors 
$M_i = \ketbra{\widetilde{\mu}_i}{\widetilde{\mu}_i}$ with $\ket{\widetilde{\mu}_i} = \sqrt{\eta_i} \rho_\out^{-\onehalf} \ket{\psi_i}$. 
The vectors $\ket{\widetilde{\mu}_i}$ have the following property~\cite{Holevo78,Eldar01}, which elucidates the name given to the measurement:
they minimize the sum of the square norms 
$\| \ket{\widetilde{\mu}_i} - \sqrt{\eta_i} \ket{\psi_i} \|^2$ under the constraint
$\sum_i \ketbra{\widetilde{\mu}_i}{\widetilde{\mu}_i} = 1$.  
If the $\ket{\psi_i}$ are linearly independent and span $\Hh_\SSS$, so that $m=n$, then  
$\{ \ket{\widetilde{\mu}_i}\}$ is an \ONB of $\Hh_\SSS$. In that case $\{ M_i^\lsm \}$ is a von Neumann measurement (see Sec.~\ref{sec-generalized_meas}).

\vspace{1mm}

\begin{exercice}
The aforementioned property of a least square \meas  can be stated as follows:
\begin{equation} \label{eq-least_square_sum}
\min_{ \{ \ket{\widetilde{\mu}_i} \} } \biggl\{ \sum_{i=1}^m \bigl\| \ket{\widetilde{\mu}_i} - \sqrt{\eta_i} \ket{\psi_i} \bigr\|^2  \biggr\}
 = n_\SSS +1 - 2 \tr ( \rho_\out^\onehalf ) \quad \text{ with } \quad \rho_\out = \sum_i \eta_i \ketbra{\psi_i}{\psi_i} \;,
\end{equation}
the  minimum being over all families $\{ \ket{\widetilde{\mu}_i} \}_{i=1}^m$ in $\Hh_\SSS$ such that 
$\sum_i \ketbra{\widetilde{\mu}_i}{\widetilde{\mu}_i} = 1$. This minimum
is achieved \ifif $\ket{\widetilde{\mu}_i} = \sqrt{\eta_i} \rho_\out^{-1/2} \ket{\psi_i}$ (up to irrelevant phase factors).

\vspace{1mm}

\noindent Sketch of the proof. 
{\rm ~\cite{Eldar01,Huang05} Define $A= \sum_i \sqrt{\eta_i} \ketbra{\psi_i}{i}$
and $B = \sum_i \ketbra{\widetilde{\mu}_i}{i}$ in analogy with  (\ref{eq-square_root_rho_i}). Then 
observe that the sum to be minimized in (\ref{eq-least_square_sum}) is equal to
$\| A^\ast - B^\ast \|_2^2 = 1+n_\SSS - 2 \re \tr ( A B^\ast)$,
and use (\ref{eq-main_property_L_P_norm}). 
\finpro 
}  
\end{exercice}

\vspace{1mm}

As suggested by this result, the least square measurement plays an important role in distinguishing quantum states drawn from a given ensemble.
This point will be  discussed in Sec.~\ref{sec-QSD_and_lsm} below.

Let us recall from Sec.~\ref{sec-link_meas_QO_ensemble} that  the relation $\rho_i = \Mm ( \ketbra{i}{i} )$, where 
$\{ \ket{i} \}_{i=1}^m$ is a fixed \ONB of $\Hh_\PP$, can be used to associate to  a \QO  $\Mm : \Bb(\Hh_\PP) \rightarrow \Bb(\Hh_\SSS)$ an ensemble  $\{ \rho_i  , \eta_i \}_{i=1}^m$ 
on $\Hh_\SSS$.
Conversely, if $\{ \rho_i , \eta_i\}$ is an ensemble on $\Hh_\SSS$, the operation $\Mm$ with
Kraus operators $A_{ik} = A_i \ket{k} =  \sqrt{p_{ik}} \ketbra{\Psi_{ik}}{i}$ satisfies this relation (here $A_i$ is the operator (\ref{eq-square_root_rho_i})).
Similarly, the relation  (\ref{eq-correspondance_POVM_QO}) establishes a one-to-one correspondence 
between POVMs $\{ M_i\}$ on $\Hh_\SSS$ and \QOs $\Rr : \Bb(\Hh_\SSS) \rightarrow \Bb(\Hh_\PP)$.
It was recognized by Barnum and Knill~\cite{Barnum02} that  
{\it the least square \meas associated to the ensemble $\{ \rho_i = \Mm ( \ketbra{i}{i}), \eta_i\}$
is nothing but the \meas corresponding to the transpose operation $\Rr_{\Mm,\rho_\inp }$ of $\Mm$ for the state  $\rho_\inp = \sum_i \eta_i \ketbra{i}{i}$}.
Actually, since $\Mm ( \rho_\inp ) = \rho_\out$, according to the Definition~\ref{def-transpose_operation},
\begin{equation}
R_{ik} = \rho_\inp^\onehalf A_{ik}^\ast \rho_\out^{-\onehalf} = \sqrt{\eta_i p_{ik}} \ketbra{i}{\psi_{ik}} \rho_\out^{-\onehalf} 
\end{equation}
are Kraus operators for $\Rr_{\Mm,\rho_\inp }$. Thus
\begin{equation}
 M_i^\lsm   = \eta_i \rho_\out^{-\onehalf} \rho_i \rho_\out^{-\onehalf}  = \sum_{k} R_{ik}^\ast R_{ik}  = \Rr_{\Mm,\rho_\inp }^\ast ( \ketbra{i}{i} )\;.
\end{equation}
Conversely, it is immediate to verify that  $\Rr_{\Mm,\rho_\inp }(\sigma) = \sum_{ik} R_{ik} \sigma R_{ik}^\ast = \sum_i \tr ( M_i^\lsm \sigma) \ketbra{i}{i}$,
hence  $\Rr_{\Mm,\rho_\inp }$ is the operation associated to  $\{ M_i^\lsm \}$ by the relation (\ref{eq-correspondance_POVM_QO}).

\newpage
\section{Quantum state discrimination} \label{sec_qsd}

 The carriers of information in quantum communication and quantum computing are quantum systems, and the information is encoded in the
 states of those systems. After processing the information, it is
 necessary to perform measurements  in order to read out the result of the computation. In other words,
one has to determine the output state of the system. If these possible outputs form a set of orthogonal states, that is, if
they are given by $m$ known density matrices $\rho_i$ with orthogonal supports, then it is easy to devise a measurement
which discriminates them without any error (a von Neumann measurement with projectors $\Pi_i$ 
 onto $\range (\rho_i)$ will do the job). However, when the $\rho_i$ are non-orthogonal a perfect discrimination is 
 impossible. Indeed, if two non-orthogonal states
$\ket{\psi_1}$ and $\ket{\psi_2}$ could be discriminated perfectly then one could duplicate those states
by producing copies of  $\ket{\psi_i}$ if the measurement outcome is $i=1,2$, without prior knowledge on which of the two states one actually possesses. This would
contradict the no-cloning theorem\footnote{
No unitary evolution on a system $\SSS$ initially in state $\ket{\psi}$ and a 
register $\RR$ initially in state $\ket{\phi}$ can transform $\ket{\Psi}= \ket{\psi}\ket{\phi}$ into $\ket{\Psi '}= \ket{\psi} \ket{\psi}$ for any
$\ket{\psi}$ belonging to a set of distinct non-orthogonal states, e.g. $\ket{\psi} \in \{ \ket{\psi_1} ,\ket{\psi_2} \}$. Actually, 
the scalar products $\braket{\Psi_1}{\Psi_2}= \braket{\psi_1}{\psi_2}$ and $\braket{\Psi_1'}{\Psi_2'} = \braket{\psi_1}{\psi_2}^2$
are different if $\braket{\psi_1}{\psi_2} \not=0,1$. More generally, the no-cloning theorem tells us that one cannot duplicate 
unknown states by using any (not necessarily unitary) quantum evolution, except when these states pertain to a family of 
orthogonal states~\cite{Barnum96}.
}.
Consequently, one can extract less information from an ensemble of non-orthogonal states than from 
an ensemble of orthogonal ones. 

It is of interest to find the best measurement to  distinguish non-orthogonal states $\rho_i$  with 
the smallest possible failure probability.  We study this state discrimination problem in this section.  
This is a quite important issue in   quantum cryptography and  in
quantum communication in general. As emphasized in the introduction of this article, we aim  at  explaining some typical questions, providing examples, and  
 establishing basic general results that will be used in the next sections, rather than giving a full account on the subject. We refer the reader  to the 
review articles~\cite{Chefles_review,Bergou_review,Bergou_tutorial} for more complete presentations.
Measurements for distinguishing quantum states can  also be optimized using other criteria than the minimal probability of equivocation. For instance,
one can try to maximize the mutual information between the initial distribution of the state ensemble and the distribution of the \meas outcomes.
This optimization problem, which
plays an important role in the transmission of information in quantum channels, is briefly discussed at the end of this section.

Before entering into the detail of the theory, let us make a philosophical remark concerning the quantum-classical differences.
Let us inquire about the quantum analog of the celebrated experiment in classical probability which consists of  picking up  randomly
colored balls contained in an urn.
In quantum mechanics, the readout of the system's state  (the color of the ball in the classical analogy) is performed by a \meas perturbing the system.
If the urn contains an ensemble of non-orthogonal states, we have just seen above that there is no way to identify with certainty
which state from the ensemble has been picked up. Therefore, the starting  assumption that  the color of the ball is known once 
it has been extracted from the urn is not fulfilled in the quantum world and identifying these colors  is already a non-trivial task!

\subsection{Discriminating quantum states drawn from a given ensemble}  \label{sec_QSD}

We review in this subsection two strategies for discriminating non-orthogonal states, known as the ambiguous and unambiguous state
discriminations.
Let us consider an ensemble $\{ \rho_i, \eta_i \}_{i=1}^m$ of states $\rho_i$ with prior probabilities $\eta_i$.
For instance, the $\rho_i$ can be some states
of the electromagnetic field encoding $m$ symbols of a given alphabet, the $i$th symbol occurring with
frequency $\eta_i$. In order to send a message, a sender prepares random states
 drawn from the ensemble and gives them to a receiver.
To  decode the message the latter must  identify these states by performing measurements.
He wants to find the measurement that minimizes the failure probability.

A first strategy, called  {\it ambiguous (or minimal error) quantum state discrimination}, 
consists in looking for a generalized  measurement with $m$ outcomes yielding the maximal success probability
$P_{\rm S}= \sum_i \eta_i p_{i|i} $, $p_{i|i}$ being the probability of   
the measurement outcome $i$ given that the state is $\rho_i$.
Here, the number of possible outcomes is chosen to be equal to the number of states in the ensemble.  
The conditional probability of the outcome $j$ given the state $\rho_i$ is (see Sec.~\ref{sec-generalized_meas})
\begin{equation} \label{eq-cond_proba_for_outcome_j}
p_{j|i} = \tr ( M_j \rho_i )
\end{equation}
so that the maximal success probability reads
\begin{equation} \label{eq-max_success_proba_POVM}
P_{\rm S}^{\,\rm{opt}} ( \{ \rho_i,\eta_i\}) =\max_{{\rm POVM}\; \{ M_i \} } \biggl\{ \sum_{i=1}^{m} \eta_i \tr ( M_i \rho_i) \biggr\} \;,
\end{equation}
where the maximum is over all POVMs $\{ M_i \}_{i=1}^m$.

A second strategy consists in seeking for a generalized measurement with $(m+1)$ outcomes enabling  
to identify perfectly each state $\rho_i$,  but such that one of the outcomes leads  to an inconclusive result.
This strategy, originally proposed by Ivanovic~\cite{Ivanovic87} and further investigated by Dieks and Peres~\cite{Dieks88,Peres88},
is called {\it unambiguous quantum state discrimination}. In other words, 
if the measurement outcome is $j \in \{ 1,\ldots, m\}$ then the receiver is certain that the state is $\rho_j$, whereas if $j=0$ he does not know. 
This means that $p_{j|i} =  p_{i|i} \delta_{ij}$ with $p_{i|i}>0$, for any $i,j = 1,\ldots, m$.
The probability of occurrence of the inconclusive outcome, $P_{0} = \sum_i \eta_i p_{0|i}$,  must be
minimized. Since $p_{0|i} = 1 - p_{i|i}$, the success probability is obtained from the same formula (\ref{eq-max_success_proba_POVM}) as for ambiguous discrimination,
but with a maximum over all POVMs $\{ M_j\}_{j=0}^m$ such that $\tr ( M_j \rho_i ) = p_{i|i} \delta_{ij}$ for $j\not= 0$. 
For pure states $\rho_i = \ketbra{\psi_i}{\psi_i}$, the rank-one measurement operators $M_j$ satisfying this condition are  
\begin{equation} \label{eq-meas_op_unambiguousQDS}
M_j = \frac{p_{j|j}}{|\braket{\psi_j^\ast}{\psi_j}|^2} \ketbra{\psi_j^\ast}{\psi_j^\ast}\quad , \quad j = 1,\ldots, m\;,
\end{equation}
with the dual normalized vectors $\ket{\psi_j^\ast}$ defined by $\braket{\psi^\ast_j}{\psi_i} = \delta_{ij} \braket{\psi^\ast_i}{\psi_i}$. 
The remaining problem is to find the values of the probabilities $p_{j|j}$ 
which maximize the success probability (\ref{eq-max_success_proba_POVM}) under
the constraint that $\{ M_j\}_{j=0}^m$ is a POVM, that is,
\begin{equation} \label{eq-cond_postitivity_M_0}
M_0 =  1 - \sum_{j=1}^m \frac{p_{j|j}}{|  \braket{\psi^\ast_j}{\psi_j}|^2} \ketbra{\psi_j^\ast}{\psi_j^\ast} \;\geq \; 0\;.
\end{equation}
This is a non-trivial problem, which has been solved so far  in particular cases only. Upper and lower bounds on the maximal success probability
can be found in terms of the scalar products 
$\braket{\psi_i}{\psi_j}$ (see e.g.~\cite{Bergou_review}).

It is worth noting that unambiguous discrimination is not always possible. For instance, a pure state ensemble 
$\{ \ket{\psi_i} , \eta_i \}$ with linearly dependent vectors $\ket{\psi_i}$ cannot be discriminated unambiguously~\cite{Chefles98}.  
Indeed, assume that $\ket{\psi_{i_0}}$ is a linear combination of the other states $\ket{\psi_i}$. Together with the no-error condition
$p_{j|i} =  p_{i|i} \delta_{i j}$, which is equivalent to $\ket{\psi_i} \in \ker M_j$ for any $j \notin \{ 0, i\}$, this means that
$\ket{\psi_{i_0}} \in \ker ( M_{i_0} )$ and thus $p_{i_0 | i_0} =0$, in contradiction with the requirement $p_{i_0 | i_0}>0$.
The same argument shows that one cannot discriminate unambiguously an ensemble of mixed states $\{ \rho_i, \eta_i\}$ 
such that one state $\rho_{i_0}$ has its support $\range ( \rho_{i_0})$ contained in the sum of the supports of the other states. 

Ambiguous and unambiguous quantum state discriminations have many applications. For instance,
the discrimination of two non-orthogonal states plays a central role 
in the  quantum cryptography protocol proposed by Bennett in 1992 to distribute 
a secrete key between two parties~\cite{Bennett92}. We will not elaborate further on these applications. 
Let us also mention that other optimization schemes
than those discussed above have been worked out~\cite{Bergou_review,Bergou_tutorial}. 
State discriminations have been implemented experimentally by  using 
polarized photons in pure states (see~\cite{Clarke01} and references therein) and, more recently, in mixed states~\cite{Bergou04}.

\subsection{Ambiguous and unambiguous discriminations of two states} \label{sec-qsd_2_states}
\subsubsection{Ambiguous discrimination}

The simplest example of ambiguous  discrimination is the case of $m=2$ states $\rho_1$ and $\rho_2$.
Then the optimal success probability and measurement are easy to determine~\cite{Helstrom}.
One starts by writing the measurement operator $M_2$ as $1-M_1$ in the expression of the success probability,
\begin{equation}  \label{eq-success_proba}
P_{\rm S,a}^{\{ M_i \} } ( \{ \rho_i,\eta_i\}) 
 =  \eta_1 \tr ( M_1 \rho_1 ) + \eta_2 \tr ( M_2 \rho_2 )
 =   
\frac{1}{2} \bigl( 1 - \tr  \Lambda   \bigr)+ \tr ( M_1 \Lambda )  
\end{equation}
with $\Lambda = \eta_1 \rho_1 - \eta_2 \rho_2$.  The maximum of $\tr ( M_1 \Lambda)$ 
 over all $M_1$ satisfying $0 \leq M_1 \leq  1$ is achieved when $M_1$  is the spectral projector $\Pi_1$ 
associated to   the positive eigenvalues $\lambda_1  \geq \cdots \geq \lambda_{p} > 0$
of the Hermitian matrix $\Lambda$.
Consequently, the maximal success probability is given by the Helstrom formula
\begin{equation}
\label{eq-opt_success_proba}
P_{\rm S,a}^{\,\opt} ( \{ \rho_i,\eta_i\}) 
= \frac{1}{2}\bigl( 1 + \tr | \Lambda | \bigr)
\quad , \quad \Lambda = \eta_1 \rho_1 - \eta_2 \rho_2 
\;.
\end{equation}
The optimal \meas is a von Neumann measurement $\{ \Pi_1^\opt, 1 - \Pi_1^\opt \}$ with $\Pi_1^\opt$ the projector
onto the support of $\Lambda_+ = (\Lambda + | \Lambda|)/2$. 
If $\Lambda \geq 0$ the optimal measurement is $\{ \Pi_1^\opt = 1, \Pi_2^\opt =0\}$, meaning that
no measurement can outperform the simple guess that the state is $\rho_1$ (a similar statement holds for $\rho_2$ if $\Lambda \leq 0$). 
For pure states $\rho_i = \ketbra{\psi_i}{\psi_i}$, (\ref{eq-opt_success_proba}) reduces to 
\begin{equation}
P_{\rm S,  a}^{\,\opt} ( \{ \ket{\psi_i} ,\eta_i\}) =  
\frac{1}{2}\Bigl( 1 + \sqrt{ 1 - 4 \eta_1 \eta_2 | \braket{\psi_1}{\psi_2} |^2} \Bigr) 
\end{equation}
and the optimal \meas consists of the rank-one eigenprojectors of $\Lambda$ 
for the positive and negative eigenvalues. When $\eta_1 = \eta_2$, these are the projections onto the two orthogonal subspaces placed symmetrically
with respect to $\ket{\psi_1}$ and $\ket{\psi_2}$, as represented in Fig.~\ref{fig1}.

\begin{figure}
\centering
\begin{minipage}{6cm}
\begin{center}
\includegraphics[width=5cm]{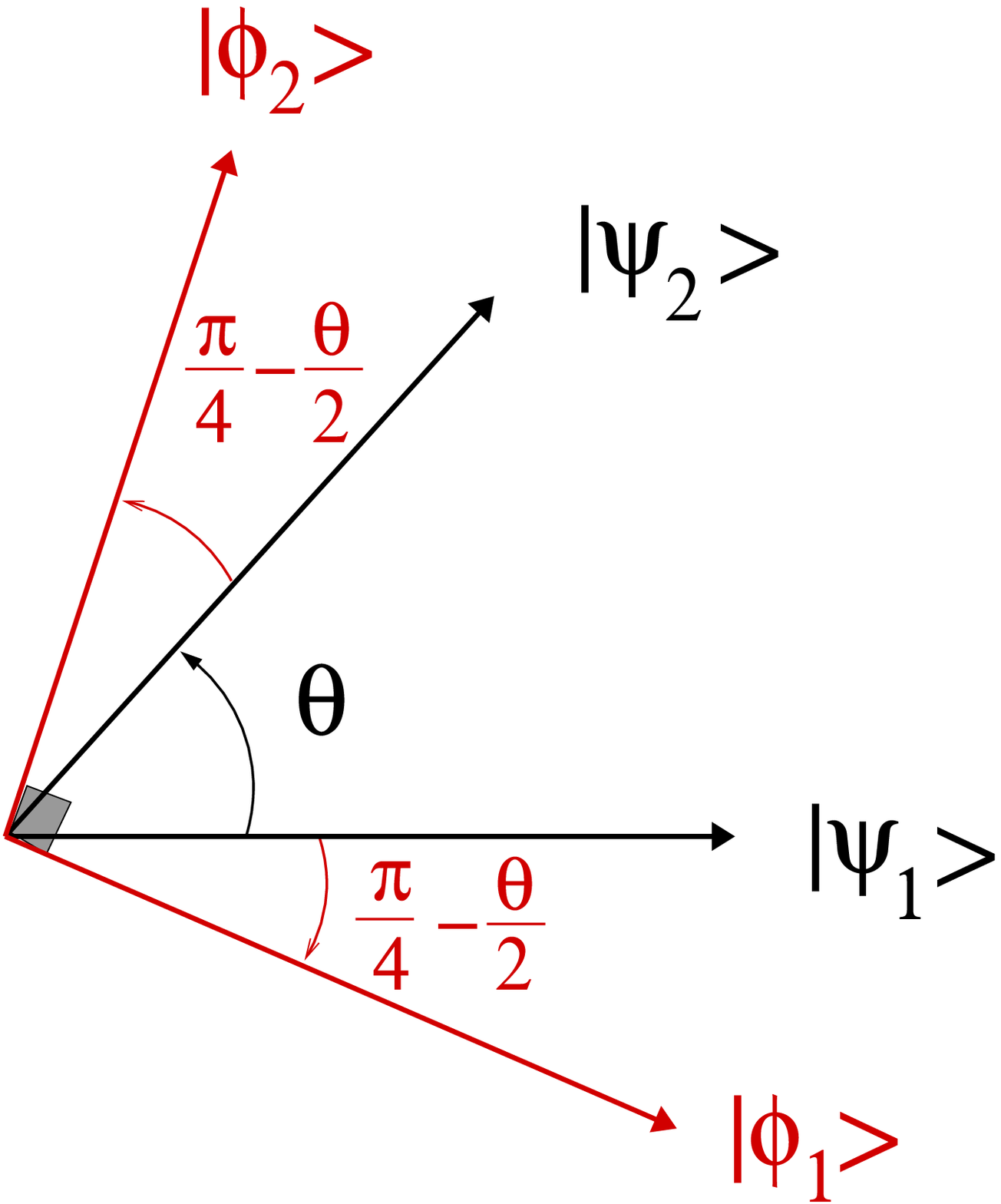}
\end{center}
\end{minipage}
\hspace{0.5cm}
\begin{minipage}{7cm}
\begin{center}
\vspace*{10mm}
\includegraphics[width=7cm]{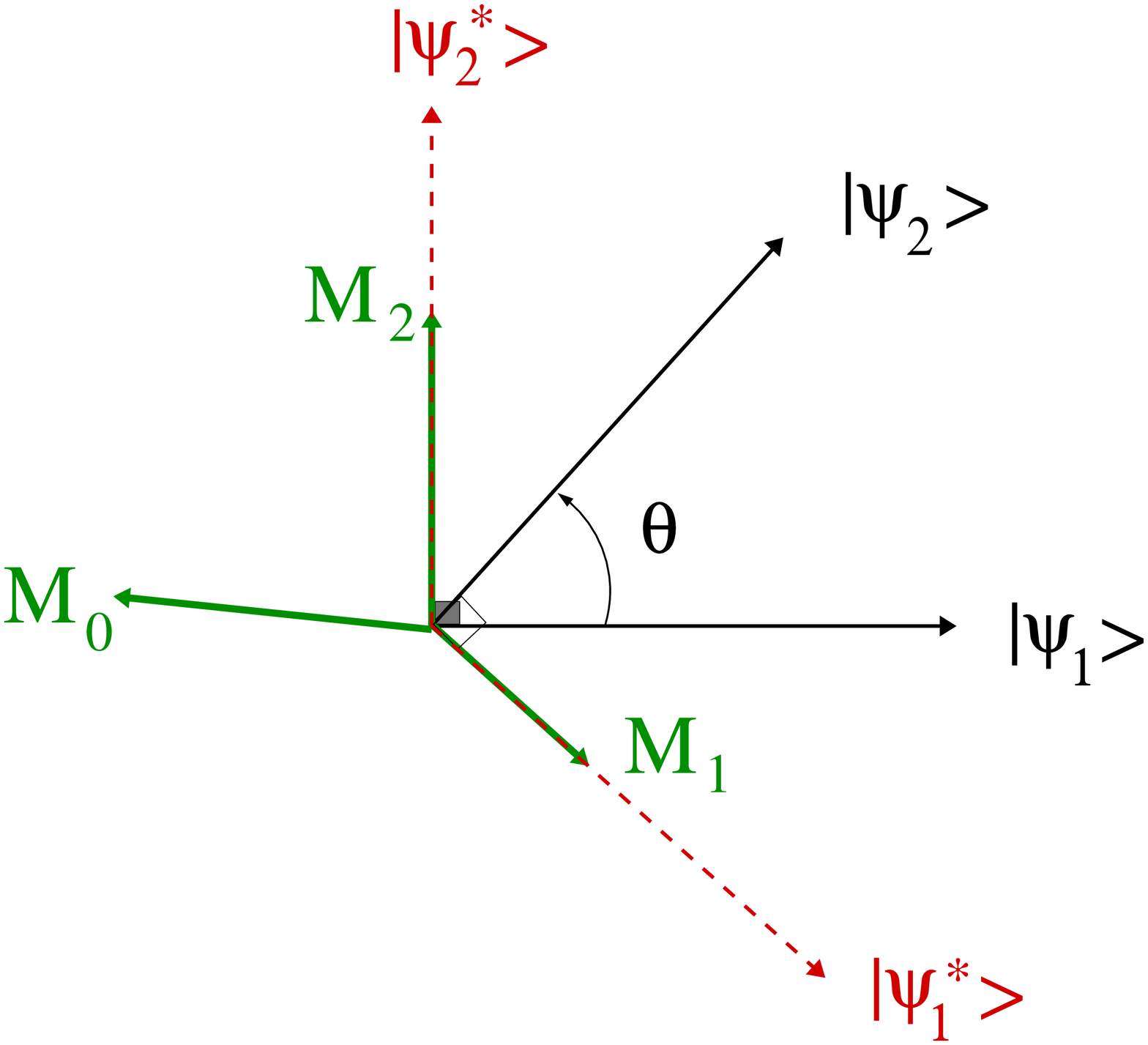}
\end{center}
\end{minipage}
\caption{Optimal \meas $\{ M_i^\opt \}$ in the discrimination of two non-orthogonal pure states $\ket{\psi_1}$ 
and $\ket{\psi_2}$ with equal prior probabilities $\eta_i = 1/2$. (a)~For ambiguous discrimination, $\{ M_i^\opt \}$ is the 
von Neumann \meas in the two orthogonal states $\ket{\phi_1}$ and $\ket{\phi_2}$ with 
$| \braket{\phi_i}{\psi_i} | = \cos ( \frac{\pi}{4}-\frac{\theta}{2} )$, that is, it is the least square \meas associated to $\{ \ket{\psi_i}, \eta_i\}$
(b) For unambiguous discrimination,  
if the maximal prior probability $\eta_\mmax$ is larger than  $q_1 = 1/ ( 1 + \cos^2 \theta)$, then 
the von Neumann \meas in the \ONB $\{ \ket{\psi_1}, \ket{\psi_2^\ast} \}$ (if $\eta_\mmax = \eta_2 > \eta_1$) or
$\{ \ket{\psi_2}, \ket{\psi_1^\ast} \}$ (if $\eta_\mmax = \eta_1 > \eta_2$) indicated by the red dashed vectors yields the smallest failure probability. 
Failure occurs when the outcome corresponds to the first vector in these two bases (inconclusive result). 
If
$1-q_1 < \eta_1 < q_1$, a smaller failure probability is obtained by using the generalized \meas 
with rank-one operators $M_i$ indicated schematically by the green vectors. 
}
\label{fig1}
\end{figure}

\subsubsection{Unambiguous discrimination of two pure states}

The power of generalized measurements is illustrated in the unambiguous discrimination of two pure states $\ket{\psi_1}$ and $\ket{\psi_2}$. 
Indeed, we will show that such measurements enable to distinguish quantum states better
than von Neumann measurements\footnote{
This can be considered as the main physical motivation to introduce generalized measurements~\cite{Peres}.
}.
Clearly, the Hilbert space $\Hh$ can be restricted to its two-dimensional subspace spanned by $\ket{\psi_1}$ and $\ket{\psi_2}$. 
The unambiguity condition implies $\ket{\psi_1} \in \ker M_2$ and $\ket{\psi_2} \in \ker M_1$, so that
the \meas operators $M_1$ and $M_2$ are of rank one and given by (\ref{eq-meas_op_unambiguousQDS}).
We can already observe at this point that the number of outcomes is larger than the space dimension, so that the unambiguous discrimination 
strategy cannot be realized with a von Neumann measurement. 

The optimal success probability is given by~\cite{Jaeger95}
\begin{equation} \label{eq-success_proba_unambiguous_2states}
P_{\rm S,u}^{\,\opt} ( \{ \ket{\psi_i}, \eta_i \} )
= 
\begin{cases}
1- 2 \sqrt{\eta_1 \eta_2} | \braket{\psi_1}{\psi_2} |
& \text{ if $1- q_1 \leq \eta_1 \leq q_1$}
\\
\eta_\mmax ( 1 -  | \braket{\psi_1}{\psi_2} |^2 ) &  \text{ if $\eta_\mmax \geq q_1$}
\end{cases}
\end{equation}
with $\eta_\mmax = \max \{ \eta_1, \eta_2\}$ and $q_1 = 1/(1+| \braket{\psi_1}{\psi_2} |^2 )$. 
It is instructive to establish this formula by using  the Neumark extension theorem~\cite{Bergou_review}.
Thanks to  Theorem~\ref{eq-th-Neumark}, one can represent $\{ M_j\}$  as
a von Neumann measurement on the larger space $\Hh \otimes \Hh_\EE$, with  $\Hh_\EE \simeq \complex^3$.
Let $\{A_j\}_{j=0}^2$ be the Kraus operators for the \meas and $\ket{\epsilon_0}$, $U$, and 
 $\pi_j^\EE$ be as in this theorem.
We may assume that  $\pi_j^\EE  = \ketbra{j}{j}$ are of rank one, where $\{ \ket{j} \}_{j=0}^2$ is an \ONB of $\Hh_\EE$ 
(see the proof of Theorem~\ref{eq-th-Neumark}). One writes
\begin{equation} \label{eq-QSD_m=2_Neumark}
\ket{\Psi_i '} = U \ket{\psi_i} \ket{\epsilon_0} = \sum_{j=0}^2 \sqrt{p_{j|i}} \ket{\varphi_{j|i}} \ket{j}
\end{equation}
for $i=1,2$, where $\sqrt{p_{j|i}} \ket{\varphi_{j|i}} = \braket{j}{\Psi_i '} \in \Hh$ are in general non-orthogonal for distinct $j$'s
and $\| \varphi_{j|i} \| = 1$.
By  (\ref{eq-Neumark_extension}) and (\ref{eq-conditional_state}) the unnormalized post-\meas states are 
$\widetilde{\rho}_{j|i}= \braket{j}{\Psi_i '} \braket{\Psi_i '}{j} =  p_{j|i} \ketbra{\varphi_{j|i}}{\varphi_{j|i}}$, hence $p_{j|i}$ and
$\ket{\varphi_{j|i}}$ can be interpreted as the  probability of outcome $j$ and the corresponding conditional state  
for the input state $\ket{\psi_i}$.
Since we require $p_{2|1}=p_{1|2}=0$, the unitarity of $U$ imposes the conditions
$p_{0|i}= 1 - p_{i|i}$  and
$\braket{\Psi_1'}{\Psi_2'} =  \sqrt{p_{0|1} p_{0|2}} \braket{\varphi_{0|1}}{\varphi_{0|2}} = \braket{\psi_1}{\psi_2}$.
The last relation implies that the probabilities $p_{0|i}$  satisfy 
\begin{equation} \label{eq-bound_on_proba_inconclusive}
p_{0|1} p_{0|2} \geq  p_{0|1} p_{0|2} |\braket{\varphi_{0|1}}{\varphi_{0|2}}|^2 = \cos^2 \theta\;,
\end{equation}
where we have set $\cos \theta=  | \braket{\psi_1}{\psi_2} |$. 
Note that this bound could have been obtained directly from (\ref{eq-cond_postitivity_M_0}), which is easy to solve since we are dealing here 
with $2 \times 2$ matrices~\cite{Bergou_review}.  

In order to maximize the success probability $P_{\rm S} = \sum_i \eta_i p_{i|i}= 1 - \sum_i \eta_i p_{0|i}$, we are 
looking for the smallest possible $p_{0|1}$ and $p_{0|2}$. For such $p_{0|i}$'s the inequality (\ref{eq-bound_on_proba_inconclusive}) is an equality.
Assuming $\cos \theta >0$, this holds whenever $\ket{\varphi_{0|2}}= \E^{\I \delta} \ket{\varphi_{0|1}}$
with $\delta = \arg \braket{\psi_1}{\psi_2}$. 
Accordingly, the conditional post-\meas state for the inconclusive outcome  is  the same irrespective of the input state $\ket{\psi_i}$.
This is physically meaningful since if this post-measurement state was depending on    
$\ket{\psi_i}$ then one could perform a new \meas on it to increase further the success probability.
In summary, for the optimal \meas one has
\begin{equation} \label{eq-derivation_opt_proba_unambiguousQSD}
\ket{\Psi_i '} = \sqrt{ p_{i|i}} \ket{\varphi_{i|i}} \ket{i} + \sqrt{p_{0|i}}\, \E^{\I \delta_i} \ket{\Phi_0}
\end{equation}
with $\ket{\Phi_0} = \ket{\varphi_{0|1}} \ket{0}$ and $\delta_1=0$, $\delta_2=\delta$.

The failure probability 
\begin{equation}
P_{0} = \eta_1  p_{0|1} + \eta_2 \frac{\cos^2 \theta}{p_{0|1}}
\end{equation}
is easy to minimize as a function of $p_{0|1}$. The minimum is achieved
for $p_{0|1}^\opt = \sqrt{\eta_2/\eta_1} \cos \theta$ and is equal to $P_0^\opt = 2 \sqrt{\eta_1\eta_2} \cos \theta$. This yields the 
upper expression in (\ref{eq-success_proba_unambiguous_2states}). 
The restrictions on the values of $\eta_1$ come from the conditions 
$p_{0|1}^\opt \leq 1$ and $p_{0|2}^\opt \leq 1$. 
When $\eta_1 \leq 1-q_1$, the minimum is achieved for $p_{0|1}^\opt = 1$ and $p_{0|2}^\opt = \cos^2 \theta$, \ie
$p_{1|1}^\opt = 0$ and $p_{2|2}^\opt = \sin^2 \theta $. 
In such a case only the state $\ket{\psi_2}$ can be identified with certainty, as
$\ket{\psi_1}$ always produces an inconclusive outcome. 
Strictly speaking this does not  correspond to an unambiguous discrimination.
One can nevertheless determine the optimal measurement, characterized by $M_1^\opt = 0$ and by
two orthogonal projectors  $M_2^\opt  = \ketbra{\psi_2^\ast}{\psi_2^\ast}$ and
$M_0^\opt = \ketbra{\psi_1}{\psi_1}$, see (\ref{eq-meas_op_unambiguousQDS}).
A similar statement holds  when  $\eta_1 \geq q_1$ by exchanging the indices $1$ and $2$.
The corresponding success probability is given by the lower expression in (\ref{eq-success_proba_unambiguous_2states}).

These results are summarized in Fig.~\ref{fig1}.
As claimed above, when $1-q_1 < \eta_1 < q_1$ generalized measurements, obtained via a coupling of the system 
with an ancilla and a measurement on the latter, do better in decoding the message than a von Neumann
measurement performed directly on the system.

\subsubsection{Unambiguous discrimination of two mixed states}

Let us now turn to the case of two mixed states $\rho_1$ and $\rho_2$. Such states cannot be unambiguously 
discriminated when $\range \rho_1$ is contained in $\range \rho_2$ or vice versa. By the unambiguity condition,
$\range M_1 \subset \ker \rho_2$ and $\range M_2 \subset \ker \rho_1$. A trivial situation is when $\ker \rho_1\, \bot \,\ker \rho_2$, in  
which case the optimal POVM is the von Neumann \meas with $M_1$ and $M_2$ equal to the projectors on $\ker \rho_2$ and $\ker \rho_1$, respectively. Then
the minimal failure probability is $P_0^\opt = \tr[ ( \eta_1 \rho_1 + \eta_2 \rho_2 ) \Pi_{0}]$, $\Pi_{0}$ being the projector onto
$\range \rho_1 \cap \range \rho_2$.
One can as before restrict the Hilbert space so that $\range \rho_1 + \range \rho_2 = \Hh$.
If $\range \rho_1$ and $\range \rho_2$ have co-dimension one in $\Hh$, then $M_1$ and $M_2$ are of rank one and take the form (\ref{eq-meas_op_unambiguousQDS})
with $\ket{\psi_1^\ast} \in \ker \rho_2$, $\ket{\psi_2^\ast} \in \ker \rho_1$, and $|\braket{\psi_i^\ast}{\psi_i} |^2$ replaced by 
$R_i = \bra{\psi_i^\ast} \rho_i \ket{\psi_i^\ast}$.
A simple generalization of (\ref{eq-success_proba_unambiguous_2states}) then yields~\cite{Rudolph03}
\begin{equation} \label{eq-success_proba_unambiguous_2_mixed_states}
P_{\rm S, u}^{\,\opt} ( \{ \rho_i, \eta_i \} ) = 
P_{\rm S}^{\,\opt} ( R_i, \eta_i ) \equiv 
\begin{cases}
\dss \frac{\eta_1 R_1 + \eta_2 R_2 - 2 \sqrt{\eta_1 \eta_2 R_1 R_2} \cos \theta}{\sin^2 \theta}
& \dss \text{ if } \cos^2 \theta < \min \Bigl\{ \frac{\eta_1 R_1}{\eta_2 R_2} , \frac{\eta_2 R_2}{\eta_1 R_1} \Bigr\}
\\[4mm]
\dss \max \{ \eta_1 R_1, \eta_2 R_2 \}
& \text{ otherwise} 
\end{cases}
\end{equation}
with $\cos \theta = | \braket{\psi^\ast_1}{\psi^\ast_2} |$.
For kernels of dimensions $d_2 \geq  d_1 > 1$, 
by a standard linear algebra argument one can construct two orthonormal bases  $\{ \ket{\psi^\ast_{2 k}} \}_{k=1}^{d_1}$ of $\ker{\rho_1}$
and $\{ \ket{\psi^\ast_{1k}} \}_{k=1}^{d_2}$ of $\ker \rho_2$ 
such that  $\braket{\psi^\ast_{1k}}{\psi^\ast_{2l}} = \delta_{kl} \cos \theta_k$, with $\theta_k \in [0,\pi/2]$. 
Let us take $M_i = \sum_k M_{ik}$ for $i=1,2$, with $M_{ik}= m_{ik} \ketbra{\psi_{ik}^\ast}{\psi_{ik}^\ast}$. 
Optimizing $P_{\rm S,u}^{\{ M_i\}}$ over the non-negative numbers $m_{ik}$ under the constraint $1-M_1 - M_2 \geq 0$ reduces to the optimization problem for rank-one
\meas operators studied before (in fact,
this constraint  is equivalent to $1-M_{1k}-M_{2k}\geq 0$ for $k=1,\ldots, d_1$ and $1-M_{1k} \geq 0$ for $d_1 < k \leq d_2$).
This gives the lower bound~\cite{Rudolph03}
\begin{equation} \label{eq-lower_bound_P_S_2_states}
P_{\rm S, u}^{\,\opt} ( \{ \rho_i, \eta_i \} ) \geq \sum_{k=1}^{d_1} P_{\rm S}^{\,\opt} ( R_{ik}, \eta_i ) + \eta_1 \sum_{d_1 < k \leq d_2} R_{1k}
\quad \text{ with $R_{ik} = \bra{\psi_{ik}^\ast} \rho_i \ket{\psi_{ik}^\ast}$.}
\end{equation}
An upper bound can be obtained in terms of the fidelity between the states $\rho_1$ and $\rho_2$ 
defined by
$F(\rho_1, \rho_2 ) = (\tr ( |  \sqrt{\rho_1} \sqrt{\rho_2} | ) )^2$ (see Proposition~\ref{prop_lowe_bound_P_S_unambiguous} and Remark~\ref{exo-bound_on_QSD} 
below)~\cite{Rudolph03},
\begin{equation} \label{eq-upper_bound_unambiguousQSD}
P_{\rm S, u}^{\,\opt} ( \{ \rho_i, \eta_i \} ) \leq 
\begin{cases}
1 - 2 \sqrt{\eta_1 \eta_2 F ( \rho_1,\rho_2)}
& \dss \text{ if } F( \rho_1,\rho_2)  < \frac{\eta_\mmin}{\eta_\mmax}
\\[2mm]
\eta_\mmax ( 1 - F (\rho_1,\rho_2 ) ) 
& \text{ otherwise.} 
\end{cases}
\end{equation}

A nice application of two mixed state discrimination is the state comparison problem~\cite{Barnett03}. 
Consider two independent copies of a given system, the state of which is drawn
from the pure state ensemble $\{ \ket{\psi_i}, 1/2 \}_{i=1,2}$. One would like to decide with the help of an appropriate measurement if the two copies are in 
the same state or not, without further information on the actual state of each copies.  If $\ket{\psi_1}$ and $\ket{\psi_2}$ are not orthogonal, this can only
be done with a probability of success $P_{\rm S, comp} < 1$. This  amounts to discriminate the two mixed states
\begin{eqnarray}  \label{eq-def_rho_eq_rho_diff}
\nn
\rho_{\rm eq} & = & \frac{1}{2} \ketbra{\psi_1 \otimes  \psi_1}{\psi_1 \otimes  \psi_1}+\frac{1}{2} \ketbra{\psi_2 \otimes \psi_2}{\psi_2 \otimes \psi_2}
\\
\rho_{\rm diff} & = & \frac{1}{2} \ketbra{\psi_1 \otimes  \psi_2}{\psi_1 \otimes  \psi_2}+ \frac{1}{2} \ketbra{\psi_2 \otimes \psi_1}{\psi_2 \otimes  \psi_1}\;.
\end{eqnarray}
It is shown in~\cite{Rudolph03} that for such mixed states of rank two, the lower and upper bounds in (\ref{eq-lower_bound_P_S_2_states}) and  
(\ref{eq-upper_bound_unambiguousQSD}) coincide. A simple calculation (see Remark~\ref{exo-bound_on_QSD} below) 
then gives the optimal success probability~\cite{Barnett03}
\begin{equation}
P_{\rm S,comp}^{\,\opt} = 1 - | \braket{\psi_1}{\psi_2} | 
\;.
\end{equation}

\subsection{Discrimination with least square measurements} \label{sec-QSD_and_lsm}

How well does the least square measurement (Sec.~\ref{sec-least_square_meas}) in  discriminating ambiguously quantum states? 
More precisely, let 
\begin{equation} 
P_{\rm S, a}^{\rm lsm} ( \{ \rho_i , \eta_i \} )=\sum_i \eta_i \tr ( \rho_i M^\lsm_i )
\end{equation}
be the success probability in discriminating the states $\rho_i$ by performing the least square measurement $\{ M_i^\lsm \}$ associated to $\{ \rho_i,\eta_i\}$.
We would like to compare $P_{\rm S, a}^{\rm lsm}$ with the optimal success probability.

Let us first observe that if  $\rho_i = \Mm ( \ketbra{i}{i} )$, $\Mm$ being  a \QO on $\Bb( \Hh)$ and
$\{ \ket{i} \}_{i=1}^n$  a fixed orthonormal basis of $\Hh$,
then $P_{\rm S,a} ( \{ \rho_i, \eta_i\})$ is related to the entanglement 
fidelity defined in  Sec.~\ref{sec-approximate_reverse_operation}. 
Recall that any ensemble $\{ \rho_i, \eta_i\}_{i=1}^m$ with $m \leq n$ states can be obtained in this way from an operation  
$\Mm: \Bb ( \Hh ) \rightarrow \Bb ( \Hh )$  
(since $m \leq n$ we can identify here the pointer space $\Hh_\PP$ with  a subspace of $\Hh$, see  Sec.~\ref{sec-link_meas_QO_ensemble}).
To establish the relation with the average fidelity (\ref{eq-average_entanglement_fidelity}), consider a POVM  $\{ M_i\}_{i=1}^m$  with $m$ \meas
 operators and let us associate to it  the \QO $\Rr$ on $\Bb ( \Hh )$ defined by $\Rr^\ast ( \ketbra{i}{j} )= M_i \delta_{ij}$. Then
\begin{equation}
P_{\rm S,a}^{ \{ M_i\} } \bigl( \{ \rho_i, \eta_i\} \bigr)
= \sum_{i=1}^m \eta_i \tr [ \Rr^\ast ( \ketbra{i}{i} ) \rho_i ] 
= \sum_{i=1}^m \eta_i \bra{i} \Rr \circ  \Mm ( \ketbra{i}{i} ) \ket{i}  
=\overline{F}_{\rm e} \bigl( \{ \ket{i} , \eta_i \} , \Rr \circ \Mm \bigr) 
\end{equation}
thanks to the equality of the entanglement fidelity with the  
input-output fidelity for pure states. In view of the one-to-one correspondence between
POVMs with $m \leq n$ operators and quantum operations on $\Bb(\Hh)$  we obtain
the following relation between $P_{\rm S, \, a}^{\,\opt}$  and the maximal fidelity over all  recovery operations $\Rr$ on $\Bb ( \Hh)$:
\begin{equation} \label{eq-link_success_proba_entanglement_fidelity}
P_{\rm S, \, a}^{\,\opt} ( \{ \rho_i , \eta_i \}_{i=1}^m ) = \max_{\Rr} \bigl\{  \overline{F}_{\rm e} ( \{ \ket{i} , \eta_i \}_{i=1}^m , \Rr \circ \Mm ) \bigr\}
\quad , \quad m \leq n \;.
\end{equation}
Furthermore, the optimal \meas operators are given in terms of the optimal recovery operation $\Rr^\opt$
by $M_i^\opt = ( \Rr^\opt )^\ast ( \ketbra{i}{i} )$.
According to Proposition~\ref{prop_transpose_operation}, taking $\Rr$ to be the transpose operation 
$\Rr_{\Mm, \rho_\inp}$ of $\Mm$ for the state $\rho_\inp = \sum_i \eta_i \ketbra{i}{i}$ gives 
an entanglement  fidelity larger than the square of the \RHS of (\ref{eq-link_success_proba_entanglement_fidelity}).
But the \meas associated to $\Rr_{\Mm, \rho_\inp}$ is the least square measurement, 
\ie $M_i^\lsm = \Rr_{\Mm, \rho_\inp}^\ast ( \ketbra{i}{i})$ (see Sec.~\ref{sec-least_square_meas}). 
As a result,  Proposition~\ref{prop_transpose_operation}  yields the following inequality.

\vspace{2mm}

\begin{corollary} If $m \leq n = \dim \Hh$, then
\begin{equation} \label{eq-upper_bound_P_S_in_term_of_P_lsm}
P_{\rm S, a}^{\,\opt} ( \{ \rho_i , \eta_i \}_{i=1}^m ) \leq \Bigl( P_{\rm S, a}^{\rm lsm} ( \{ \rho_i , \eta_i \}_{i=1}^m ) \Bigr)^\onehalf \;.
\end{equation}
\end{corollary}

\vspace{1mm}

Thus, if the error probability for discriminating $\{ \rho_i, \eta_i\}$ using the least square \meas is small, then
it is at most twice  the minimal error probability  $P_{\rm err,a}^{\,\opt}= 1 - P_{\rm S,a}^{\,\opt}$, up to a small correction of the order of $(P_{\rm S,a}^{\rm lsm} )^2$. 
Small error probabilities occur for almost orthogonal states. Therefore, for such states least square measurements are nearly 
optimal~\cite{Hausladen94,Barnum02}.

It is worth  mentioning that least square measurements  are also asymptotically optimal for 
discriminating  ambiguously equiprobable linearly independent pure states~\cite{Holevo78}.
In addition, they optimally discriminate equiprobable states drawn from 
a symmetric ensemble, like for instance 
the states  $\rho_i = U^{i-1} \rho_1 (U^{i-1})^\ast$ related between themselves  
through conjugations by powers of a single unitary operator $U$ satisfying $U^m=\pm 1$ (see~\cite{Ban97,Barnett01,Chou03,Eldar01} and references therein).
Necessary and sufficient conditions for the optimality of least square measurements in state discrimination
have been investigated in~\cite{Eldar04,Sasaki98}.

\subsection{General results on ambiguous discrimination} \label{sec-gen_result_QSD}

Let $\{ \rho_i, \eta_i\}_{i=1}^m$ be an ensemble of $m$ states of a system with a $n$-dimensional Hilbert space $\Hh$.
Hereafter we assume that $\eta_i >0$ for all $i=1,\ldots , m$, so that $m$ is the actual number of states to discriminate.
We denote by $\widetilde{\rho}_i = \eta_i \rho_i$ the unnormalized states with trace equal to the prior probability $\eta_i$. 
To shorten notation, the dependence
of the success probability $P_{\rm S}$ on the ensemble is not written explicitly.
The following proposition contains one of the few results in ambiguous discrimination applying to arbitrary  
ensembles.

\vspace{2mm}

\begin{proposition} \label{prop-Holevo_QSD}{\rm \cite{Holevo73,Yuen75,Eldar_Megretski03}}
The optimal success probability in ambiguous state discrimination is given by 
\begin{equation} \label{eq-max_success_proba_as_minumum_Upsilon}
P_{\rm S,a}^{\,\opt}  = \inf_{\Upsilon \geq \widetilde{\rho}_i} \bigl\{ \tr ( \Upsilon) \bigr\} \;,
\end{equation}
where the infimum is over all self-adjoint operators $\Upsilon$ satisfying $\Upsilon \geq \widetilde{\rho}_i$ for any $i=1,\ldots , m$.
Moreover,  the POVM $\{ M_i^\opt \}_{i=1}^m$ is optimal 
\ifif the operator 
$\Upsilon^\opt = \sum_i \widetilde{\rho}_i M_i^\opt$ satisfies the two conditions
\begin{itemize}
\item[(i)] $\Upsilon^\opt$ is self-adjoint;
\item[(ii)] $\Upsilon^\opt \geq \widetilde{\rho}_i$ for any $i=1,\ldots, m$.
\end{itemize}
In such a case, the infimum in the \RHS of (\ref{eq-max_success_proba_as_minumum_Upsilon}) 
is attained for $\Upsilon=\Upsilon^\opt$.
\end{proposition}

\vspace{1mm}

The fact that (ii) is sufficient to ensure the optimality of $\{ M_i^\opt\}$ is obvious from the relation
\begin{equation}
P_{\rm S,a}^{\,\opt} - P_{\rm S,a}^{ \{ M_i\} } = \sum_{i=1}^m \tr [ (\Upsilon^\opt - \widetilde{\rho}_i ) M_i ] \;.
\end{equation}
The necessary and sufficient conditions (i) and (ii)
are due to Holevo~\cite{Holevo73}, who derived them by considering a specific one-parameter family $\{ M_i (\varepsilon)\}$ 
of POVMs such that  $M_i ( 0)=M_i^\opt$ and by exploiting the fact that 
$\partial P_{\rm S,a}^{ \{ M_i (\varepsilon)\} }/\partial \varepsilon=0$ for $\varepsilon=0$ 
(see~\cite{Helstrom}, chapter 4).  
Yuen, Kennedy, and Lax ~\cite{Yuen75} proposed another derivation based on a duality argument in vector space optimization. We shall present below 
the related  proof of Eldar, Megretski and Verghese~\cite{Eldar_Megretski03}.

Let us note that (i) and (ii) imply
\begin{equation} \label{eq-CNS_optimal_meas_QSD}
(\Upsilon^\opt - \widetilde{\rho}_i ) M_i^\opt = M_i^\opt (\Upsilon^\opt - \widetilde{\rho}_i ) = 0
\quad , \quad i=1,\ldots,m\;.
\end{equation}
In fact, since $\sum_i \tr [(\Upsilon^\opt - \widetilde{\rho}_i  ) M_i^\opt ] =0$ and $ \Upsilon^\opt  - \widetilde{\rho}_i \geq 0$ by  (ii), one deduces that $| ( \Upsilon^\opt - \widetilde{\rho}_i )^{1/2} (M_i^\opt)^{1/2}|^2 =0$
(recall that $A \geq 0$ and $\tr(A) = 0$ imply $A=0$). One concludes from  this equality that $( \Upsilon^\opt - \widetilde{\rho}_i ) M_i^\opt=0$.
It is easy to see by eliminating $\Upsilon^\opt$ that (\ref{eq-CNS_optimal_meas_QSD}) is equivalent to
\begin{equation} \label{eq-CNS_optimal_meas_QSD_bis}
M_i^\opt ( \widetilde{\rho}_i - \widetilde{\rho}_j ) M_j^\opt = 0 \quad ,\quad  i,j = 1,\ldots, m\;.
\end{equation}
The condition (\ref{eq-CNS_optimal_meas_QSD_bis}) automatically implies that $\Upsilon^\opt$ is self-adjoint.
Hence a necessary and sufficient condition for $\{ M_i^\opt \}$ to be optimal is given by 
conditions (ii) and (\ref{eq-CNS_optimal_meas_QSD_bis}). 

Except in special cases such as ensembles of equiprobable states related by a symmetry~\cite{Ban97,Barnett01,Eldar01,Chou03}, it
is  difficult in practice to obtain the optimal measurement and success probability from the above necessary and sufficient conditions. 
Nevertheless, the formulas (\ref{eq-max_success_proba_as_minumum_Upsilon}) and (\ref{eq-CNS_optimal_meas_QSD}) are 
helpful for computing these quantities numerically. For indeed, the minimization task
in (\ref{eq-max_success_proba_as_minumum_Upsilon}) is simpler than the maximization in 
(\ref{eq-max_success_proba_POVM}) and can be solved efficiently with the help of convex semidefinite programs~\cite{Eldar_Megretski03}.   

\vspace{1mm}

\proof The main idea is to show that the minimization problem in (\ref{eq-max_success_proba_as_minumum_Upsilon}) is
dual to the maximization problem in (\ref{eq-max_success_proba_POVM}). More precisely, there exists a convex set $\Gamma \subset \Bb(\Hh)_{\rm s.a.}$
such that
\begin{equation} \label{eq-equality_proof_NSC_QSD_2}
P_{\rm S,a}^{\{ M_i\} } \leq \tr ( \Upsilon) \quad , \quad  \; \forall\; \{ M_i \}\;{\rm POVM}, \;\forall\; \Upsilon \in \Gamma\;,
\end{equation}
and the maximum of the left-hand member is equal to the minimum of the right-hand member, \ie $P_{\rm S,a}^{\,\opt} = \min_{\Upsilon \in \Gamma} \tr ( \Upsilon) $.
The set $\Gamma$ is defined by
\begin{equation}
\Gamma = \bigl\{ \Upsilon \in \observables_{\rm s.a.} \,;\,  \Upsilon \geq  \widetilde{\rho}_i \, , \, i=1,\ldots, m \bigr\}\;.
\end{equation}
Then 
$\tr ( \Upsilon ) - P_{\rm S,a}^{\{ M_i\} } = \sum_i \tr [ ( \Upsilon - \widetilde{\rho}_i ) M_i ] \geq 0$ for any $\Upsilon \in \Gamma$, so that 
(\ref{eq-equality_proof_NSC_QSD_2}) holds true.
Let us now
define the following convex subset $\Omega$ of the real vector space $\Bb (\Hh)_{\rm s.a.} \times \real$:
\begin{equation}
(B, x) \in \Omega \quad \Leftrightarrow  \quad B= \sum_{i=1}^m B_i - 1 \;,\;x= r - \sum_{i=1}^m \tr ( B_i \widetilde{\rho}_i )
\;\text{ with } \; B_i \geq 0\; \text{ and }\; r > P_{\rm S,a}^{\,\opt}\;.
\end{equation}
This space is endowed with the scalar product $\langle (B,x)\, ,\, (C, y)\rangle = \tr(BC) + x y$.
Since $\Omega$ is convex and does not contain $(0,0)$, by the separating hyperplane theorem one can find 
a non-vanishing vector $(\Upsilon_a ,a) \in \Bb (\Hh)_{\rm s.a.} \times \real$ such that
 $\langle  (\Upsilon_a ,a)  \, , \, (B, x) \rangle \geq 0$ for any $(B,x) \in \Omega$, that is
\begin{equation} \label{eq-bould_from_separating_hyperplane}
 \tr \Bigl[ \Upsilon_a  \Bigl(\sum_{i=1}^m B_i -1 \Bigr) \Bigr] + a \Bigl( r - \sum_{i=1}^m \tr ( B_i \widetilde{\rho}_i ) \Bigr) \geq 0\;.
\end{equation}
Taking $B_i = t \ketbra{\varphi}{\varphi}$ if $i=k$ and zero otherwise, with $\ket{\varphi} \in \Hh$ and $t>0$, 
and letting $t \rightarrow \infty$, we obtain
 $\bra{\varphi} \Upsilon_a \ket{\varphi} - a \bra{\varphi} \widetilde{\rho}_k \ket{\varphi} \geq 0$.
But $\ket{\varphi}$ and $k$ are arbitrary, hence
\begin{equation} \label{eq-first_bound_proof_NSC_QSD_1}
\Upsilon_a \geq a \widetilde{\rho}_i \quad , \quad i=1,\ldots,m \;.
\end{equation}
Similarly, taking $B_i=0$ for all $i$ and $r \rightarrow P_{\rm S,a}^{\,\opt}$, (\ref{eq-bould_from_separating_hyperplane}) yields 
\begin{equation} \label{eq-first_bound_proof_NSC_QSD_2}
a P_{\rm S,a}^{\,\opt} \geq \tr ( \Upsilon_a )  \;.
\end{equation}
From the same choice of $B_i$ and $r \rightarrow \infty$ one gets $a \geq 0$. If $a = 0$ then $\Upsilon_a \geq 0$ and $\tr ( \Upsilon_a) = 0$ 
by (\ref{eq-first_bound_proof_NSC_QSD_1}) and (\ref{eq-first_bound_proof_NSC_QSD_2}). This would imply $\Upsilon_a =0$, in contradiction with
$(\Upsilon_a, a ) \not= (0,0)$. Thus $a>0$. The self-adjoint operator $\Upsilon^\opt = \Upsilon_a/a$ satisfies 
$\Upsilon^\opt \geq  \widetilde{\rho}_i$ for all $i$ (\ie $\Upsilon^\opt  \in \Gamma$) 
and $\tr ( \Upsilon^\opt ) \leq  P_{\rm S,a}^{\,\opt}$, see  (\ref{eq-first_bound_proof_NSC_QSD_1}) and 
(\ref{eq-first_bound_proof_NSC_QSD_2}). The converse of the last inequality follows from (\ref{eq-equality_proof_NSC_QSD_2}). Whence
$P_{\rm S,a}^{\,\opt} = \tr ( \Upsilon^\opt ) = \min_{\Upsilon \in \Gamma} \tr ( \Upsilon)$,
as claimed in the proposition.
This identity implies $\sum_i \tr [( \Upsilon^\opt - \widetilde{\rho}_i) M_i^\opt ] = 0$ if  $\{ M_i^\opt\}$ is an
optimal POVM. But all traces in the sum are non-negative, thus they vanish and (\ref{eq-CNS_optimal_meas_QSD}) is satisfied by 
the arguments given above to derive this equation.
It results from (\ref{eq-CNS_optimal_meas_QSD}) that 
$ \Upsilon^\opt = \sum_i  \widetilde{\rho}_i M_i^\opt =   \sum_i  M_i^\opt  \widetilde{\rho}_i$.
This concludes the proof. 
\finpro

\vspace{3mm}

Let us consider the success probability
\begin{equation} \label{eq-max_success_proba_von_Neumann}
P_{\rm S,a}^{\rm{opt\,v.N.}} ( \{ \rho_i,\eta_i \})
= \max_{ \{ \Pi_i \} } \biggl\{ \sum_{i=1}^{m} \eta_i \tr ( \Pi_i \rho_i) \biggr\}\;,
\end{equation}
where the maximum is over all von Neumann measurements $\{ \Pi_i\}_{i=1}^m$. 
A natural question is whether this probability may be equal to $P_{\rm S,a}^{\,\opt}$, \ie whether  the states $\rho_i$  may be
 discriminated optimally with a von Neumann measurement.
We have already argued above that this is not always the case, even for pure states. 
A simple consequence of Proposition~\ref{prop-Holevo_QSD}  is that the equality holds for
{\it linearly independent states}. The states $\rho_i$ are called linearly independent if their eigenvectors $\ket{\zeta_{ij}}$
with non-zero eigenvalues 
form a linearly independent family $\{ \ket{\zeta_{ij}} \}_{i=1, \ldots , m}^{j=1,\ldots, r_i}$
in $\Hh$ (here $r_i$ is the rank of $\rho_i$).
We say that they span the Hilbert space $\Hh$ if $\Hh= \Span \{ \ket{\zeta_{ij}} \}_{i=1, \ldots , m}^{j=1,\ldots, r_i}$.
Without loss of generality one can restrict  $\Hh$  to a subspace $\Hh'$ spanned by the $\rho_i$.

\vspace{2mm}

\begin{corollary} {\rm \cite{Eldar_Megretski03}} 
Let  $\{ \ket{\psi_i} , \eta_i\}_{i=1}^m$ be an ensemble of pure states spanning $\Hh$. Then the optimal measurement operators 
$M_i^\opt$ in ambiguous state discrimination are of rank one. More generally, for any ensemble  $\{ \rho_{i}  , \eta_i\}_{i=1}^m$ spanning $\Hh$, the optimal 
measurement operators have ranks $\rank ( M_i^\opt ) \leq \rank ( \rho_i )$ for all $i=1,\ldots , m$.
\end{corollary}

\vspace{1mm}

\begin{corollary} {\rm \cite{Eldar03}} \label{prop_Eldar_Megretski}
Let $\{ \rho_i , \eta_i\}_{i=1}^m$ be an ensemble of linearly independent states spanning $\Hh$. 
Then an optimal measurement in ambiguous state discrimination is a von Neumann measurement
with orthogonal projectors $M_i^\opt = \Pi_i^\opt$ of rank $r_i=\rank ( \rho_i)$. In particular, the probabilities 
(\ref{eq-max_success_proba_POVM}) and (\ref{eq-max_success_proba_von_Neumann}) are equal.

\end{corollary}

\vspace{1mm}

\proof Let us set $N_i^\opt = \Upsilon^\opt - \widetilde{\rho}_i$. The relation (\ref{eq-CNS_optimal_meas_QSD}) implies $\range M_i^\opt \subset \ker N_i^\opt$, hence
$\rank (   M_i^\opt ) \leq \dim ( \ker N_i^\opt)$. Since the rank of the sum of two matrices is smaller or equal to the sum of their ranks,
$\rank ( \Upsilon^\opt ) \leq \rank ( N_i^\opt ) + r_i$ and thus
$\dim ( \ker  N_i^\opt ) \leq \dim ( \ker  \Upsilon^\opt ) + r_i$.
But $\ker   \Upsilon^\opt   \subset [\range ( \rho_i)]^\bot$ for all $i$ according to the condition (ii) of Proposition~\ref{prop-Holevo_QSD}.   
Consequently, if the states $\rho_i$ span $\Hh$ then $\ker   \Upsilon^\opt = \{ 0\}$. This shows that $\rank (   M_i^\opt ) \leq r_i$.
If furthermore the $\rho_i$ are linearly independent, then $\sum_i r_i = n=\dim \Hh$. Introducing the spectral 
decomposition  $M_i^\opt = \sum_k \ketbra{\widetilde{\mu}_{ik}}{\widetilde{\mu}_{ik}}$ with unnormalized vectors $\ket{\widetilde{\mu}_{ik}}$, $k=1,\ldots, r_i$, 
and noting that 
the sum $\sum_{i,k} \ketbra{\widetilde{\mu}_{ik}}{\widetilde{\mu}_{ik}}=1$ contains at most $n$ terms, it follows that   
$\{ \ket{\widetilde{\mu}_{ik}}\}$ is an orthonormal
basis of $\Hh$. Thus $M_i^\opt$ are orthogonal projectors of rank $r_i$. 
\finpro

\subsection{Bounds on the maximal success probability} \label{sec-bound_success_proba}

We now establish some inequalities satisfied by $P_{\rm S}^{\,\opt}$ for any number  $m$ of states to discriminate.
A review of various upper bounds for ambiguous discrimination can be found in~\cite{Qiu10}. We only discuss here the bounds involving the 
fidelity
\begin{equation} \label{eq-def_Ulhmann_fidelity}
F(\rho,\sigma)= \| \sqrt{\rho} \sqrt{\sigma} \|_1^2 = \Bigl( \tr [ ( \sqrt{\sigma} \rho \sqrt{\sigma} )^\onehalf ] \Bigr)^2\;.
\end{equation}
The properties of this fidelity will be analyzed in the forthcoming Sec.~\ref{sec-Bures_distance}. Let us only mention
here that $F(\rho,\sigma)$ is symmetric under the exchange of $\rho$ and $\sigma$ 
(actually, $ \sqrt{\sigma} \rho \sqrt{\sigma}$ and  $ \sqrt{\rho} \,\sigma \sqrt{\rho}$
have the same non-zero eigenvalues) and reduces for pure states $\rho_\psi = \ketbra{\psi}{\psi}$ and  
$\sigma_\phi = \ketbra{\phi}{\phi}$ to the square modulus of the scalar product $\braket{\psi}{\phi}$, \ie $F(\rho_\psi,\sigma_\phi)=| \braket{\psi}{\phi}|^2$. 
More generally, $F (\rho, \sigma )$ can be seen as a measure of non-orthogonality of $\rho$ and $\sigma$.
 
The following lower and upper bounds on  the maximum success probability $P_{\rm S,a}^{\,\opt}$ for ambiguous state discrimination 
are taken from Refs.~\cite{Barnum02} and~\cite{Montanaro08}, respectively\footnote{
The upper bound is established in~\cite{Barnum02} (and is often reported in subsequent works) with an unnecessary extra factor of two in front of the sum
(after correcting the obvious misprints in this reference).
}.

\vspace{2mm}

\begin{proposition} {\rm (Barnum and Knill~\cite{Barnum02}, Montanaro~\cite{Montanaro08}).} \label{prop_lower_and_upper_bounds_on_P_S_amb}
For any ensemble  $\{ \rho_i, \eta_i \}_{i=1}^m$, one has
\begin{equation} \label{eq-lower_and_upper_bounds_on_P_S_amb}
1 - \sum_{i > j} \sqrt{\eta_i \eta_j F( \rho_i, \rho_j )}
\leq P_{\rm S,a}^{\,\opt} ( \{ \rho_i, \eta_i \} )  \leq
1 - \sum_{i > j} \eta_i \eta_j F( \rho_i, \rho_j )\;.
\end{equation}
\end{proposition}

\vspace{1mm}

The inequalities (\ref{eq-lower_and_upper_bounds_on_P_S_amb}) make quantitative the intuitive fact that the more  pairwise orthogonal are
the states $\rho_i$, the larger is the success probability to discriminate them, and conversely.

\vspace{1mm}

\proof
Let $\rho_i = A_i A_i^\ast$, the operators $A_i$ being, for instance, given  by (\ref{eq-square_root_rho_i}). 
Given a POVM $\{ M_i \}$ with Kraus operators $R_i$ (\ie $M_i = R_i^\ast R_i$), we set 
\begin{equation} \label{eq-def_matrices_S_and_B}
S_{ij} = \sqrt{\eta_j} R_i A_j \quad , \quad 
B_{ij} = \sqrt{\eta_i \eta_j} A_i^\ast A_j \;.
\end{equation}
We view $S=(S_{ij})_{i,j=1}^m$ and $B=( B_{ij})_{i,j=1}^m$ as $m \times m$ matrices with values in $\observables$, which
are related by $S^\ast S = B \geq 0$ (this follows from $\sum_i R_i^\ast R_i = 1$).
Observe that
\begin{equation} \label{eq-proof_bounds_on_opt_success_proba}
P_{\rm S,a}^{\{ M_i\}} = \sum_{j} \eta_j \tr ( M_j {\rho}_j ) = 1 - \sum_{i \not= j}  \eta_j \tr ( M_i {\rho}_j )
 = 1 - \sum_{i \not= j}  \| S_{ij} \|^2_2 
\end{equation}
and
\begin{equation} \label{eq-proof_bounds_on_opt_success_proba2}
\eta_i \eta_j F ( \rho_i , \rho_j ) = \eta_i \eta_j \| \sqrt{{\rho}_i} \sqrt{{\rho}_j} \|_1^2 
 = \eta_i \eta_j \| U_i^\ast \sqrt{{\rho}_i} \sqrt{{\rho}_j} U_j \|^2_1   
 = \| B_{ij} \|_1^2 \;,
\end{equation}
where $\| \cdot \|_{1,2}$ are the trace and Hilbert Schmidt norms. We  have used in (\ref{eq-proof_bounds_on_opt_success_proba2}) 
 the polar decomposition $A_i = \sqrt{{\rho}_i} U_i$ and the unitary invariance of these norms.
The main idea  
to prove the first inequality in (\ref{eq-lower_and_upper_bounds_on_P_S_amb}) is to bound from below the optimal success probability $P_{\rm S,a}^{\,\opt}$  
by the success probability $P_{\rm S,a}^\lsm$ for discriminating the states with the least square measurement~\cite{Barnum02}.
For the latter, the matrix $S$ in (\ref{eq-def_matrices_S_and_B}) is the square root of  $B$
(in fact, according to (\ref{eq-definition_LSM}), $S^\lsm_{ij}= \sqrt{\eta_i \eta_j} A_i^\ast \rho_\out^{-1/2} A_j$ so that
$S^\lsm \geq 0$, and it has been argued above that $|S|^2 = B$). For instance, if the $\rho_i$  are pure states $\ket{\psi_i}$,
$B$ and $S^\lsm$ can be identified with the scalar product matrices $( \braket{\widetilde{\psi}_i}{\widetilde{\psi}_j} )_{i,j=1}^m$ and 
$( \braket{\widetilde{\mu}_i}{\widetilde{\psi}_j} )_{i,j=1}^m$, respectively, with  $\ket{\widetilde{\psi_i}} = \sqrt{\eta_i} \ket{\psi_i}$ and
$\ket{\widetilde{\mu}_i} = \sqrt{\eta_i} \rho_\out^{-1/2} \ket{\psi_i}$, the latter  being the vectors describing 
the least square measurement (Sec.~\ref{sec-least_square_meas}). 
The identity  $S^\lsm = \sqrt{B}$ then becomes evident from the definition of a POVM\footnote{
This remarkable identity has been singled out for pure states in~\cite{Hausladen96}. The authors of this reference suggest to use it
as a definition of the least square measurement.
}.
Therefore, in view of (\ref{eq-proof_bounds_on_opt_success_proba}), $P_{\rm S,a}^{\,\opt} \geq P_{\rm S,a}^\lsm = 1 - \sum_{i\not= j} \| ( \sqrt{B})_{ij} \|_2^2$.
The lower bound in (\ref{eq-lower_and_upper_bounds_on_P_S_amb}) comes from 
the following norm inequality proven in  Appendix~\ref{app-norm_inequalities}:
for any fixed $j=1,\ldots, m$,
\begin{equation} \label{eq-lower_and_upper_bounds_on_P_S_amb_ter}
 \sum_{i,i \not= j} \| (\sqrt{B})_{ij} \|_2^2 \leq  \onehalf \sum_{i,i \not= j} \| B_{ij} \|_1\;,
\end{equation}
where the last sum is related to the fidelities by (\ref{eq-proof_bounds_on_opt_success_proba2}).

It remains to establish  the upper bound. With the notation above, this bound takes the form
\begin{equation} \label{eq-lower_and_upper_bounds_on_P_S_amb_bis}
\onehalf \sum_{i \not= j} \| B_{ij} \|_1^2 \leq  \sum_{i \not= j} \| S_{ij} \|_2^2 \;.
\end{equation}
Fixing $j$ again and introducing the notation $\| \cdot \|_{1/2}$ as in (\ref{eq-Lp_norm}) (note that this is not a norm),  if one can show that
\begin{equation} \label{eq-proof_prop_bounds_on_P_S}
\Bigl\| \sum_{i,i \not= j}  |B_{ij}|^2 \Bigr\|_\onehalf  \leq \sum_{i,i \not= j} \Bigl( \| S_{ij} \|_2^2 +  \| S_{ji} \|_2^2 \Bigr)
\end{equation}
then the required inequality (\ref{eq-lower_and_upper_bounds_on_P_S_amb_bis}) will be proven. Actually,
$\sum_{i} \| B_{ij} \|_1^2 = \sum_i \| |B_{ij}|^2 \|_{1/2} \leq \| \sum_i | B_{ij} |^2 \|_{1/2}$ by the   inverse Minkowski inequality
(\ref{eq-Minkowski_inequality}) in Appendix~\ref{app-norm_inequalities}.
In order to show (\ref{eq-proof_prop_bounds_on_P_S}), let us 
introduce the following  $(m-1)\times( m-1)$ matrices with values in $\observables$:
\begin{equation}
\begin{array}{ccccccc}
C^{(j)} & = & \dss \sum_{i,i\not= j} (S_{ji})^\ast \otimes  \ketbra{i}{1} 
& , &  D^{(j)} & = & \dss  S_{jj} \otimes \ketbra{1}{1}
\\
E^{(j)} & = & \dss \sum_{i\not= j} \sum_{k \not= j} (S_{ki} )^\ast \otimes \ketbra{i}{k}
& , & F^{(j)}  & = & \dss \sum_{k,k\not= j} S_{kj}\otimes  \ketbra{k}{1} 
\end{array}
\end{equation}
(here $\ketbra{i}{k}$ stands for the matrix with  vanishing entries except in the $i$th raw and $k$th column, which has a unit entry).
An explicit calculation leads to 
\begin{equation} \label{eq-proof_prop_bounds_on_P_S_ter}
\bigl\| C^{(j)} D^{(j)} +  E^{(j)}  F^{(j)} \bigr\|_1^2 = \Bigl\| \sum_{i,i \not= j} | B_{ij}|^2 \Bigr\|_\onehalf
\quad , \quad   \bigl\| C^{(j)} \bigr\|_2^2 = \sum_{i,i \not=j}  \| S_{ji} \|_2^2 \quad , \quad  \bigl\| F^{(j)} \bigr\|_2^2 = \sum_{k,k \not=j} \| S_{kj}\|_2^2 \;.
\end{equation}
Furthermore,
\begin{equation} \label{eq-proof_prop_bounds_on_P_S_bis}
\bigl\| C^{(j)}  \bigr\|_2^2 + \bigl\| D^{(j)}  \bigr\|_2^2 + \bigl\| E^{(j)}  \bigr\|_2^2 + \bigl\| F^{(j)}  \bigr\|_2^2 
= \sum_{i,k} \| S_{ik} \|_2^2
 =  \sum_{k} \eta_k \tr ( {\rho}_k ) = 1\;.
\end{equation}
We can now take advantage of the norm inequality (\ref{eq-norm_1_and_2_inequality})
of Appendix~\ref{app-norm_inequalities}.
Because of (\ref{eq-proof_prop_bounds_on_P_S_bis}), this gives
\begin{equation}
\bigl\| C^{(j)} D^{(j)} +  E^{(j)}  F^{(j)} \bigr\|_1^2  \leq \bigl\| C^{(j)} \bigr\|_2^2 + \bigl\| F^{(j)} \bigr\|_2^2\;.
\end{equation}
We plug the equalities (\ref{eq-proof_prop_bounds_on_P_S_ter}) into this result to obtain (\ref{eq-proof_prop_bounds_on_P_S}).  
This concludes the proof.
\finpro

\vspace{3mm}

Let us now turn to unambiguous discrimination. The following easy-to-derive bound generalizes the upper line in
(\ref{eq-upper_bound_unambiguousQSD}).

\vspace{2mm}

\begin{proposition} {\rm \cite{Feng04}} \label{prop_lowe_bound_P_S_unambiguous}
The maximum success probability for unambiguous state discrimination is bounded by
\begin{equation} \label{eq-lower_bound_P_S_unamb}
 P_{\rm S,u}^{\,\opt} ( \{ \rho_i, \eta_i \}) \leq 1 - \biggl( \frac{2m}{m-1}  \sum_{i>j} \eta_i \eta_j F (\rho_i, \rho_j ) \biggr)^\onehalf\;.
\end{equation}
\end{proposition}

\vspace{1mm}

\proof
The failure probability $P_0 = 1 - P_{\rm S,u}$ satisfies
\begin{equation}
P_0^2 =  \biggl( \sum_{i=1}^m \eta_i \tr ( M_0 \rho_i ) \biggr)^2 
\geq 
\frac{m}{m-1}  \sum_{i\not= j} \eta_i \eta_j \tr ( M_0 \rho_i )\tr ( M_0 \rho_j ) 
\geq \frac{m}{m-1}   \sum_{i\not= j} \eta_i \eta_j  \bigl| \tr ( U_{ij} \sqrt{\rho_i} M_0 \sqrt{\rho_j} )\bigr|^2 \;,
\end{equation}
where $U_{ij}$  are arbitrary unitary operators and the first and second bounds follow from the Cauchy-Schwarz 
inequality. 
Expressing $M_0$ as $1 - \sum_i M_i$ and using $\range M_i \subset \ker \rho_j$ for $i \not=j$,
one gets $\tr ( U_{ij} \sqrt{\rho_i} M_0 \sqrt{\rho_j} )= \tr ( U_{ij} \sqrt{\rho_i} \sqrt{\rho_j} )$. 
Maximizing over all unitaries $U_{ij}$ and using the formula $F( \rho_i,\rho_j) = \max_{U} | \tr ( U \sqrt{\rho_i} \sqrt{\rho_j} )|^2$, one
obtains (\ref{eq-lower_bound_P_S_unamb}).
\finpro

\vspace{3mm}

One infers from the last two propositions and the Cauchy-Schwarz inequality that

\vspace{2mm}

\begin{corollary}
The minimal failure probabilities $P_{\rm err,a}^{\,\opt} = 1 - P_{\rm S,a}^{\,\opt}$ and $P_0^{\,\opt}$ for  discriminating 
$m$ states ambiguously and unambiguously satisfy $P_0^{\,\opt} \geq 2 P_{\rm err,a}^{\,\opt} /(m-1)$.   
\end{corollary}

\vspace{1mm}

In particular, as noted in~\cite{Bergou_review}, for two states $P_0^{\,\opt}$ is at least twice
larger than $P_{\rm err,a}^{\,\opt}$.
 
\subsection{The Holevo bound} \label{sec_Holevo_bound}

Let us come back to the issue of encoding an input message $A$ in an ensemble  $\{ \rho_i , \eta_i\}$ of quantum states 
and transmitting it to a receiver.
From an information point of view, it makes sense to optimize the measurement  in
such a way as to maximize the  mutual information between the input message $A$ and the
output message $B$ reconstructed by the receiver (that is, $B$ is the set of measurement outcomes).
 This mutual information is defined as~\cite{Shannon48}
\begin{equation}
I_{A : B} = H (A) + H( B ) -  H(A,B) \;,
\end{equation}
where $H(A)= -
\sum_i \eta_i \ln \eta_i$ is the Shannon entropy of the input message,
$H(B) = - \sum_j p_{j} \ln p_{j}$ is the Shannon entropy of the \meas outcomes $B$ with probabilities
$p_{j} = \sum_i \eta_i \tr ( M_j \rho_i)$, and 
$H(A,B)= - \sum_{i,j} p_{ij} \ln p_{ij}$ is the Shannon entropy
of the joint process $(A,B)$ with probabilities 
$p_{ij}=   \eta_i p_{j|i} = \eta_i \tr ( M_j \rho_i )$, see~(\ref{eq-cond_proba_for_outcome_j}).
One can show from the concavity of the logarithm that $I_{A:B} \geq 0$ and $I_{A:B}=0$ if and only if $A$ and $B$ are independent.

The conditional Shannon entropies are defined by
\begin{equation} \label{eq-class_cond_entropy}
H(B|A)= - \sum_i \eta_i \sum_j p_{j|i} \ln p_{j|i}  \quad , \quad 
H(A|B) = - \sum_j p_j \sum_i \eta_{i|j} \ln \eta_{i|j}
\;,
\end{equation}
where $p_{j|i} = \tr ( M_j \rho_i )$ is the conditional probability of the \meas outcome $j$ given the state $\rho_i$ and
$\eta_{i|j}$ the conditional ({\it a posteriori}) probability that the state is $\rho_i$  given the outcome $j$. The latter is given by the Bayes rule
$\eta_{i|j}  = \eta_i p_{j|i}/p_j$.
The conditional entropy $H( A | B)$ represents the lack of knowledge of the receiver  on the state of the ensemble 
that was sent to him, after he has performed the measurement. In general the \meas producing the lowest value of 
$H( A | B)$ is not a von Neumann measurement~\cite{Davies78}. Thanks to the well-known relation $H(A,B)= H(A) + H(B|A)=H(B)+ H(A|B)$, the mutual information
can be expressed in terms of these conditional entropies as~\cite{Shannon48},
\begin{equation} \label{eq-def_mutual_info}
I_{A:B} = H(A) - H(A|B) = H(B) - H(B|A)\;.
\end{equation}
As $H(A|B) \geq 0$ one has $I_{A:B} \leq H (A)$, with equality \ifif $B$ is a function of $A$.
This means that  if $I_{A:B}$ is maximal, \ie $I_{A:B} = H(A)$, the receiver can reconstruct without any error
the message $A$ from his measurement outcomes. As stressed at the beginning of this section, 
this is never the case if $A$ is encoded
using non-orthogonal states $\rho_i$. Hence $I_{A:B} < H (A)$ for non-orthogonal states. The maximum 
\begin{equation} \label{eq-max_mutual_info}
\max_{{\rm POVM}\;\{ M_i\}} \bigl\{ I_{A:B} \bigr\} 
\end{equation}
measures the maximal amount of information accessible to the receiver, that is, how well can he reconstruct the message.
The determination of  the optimal measurement maximizing  $I_{A:B}$  appears to be a 
more difficult task than the minimization of the probability of error in state discrimination. However, one can
place an upper bound on the maximal information (\ref{eq-max_mutual_info}) by means of the Holevo inequality
\begin{equation} \label{eq-Holevo_bound}
I_{A:B} \; \leq \; \chi_{\rm Holevo} = S (\rho) - \sum_i \eta_i S ( \rho_i)
\quad , \quad \rho = \sum_i \eta_i \rho_i \;,
\end{equation}
where $S ( \rho) = - \tr ( \rho \ln \rho)$ is the von Neumann entropy of $\rho$. The proof of this important result relies on the 
monotonicity of the quantum mutual information under certain \QOs (see Remark~\ref{exo-Holevo_bound} below).
The positive number $ \chi_{\rm Holevo}$ is called the Holevo quantity.
We will show below  that $\chi_{\rm Holevo} \leq H ( \{ \eta_i\})$ with equality \ifif the $\rho_i$ have
orthogonal supports (see (\ref{eq_entropy_as_min_over_pure_state_dec2_gen})). We thus recover the aforementioned fact  that for non-orthogonal states $\rho_i$
the maximum (\ref{eq-max_mutual_info}) is smaller than 
the entropy $H (A)$ of the input message.

\newpage
\section{Quantum entropies} \label{sec-entropies}

In this Section we
give the definitions and main properties of the von Neumann entropy, the corresponding relative entropy, and the quantum R\'enyi relative entropies.
For classical systems these entropies reduce to the Shannon entropy, the Kullback-Leibler divergence, and the R\'enyi divergences, respectively, 
which are central objects in classical information theory.
To begin with we recall   in Sec.~\ref{sec-von_Neumann_wentropy} the standard  properties of the von Neumann entropy.
The most important result for our purpose is the
monotonicity of the corresponding relative 
entropy  with respect to quantum operations and the characterization of pairs of states which have the same relative entropy than their transformed states
under a given operation. The proof of this result, which will be used later in  Sec.~\ref{sec-discord}, is given  in Sec.~\ref{sec-relative_entropy}. 
We finally present in Sec.~\ref{sec-cond_Renyi_entropy} the quantum version of the
R\'enyi divergences introduced recently
in~\cite{Mueller-Lennert13,Wilde13,Frank13}. This quantum version contains as special cases the 
von Neumann relative entropy and the logarithm of the fidelity (\ref{eq-def_Ulhmann_fidelity}). The fidelity and
the closely related Bures distance will be the subject of
Sec.~\ref{sec-Bures_distance}. Together with the von Neumann relative entropy, it  plays a major role in our
geometrical approach of quantum correlations 
(Sec.~\ref{sec-geometric_measures}). The generalization  of this approach 
to  the whole family formed by the relative R\'enyi entropies constitutes an interesting open problem that will not be deeply explored in this article.
The reader may thus skip Sec.~\ref{sec-cond_Renyi_entropy} in a first reading.

\subsection{The von Neumann entropy} \label{sec-von_Neumann_wentropy}

The entropy $H(\{ p_k\} )= - \sum_k p_k \ln p_k$ introduced by Shannon in his two celebrated
 1948 papers~\cite{Shannon48} quantifies the amount of information at our disposal on the state of a classical system. It 
vanishes when the state is 
perfectly known and takes its maximum value (equal to
$\ln n$ if the system has  $n$ distinct possible states) when one has no
information on this state at all, that is, if all possible
states are equiprobable.  The quantum analog of the Shannon entropy is
the von Neumann entropy
\begin{equation} \label{eq-def_S}
S (\rho) = - \tr ( \rho \ln \rho )\;.
\end{equation}
This is a unitary invariant quantity, \ie $S( U \rho U^\ast ) = S ( \rho)$ for $U$ unitary. Moreover, $S$ is
additive for composite systems, \ie $S ( \rho_\AAA \otimes \rho_\BB ) = S ( \rho_\AAA ) + S ( \rho_\BB )$ 
for any states $\rho_\AAA$ and $\rho_\BB$ of the systems $\AAA$ and
$\BB$.  Another important property of $S$ is its strictly
concavity\footnote{
This comes from the strict convexity of $f(x)=x \ln x$. Actually, it is not hard to prove that if $f$ is strictly convex then
the map $\rho \in \states (\Hh) \mapsto \tr [ f(\rho)]$ is strictly
convex~\cite{Carlen}.
},
\ie for any states $\rho_0, \rho_1$ and $0 \leq
\eta \leq 1$ it holds $S( (1-\eta) \rho_0 + \eta \rho_1) \geq (1-\eta)
S ( \rho_0) + \eta S ( \rho_1)$, with equality \ifif $\rho_0=\rho_1$
or $\eta \in \{ 0,1\}$.

A much less trivial property of importance in quantum information theory is the so-called
strong subadditivity
\begin{equation} \label{eq-strong_subadditivity_S}
S ( \rho_{\sf{AB}} ) + S ( \rho_{\sf{BC}} ) - S (\rho_{\sf{ABC}} ) - S
(\rho_{\sf{B}} ) \geq 0\;,
\end{equation}
where $\rho_{\sf{ABC}}$ is a state of $\ABC$ with marginals 
$\rho_{\sf{AB}} = \tr_{\sf{C}} (\rho_{\sf{ABC}})$, $\rho_{\sf{BC}} = \tr_A ( \rho_{\sf{ABC}})$, and $\rho_{\sf{B}} = \tr_{\sf{AC}}
(\rho_{\sf{ABC}})$.  The inequality (\ref{eq-strong_subadditivity_S})
was first proven by Lieb and Ruskai~\cite{Lieb73} by using a former
work of Lieb~\cite{Lieb73a} on the concavity of the map
$\rho \mapsto \tr (K^\ast \rho^{1+\beta}  K \rho^{-\beta})$ for $-1 \leq \beta \leq 0$ (see Lemma~\ref{lemma_Lieb_concavity_Ando_convexity} below).  
Alternatively, (\ref{eq-strong_subadditivity_S}) is a direct consequence of the monotonicity of the relative
entropy   (Theorem \ref{theo-Petz} below), which can be established by other means than Lieb's concavity theorem.  
Choosing $\Hh_\BB = \complex$,  the strong subadditivity (\ref{eq-strong_subadditivity_S}) implies that $S$ is
subadditive, \ie $S (\rho_{\sf{AC}}) \leq S ( \rho_\AAA ) + S(\rho_\CC )$.

As is well know in statistical physics, the von Neumann entropy $S(\rho)$ is the Legendre transform of the
free energy $\Phi (\beta, H ) = -\beta^{-1} \ln \tr (\E^{-\beta H})$.  
More precisely, one has (see~\cite{Carlen}, Theorem 2.13)
\begin{equation} \label{eq-S_Legendre_transform_F}
S (\rho) = \inf_{H \in \observables_{\rm s.a.}} \bigl\{ \beta \tr ( H \rho ) - \beta \Phi
(\beta,H) \bigr\} \quad , \quad \Phi (\beta, H ) = \inf_{\rho \in
  \states (\Hh)} \bigl\{ \tr ( H \rho ) - \beta^{-1} S (\rho) \bigr\} \;,
\end{equation}
and the last infimum is attained \ifif $\rho$ is the Gibbs state
$\rho_\beta = \E^{-\beta H}/\tr ( \E^{-\beta H} )$. The free energy is a concave function of the energy observable $H$.

 The following identity will be  used repeatedly in Secs.~\ref{sec-entanglement} and~\ref{sec-discord}:
\begin{equation} \label{eq-equality_entropy_reduce_states}
S( \rho_\AAA ) = S ( \rho_\BB)
\quad \text{ if $\rho_\AAA$ and $\rho_\BB$ are the reduced states of the pure state $\ket{\Psi_\AB}$ of $\AB$.}
\end{equation} 
It is a consequence of Theorem~\ref{theo_Schmidt_dec}, since if $\ket{\Psi_\AB}$ has Schmidt coefficients  $\mu_i$ then
$S(\rho_\AAA ) = S ( \rho_\BB) = -\sum_i \mu_i \ln \mu_i$.

A last identity worthwhile mentioning here is
\begin{equation} \label{eq_entropy_as_min_over_pure_state_dec}
S ( \rho) = \min_{ \{ \ket{\psi_i}, \eta_i\}} H ( \{ \eta_i\}) = \min_{ \{ \ket{\psi_i}, \eta_i\}}  \Bigl\{ - \sum_{i=1}^m \eta_i \ln \eta_i \Bigr\}\;,
\end{equation}
where the minimum is over all pure state decompositions of $\rho$. Furthermore, a decomposition minimizes
$H ( \{ \eta_i\})$ 
\ifif  it is a spectral decomposition of $\rho$. These statements can be justified as follows\footnote{
An alternative proof can be found in~\cite{Nielsen}.
}.
Let $\{ \ket{k}, p_k\}_{i=1}^r$ be a spectral decomposition of $\rho$, with $ r= \range (\rho)$.
An arbitrary pure state decomposition  $\{ \ket{\psi_i}, \eta_i\}_{i=1}^m$ of $\rho$ has the form
$\sqrt{\eta_i} \ket{\psi_i} = \sum_k u_{ik} \sqrt{p_k} \ket{k}$, where 
$(u_{ik})$ is a $m \times m$ unitary matrix and $m \geq r$ 
(see (\ref{eq-link_between_pure_state_dec})). Setting $p_k=0$ for $r < k \leq m$ one gets
$\eta_i = \sum_k | u_{ik} |^2 p_k$.
Since $f(x) = x \ln x$ is  strictly convex, one finds
\begin{equation}  \label{eq_entropy_as_min_over_pure_state_dec2}
- H ( \{ \eta_i\} ) = \sum_{i=1}^m \eta_i \ln \eta_i \leq \sum_{i,k=1}^m | u_{ik} |^2 p_k \ln p_k = \sum_{k=1}^r p_k \ln p_k = - S ( \rho)\;,
\end{equation}
so that $S ( \rho) \leq H ( \{ \eta_i\} )$. 
By strict convexity, the inequality in (\ref{eq_entropy_as_min_over_pure_state_dec2}) is an equality \ifif for any $i$, there exists 
 some $k_i \in \{ 1, \ldots , r+1\}$ such that 
$u_{ik} = 0$ when $k \notin I_{i} = \{ k =1,\ldots, m ; p_k = p_{k_i} \}$. Thus $S ( \rho) = H ( \{ \eta_i\} )$ \ifif
\begin{equation} \label{eq-proof_entropy_as_min_over_pure_state_dec}
\sqrt{\eta_i} \ket{\psi_i} = \sqrt{p_{k_i}} \sum_{k \in I_i} u_{ik} \ket{k} 
\end{equation}
are eigenvectors of $\rho$ with eigenvalue $\eta_i =p_{k_i}$ (if $p_{k_i} \not= 0$). It remains to check that $\braket{\psi_i}{\psi_j}=0$ when $p_{k_i} = p_{k_j}\not= 0$.
This comes from the unitarity of  $(u_{ik})$. This yields the desired result.
The inequality (\ref{eq_entropy_as_min_over_pure_state_dec2}) can be easily generalized to get\footnote{
This follows from  (\ref{eq_entropy_as_min_over_pure_state_dec}) by writing the spectral decompositions of the $\rho_i$ (see~\cite{Nielsen}, Sec.~11.3). 
}
%
\begin{equation} \label{eq_entropy_as_min_over_pure_state_dec2_gen}
S ( \rho ) \leq H ( \{ \eta_i\} ) + \sum_i \eta_i S ( \rho_i ) 
\end{equation}
for any ensemble $\{ \rho_i , \eta_i \}$ forming a convex decomposition of $\rho$.
Moreover, one has equality \ifif the $\rho_i$ have orthogonal
supports.

\subsection{Relative entropy} \label{sec-relative_entropy}

A related quantity to the von Neumann entropy is the relative entropy
introduced by Umegaki \cite{Umegaki24} and later extended by
Araki~\cite{Araki76} in the von Neumann algebra setting,
\begin{equation} \label{eq-def_relative_S}
S ( \rho || \sigma ) =
\begin{cases}
 \tr \bigl( \rho ( \ln \rho - \ln \sigma ) \bigr) & \text{ if $\ker
   (\sigma) \subset \ker ( \rho )$} \\ +\infty & \text{ otherwise.}
\end{cases} 
\end{equation}
Note that by taking $\sigma=1/n$ proportional to the identity operator, $S(\rho || 1/n) = \ln n - S(\rho)$ is the difference 
between the maximal and the von Neumann entropy of $\rho$.
The relative entropy has the following properties: 
\begin{itemize}
\item[(i)] $S ( \rho || \sigma ) \geq 0$ with equality \ifif $\rho =
  \sigma$;
\item[(ii)] unitary invariance $S ( U \rho U^\ast || U \sigma U^\ast )
  = S (\rho || \sigma)$ for any unitary $U$;
\item[(iii)] additivity for composite systems: $S( \rho_\AAA \otimes \rho_\BB || \sigma_\AAA \otimes \sigma_\BB ) = S( \rho_\AAA || \sigma_\AAA ) 
+ S ( \rho_\BB || \sigma_\BB )$;
\item[(iv)] joint convexity: if $0 \leq \eta \leq 1$ then $S ( (1-\eta) \rho_0 + \eta \rho_1 || (1-\eta) \sigma_0 + \eta \sigma_1 )
  \leq (1-\eta ) S ( \rho_0 || \sigma_0 ) + \eta S ( \rho_1 ||
  \sigma_1 ) $.
\end{itemize}
The first property (i) follows from Klein's inequality, which states
that if $f$ is continuous and strictly convex, then $\tr [ f(A) - f(B)
  - (A-B) f'(B)] \geq 0$, with equality \ifif $A=B$. Its proof can be
found for instance in the excellent lecture notes of E.A. Carlen~\cite{Carlen}.
The properties (ii) and (iii) are immediate consequences of the cyclicity
of the trace and the relation $\ln ( \rho_\AAA \otimes \rho_\BB ) = \ln \rho_\AAA \otimes 1 + 1 \otimes \ln \rho_\BB$, as in the case of the
von Neumann entropy.  The last property (iv) can be deduced from the strong subadditivity (\ref{eq-strong_subadditivity_S})~\cite{Lindblad74,Lindblad75}.
It will be proven in Sec.~\ref{sec-cond_Renyi_entropy}.  Let us point out that
(i) implies the aforementioned subadditivity $S (\rho_{\sf{AC}}) \leq S ( \rho_\AAA ) + S(\rho_\CC )$ of the von Neumann entropy, with
equality \ifif $\rho_{\sf{AC}}= \rho_\AAA \otimes \rho_\CC$ is a
product state (in fact, $S( \rho_{\sf{AC}} || \rho_\AAA \otimes \rho_\CC ) = S ( \rho_\AAA) + S ( \rho_\CC) - S( \rho_{\sf{AC}} )$).

Another fundamental property of $S ( \rho || \sigma ) $ is its monotonicity with respect to CP trace-preserving mappings. This monotonicity
means that if one performs the same measurement on two
states without readout of the outcomes, the pair of post-measurement states has a lower relative entropy  
than the pair of states before the measurement.  This fact was first proven by
Lindblad~\cite{Lindblad75} (see also \cite{Araki76} and~\cite{Uhlmann77}). Notice that unlike the relative entropy, 
the von Neumann entropy is not monotonous with respect to non-projective measurements (see~\cite{Nielsen}, Exercise 11.15).
 The following theorem provides a necessary and sufficient
condition on the two states such that the monotonicity of the relative entropy is satisfied
with equality. It is due to Petz~\cite{Petz03}.

\vspace{2mm}

\begin{theorem} \label{theo-Petz} {\rm (Monotonicity of the relative entropy~\cite{Petz03,Hayden04})} 
For any quantum operation $\Mm : \Bb( \Hh) \rightarrow \Bb ( \Hh')$ one
has $S ( \rho || \sigma ) \geq S ( \Mm (\rho) || \Mm (\sigma))$ for
all states $\rho$, $\sigma \in \states (\Hh)$. The inequality is an
equality \ifif there exists a quantum operation $\Rr : \Bb(\Hh') \rightarrow \Bb ( \Hh)$ such that $\Rr \circ \Mm ( \sigma) = \sigma$
and $\Rr \circ \Mm ( \rho) = \rho$. This \QO is the transpose
operation $\Rr = \Rr_{\Mm,\sigma}$ defined in (\ref{eq-def_widehat_Mm}).
\end{theorem}

\vspace{1mm}

Let us recall from Sec.~\ref{sec-transpose_operation} that the transpose operation $\Rr_{\Mm,\sigma}$ is the \QO with Kraus operators
\begin{equation} \label{eq-transpose_operation}
R_i = \sqrt{\sigma} A_i^\ast  \Mm ( \sigma )^{-1/2}\;,
\end{equation}
where $\{ A_i\}$ are some Kraus operators for $\Mm$. The conditions $\Rr \circ \Mm ( \sigma) = \sigma$ and $\Rr \circ \Mm ( \rho) = \rho$, which mean
that $\rho$ and $\sigma$ can be recovered respectively from $\Mm(\rho)$ and $\Mm(\sigma)$ by
means of the same \QO  $\Rr$, is clearly sufficient to ensure the equality
$S ( \rho || \sigma ) = S ( \Mm (\rho) || \Mm (\sigma))$ if monotonicity holds true. It is remarkable  that this is also a necessary condition, with  
$\Rr=\Rr_{\Mm,\sigma}$ the approximate reversal of $\Mm$ introduced in the context of quantum  error correction (Sec.~\ref{sec-tranpose_op_and_lsm}).

We present below the derivation  of this result given by Petz in Ref.~\cite{Petz03}, which also provides a nice and simple proof
of the monotonicity. A completely different proof of the monotonicity, based on Lieb's concavity theorem
as in Ref.~\cite{Lindblad75,Carlen,Frank13}, will be given in
Sec.~\ref{sec-cond_Renyi_entropy} in the more general setting of the R\'enyi entropies.  It is
noteworthy that Petz's derivation does neither rely on
the Stinespring theorem nor on the Kraus decomposition (albeit it takes advantage of one of its consequence, 
namely, the Kadison-Schwarz inequality).  It makes use of the theory of operator convex functions and of Araki's
relative modular operators~\cite{Araki82}. Let $\Mm$ be a
\QO $\Bb( \Hh) \rightarrow \Bb ( \Hh')$ and $\rho$ and $\sigma$ be two
states of $\states (\Hh)$ such that $\rho$ and $\Mm (\rho)$ are invertible. One can
define two relative modular operators by (see Sec.~\ref{sec-states})
\begin{equation} \label{eq-def_modular_op2}
\Delta_{\sigma|\rho} (B ) = \sigma B \rho^{-1} \quad , \quad
\Delta_{\Mm(\sigma)| \Mm(\rho)} ( B') = \Mm (\sigma ) B' \Mm (\rho)^{-1}
\quad , \quad B \in \observables \; , \; B' \in \Bb ( \Hh') \;.
\end{equation}

\proof Let us set $\rho_\Mm = \Mm (\rho)$ and $\sigma_\Mm = \Mm ( \sigma)$ and assume that $\rho$, $\sigma$, $\rho_\Mm$, and
$\sigma_\Mm$ are invertible.  In the whole proof these states are fixed, so to simplify notation we write $\Delta$ instead of $
\Delta_{\sigma|\rho}$ and $\Delta_\Mm$ instead of $ \Delta_{\sigma_\Mm |\rho_\Mm}$. We set $\xi= \rho^\onehalf$ and $\xi_\Mm= \rho_\Mm^\onehalf$. 
One can view these two operators as
unit vectors  in $\Bb(\Hh)$ and $\Bb(\Hh')$, respectively, for the 
Hilbert-Schmidt scalar product $\langle \cdot , \cdot \rangle$. 
The first observation is that
\begin{equation} \label{eq-proof_Petz_theo0}
S ( \rho || \sigma )
= \langle \xi \, , \, ( \ln \rho - \ln \sigma ) \xi \rangle = - \langle \xi \, , \, \ln (\Delta ) \xi  \rangle
= \int_0^\infty \D t \, \Bigl( \bigl\langle \xi \, , \,
(\Delta + t)^{-1} \xi \bigr\rangle - (1+ t)^{-1} \Bigr)\;.
\end{equation}
The third equality can be established, for instance, with the help of the first 
identity in (\ref{eq-log_as_limit}) (see Appendix~\ref{app_operator_convex_functions}).
Therefore,  in order to prove
that $S ( \rho || \sigma ) \geq S ( \rho_\Mm || \sigma_\Mm )$, it suffices to show that for any $t >0$,
\begin{equation} \label{eq-proof_monotonocity_Petz0}
 \bigl\langle \xi_\Mm \, , \, (\Delta_\Mm + t)^{-1}
 \xi_\Mm \bigr\rangle \leq \bigl\langle \xi \, ,
 \, (\Delta + t)^{-1} \xi \bigr\rangle \;.
\end{equation}
To this end, let us consider the operator $\Cc_\Mm$  defined by
\begin{equation} \label{eq-def_VV_m}
\Cc_\Mm ( B ' \xi_\Mm ) = \Mm^\ast ( B' ) \xi \quad ,
\quad B \in \Bb ( \Hh' )\;.
\end{equation}
Since $\{ B' \xi_\Mm ; B' \in \Bb ( \Hh')\}$ is equal to\footnote{
In the  theory of  C$^\ast$-algebras, if this equality is true upon completion of $\{ B' \xi_\Mm ; B' \in \Bb ' \}$ for the Hilbert-Schmidt norm one says  
that $(  B' \in \Bb ' \mapsto \Ll_{B'} , \xi_\Mm)$ defines a cyclic representation of the algebra $\Bb '$ on the 
Hilbert space $\Bb ( \Hh')$~\cite{Bratteli}.
}
 $\Bb ( \Hh')$ by the invertibility of $\rho_\Mm$, (\ref{eq-proof_monotonocity_Petz1}) defines
an operator $\Cc_\Mm$ from $\Bb ( \Hh')$ to $\Bb ( \Hh )$. 
Then
\begin{equation} \label{eq-proof_monotonocity_Petz1}
\Cc_\Mm^\ast \Delta \Cc_\Mm \leq \Delta_\Mm \;.
\end{equation}
Actually, thanks to the Kadison-Schwarz inequality
(\ref{eq-Kadison-Schwarz}) and  the relation $(\Mm^\ast ( {B'}^\ast ))^\ast =
\Mm^\ast ( B' )$, one has
\begin{eqnarray}
\nn
\bigl\langle \Cc_\Mm ( B' \xi_\Mm ) \, , \, \Delta \, \Cc_\Mm
( B' \xi_\Mm ) \bigr\rangle
& = & \nn
\tr \Bigl( | \Mm^\ast ( {B'}^\ast)|^2  \sigma \Bigr)
\\
& \leq &
\tr \Bigl( \Mm^\ast ( B' {B'}^\ast ) \sigma \Bigr)
= \bigl\langle B' \xi_\Mm \, , \, \Delta_\Mm B'
\xi_\Mm \bigr\rangle \;.
\end{eqnarray}
%
One shows similarly that $\| \Cc_\Mm ( B' \xi_\Mm ) \|_2 \leq \| B' \xi_\Mm \|_2$ for any $B '\in \Bb (\Hh')$, hence 
$\| \Cc_\Mm \| \leq 1$.

We now use the fact that the function $f(x) = (x+t)^{-1}$ is operator
monotone-decreasing and operator convex. The definitions of operator monotone and operator convex functions are given in
 Appendix~\ref{app_operator_convex_functions}.
Together with the bound
(\ref{eq-proof_monotonocity_Petz1}), this implies\footnote{
In~\cite{Petz03} the last term in the \RHS is omitted. This is a not correct as the Jensen-type inequality
(\ref{eq-Jensen_for_operator_convex}) cannot be applied for the
function $f(x)=(x+t)^{-1}$, because it does not satisfy the condition $f(0)\leq 0$.  Fortunately, this term disappears in
(\ref{eq-proof_monotonocity_Petz0}) due to the trace-preserving property of $\Mm$ and the proof goes through.
}
%
\begin{equation} \label{eq-proof_monotonocity_Petz2}
(\Delta_\Mm + t )^{-1} \leq ( \Cc_\Mm^\ast \Delta \Cc_\Mm + t )^{-1}
  \leq \Cc_\Mm^\ast (\Delta + t )^{-1} \Cc_\Mm + t^{-1} ( 1 -
  \Cc_\Mm^\ast \Cc_\Mm ) \;.
\end{equation}
The last inequality follows by applying the Jensen-type inequality
(\ref{eq-Jensen_for_operator_convex}) for the operator convex function
$g(x) = ( x + t)^{-1} - t^{-1}$ satisfying $g(0) = 0$ and the contraction $\Cc_\Mm$.  Since
$\Cc_\Mm ( \xi_\Mm ) = \xi$ by (\ref{eq-def_VV_m})
and $\Mm^\ast ( 1) = 1$, the inequality (\ref{eq-proof_monotonocity_Petz2}) entails
\begin{equation}
\bigl\langle \xi_\Mm \, , \, (\Delta_\Mm + t )^{-1}
\xi_\Mm \bigr\rangle \leq \bigl\langle \xi \, ,\,
(\Delta + t )^{-1} \xi \bigr\rangle + t^{-1} \bigl( \tr (
\rho_\Mm ) - \tr ( \rho ) \bigr) \;.
\end{equation}
The term proportional to $t^{-1}$
vanishes because $\Mm$ is trace preserving, hence one obtains the desired bound
(\ref{eq-proof_monotonocity_Petz0}). We have thus proven the monotonicity of
the relative entropy.

In addition to its simplicity, the above proof offers the
advantage that it easily yields a necessary and sufficient
condition for having $S ( \rho || \sigma ) = S ( \rho_\Mm || \sigma_\Mm)$.  Actually, this equality holds \ifif
(\ref{eq-proof_monotonocity_Petz0}) is an equality, \ie
\begin{equation} \label{eq-proof_monotonocity_Petz3}
\bigl\langle \xi_\Mm \, , \, (\Delta_\Mm + t )^{-1}
\xi_\Mm \bigr\rangle = \bigl\langle \xi_\Mm \, ,
\, \bigl( \Cc_\Mm^\ast (\Delta + t )^{-1} \Cc_\Mm + t^{-1} ( 1 -
\Cc_\Mm^\ast \Cc_\Mm ) \bigr) \xi_\Mm \bigr\rangle
\end{equation}
 for all $t >0$. But for any operators $X$, $Y$, and $Z$ with $Z$
 invertible and $X \leq Y$, $\langle Z , X Z \rangle = \langle Z , Y Z \rangle$ implies $X Z  = Y Z$.  Hence we can infer from
(\ref{eq-proof_monotonocity_Petz2}) and
(\ref{eq-proof_monotonocity_Petz3}) that
\begin{equation} \label{eq-proof_monotonocity_Petz3bis}
(\Delta_\Mm + t )^{-1} \xi_\Mm = \Cc_\Mm^\ast (\Delta + t   )^{-1} \xi\;\quad , \quad t>0\;,
\end{equation}
where we have used the identity $\Cc_\Mm^\ast \Cc_\Mm (\xi_\Mm ) =
\xi_\Mm$ (in fact, one easily finds that
$\langle \Cc_\Mm (B' \xi_\Mm) \,,\,\Cc_\Mm \xi_\Mm \rangle  = \langle B' \xi_\Mm\,,\,\xi_\Mm \rangle$ for any
$B '\in \Bb (\Hh')$). 
Therefore,
%
\begin{equation} \label{eq-proof_monotonocity_Petz4}
\bigl\| \Cc_\Mm^\ast (\Delta + t )^{-1} \xi \bigr\|^2_2
= \bigl\langle (\Delta_\Mm + t )^{-2} \xi_\Mm ,
\xi_\Mm \bigr\rangle
= \bigl\langle \Cc_\Mm^\ast (\Delta + t )^{-2} \xi ,
\xi_\Mm \bigr\rangle = \bigl\| (\Delta + t )^{-1}
\xi \bigr\|^2_2\;,
\end{equation}
%
where the second equality is obtained by differentiating 
(\ref{eq-proof_monotonocity_Petz3bis}) with respect to $t$.  Now, the
identity $\| \Cc^\ast (X) \|_2 = \| X \|_2$ for $\Cc$ a contraction
implies that $\Cc \Cc^\ast (X) = X$ (in fact, then the Cauchy-Schwarz
inequality $\langle X \, , \Cc \Cc^\ast (X) \rangle \leq \|X \|_2 \| \Cc \Cc^\ast (X) \|_2 \leq \| X \|^2_2$ is an equality, so that 
$\Cc \Cc^\ast (X)$ must be proportional to $X$).  We conclude that
\begin{equation}
\Cc_\Mm ( \Delta_\Mm + t )^{-1} \xi_\Mm = \Cc_\Mm
\Cc_\Mm^\ast ( \Delta + t )^{-1} \xi = (\Delta + t )^{-1}
\xi
\end{equation}
for any $t>0$. By means of the functional calculus, one deduces from this identity that
\begin{equation}
\Cc_\Mm \Delta_\Mm^{-\onehalf} \xi_\Mm = \Delta^{-\onehalf}
\xi\;.
\end{equation}
In view of the definitions (\ref{eq-def_modular_op2}) and
(\ref{eq-def_VV_m}) and the invertibility of $\rho$, the last formula
gives $\Mm^\ast ( \sigma_\Mm^{-\onehalf} \xi_\Mm ) = \sigma^{-\onehalf} \xi$. By multiplying by the adjoint
and using the Kadison-Schwarz inequality, we arrive at
\begin{equation}
\sigma^{-\onehalf} \rho\, \sigma^{-\onehalf} \leq \Mm^\ast ( \sigma^{-\onehalf}_\Mm \rho_\Mm \sigma^{-\onehalf}_\Mm)\;,
\end{equation}
that is, $\rho \leq {\Rr}_{\Mm,\sigma} ( \rho_\Mm )$ with
${\Rr}_{\Mm,\sigma}$ defined in (\ref{eq-def_widehat_Mm}).  But 
$\tr [\rho] =  \tr [ \rho_\Mm ]= \tr [ {\Rr}_{\Mm,\sigma} ( \rho_\Mm )]$, whence $\rho = {\Rr}_{\Mm,\sigma} ( \rho_\Mm )$.  The other
equality $\sigma = {\Rr}_{\Mm,\sigma} ( \sigma_\Mm )$ is obvious.
Reciprocally, as stressed above, these two identities imply $S ( \rho || \sigma ) = S( \rho_\Mm || \sigma_ \Mm)$ thanks to 
the monotonicity of the relative entropy and the fact that ${\Rr}_{\Mm,\sigma} $ is a quantum operation.  \finpro

\vspace{2mm}

Let us end this subsection by pointing out that the strong subadditivity of the von Neumann entropy, the joint convexity  of the relative entropy, and its
monotonicity can be deduced from each other.  For instance, the strong subadditivity
(\ref{eq-strong_subadditivity_S}) is a simple consequence of the monotonicity. Actually, one  checks that
\begin{equation} \label{eq-link_strong_addtivity_monotonicity}
S ( \rho_{\sf{AB}} ) + S ( \rho_{\sf{BC}} ) - S (\rho_{\sf{ABC}} ) - S (\rho_\BB) = S (\rho_{\sf{ABC}} || \rho_\AAA \otimes \rho_{\sf{BC}} )
- S ( \Mm_C ( \rho_{\sf{ABC}}) || \Mm_C ( \rho_\AAA \otimes \rho_{\sf{BC}}) )
\end{equation}
with $\Mm_\CC : \rho \mapsto \tr_\CC (\rho)$. It is easy to show that
$\Mm_\CC$ is a CP and trace-preserving map $\Bb (\Hh_{\ABC}) \rightarrow \Bb ( \Hh_\AB)$, therefore (\ref{eq-strong_subadditivity_S}) follows from
Theorem~\ref{theo-Petz}. With the help of this theorem it is  also possible to characterize
all states $\rho_{\sf{ABC}}$ such that
(\ref{eq-strong_subadditivity_S}) becomes an equality \cite{Hayden04}.

Conversely, Lindblad~\cite{Lindblad74,Lindblad75} proves the monotonicity
inequality from the strong subadditivity.
The basic idea is to show that the strong subadditivity of the von
Neumann entropy or the closely related Lieb concavity theorem imply the
joint convexity (iv) of the relative entropy. The corresponding arguments are given in Sec.~\ref{eq-main_prop_rel_alpha_ent}  below. 
 One can then deduce the
monotonicity of the relative entropy from its joint convexity (iv)
with the help of Stinespring's theorem as
follows~\cite{Uhlmann73a,Wolf,Frank13}. Recall that if $\mu_H$ is the
normalized Haar measure on the group $U(n)$ of $n \times n$
unitary matrices, then $\int \D \mu_H ( U) \,U B U^\ast = n^{-1} \tr ( B)$ for any $B \in \observables$ (in fact, all diagonal matrix
elements of the \LHS in an arbitrary basis are equal, as follows from the
left-invariance $\D \mu_H ( V U) = \D \mu_H ( U)$ for $V\in U (n)$; as a result, this \LHS is proportional to the identity
matrix).  We infer from Stinespring theorem~\ref{theo-Stinespring}
that
\begin{equation}
\Mm ( \rho) \otimes ( 1 /n_\EE ) = \int_{U(n_\EE)} \D \mu_H ( U_\EE ) \,(1 \otimes U_\EE) U \rho \otimes
\ketbra{\epsilon_0}{\epsilon_0} U^\ast (1 \otimes U_\EE^\ast)
\end{equation}
with $U$ unitary on $\Hh_{\SE}$. Thanks to the additivity (iii), the
joint convexity convexity (iv), and the unitary invariance (ii), we get
\begin{eqnarray}
\nn & & S ( \Mm(\rho) || \Mm( \sigma) ) = S \bigl( \Mm ( \rho) \otimes (1 /n_\EE ) \bigl| \bigr| \Mm (\sigma) \otimes (1 /n_\EE ) \bigr) 
\\ \nn 
&  & \hspace*{1cm} 
\leq \int_{ U(n_\EE)} \D \mu_H ( U_\EE ) S \bigl( (1 \otimes U_\EE) U \rho \otimes \ketbra{\epsilon_0}{\epsilon_0} U^\ast (1 \otimes U_\EE^\ast) \bigl|\bigr|
 (1 \otimes U_\EE) U \sigma \otimes \ketbra{\epsilon_0}{\epsilon_0} U^\ast (1 \otimes U_\EE^\ast) \bigr) 
\\ & & \hspace*{1cm} =
\int_{ U (n_\EE)} \D \mu_H ( U_\EE ) S ( \rho || \sigma ) = S ( \rho || \sigma ) \;.
\end{eqnarray}
By the same argument, one can show a slightly more general result. 

\vspace{2mm}

\begin{proposition} \label{prop-convexity_implies_monotonicity}
Let $f: \states ( \Hh ) \times \states ( \Hh) \rightarrow \real$ be a unitary-invariant jointly 
convex function for any finite Hilbert space $\Hh$, which satisfies 
$f(\rho \otimes \tau, \sigma\otimes \tau) = f(\rho, \sigma)$ for all $\rho, \sigma \in \states ( \Hh)$ and $\tau \in \states (\Hh')$. 
Then $f$ is monotonous with respect to quantum operations.
\end{proposition}

\subsection{Quantum relative R\'enyi entropies} \label{sec-cond_Renyi_entropy}
\subsubsection{Definitions}

In the classical theory of information, other entropies than the Shannon entropy play a role when ergodicity breaks down or outside
the asymptotic regime. The R\'enyi entropy depending on a parameter
$\alpha >0$ unifies these different entropies.  In the quantum setting, it is defined as
\begin{equation}
S_\alpha (\rho) = (1-\alpha)^{-1} \ln \tr ( \rho^\alpha ) \;.
\end{equation}
It is easy to show that $S_\alpha (\rho)$ converges to the von Neumann
entropy $S(\rho)$ when $\alpha \rightarrow 1$ and that $S_\alpha
(\rho)$ is a non-increasing function of $\alpha$.

A first definition of the  quantum relative R\'eyni entropy is
\begin{equation}  \label{eq-relative_Renyi_entropy_normal}
{S}_\alpha^{\rm (n)} ( \rho || \sigma)
= (\alpha-1)^{-1} \ln ( \tr [\rho^\alpha \sigma^{1-\alpha} ])  \quad , \quad \alpha > 0\;,\;\alpha \not= 1 \;.
\end{equation}
This entropy appears naturally in the context of the quantum  hypothesis testing (Sec.~\ref{eq-Q_hyp_testing} below). 
We shall discuss here a symmetrized version proposed recently by M\"uller-Lennert {\it et al.}~\cite{Mueller-Lennert13} and 
by Wilde, Winter, and Yang~\cite{Wilde13}. It is given by
\begin{equation} \label{eq-relative_Renyi_entropy}
S_\alpha ( \rho || \sigma ) = (\alpha-1)^{-1} \ln \tr \bigl[ \bigl(
  \sigma^{\frac{1-\alpha}{2 \alpha}} \rho \,\sigma^{\frac{1-\alpha}{2
      \alpha}} \bigr)^\alpha \bigr] 
\end{equation}
if $\alpha \in (0,1)$ and $\tr( \sigma \rho ) > 0$  or if  $\alpha>1$  and $\ker \sigma \subset \ker \rho$ (if none of these conditions are
satisfied, one sets $S_\alpha (  \rho || \sigma ) = +\infty$).  This
relative entropy has been used in Ref.~\cite{Wilde13} to solve an important open problem
related to the transmission of information in noisy quantum channels. It seems likely that
much more applications in quantum information theory  will be encountered in the future. 
The  entropies $S_\alpha$ appeared recently as central objects in a very different context,
namely, the quantum fluctuation relations in out-of-equilibrium  statistical physics~\cite{Jaksic13,Jaksic14}. A nice feature
of the family $\{ S_\alpha \}_{\alpha > 0}$ is that it contains
the von Neumann relative entropy, the fidelity entropy, and the max-entropy as
special cases. Furthermore, $S_\alpha$ depends continuously and monotonously on $\alpha$.  The fidelity-entropy is obtained for
$\alpha=1/2$. It is given by 
$S_{1/2} ( \rho || \sigma ) = -\ln F (\rho, \sigma)$, where  
$F(\rho,\sigma)$ is the fidelity (\ref{eq-def_Ulhmann_fidelity}).
The max-entropy is defined by
\begin{equation} \label{eq-max_entropy}
S_\infty ( \rho || \sigma ) = \lim_{\alpha \rightarrow \infty} S_\alpha ( \rho || \sigma ) = \ln \| \sigma^{-\onehalf} \rho
\sigma^{-\onehalf} \|\;,
\end{equation}
where $\| \cdot \|$ is the operator norm. The second equality follows from $\| A\|_\alpha \rightarrow \| A \|$ as 
$\alpha \rightarrow \infty$ (see Sec.~\ref{sec-states_and_observables}).
Finally, one recovers the von Neumann relative
entropy (\ref{eq-def_relative_S}) by letting $\alpha \rightarrow 1$,
\begin{equation} \label{eq-relative_Reyni_ent_alpha=1}
S ( \rho || \sigma ) = \lim_{\alpha \rightarrow 1} S_\alpha ( \rho || \sigma )\;.
\end{equation}
To justify this statement, let us set $A(\alpha) = \sigma^{\frac{1-\alpha}{2\alpha}} \rho \sigma^{\frac{1-\alpha}{2\alpha}}$. 
Explicit calculations show that
\begin{eqnarray}
\nn
\frac{\D  \tr [ A(\alpha)^\alpha ] }{\D \alpha} 
& = & 
\tr [ A(\alpha)^\alpha \ln A(\alpha)]  
  + \alpha \tr \Bigl[ A(\alpha)^{\alpha-1} \frac{\D A}{\D \alpha}  \Bigr]\
\\ 
\frac{\D A}{\D \alpha}
& = & 
-\frac{1}{2 \alpha^2} \Bigl(\ln (\sigma)  A (\alpha) + A(\alpha) \ln (\sigma) \Bigr)\;.
\end{eqnarray}
Consequently, 
$ S_\alpha ( \rho || \sigma ) \rightarrow  
(\D \ln \tr [ A(\alpha)^\alpha ]/\D \alpha )_{\alpha=1} = \tr ( \rho \ln \rho - \rho \ln \sigma )$ as 
$\alpha \rightarrow 1$.
Note that a similar result holds for the unsymmetrized R\'enyi entropy (\ref{eq-relative_Renyi_entropy_normal}), \ie
$ S ( \rho || \sigma ) = \lim_{\alpha \rightarrow 1} S_\alpha^{\rm (n)} ( \rho || \sigma )$.
Let us also emphasize that
\begin{equation}
S_\alpha ( \rho || \sigma ) \leq {S}_\alpha^{\rm (n)}  ( \rho || \sigma )
\end{equation}
by the Lieb-Thirring trace
inequality (\ref{eq-Lieb-Thirring}).

For commuting matrices $\rho = \sum p_k \ketbra{k}{k}$ and $\sigma = \sum_k q_k \ketbra{k}{k}$, both $S_\alpha ( \rho || \sigma )$ and
${S}_\alpha^{\rm (n)}  ( \rho || \sigma )$ reduce to the classical R\'eyni divergence
\begin{equation} \label{eq-Reyni_divergence}
S_\alpha^{\rm clas} ( \pv || \qv ) = (\alpha-1)^{-1} \ln \Bigl( \sum_{k=1}^n p_k^\alpha q_k^{1-\alpha} \Bigr) \;,
\end{equation}
which is non-negative for $\alpha>0$ by the H\"older inequality.

\subsubsection{Main properties} \label{eq-main_prop_rel_alpha_ent}

It is shown in this subsection that the R\'enyi relative entropy
$S_\alpha ( \rho || \sigma)$ satisfies the same properties (i-iv) as
the von Neumann relative entropy in Sec.~\ref{sec-relative_entropy}
for any $\alpha \in [ 1/2,1]$.  For $0 < \alpha < \infty$ we define the $\alpha$-fidelity by
\begin{equation} \label{eq-def_alpha_fidelity}
F_\alpha (\rho || \sigma ) = \| \rho^\onehalf \sigma^{\frac{\beta}{2}} \|_{2 \alpha}^2 = 
\| \sigma^{\frac{\beta}{2}} \rho \sigma^{\frac{\beta}{2}}\|_\alpha 
= \E^{-\beta S_\alpha ( \rho || \sigma)} 
\quad \text{ with } \quad \beta = \frac{1-\alpha}{\alpha}\;.
\end{equation}
Here, we have used the notation $\| A \|_{2 \alpha} = (\tr [ (A^\ast A)^\alpha] )^{\frac{1}{2\alpha}}$ even if this does not correspond to a norm when
$0< \alpha < 1/2$.

\vspace{2mm}

\begin{theorem} \label{theo-Renyi_relative_entropy} For any $\alpha>0$, one has 
\begin{itemize}
\item[(i)] $S_\alpha ( \rho || \sigma ) \geq 0$ with equality \ifif $\rho = \sigma$;
\item[(ii)] $S_\alpha ( \rho|| \sigma)$ is unitary invariant;
\item[(iii)]$S_\alpha ( \rho|| \sigma)$ is additive for composite systems;
\item[(iv)] $F_\alpha ( \rho || \sigma)^\alpha$ is jointly concave for $\alpha \in [ 1/2, 1)$ and jointly convex for $\alpha > 1$.
    In particular, $S_\alpha ( \rho|| \sigma)$ is jointly convex for   $\alpha \in [ 1/2, 1]$;
\item[(v)] if $\alpha \geq 1/2$ then $S_\alpha ( \rho|| \sigma) \geq S_\alpha ( \Mm (\rho) || \Mm ( \sigma))$ for any \QO $\Mm$ on
  $\observables$.
\end{itemize}
\end{theorem} 

\vspace{1mm}

The statements (i-iii), as well as (iv-v) for a restricted range of $\alpha$, namely $\alpha \in (1,2]$, have been established in
  \cite{Mueller-Lennert13,Wilde13}.  
The justification of (iv-v) in full  generality is due to Frank and Lieb~\cite{Frank13}.

\vspace{1mm}

\proof The unitary invariance (ii) and additivity (iii) are evident
and also hold for the $\alpha$-fidelity. We now argue that the non-negativity (i) and the monotonicity (iv) can be deduced
from the convexity/concavity property~(iv). 
Thanks to Proposition~\ref{prop-convexity_implies_monotonicity}, (iv) 
 implies that if $\alpha \in [1/2,1)$ then $F_\alpha ( \Mm ( \rho) || \Mm ( \sigma ) ) \geq
F_\alpha ( \rho || \sigma)$ for any \QO $\Mm$, and the reverse
  inequality holds true if $\alpha>1$. The monotonicity of $S_\alpha$ for
  $\alpha \geq 1/2$ then follows immediately (the case $\alpha=1$ is obtained by continuity, see (\ref{eq-relative_Reyni_ent_alpha=1})).
Let $\{ \ket{k}\}$ be an \ONB of $\Hh$ and  
$\Mm_{\Pi}$ be the \QO (\ref{eq-von_Neumann_measurement})
  associated to the von Neumann measurement $\{ \Pi_k=\ketbra{k}{k}  \}$. The
  monotonicity entails
\begin{equation} \label{eq-proof_positivity_S_alpha}
S_\alpha ( \rho || \sigma ) \geq S_\alpha ( \Mm_{\Pi} ( \rho) ||
\Mm_{\Pi} ( \sigma) ) = S_\alpha^{\rm clas} ( \pv || \qv)\;,
\end{equation}
where $\pv$ and $\qv$ are the vectors with components $p_k = \bra{k} \rho \ket{k}$ and $q_k = \bra{k} \sigma \ket{k}$.  
Since the classical R\'enyi divergence (\ref{eq-Reyni_divergence}) is non-negative and vanishes \ifif 
$\pv = \qv$, we deduce from (\ref{eq-proof_positivity_S_alpha}) that $S_\alpha ( \rho || \sigma )\geq 0$, with equality \ifif 
$\bra{k} \rho \ket{k} = \bra{k} \sigma \ket{k}$ for all $k$. The \ONB $\{ \ket{k}\}$
being arbitrary, this justifies the assertion (i) for $\alpha \geq 1/2$.  To show this assertion for $\alpha \in (0, 1/2)$, 
we argue as
in~\cite{Mueller-Lennert13} that
\begin{equation} \label{eq-proof_theorem_Reyni_ent0}
S_\alpha ( \rho || \sigma ) \geq S_\alpha ( \Mm_\Pi (\rho) || \sigma ) = S_\alpha^{\rm clas} ( \pv || \qv) 
\end{equation}
with $0 < \alpha < 1$, $\Mm_\Pi$ being as before associated with the von Neumann $\{ \Pi_k = \ketbra{k}{k} \}$ but with $\{ \ket{k} \}$ an 
orthonormal eigenbasis of $\sigma$.
Actually, let $\alpha \in (0,1)$ and let us  set $A(\beta) = \sigma^{\frac{\beta}{2}} \rho \sigma^{\frac{\beta}{2}}$ with $\beta = \alpha^{-1} - 1$.
By virtue of the Jensen type inequality (\ref{eq-Jensen_type_ineq}) of Appendix~\ref{app_operator_convex_functions},
one has
\begin{equation} \label{eq-proof_theorem_Reyni_ent}
\bigl( \Mm_\Pi (  A(\beta) ) \bigr)^\alpha 
\geq 
 \Mm_\Pi \bigl( A(\beta)^\alpha  \bigr)
\end{equation}
due to the operator concavity of $f(x) = x^\alpha$.  Hence, by the trace-preserving property of $\Mm_\Pi$
and the identity $\sigma^{\frac{\beta}{2}}  \Mm_\Pi ( \rho) \sigma^{\frac{\beta}{2}}  = \Mm_\Pi ( A ( \beta))$,
\begin{eqnarray} \label{eq-proof_positivity_S_alpha2}
\nn
 S_\alpha ( \rho || \sigma ) 
& = & 
(\alpha-1 )^{-1} \ln \tr \bigl[ \Mm_\Pi \bigl( A(\beta)^\alpha \bigr)\bigr] 
\\
& \geq & 
(\alpha-1)^{-1} \ln \tr \bigl[ \bigl( \Mm_\Pi \bigl( A(\beta) \bigr) \bigr)^\alpha \bigr] 
=  S_\alpha ( \Mm_\Pi (\rho) || \sigma ) \;.
\end{eqnarray}
This proves (\ref{eq-proof_theorem_Reyni_ent0}) and thus the non-negativity of $S_\alpha$ for $\alpha \in (0,1)$. 
Observe that $S_\alpha ( \rho || \sigma ) =  S_\alpha ( \Mm_\Pi (\rho) || \sigma )$ \ifif
(\ref{eq-proof_theorem_Reyni_ent}) holds with equality, that is,
$\bra{k} A ( \beta) \ket{k}^\alpha = \bra{k} A(\beta)^\alpha \ket{k}$ for all $k$.
By the strict concavity of $f(x) = x^\alpha$, $\{ \ket{k}\}$ must then be an eigenbasis of $A(\beta)$, and thereby also of $\rho$.  
Thus $\rho$ and $\sigma$ commute and $S_\alpha ( \rho || \sigma)$ coincides with the classical R\'enyi divergence $S_\alpha^{\rm clas} ( \pv || \qv )$.
By the aforementioned properties of  $S_\alpha^{\rm clas} ( \pv || \qv )$, it follows from (\ref{eq-proof_theorem_Reyni_ent0}) that
$S_\alpha ( \rho || \sigma)=0$ implies $\pv = \qv$ and thus $\rho = \sigma$.

It remains to show the statement (iv) of the theorem.
Following~\cite{Frank13}, we obtain (iv) with the help of a duality
formula for $F_\alpha ( \rho, \sigma)$ and of Lieb's concavity and
Ando's convexity theorems. We omit here the proof of these two important theorems, which can be found 
in~\cite{Carlen} (see also~\cite{Nielsen} for the Lieb theorem). The
duality formula will be shown at the end this subsection.

\vspace{2mm}

\begin{lemma} \label{lemma_Lieb_concavity_Ando_convexity}
{\rm (Lieb's concavity and Ando's convexity theorem~\cite{Ando79,Lieb73a})}  For any $K \in \observables$ and
any $\beta \in [-1,1]$, the function $ ( R, S ) \mapsto  \tr (
K^\ast R^q K S^{-\beta} )$ on $\observables_{+} \times
\observables_{+}$ is jointly concave in $( R,S)$ if $-1 \leq \beta
\leq 0$ and $0 \leq q \leq 1+\beta$ and is jointly convex in
$(R,S)$ if $0 \leq \beta \leq 1$ and $1 + \beta \leq q \leq
2$.
\end{lemma}

\vspace{1mm}

\begin{lemma} \label{lemma_duality_formula}
{\rm  (Duality formula for the $\alpha$-fidelity~\cite{Frank13})} If $\alpha \in
(0,1)$ (that is, $\beta = \alpha^{-1} - 1 >0$) then
\begin{equation} \label{eq-duality_S_alpha}
F_\alpha ( \rho, \sigma )^\alpha = \inf_{H \geq 0} \Bigr\{ \alpha \tr
( H \rho ) + (1-\alpha) \tr \bigr[ ( \sqrt{H} \sigma^{-\beta} \sqrt{H}
  )^{-\frac{1}{\beta}} \bigr] \Bigr\} \;.
\end{equation}
If $\alpha>1$ (that is, $-1 < \beta < 0$), the same identity holds but with
the infimum replaced by a supremum.
\end{lemma}

\vspace{1mm}

Given Lemma~\ref{lemma_duality_formula}, if one can show that, for a
fixed operator $B \in \observables$, the function
\begin{equation}
g_{B,\beta} (\sigma) = \tr \bigl[ ( B^\ast \sigma^{-\beta} B
  )^{-\frac{1}{\beta}} \bigr]
\end{equation}
is concave in $\sigma$ when $-1 \leq \beta \leq 1$, $\beta \not= 0$, it will follow that $F_\alpha (\rho || \sigma)^\alpha$ is jointly
concave for $\alpha \in [1/2,1)$ (\ie $0 < \beta \leq 1$) and jointly convex for $\alpha >1$ (\ie $-1 < \beta < 0$), thereby proving
Theorem~\ref{theo-Renyi_relative_entropy}.  We first assume $-1 \leq \beta < 0$.
  For any operator $Y \geq 0$, let us set
\begin{equation}
h_Y (X) = \tr ( Y X^{1 + \beta} ) - (1+\beta) \tr (X)
\end{equation}
with $X\in \observables_+$. Given two self-adjoint matrices $Y$
and $Z$, it is known that (see~\cite{Bhatia}, Problem III.6.14)
\begin{equation} \label{eq-proof_Frank_Lieb1}
\sum_{i=1}^n y_{n-i} z_{i} \leq \tr ( Y Z ) \leq \sum_{i=1}^n y_{i}
z_{i}\;,
\end{equation}
where $y_1 \geq y_2 \geq \cdots \geq y_n$ and $z_1 \geq z_2 \geq \cdots \geq z_n$ are the eigenvalues of $Y$ and $Z$ in non-increasing
order. Therefore,
\begin{equation}
\sup_{X \geq 0} \{ h_Y ( X) \} = \max_{ \xv} \Bigl\{ \sum_{i=1}^n
\bigl( y_{i} x_i^{1+\beta} - ( 1 + \beta) x_i \bigr) \Bigr\} = - \beta
\sum_{i=1}^n y_i^{-\frac{1}{\beta}} = -\beta \tr \bigl(
Y^{-\frac{1}{\beta}} \bigr) \;,
\end{equation}
the maximum in the second member being over all vectors $\xv \in \real_+^n$.  Similarly, it follows from (\ref{eq-proof_Frank_Lieb1})
that if $0 < \beta \leq 1$ then $\inf_{X \geq 0} \{ h_Y ( X) \} = -\beta \tr ( Y^{-\frac{1}{\beta}} )$.  Plugging $Y = B^\ast \sigma^{-\beta} B$ 
into these identities, one finds
\begin{equation} \label{eq-proof_Frank_Lieb2}
g_{B,\beta} (\sigma) =\sup_{X \geq 0} \Bigl\{  - \beta^{-1}  \bigl( \tr (
B^\ast \sigma^{-\beta} B X^{1 + \beta} ) - (1+\beta) \tr (X) \bigr) \Bigr\}
\quad , \quad -1 \leq \beta < 0\;\text{ or }\; 0 < \beta \leq 1.
\end{equation}
Let us introduce the $2 \times 2$ block matrices
\begin{equation}
K = \left( \begin{array}{ll} 0 & 0 \\ B^\ast & 0 \end{array} \right)
\quad , \quad S = \left( \begin{array}{ll} \sigma & 0 \\ 0 &
  X \end{array} \right) \;.
\end{equation}
A simple calculation gives
\begin{equation} \label{eq-proof_Frank_Lieb3}
\tr ( B^\ast \sigma^{-\beta} B X^{1 + \beta} ) = \tr_{\Hh \otimes
  \complex^2} ( K^\ast S^{1+\beta} K S^{-\beta} ) \;.
\end{equation}
By Lemma~\ref{lemma_Lieb_concavity_Ando_convexity}, the \RHS of (\ref{eq-proof_Frank_Lieb3}) is concave (respectively convex) in $S$ 
when $-1 \leq \beta < 0$ (respectively $0 < \beta \leq 1$). As a result, the \LHS is jointly concave (convex) in $(\sigma,X)$. 
But the maximum over $X$ of a jointly concave  function $f(\sigma, X)$ is concave  in $\sigma$. Thanks to (\ref{eq-proof_Frank_Lieb2}), 
we may conclude that $g_{B,\beta} (\sigma)$ is concave in $\sigma$ for all $\beta \in [-1,1]$, $\beta \not=0$. The proof of 
 Theorem~\ref{theo-Renyi_relative_entropy} is now complete.  \finpro

\vspace{2mm}

Let us come back to the duality formula
(\ref{eq-duality_S_alpha}). We observe in passing that this formula
bears some similarity with the variational formula
(\ref{eq-S_Legendre_transform_F}) for the von Neumann entropy.

\vspace{2mm}

\Proofof{lemma~\ref{lemma_duality_formula}} Since  $\sigma^{-\frac{\beta}{2}} H \sigma^{-\frac{\beta}{2}}$ has
the same non-zero eigenvalues as $\sqrt{H} \sigma^{-\beta} \sqrt{H}$,
the quantity inside the infimum in (\ref{eq-duality_S_alpha}) is equal
to
\begin{equation}
g ( H ) = \alpha \tr ( H \rho ) + (1-\alpha) \tr \bigr[ (
  \sigma^{-\frac{\beta}{2}} H \sigma^{-\frac{\beta}{2}}
  )^{-\frac{1}{\beta}} \bigr] \;.
\end{equation}
Differentiating the \RHS with respect to the matrix elements of $H$ in
the some \ONB $\{ \ket{i}\}$  and using the relation $\partial \tr [ f(B)]/\partial B_{ij} = f' ( B)_{ji}$ with $f(x)$ a $C^1$-function,
we get
\begin{equation}
\frac{\partial g ( H)}{\partial H_{ij}} = \alpha \Bigl( \rho - \sigma^{-\frac{\beta}{2}} ( \sigma^{-\frac{\beta}{2}} H
\sigma^{-\frac{\beta}{2}} )^{-\frac{1}{\beta}-1} \sigma^{-\frac{\beta}{2}} \Bigr)_{ji}\;.
\end{equation}
Hence $g(H)$ has an extremum \ifif $H = \widehat{H} = \sigma^{\frac{\beta}{2}} ( \sigma^{\frac{\beta}{2}} \rho \sigma^{\frac{\beta}{2}} )^{\alpha-1} \sigma^{\frac{\beta}{2}} \geq
0$.  But
\begin{equation}
g ( \widehat{H} ) = \tr [( \sigma^{\frac{\beta}{2}} \rho  \sigma^{\frac{\beta}{2}})^\alpha ]
= F_\alpha ( \rho || \sigma )^\alpha\;.
\end{equation}
As $B \in \observables_{+}
\mapsto \tr (B^p)$ is convex for $p \geq 1$ or $p \leq 0$, 
$g(H)$ is convex if $\alpha \in (0,1)$
(\ie $- \beta^{-1} < 0$) and concave if $\alpha>1$ (\ie $- \beta^{-1} > 1$).  It follows that $g ( \widehat{H} )$ is
a minimum for $\alpha \in (0,1)$ and a maximum for $\alpha>1$.
\finpro

\vspace{2mm}

Let us point out that it follows from Lemma~\ref{lemma_Lieb_concavity_Ando_convexity} that the normal-ordered R\'enyi entropy
(\ref{eq-relative_Renyi_entropy_normal}) is also jointly convex for $\alpha \in (0,1)$. 
Taking $\alpha \rightarrow 1$ and recalling that $S_\alpha^{\rm (n)} ( \rho || \sigma) \rightarrow S(\rho||\sigma)$,
this gives a direct proof the joint convexity of the  relative von Neumann entropy $S(\rho||\sigma)$ 
from the Lieb concavity theorem, as noted by Lindblad~\cite{Lindblad74,Lindblad75}. 
Combined with  Proposition~\ref{prop-convexity_implies_monotonicity}, this leads to a  
 completely different justification of the monotonicity of $S(\rho||\sigma)$ in Theorem~\ref{theo-Petz}
than that presented in Sec.~\ref{sec-relative_entropy}.
It would be interesting to look for a generalization of the arguments of Petz in Sec.~\ref{sec-relative_entropy} to the case of the
$\alpha$-entropies.

\subsubsection{Monotonicity in $\alpha$}

As stated above, a very nice feature of the $\alpha$-entropy (\ref{eq-relative_Renyi_entropy}) is that,  like the classical
R\'enyi divergence, it is monotonous in $\alpha$. This leads in particular to some 
bound between the relative von Neumann entropy and the fidelity
(see (\ref{eq-rel_ent_bound}) below).

\vspace{2mm}

\begin{proposition} {\rm \cite{Mueller-Lennert13}} \label{prop-monotonicity_in_alpha_of_S_alpha}
For any $\rho, \sigma \in \states (\Hh)$, $S_\alpha ( \rho || \sigma)$ is a non-decreasing function of $\alpha$ on $(0,\infty)$.
\end{proposition}

\vspace{1mm}

\proof One first derive the following identity similar to
(\ref{eq-proof_Frank_Lieb2}):
\begin{equation} \label{proof_monotonicty_in_alpha_S_alpha}
\bigl( g_{B,-\alpha^{-1}} (\sigma) \bigr)^{\frac{1}{\alpha}} = \bigl\|
B^\ast \sigma^{1/\alpha} B \bigr\|_{\alpha} = \sup_{\tau \geq 0,
  \tr(\tau)=1} \tr \bigl( B^\ast \sigma^{1/\alpha} B \,
\tau^{1-1/\alpha} \bigr)\quad, \quad \alpha\geq 1\;.
\end{equation}
If $0 < \alpha \leq 1$ the supremum has to be
replaced by an infimum.   When $\alpha \geq 1$ this identity is nothing but a rewriting of
the H\"older's inequality (\ref{eq-main_property_L_P_norm}). 
The derivation for $\alpha \in (0,1)$ relies on (\ref{eq-proof_Frank_Lieb1}) and
follows the same lines as for the derivation of (\ref{eq-proof_Frank_Lieb2})
(apart from the fact that we substituted
$\beta$ by $-1/\alpha$), but one must introduce a Lagrange
multiplier to account for the constraint $\tr ( \tau)=1$. 
Applying the relation (\ref{proof_monotonicty_in_alpha_S_alpha}) for $B= \sigma^{-\onehalf} \rho^{\onehalf}$ and plugging 
the identity $\| \sigma^{\frac{\beta}{2}} \rho \sigma^{\frac{\beta}{2}} \|_{\alpha} = \| \rho^\onehalf \sigma^{\beta} \rho^\onehalf \|_{\alpha}$ 
into (\ref{eq-def_alpha_fidelity}), we are led to
\begin{equation} \label{eq-variational_formula_for_S_alpha}
S_\alpha ( \rho || \sigma) = \sup_{\tau \in \states (\Hh)} \bigl\{ -\beta^{-1} \ln F_{\alpha} ( \rho || \sigma ; \tau ) \bigr\} 
\quad , \quad 
F_{\alpha} ( \rho || \sigma ; \tau ) = \tr \bigl( \rho^{\onehalf} \sigma^\beta \rho^{\onehalf} \tau^{-\beta} \bigr) =
\bigl\langle \xi\, , \, \Delta_{\sigma| \tau}^\beta \xi \bigr\rangle\;,
\end{equation}
for any $\alpha>0$, $\alpha \not= 1$. In the last identity $\xi = \rho^\onehalf$ and
we have introduced the relative modular operator, see (\ref{eq-def_modular_op}).  
For any fixed $\tau \in \states ( \Hh)$, one finds
\begin{equation}
\frac{\D }{\D \beta} \Bigl( - \beta^{-1} \ln F_{\alpha} ( \rho || \sigma ; \tau ) \Bigr) =
- \frac{1}{\beta^2 F_{\alpha} ( \rho || \sigma ; \tau)} 
\Bigl( \langle \xi\, ,\, \Delta^\beta_{\sigma| \tau} \ln ( \Delta^\beta_{\sigma| \tau}) \xi \rangle
 - \langle \xi \, ,\, \Delta^\beta_{\sigma| \tau} \xi \rangle 
\ln \langle \xi \, ,\, \Delta^\beta_{\sigma| \tau} \xi \rangle \Bigr) \;.
\end{equation}
The Jensen inequality applied to the convex function $f(x) = x \ln x$ implies that the quantity inside the parenthesis in the \RHS 
is non-negative. Thus $-\beta^{-1} F_{\alpha} ( \rho || \sigma; \tau)$ is a non-increasing function of $\beta$.  This is true for any density matrix $\tau$,
thus one infers from (\ref{eq-variational_formula_for_S_alpha}) that $\alpha \mapsto S_\alpha ( \rho || \sigma)$ is non-decreasing.
\finpro

\newpage
\section{The Bures distance and Uhlmann fidelity}
\label{sec-Bures_distance}

In this section we study the Bures distance  on the set of quantum states $\states ( \Hh)$.
This distance  is Riemannian and  monotonous with respect to quantum operations. It is a simple function of the fidelity (\ref{eq-def_Ulhmann_fidelity}).
Its metric coincides with the quantum Fisher information quantifying the best achievable precision in the parameter estimation problem discussed 
in Sec.~\ref{sec-interferometry}. The material of this section (as well as of Sec.~\ref{sec-QHT_and_parameter_est}) is completely independent from that
 of sections~\ref{sec-entanglement} and \ref{sec-discord}, so it is possible at this point to proceed  directly to Sec.~\ref{sec-entanglement}. The reading of 
Secs.~\ref{sec-contractive_distances}--\ref{sec_bounds_Bures_dist_and_trace} is, however, recommended before 
going through Sec.~\ref{sec-geometric_measures} devoted to the geometrical measures of quantum correlations, where the  
 Bures distance plays the key role. 
The section is organized as follows. 
Sec.~\ref{sec-contractive_distances} contains a short discussion on contractive (\ie monotonous) distances.
It is argued there that the distances induced by the $\| \cdot \|_p$-norm are not contractive save for $p=1$.  
The definition and main properties of the Bures distance are given in Secs.~\ref{sec-def-Bures_distance}--\ref{sec_bounds_Bures_dist_and_trace}. 
The Bures metric is determined in Sec.\ref{sec-Bures_metric}. Finally, Sec.~\ref{sec-charac_contractive_dist} contains 
the proof of an important result of Petz on the characterization of all Riemannian contractive metrics on $\states ( \Hh)$ 
for finite-dimensional Hilbert spaces $\Hh$.

\subsection{Contractive and convex distances} \label{sec-contractive_distances}

In order to quantify how far are two  states $\rho$ and $\sigma$ it is necessary to define a distance on the set $\Ee ( \Hh )$ of
quantum states.  One has a priori the choice between many distances.  The most common ones are
the $L^p$-distances defined by (\ref{eq-Lp_norm}).  In quantum
information theory it seems, however, natural to impose the following requirement.

\vspace{2mm}

\begin{definition} \label{def-contractive_dist} 
A distance $d$ on the sets of quantum states is contractive if
for any finite Hilbert spaces $\Hh$ and $\Hh'$, any quantum operation $\Mm : \Bb(\Hh) \rightarrow \Bb(\Hh' )$, and any $\rho$, $\sigma \in \Ee (\Hh)$,
it holds
\begin{equation}
d ( \Mm (\rho), \Mm (\sigma) ) \leq d (\rho,\sigma)\;.
\end{equation}
\end{definition}

\vspace{1mm}

A contractive distance is in particular invariant under unitary
conjugations, \ie
\begin{equation}
d \bigl( U \rho\, U^\ast, U \sigma\, U^\ast \bigr) = d(\rho,\sigma)
\quad \text{if $U$ is unitary}
\end{equation}
(in fact, $\rho \mapsto U \rho \, U^\ast$ is an invertible \QO on $\Bb(\Hh)$).
For such a distance, if a generalized measurement is performed on a
system, two states are closer from each other after  the measurement than before it,
and if the system is subject to a unitary evolution the distance between the time-evolved states remains unchanged.

For $p>1$, the distances $d_p$ (in particular, the Hilbert-Schmidt
distance $d_2$) are not contractive. A counter-example for two qubits
is obtained~\cite{Ozawa00} by taking $\Mm ( \rho) = A_1 \rho A_1^\ast + A_2 \rho A_2^\ast$ with 
\begin{equation}
A_1 = \sigma_+ \otimes 1 \quad , \quad A_2 = \sigma_+ \sigma_- \otimes 1
\quad , \quad  \rho= \onehalf \otimes \sigma_+ \sigma_- \quad , \quad \sigma= \onehalf \otimes \sigma_- \sigma_+
\end{equation}
(here $\sigma_+ = \ketbra{1}{0}$ is the raising operator and $\sigma_-= \sigma_+^\ast$).  Then $\| \Mm (\rho)
- \Mm (\sigma) \|_p = 2^{1/p}$ is larger than $\| \rho-\sigma\|_p =
2^{2/p-1}$.

\vspace{2mm}

\begin{proposition} \label{prop-trace_distance} {\rm \cite{Ruskai94}}
The trace distance $d_1$ is contractive.
\end{proposition}

\vspace{1mm}

\proof: Let $R=\rho-\sigma=R_+-R_-$
with $R_{\pm} = (|R | \pm R)/2 = \pm R P_\pm \geq 0$ the positive and negative parts of $R$ (here $P_+$ and $P_-$ are the spectral projectors of $R$ on
$[0,\infty) $ and $(-\infty,0)$).  Then $\| R \|_1 = \tr(R_++R_-)= 2 \tr ( R_+)$ because $\tr (R)=\tr(R_+)-\tr(R_-)=0$.  Since $\Mm$ is
trace preserving and CP, one has $\| \Mm (R)\|_1 = 2 \tr [ \Mm (R)_+  ]$ and 
$\Mm ( R)_+ = ( \Mm (R_+)-\Mm (R_-))_+ \leq \Mm ( R_+  )$. 
Thus $ \| \Mm (R)\|_1 \leq 2 \tr [ \Mm (R_+) ] = 2 \tr [ R_+] =  \| R \|_1 $.
\finpro

\vspace{2mm}

A distance $d$ on $\Ee (\Hh)$ is {\it jointly convex} if for any state ensembles 
$\{\rho_i, p_i \}$ and $\{\sigma_i ,  p_i \}$ with the same probabilities $p_i$,
\begin{equation}
d \Bigl( \sum_i p_i \rho_i , \sum_i p_i \sigma_i \Bigr) \leq \sum_i
p_i d(\rho_i,\sigma_i)\;.
\end{equation}
Since they are associated to a norm, the distances $d_p$ are
jointly convex for any $p \geq 1$.

\subsection{The Bures distance} \label{sec-def-Bures_distance}

We now introduce the Bures distance $d_{\rm B}$.
This distance is contractive like $d_1$. It was first considered by Bures in
the context of infinite products of von Neumann
algebras~\cite{Bures69} (see also~\cite{Araki70}) and  was later
studied in a series of papers by
Uhlmann~\cite{Uhlmann76,Uhlmann86,Uhlmann95}. Uhlmann used it to define
parallel transport and related it to the fidelity generalizing the usual fidelity $| \braket{\psi}{\phi} |^2$
between pure states.  Indeed, $d_{\rm B}$ is a extension to mixed states
of the Fubini-Study distance on the projective space $P \Hh$ of pure
states,
\begin{equation} \label{eq-def_Fubini_study}
d_{\rm FS} \bigl(  \rho_{\psi}  , \sigma_{\phi}  \bigr) =
\inf_{\ket{{\psi}},\ket{{\phi}} } \bigl\|\ket{{\psi}} -
\ket{{\phi}} \bigr\| = \bigl( 2 - 2 | \braket{\psi}{\phi} |
\bigr)^{\frac{1}{2}} \;,
\end{equation}
where the infimum in the second member is over all representatives $\ket{\psi}$ of $\rho_{\psi}  \in P \Hh$ and 
 $\ket{\phi}$ of $\sigma_{\phi} \in P \Hh$ (\ie $\rho_\psi = \ketbra{\psi}{\psi}$ and  $\sigma_\phi = \ketbra{\phi}{\phi}$). 
Observe that the 
third member is independent of these representatives.  For two mixed states $\rho$ and $\sigma$ in
$\states ( \Hh)$, one can define analogously~\cite{Uhlmann86,Hubner92}
\begin{equation} \label{eq-def_Bures_dist_1}
d_{\rm B} ( \rho,\sigma) = \inf_{A,B} d_2 ( A-B)\;,
\end{equation}
where the infimum is over all Hilbert-Schmidt matrices $A$ and $B$ satisfying $A A^\ast =\rho$ and $B B^\ast =\sigma$. Such matrices are
given by $A= \sqrt{\rho} V$ and $B=\sqrt{\sigma} W$ for some unitaries $V$ and $W$
(polar decompositions).  If $\rho = \rho_{\psi}$ and $\sigma= \sigma_{\phi}$ are pure
states, then $A = \ketbra{\psi}{\mu}$ and $B =\ketbra{\phi}{\nu}$ with $\| \mu \|=\|\nu \|=1$, so that
(\ref{eq-def_Bures_dist_1}) reduces to the Fubini-Study distance (\ref{eq-def_Fubini_study}).

For mixed states $\rho$ and $\sigma$, the \RHS of
(\ref{eq-def_Bures_dist_1}) is given by 
\begin{equation}
\bigl( 2 - 2 \sup_{U} \re \tr ( U
\sqrt{\rho} \sqrt{\sigma} ) \bigr)^{\frac{1}{2}}
\end{equation}
with a supremum over all unitaries $U = W V^\ast$.   This supremum is equal to $\| \sqrt{\rho} \sqrt{\sigma}\|_1$ and is attained
\ifif $U {U}_0 | \sqrt{\rho} \sqrt{\sigma} |^{\frac{1}{2}}=
|\sqrt{\rho} \sqrt{\sigma}|^{\frac{1}{2}}$, where ${U}_0$ is such that $\sqrt{\rho} \sqrt{\sigma}= {U}_0 | \sqrt{\rho} \sqrt{\sigma}|$
(see Sec.~\ref{sec-states_and_observables}). Equivalently, the infimum in
(\ref{eq-def_Bures_dist_1}) is attained \ifif the parallel transport
condition $A^\ast B \geq 0$ holds.  We obtain the following equivalent definition of $d_{\rm B}$.

\vspace{2mm}

\begin{definition} \label{def-Bures_distance_and_fidelity}
For any states $\rho,\sigma \in \states ( \Hh)$,
\begin{equation} \label{eq-def_Bures_dist_2}
d_{\rm B} ( \rho,\sigma) = \bigl( 2 - 2 \sqrt{F(\rho,\sigma )}
\bigr)^{\frac{1}{2}}
\end{equation}
where the Uhlmann fidelity  is defined by
\be \label{eq-fidelity} 
F(\rho,\sigma) = \| \sqrt{\rho} \sqrt{\sigma} \|_1^2 = \Bigl( \tr \bigl[ (  \sqrt{\sigma} \rho \sqrt{\sigma} )^{\onehalf} \bigr] \Bigr)^2 \;.  
\ee
\end{definition}

\vspace{1mm}

The fidelity $F(\rho,\sigma)$ is symmetric in $(\rho,\sigma)$ and belongs to the interval $[0,1]$.  It is clearly a generalization of
the usual pure state fidelity $F (\ket{\psi}, \ket{\phi})= |\braket{\psi}{\phi} |^2$. If $\sigma_\phi$ is pure, then
\begin{equation} \label{eq-fidelity_pure_state}
F(\rho , \sigma_\phi) = \bra{\phi} \rho \ket{\phi} 
\end{equation}
for any $\rho \in \states ( \Hh)$.

It is immediate on (\ref{eq-def_Bures_dist_1}) that $d_{\rm B}$ is
positive and symmetric, and $d_{\rm B} (\rho,\sigma)=0$ if and only if
$\rho=\sigma$. The triangle inequality is more difficult to show. It can be established with the help 
of the following astonishing theorem.

\vspace{2mm}

\begin{theorem} { \rm  (Uhlmann~\cite{Uhlmann76})}  \label{theo-Uhlmann}
Let $\rho$, $\sigma \in \states ( \Hh)$ and $\ket{\Psi}$ be a purification of $\rho$ on the space $\Hh \otimes
\Kk$, with $\dim \Kk \geq \dim \Hh$. Then
\begin{equation} \label{eq-theo_Uhlmann}
F(\rho,\sigma) = \max_{\ket{\Phi}} | \braket{\Psi}{\Phi} |^2
\end{equation}
where the maximum is over all purifications $\ket{\Phi}$ of $\sigma$ on $\Hh \otimes \Kk$.
\end{theorem}

\vspace{1mm}

\proof We give here a simple proof due to Josza~\cite{Jozsa94}. Let us first assume $\Kk \simeq \Hh$.  Let $\ket{\Psi}$ and $\ket{\Phi}$ be purifications of $\rho$ and $\sigma$
on $\Hh \otimes \Hh$, respectively.  As it has been noticed in Sec.~\ref{sec-purification}, by the Schmidt decomposition
these purifications can always be written as
\begin{equation} \label{eq-proof_Ulhmann_th1}
\ket{\Psi} = \sum_{k=1}^n \sqrt{p_k} \ket{k} \ket{f_k} \quad , \quad
\ket{\Phi} = \sum_{k=1}^n \sqrt{q_k} (U \ket{k}) \ket{g_k}\;,
\end{equation}
where $\rho = \sum_k p_k \ketbra{k}{k}$ and $\sigma = \sum_k q_k U \ketbra{k}{k} U^\ast$ are spectral decompositions
 of $\rho$ and
$\sigma$, $U$ is a unitary operator on $\Hh$, and $\{ \ket{f_k} \}_{i=1}^n$ and $\{ \ket{g_k} \}_{i=1}^n$ are two \ONBs of $\Hh$.
Defining the unitaries $V$ and $W$ on $\Hh$ by $\ket{f_k}= V \ket{k}$
and $\ket{g_k}= W \ket{k}$ for any $k=1,\ldots, n$, we have
\begin{equation}
\ket{\Psi} = \sqrt{\rho} \otimes V \ket{\Sigma} \quad , \quad
\ket{\Phi} = \sqrt{\sigma} \,U \otimes W \ket{\Sigma} \quad
\text{with} \quad \ket{\Sigma}= \sum_{k=1}^n \ket{k} \ket{k}\;.
\end{equation}
The vector $\ket{\Sigma}$ is the vector associated to the identity operator on $\Bb ( \Hh)$ by
the isomorphism (\ref{eq-ismoemtry_operators_vectors}).
For any $X, Y \in \observables$, one obtains by setting $O = X^T \otimes Y$ in (\ref{eq-reshuffling_op}) and 
noting that $\tr ( O^\Rr ) = \tr ( X Y)$ that
\begin{equation}
\tr ( X Y  ) = \bra{\Sigma} X^{T} \otimes Y \ket{\Sigma}
\end{equation}
(here $ X^{T}$ is the transpose of $X$ in the basis $\{ \ket{k} \}$).  Introducing the unitary $U_0= V^\ast W U^T$, this
gives
\begin{equation}
\sup_{\ket{\Phi}} | \braket{\Phi}{\Psi} | = \sup_{W} | \bra{\Sigma} U^\ast \sqrt{\sigma} \sqrt{\rho} \otimes W^\ast V \ket{\Sigma} | 
 =
\sup_{U_0} | \tr ( \sqrt{\rho} \sqrt{\sigma} \,U_0^\ast ) | = \| \sqrt{\rho}
\sqrt{\sigma} \|_1 \;.
\end{equation}
The last equality comes from (\ref{eq-main_property_L_P_norm}). This proves the desired result.
 The supremum is achieved by choosing
$\ket{\Phi}$ as in (\ref{eq-proof_Ulhmann_th1}) with $U = U_0^T  ( W^\ast)^T V^T$, $U_0$ being a unitary in the polar
decomposition of $\sqrt{\rho} \sqrt{\sigma}$. 

If $\Kk$ has a dimension $m$ larger than $n$, we extend $\rho$ and $\sigma$ to a space $\Hh' \simeq \Kk$ by adding to them new
orthonormal eigenvectors $\ket{k}$ and $U \ket{k}$ with zero
eigenvalues $p_k = q_k = 0$, $k=n+1,\ldots, m$.  This does not change the fidelity $F(\rho,\sigma)$, thus 
$F(\rho,\sigma) = \max_{\ket{\Phi'}} | \braket{\Psi'}{\Phi'}|^2$, where $\ket{\Psi'}$ is
a purification of $\rho'=\sum_{k=1}^{m} p_k \ketbra{k}{k}=\rho$ on $\Hh' \otimes \Hh'$, and similarly for $\ket{\Phi'}$.  
But $\ket{\Psi'}$ and $\ket{\Phi'}$ have the form (\ref{eq-proof_Ulhmann_th1}), hence they belong to $\Hh \otimes \Kk$.  
\finpro

\vspace{3mm}

Let $\rho$, $\sigma$, and $\tau$ be three states of $\states ( \Hh)$ and $\ket{\Psi}$ be a purification of $\rho$ on $\Hh \otimes \Hh$.
According to Theorem~\ref{theo-Uhlmann}, there exists a purification $\ket{\Phi}$ of $\sigma$ on $\Hh \otimes \Hh$ such that $F
(\rho,\sigma) = | \braket{\Psi}{\Phi} |^2$. One can choose the
arbitrary phase factor of $\ket{\Phi}$ in such a way that
$\braket{\Psi}{\Phi} \geq 0$, whence $\sqrt{F (\rho,\sigma)} =
\braket{\Psi}{\Phi}$. Similarly, there exists a purification
$\ket{\chi}$ of $\tau$ such that $\sqrt{F(\sigma, \tau)} =
\braket{\Phi}{\chi} \geq 0$.  In view of (\ref{eq-def_Bures_dist_2}) and (\ref{eq-theo_Uhlmann}),
\begin{eqnarray}
\nn d_{\rm B} ( \rho, \tau) & \leq & \bigl( 2 - 2 | \braket{\Psi}{\chi} | \bigr)^\onehalf 
\\
\nn
& \leq & \bigl( 2 - 2 \re \braket{\Psi}{\chi} | \bigr)^\onehalf
  = \bigl\| \ket{\Psi} - \ket{\chi} \bigr\|
\\ & \leq & \bigl\| \ket{\Psi} - \ket{\Phi} \big\| + \bigl\| \ket{\Phi} - \ket{\chi} \big\| 
= \bigl( 2 - 2 \braket{\Psi}{\Phi} | \bigr)^\onehalf + \bigl( 2 - 2 \braket{\Phi}{\chi} | \bigr)^\onehalf \;,
\end{eqnarray}
showing that $d_{\rm B}$ satisfies the triangle inequality $d_{\rm B}
( \rho, \tau) \leq d_{\rm B} ( \rho, \sigma) + d_{\rm B} ( \sigma,
\tau)$.

\vspace{2mm}

\begin{corollary} \label{prop-d_B_contractive}
The map $(\rho,\sigma) \mapsto d_{\rm B} ( \rho, \sigma)$ defines a distance $d_{\rm B}$ on  quantum states, with values in
$[0,1]$. This distance is contractive. Moreover, $d_{\rm B}^2$ is jointly convex.
\end{corollary}

\vspace{1mm}

Note that $d_{\rm B}$ is not jointly convex.  One gets a
counter-example by choosing $\rho_0 = \sigma_0 = \ketbra{0}{0}$, $\rho_1 = \ketbra{1}{1}$, $\sigma_1 = \ketbra{2}{2}$, and
$p_0=p_1=1/2$, $\{ \ket{0}, \ket{1}, \ket{2} \}$ being an orthonormal
family in $\Hh$.

It is clear on (\ref{eq-fidelity}) that $F(\rho,\sigma)=0$ \ifif
$\rho$ and $\sigma$ have orthogonal supports, $\range \rho \,{\bot}\, \range \sigma$. Therefore, two states $\rho$ and $\sigma$ have a
maximal distance $d_{\rm B} ( \rho,\sigma)=1$ if they are  orthogonal and thus perfectly distinguishable.

\vspace{1mm}

\proof We have already established above that $d_{\rm B}$ satisfies all the axioms of a distance. To show the contractivity, it is enough to check
that for any \QO $\Mm : \Bb ( \Hh) \rightarrow \Bb ( \Hh' ) $ and any states $\rho, \sigma \in \states (\Hh)$,
\begin{equation} \label{eq-fidelity_is_monotonous}
F ( \Mm ( \rho) , \Mm ( \sigma)) \geq F ( \rho,\sigma)\;.
\end{equation}
This property of the fidelity is a consequence of the contractivity of the relative R\'enyi entropy for  $\alpha=1/2$
(Theorem~\ref{theo-Renyi_relative_entropy}{\it (v)}).  It is, however,  instructive to re-derive this result
from Theorem~\ref{theo-Uhlmann}. According to this theorem, there exist some purifications
$\ket{\Psi}$ and $\ket{\Phi}$ of $\rho$ and $\sigma$ on $\Hh \otimes \Kk$ such that $F ( \rho,\sigma ) = | \braket{\Psi}{\Phi} |^2$. 
Now, thanks to (\ref{eq-purification_Mm_rho}) one obtains some purifications $\ket{\Psi_\Mm} = 1_\Kk \otimes U \ket{\Psi} \ket{\epsilon_0}$ 
 of $\Mm (\rho)$ and 
$\ket{\Phi_\Mm} = 1_\Kk \otimes U \ket{\Phi} \ket{\epsilon_0}$ of $\Mm (\sigma)$ on 
$\Kk \otimes \Hh' \otimes \Hh_\EE'$, with $\ket{\epsilon_0} \in \Hh_\EE$ and
$U : \Hh \otimes \Hh_\EE \rightarrow \Hh' \otimes \Hh_\EE'$ unitary. Thus
\begin{equation}
F ( \Mm ( \rho) , \Mm ( \sigma)) \geq | \braket{\Psi_\Mm}{\Phi_\Mm}
|^2 = | \braket{\Psi}{\Phi} |^2 = F ( \rho, \sigma) \;.
\end{equation}  
The joint convexity of $d_{\rm B}^2$ is a consequence of the bound\footnote{
Note that one cannot replace $\sqrt{F}$ by $F$ in this inequality,
that is, $F(\rho,\sigma)$ is not jointly concave (one can take the same
counter-example as that given above for $d_{\rm B}$). 
However, by a slight modification of the proof of Corollary~\ref{prop-d_B_contractive} one can show that $\rho \mapsto
F(\rho,\sigma)$ and $\sigma \mapsto F(\rho,\sigma)$ are concave.  In
their book~\cite{Nielsen}, Nielsen and Chuang define the fidelity as
the square root of (\ref{eq-fidelity}). This must be kept in mind when
comparing the results in this monograph with those of this article.  
}
%
\begin{equation} \label{eq-concavity_fidelity}
\sqrt{F \Bigl( \sum_i p_i \rho_i , \sum_i q_i \sigma_i \Bigr)} \geq
\sum_i \sqrt{p_i q_i} \sqrt{F ( \rho_i, \sigma_i )}\;,
\end{equation}
where $\{  \rho_i, p_i \}$ and $\{\sigma_i,  q_i  \}$ are arbitrary ensembles in $\states ( \Hh)$.  Note that the statement
(\ref{eq-concavity_fidelity}) is slightly more general than the joint concavity of the square root of $F(\rho, \sigma)$ proven in Sec.~\ref{sec-cond_Renyi_entropy}
(Theorem~\ref{theo-Renyi_relative_entropy}{\it (iv)}).  To show that (\ref{eq-concavity_fidelity}) is true, we introduce as before some
purifications $\ket{\Psi_i}$ of $\rho_i$ and $\ket{\Phi_i}$ of $\sigma_i$ on $\Hh \otimes \Hh$ such that 
$\sqrt{F ( \rho_i, \sigma_i )} = \braket{\Psi_i}{\Phi_i}$.  Let us define the vectors
\begin{equation}
\ket{\Psi} = \sum_i \sqrt{p_i} \ket{\Psi_i} \ket{\epsilon_i} \quad ,
\quad \ket{\Phi} = \sum_i \sqrt{p_i} \ket{\Phi_i} \ket{\epsilon_i}
\end{equation}
in $\Hh \otimes \Hh \otimes \Hh_\EE$, where $\Hh_\EE$ is an auxiliary Hilbert space and $\{ \ket{\epsilon_i} \}$ is an \ONB of $\Hh_\EE$. Then
$\ket{\Psi}$ and $\ket{\Phi}$ are purifications of $\rho = \sum_i p_i \rho_i$ and $\sigma = \sum_i q_i \sigma_i $, respectively. One infers
from Theorem~\ref{theo-Uhlmann} that
\begin{equation}
\sqrt{F ( \rho, \sigma )} \geq | \braket{\Psi}{\Phi} | = \sum_i \sqrt{p_i   q_i} \braket{\Psi_i}{\Phi_i} 
= \sum_i \sqrt{p_i q_i} \sqrt{F ( \rho_i, \sigma_i )}\;.
\end{equation}
This complete the proof of the corollary. 
\finpro

\vspace{1mm}

\begin{exercice}
A consequence of  (\ref{eq-def_entanglement_fidelity}) and (\ref{eq-fidelity_pure_state}) and of the monotonicity  of the fidelity $F$ with respect to
 partial trace operations (see (\ref{eq-fidelity_is_monotonous})) is that
the entanglement fidelity $F_{\rm e} ( \rho, \Mm)$ of a state $\rho$ with respect to a \QO $\Mm$ satisfies
\begin{equation}
F_{\rm e} ( \rho, \Mm ) \leq F ( \rho, \Mm ( \rho))\;.
\end{equation}
\end{exercice}

\vspace{1mm}

\begin{exercice}
As the fidelity satisfies $F ( \rho \otimes \rho' , \sigma \otimes \sigma' ) = F( \rho, \sigma) F ( \rho', \sigma')$, the Bures distance
increases by taking tensor products,  $d_{\rm B} ( \rho \otimes \rho' , \sigma \otimes \sigma')  \geq d_{\rm B} ( \rho, \sigma)$
for any $\rho,\sigma \in \states ( \Hh)$, $\rho', \sigma' \in \states ( \Hh')$, with equality \ifif $\rho'=\sigma'$. This has to be contrasted with
the trace distance, which does not enjoy this property.
\end{exercice}

In the two following subsections we collect some important properties of the Bures distance.
We refer the reader to the monographs~\cite{Bengtsson,Nielsen} for a list of names to which these properties
should be attached.

\subsection{Bures distance and statistical distance in classical probability} \label{sec_Bures_dist_and_stat_dist}

The restriction of a distance $d$ on $\states ( \Hh)$ to all  density matrices commuting with a given state $\rho_0$ defines 
a distance on the simplex $\states_{\rm clas} = \{ \pv \in \real_{+}^n ; \sum_{i} p_i = 1\}$ of classical probabilities on the finite space 
$\{ 1, 2,\ldots, n\}$.  In particular, if $\rho$ and $\sigma$ are two commuting states with spectral decompositions
  $\rho = \sum_k p_k \ketbra{k}{k}$ and $\sigma = \sum_k q_k \ketbra{k}{k}$, then 
\begin{equation} \label{eq-classical_distance_l1}
\nn d_1 ( \rho, \sigma) = d_1^{\rm \,clas} ( \pv , \qv ) = \sum_{k=1}^n | p_k - q_k | 
\end{equation}
is the $\ell^1$-distance, and
\begin{equation} \label{eq-classical_distance}
d_{\rm B} ( \rho, \sigma) = d_{\rm H}^{\rm \,clas} ( \pv ,\qv)
  = \biggl( \sum_{k=1}^n ( \sqrt{p_k} - \sqrt{q_k} )^2 \biggr)^\onehalf = \Bigl( 2 - 2 \sum_{k=1}^n \sqrt{p_k q_k}
\Bigr)^\onehalf
\end{equation}
is the Hellinger distance.
A distance closely related  to $d_{\rm H}^{\rm \,clas}$ is the so-called statistical distance 
$\Theta^{\rm\, clas} ( \pv, \qv) = \arccos ( 1 - d_{\rm H}^{\rm \,clas} ( \pv, \qv)^2/2)$, \ie the angle between the vectors $\xv = ( \sqrt{p_k} )_{k=1}^{n}$
and $\yv  = ( \sqrt{q_k} )_{k=1}^{n} $ on the unit sphere. Given two non-commuting states $\rho$ and $\sigma$, one can consider the
distance $d^{\rm\, clas}( \pv,\qv)$ between the outcome probabilities $\pv$ and $\qv$ of a measurement performed on the system in states $\rho$ and $\sigma$,
respectively. It is natural to ask whether there is a relation between $d( \rho, \sigma)$ and the supremum of $d^{\rm\, clas}( \pv,\qv)$ over all measurements.

\vspace{2mm}

\begin{proposition} \label{prop-link_classical_fidelity}
For any $\rho, \sigma \in \states ( \Hh)$,
\begin{equation} 
\label{eq-link_Bures_distance_classical}
d_1 ( \rho, \sigma) = \sup_{\{ M_i\}} d_1^{\rm \,clas} ( \pv , \qv ) \quad , \quad
d_{\rm B} ( \rho, \sigma ) = \sup_{\{ M_i\}} d_{\rm H}^{\rm \,clas} ( \pv ,\qv)\;,
\end{equation}
where the suprema are over all POVMs $\{ M_i\}$ and $p_i = \tr ( M_i
\rho)$ (respectively $q_i = \tr ( M_i \sigma)$) is the probability of
the measurement outcome $i$ in the state $\rho$ (respectively
$\sigma$). Moreover, the suprema are achieved for von  Neumann measurements with rank-one projectors $M_i = \ketbra{i}{i}$. 
\end{proposition}

\vspace{1mm}

\proof We leave the justification of the first identity to the reader. It can be obtained by following similar arguments as 
in the proof of Proposition~\ref{prop-trace_distance} (see~\cite{Nielsen}).  Let us show the second identity.  
Given a  POVM $\{ M_i \}$, by taking advantage of the definition (\ref{eq-fidelity}) of the fidelity, the
polar decomposition $\sqrt{\rho} \sqrt{\sigma} = U | \sqrt{\rho} \sqrt{\sigma}|$, and the identity  $\sum_i M_i = 1$, one gets
\begin{equation} \label{eq-proof_fidelity_as_classical_F0}
\sqrt{F ( \rho, \sigma)} = \sum_i \tr ( U^\ast \sqrt{\rho} \sqrt{M_i}
\sqrt{M_i} \sqrt{\sigma} ) \;\leq \; \sum_i \sqrt{p_i q_i} \;.
\end{equation}
The upper bound comes from the
Cauchy-Schwarz inequality.  It remains to show that this bound can be attained for an appropriate choice of POVM.  
The Cauchy-Schwarz inequality holds with equality \ifif $\sqrt{M_i} \sqrt{\rho}\, U= \lambda_i \sqrt{M_i} \sqrt{\sigma}$ with $\lambda_i \in \complex$.  
Assuming $\sigma>0$ and observing that $\sqrt{\rho}\, U =
\sigma^{-\onehalf} | \sqrt{\rho} \sqrt{\sigma} |$, this identity can be recast as
\begin{equation} \label{eq-proof_fidelity_as_classical_F1}
\sqrt{M_i} ( R - \lambda_i ) =0 \quad \text{ with } \quad R=
\sigma^{-\onehalf} | \sqrt{\rho} \sqrt{\sigma} | \sigma^{-\onehalf}\;.
\end{equation}
Let $R = \sum_i r_i \ketbra{i}{i}$ be a spectral projection of the non-negative matrix $R$.  Taking $M_i$ to be the von Neumann projector
$M_i=\ketbra{i}{i}$ and $\lambda_i = r_i$, we find that
(\ref{eq-proof_fidelity_as_classical_F1}) is satisfied for all $i$. Thus $\sqrt{F( \rho, \sigma)}$ is equal to the \RHS of  
(\ref{eq-proof_fidelity_as_classical_F0}). If $\sigma$ is not invertible it can be approached by
invertible density matrices $\sigma_\varepsilon = ( 1- \varepsilon) \sigma + \varepsilon$, $\varepsilon>0$, and the result follows by continuity.  \finpro

\vspace{2mm}

Much as for the quantum relative R\'enyi entropies (Sec.~\ref{sec-cond_Renyi_entropy}), one may define another  distance on 
$\states ( \Hh)$ which also reduces to the Hellinger distance $d_{\rm H}^{\rm \, \clas}$ for commuting matrices, by setting
\begin{equation} \label{eq-Q_Hellinger_distance}
d_{\rm H} ( \rho, \sigma) = d_2 ( \sqrt{\rho}, \sqrt{\sigma} ) =  \Bigl( 2 - 2 \sqrt{F^{(\rm n)}_{\onehalf} (\rho || \sigma )} \Bigr)^\onehalf \;,
\end{equation}
where $F^{(\rm n)}_{\alpha} (\rho || \sigma )$ is the fidelity associated to the normal-ordered $\alpha$-entropy (\ref{eq-relative_Renyi_entropy_normal}),
namely,
\begin{equation} \label{eq-nomral-ordered_alpha_fidelity}
F^{(\rm n)}_{\alpha} (\rho || \sigma ) = \Bigl( \tr \bigl[ \rho^\alpha \sigma^{1-\alpha} \bigr] \Bigr)^{\frac{1}{\alpha}} = \E^{-\beta S_\alpha^{\rm (n)} ( \rho ||  \sigma)}
\quad , \quad  \beta = \frac{1-\alpha}{\alpha} \;.
\end{equation}
This distance is sometimes called the quantum Hellinger distance.
Thanks to Lieb's concavity theorem 
(Lem\-ma~\ref{lemma_Lieb_concavity_Ando_convexity}), $F^{(\rm n)}_{\alpha} (\rho ||\sigma )^\alpha$ is jointly concave in $(\rho,\sigma)$ for 
all $\alpha \in (0,1)$. Consequently, the square Hellinger distance $d_{\rm H} ( \rho, \sigma)^2$ is jointly convex, just as $d_{\rm B} (\rho,\sigma)^2$.
From Proposition~\ref{prop-convexity_implies_monotonicity} one then deduces that $d_{\rm H}$ is contractive. It is worth noting that $d_{\rm H}$ does 
not coincide with the Fubini-study distance (\ref{eq-def_Fubini_study}) for pure states (in fact, one finds
$F^{(\rm n)}_{1/2} (\rho_\psi || \sigma_\phi ) = | \braket{\psi}{\phi} |^4$).  
For any $\rho, \sigma \in \states ( \Hh)$, one finds by comparing (\ref{eq-def_Bures_dist_1}) and (\ref{eq-Q_Hellinger_distance}) that
$d_{\rm B} ( \rho, \sigma) \leq   d_{\rm H} ( \rho, \sigma) $.

\subsection{Comparison of the Bures  and trace distances} \label{sec_bounds_Bures_dist_and_trace}

The next result shows that the Bures and trace distances $d_{\rm B}$ and $d_1$ are equivalent and
gives optimal bounds of $d_1$ in terms of $d_{\rm B}$.

\vspace{2mm}

\begin{proposition} \label{prop_bounds_between_d_B_and_d_1}
For any $\rho, \sigma \in \states ( \Hh)$, one has
\begin{equation} \label{eq-bounds_on_d_1}
d_{\rm B} ( \rho,\sigma)^2 \leq d_1 ( \rho, \sigma ) \leq 2 \Bigl\{ 1 - \Bigl( 1 - \onehalf d_{\rm B} ( \rho, \sigma)^2 \Bigr)^2\Bigr\}^\onehalf \;.
\end{equation}
\end{proposition}

\vspace{1mm}

The lower bound has been first proven by Araki~\cite{Araki70} in the
$C^\ast$-algebra setting. We shall justify it from Proposition~\ref{prop-link_classical_fidelity} as in Ref.~\cite{Nielsen}.  The upper bound
is saturated for pure states, as shown in the proof below.
Note that this bound implies that $d_1 ( \rho, \sigma ) \leq 2 d_{\rm B} ( \rho, \sigma )$.

\vspace{1mm}

\proof We first argue that if $\rho_\psi = \ketbra{\psi}{\psi}$ and $\sigma_\phi = \ketbra{\phi}{\phi}$ are pure states, then 
$d_1 ( \rho_\psi, \sigma_\phi ) = 2 \sqrt{ 1 - F ( \rho_\psi, \sigma_\phi ) }$
and thus the upper bound in (\ref{eq-bounds_on_d_1}) is an equality.  Actually, let 
$\ket{\phi} = \cos \theta \ket{\psi} + \E^{\I   \delta} \sin \theta \ket{\psi^\bot}$, where 
$\theta , \delta \in [0,2\pi)$ and $\ket{\psi^\bot}$ is a unit vector orthogonal to $\ket{\psi}$.  Since $\rho_\psi - \sigma_\phi$ has
non-vanishing eigenvalues $\pm \sin \theta$, one has $d_1 ( \rho_\psi, \sigma_\phi ) = 2 |\sin \theta|$.  
But $F ( \rho_\psi, \sigma_\phi )= \cos^2 \theta$, hence the aforementioned statement is true.  It then follows from
Theorem~\ref{theo-Uhlmann} and from the contractivity of the trace distance with respect to partial trace operations
(Proposition~\ref{prop-trace_distance}) that for arbitrary $\rho$ and $\sigma \in \states ( \Hh)$,
\begin{equation}
d_1 ( \rho, \sigma) \leq 2 \sqrt{ 1 - F(\rho,\sigma)}\;.
\end{equation}
To bound $d_1 ( \rho, \sigma) $ from below, we use
Proposition~\ref{prop-link_classical_fidelity} and consider a generalized measurement $\{ M_i \}$ such that 
$\sqrt{F ( \rho, \sigma  )}= \sum_i \sqrt{p_i q_i}$ with $p_i = \tr ( \rho M_i)$ and $q_i = \tr( \sigma M_i)$. This yields
\begin{equation}
d_{\rm B} ( \rho, \sigma )^2 = \sum_i ( \sqrt{p_i} - \sqrt{q_i} )^2
\leq \sum_i | p_i -q_i | \leq d_1 ( \rho, \sigma)\;,
\end{equation}
where the last inequality comes from Proposition~\ref{prop-link_classical_fidelity}  again. \finpro

\vspace{2mm}

The following bound on  the relative entropy can be obtained from  (\ref{eq-def_alpha_fidelity}),  (\ref{eq-relative_Reyni_ent_alpha=1}), and Proposition~\ref{prop-monotonicity_in_alpha_of_S_alpha}
\begin{equation} \label{eq-rel_ent_bound}
S ( \rho || \sigma) \geq - 2 \ln \Bigl( 1 - \onehalf d_{\rm B} ( \rho, \sigma)^2 \Bigr) \geq-  \ln \Bigl( 1 - \frac{1}{4} d_1 ( \rho, \sigma)^2 \Bigr)
\;.
\end{equation}

\vspace{1mm}

\begin{exercice}
By taking advantage of the inequality $F ( \rho, \sigma) \geq  \tr ( \rho \sigma)$, which follows from (\ref{eq-fidelity}) and
the norm inequality $\| A\|_1 \geq \| A \|_2$,
one can establish another bound  on $S(\rho|| \sigma)$ in terms of the fidelity, which reads~\cite{Streltsov10} 
\begin{equation}
S ( \rho || \sigma ) \geq - S ( \rho) - \ln F ( \rho, \sigma)\;.
\end{equation} 
\end{exercice}

\vspace{1mm}

\begin{exercice}
The formula 
\begin{equation}
F (\rho,\sigma) = \frac{1}{4} \inf_{H > 0} \bigr\{ \tr ( H \rho ) + \tr ( H^{-1} \sigma ) \bigr\}^2 = \inf_{H > 0} \bigl\{ \tr ( H \rho ) \tr ( H^{-1} \sigma ) \bigr\} \;
\end{equation} 
can be easily proven with the help of Lemma~\ref{lemma_duality_formula} and Theorem~\ref{theo-Uhlmann}. The last expression is  due to  Alberti~\cite{Alberti83}.
\end{exercice}

\vspace{1mm}

\begin{exercice} \label{exo-bound_on_QSD}
We are now in position to show without much effort several results of Sec.~\ref{sec-qsd_2_states}.
{\rm 
\begin{itemize}
\item[(a)] The upper bound (\ref{eq-upper_bound_unambiguousQSD}) on the optimal success probability $P_{\rm S,u}^{\,\opt}$ in unambiguous discrimination
of two mixed states
can be established from Uhlmann's theorem, formula (\ref{eq-success_proba_unambiguous_2states}), 
and the fact that $P_{\rm S, u}^{\,\opt} ( \{ \rho_i , \eta_i\} ) \leq P_{\rm S, u}^{\,\opt} ( \{ \ket{\Psi_i} , \eta_i\} )$, where
$\ket{\Psi_i}$ is a purification of $\rho_i$ for any $i$~\cite{Rudolph03}.
\item[(b)] It is instructive to derive in the special case of $m=2$ states the lower bound  on $P_{\rm S,a}^{\,\opt}$ given in 
Proposition~\ref{prop_lower_and_upper_bounds_on_P_S_amb}  by 
using the Helstrom formula~(\ref{eq-opt_success_proba}), the fact that $\tr ( |\Lambda|) \geq \sum_i | \bra{i} \Lambda \ket{i} |$
for any \ONB $\{ \ket{i}\}$, and Proposition~\ref{prop-link_classical_fidelity}~\cite{Bergou_review}.
\item[(c)] The Uhlmann theorem gives an efficient way to calculate the fidelity between the two states (\ref{eq-def_rho_eq_rho_diff}) 
(the result is $F(\rho_{\rm eq}, \rho_{\rm diff} ) = | \braket{\psi_1}{\psi_2} |^2$).
\end{itemize}
}
\end{exercice}

\subsection{Bures and quantum Hellinger metrics, quantum Fisher information} \label{sec-Bures_metric}

Recall that a Riemannian metric on $\states ( \Hh)$ is a map $g$ which
associates to each $\rho \in \states ( \Hh)$ a scalar product $g_\rho$ on the
tangent space to $\states ( \Hh)$ at $\rho$. For any state $\rho$ on $\Hh$, this tangent space can be identified with the (real)  vector space
$\saobservables$ of self-adjoint operators on $\Hh$.  A metric $g$ defines a Riemannian distance $d$, which is such that
the square distance $\D s^2 = d ( \rho, \rho + \D \rho )^2$ between two infinitesimally close states $\rho$ and $\rho + \D \rho$ is given by
\begin{equation}
\D s^2 = g_\rho ( \D \rho , \D \rho) \;.
\end{equation}
The Hilbert-Schmidt distance $d_2$ is obviously Riemannian: its metric is constant and given by the scalar product
(\ref{eq-Hilbert_Schmidt_product}). In contrast, the trace distance $d_1$ is not Riemannian. 

Let us show that the Bures distance
$d_{\rm B}$ is Riemannian and determine its metric $g_{\rm B}$.
It is convenient to introduce a small parameter $t \in \real$. According to Definition~\ref{def-Bures_distance_and_fidelity} one has
\begin{equation} \label{eq-derivation_Bures_metric0}
d_{\rm B} ( \rho, \rho + t\, \D \rho )^2 = 2 - 2  \tr ( A ( t) ) 
\quad , \quad  A ( t ) =  \bigl( \sqrt{\rho} ( \rho + t \D \rho ) \sqrt{\rho} \bigr)^\onehalf\;.
\end{equation}
The scalar product $(g_{\rm B})_\rho$ will be given in terms of the eigenvectors $\ket{k}$ and eigenvalues $p_k$ of $\rho$ in the spectral 
decomposition $\rho= \sum_k p_k \ketbra{k}{k}$.
Using the notation  $\dot{A} (t) = \D A/ \D t$, $\ddot{A} (t) = \D^2 A /\D t^2$, and the identity  $A(t)^2 = \sqrt{\rho} ( \rho + t \D \rho ) \sqrt{\rho}$, one finds
\begin{eqnarray} \label{eq-derivation_Bures_metric}
\nn
\dot{A} (0) A ( 0) + A ( 0) \dot{A} (0) 
& = &  \sqrt{\rho} \,\D \rho \, \sqrt{\rho} \\
\ddot{A} (0) A ( 0) + 2 \dot{A} (0) \dot{A} (0) +  A ( 0) \ddot{A} (0) 
& = & 0
\end{eqnarray}
The first equation yields
\begin{equation}
( p_k + p_l ) \bra{k} \dot{A} (0) \ket{l} = \sqrt{p_k p_l}  \bra{k} \D \rho  \ket{l}\;.
\end{equation}
Since $\tr ( \D \rho ) = 0$, it follows that $\tr [ \dot{A} (0) ]= 0$. Assume that $A(0)=\rho$ is invertible. 
Multiplying the second equation in (\ref{eq-derivation_Bures_metric}) by $A(0)^{-1}$ and taking the trace, one verifies that
\begin{equation}
\tr [  \ddot{A} (0) ] 
 =  
- \tr \bigl[\dot{A} (0)^2  A(0)^{-1}  \bigr] = 
- \sum_{k,l=1}^n p_k^{-1} \bigl| \bra{k} \dot{A} (0) \ket{l} \bigr|^2
 =  
- \sum_{k,l=1}^n \frac{p_l | \bra{k} \D \rho \ket{l} |^2}{ (p_k+p_l)^2}
\;.
\end{equation}
Thus, going back to (\ref{eq-derivation_Bures_metric0}) we arrive at
\begin{equation}\label{eq-Bures_metric0}
d_{\rm B} ( \rho, \rho + t \D \rho )^2 = - \tr [ \ddot{A}(0) ] t^2 + \Oo ( t^3) 
=  ( g_{\rm B})_\rho ( \D \rho , \D \rho ) t^2 + \Oo ( t^3)
\end{equation}
with~\cite{Hubner92}
\begin{equation} \label{eq-Bures_metric}
(g_{\rm B})_\rho (A,A) = \onehalf \sum_{k,l=1}^n \frac{|\bra{k} A \ket{l} |^2}{p_k +
  p_l}
\quad , \quad A \in \saobservables\;,\;\rho >0 \;.
\end{equation}
The last formula defines a scalar product on $\saobservables$ by polarization, hence $d_{\rm B}$ is Riemannian with metric $g_{\rm B}$.
One readily obtains from this metric the infinitesimal volume element. The volume of $\states ( \Hh)$ and the area of its boundary
are determined in~\cite{Sommers03}.  

\vspace{2mm}

\begin{definition}
Given a state $\rho \in \states ( \Hh)$ and an observable $H \in \saobservables$, the non-negative number
\begin{equation} \label{eq-Fisher_info}
\Ff_Q ( \rho, H ) 
 = 4 ( g_{\rm B})_\rho \bigl( -\I [ H, \rho ], - \I [ H, \rho] \bigr)
  = 2  \sum_{k,l, p_k+p_l >0} \frac{(p_k-p_l)^2}{p_k + p_l} |\bra{k} H \ket{l} |^2
\end{equation}
is called the quantum Fisher information of $\rho$ with respect to $H$.
\end{definition}

\vspace{1mm}

The quantity $\Ff_Q ( \rho, H )$  has been introduced by Braunstein and Caves~\cite{Braunstein94} as a quantum analog of the Fisher information 
in statistics. Similarly to the definition of the Bures distance in Sec.~\ref{sec-def-Bures_distance},
these authors related it to the metric -- called the ``distinguishability metric'' by 
Wootters~\cite{Wootters81} -- extending  the Fubini-Study metric to mixed states.
For a pure state $\rho_\Psi = \ketbra{\Psi}{\Psi}$, the quantum Fisher information  reduces to the square quantum fluctuation of $H$, namely,
\begin{equation} \label{eq-Fisher_info_for_pure_states}
\Ff_Q ( \rho_\Psi, H ) = 4 \langle (\Delta H )^2 \rangle_\Psi  = 4 \bigl( \bra{\Psi} H^2 \ket{\Psi} - \bra{\Psi} H \ket{\Psi}^2 \bigr)\;.
\end{equation}
In general, $\sqrt{\Ff_Q ( \rho, H )}$ gives the speed at which a given state 
$\rho$ separates from its time-evolved state $\rho(t) = \E^{-\I t H} \rho \E^{\I t H}$ under the dynamics specified by the Hamiltonian $H$.
In fact,  by plugging $\D \rho/\D t  = - \I  [ H , \rho ]$ into 
(\ref{eq-Bures_metric0}) one checks that
\begin{equation}
\sqrt{\Ff_Q ( \rho, H )} = \biggl( 2 \frac{\D^2}{\D t^2}  \, d_{\rm B} ( \rho, \rho(t) )^2 \Bigr|_{t=0} \biggr)^\onehalf \approx \sqrt{2} 
\frac{ \delta d_{\rm B}}{\delta t}  \;.
\end{equation}
We postpone the discussion on the statistical interpretation of $\Ff_Q ( \rho, H )$ to Sec.~\ref{sec-interferometry} below. 
It will be argued there that $\Ff_Q ( \rho, H )$
measures the amount of quantum correlations in the state $\rho$  that can be used for improving precision in 
quantum metrology.

Let us now turn to the quantum Hellinger  distance (\ref{eq-Q_Hellinger_distance}).
We  proceed to determine the metric $g_\alpha$ associated to the normal-ordered relative  R\'enyi entropy (\ref{eq-relative_Renyi_entropy_normal}),
 from which the quantum Hellinger metric $g_{\rm H}$ is obtained by setting $\alpha=1/2$. We demonstrate that
the largest metric $ g_\alpha$ for $\alpha \in (0,1)$ is achieved for $\alpha=1/2$ and is equal to $g_{\rm H}/2$, a 
result that will be needed later on (Sec.~\ref{eq-Q_hyp_testing}). 
The metric $ g_\alpha$ is defined by 
\begin{eqnarray} \label{eq-infinitesimal_normal_alpha_ent}
\nn
S_\alpha^{\rm (n)} ( \rho + t \D \rho || \rho ) 
& = & (1-\alpha)^{-1} \bigl( 1- F_\alpha^{\rm (n)} ( \rho + t \D \rho || \rho )^\alpha \bigr)  + \Oo ( t^3)
\\
& = & 
t^2 (1-\alpha)^{-1} ( g_\alpha)_\rho  (\D \rho , \D \rho) + \Oo ( t^3)\;,
\end{eqnarray}
where $F_\alpha^{\rm (n)}$ is the $\alpha$-fidelity, see (\ref{eq-nomral-ordered_alpha_fidelity}). 
To determine $g_\alpha$  for all $\alpha \in (0,1)$, we use (\ref{eq-int_repres_A^alpha}) in Appendix~\ref{app_operator_convex_functions}
to write
\begin{eqnarray}
\nn
B_\alpha ( t) 
& = & 
\rho^{\alpha} -  ( \rho + t \D \rho)^{\alpha}  
  =   
 \frac{\sin (\alpha \pi )}{\pi} \int_0^\infty \D x \,x^{\alpha} \biggl(\frac{1}{x + \rho + t \D \rho} -  \frac{1}{x+\rho} \biggr)
\\
&=  & 
 \frac{\sin (\alpha \pi )}{\pi} \int_0^\infty \D x \, \,x^{\alpha} \biggl( - \frac{t}{x + \rho} \D \rho \frac{1}{x + \rho}
 + \frac{t^2}{x + \rho} \D \rho \frac{1}{x + \rho} \D \rho \frac{1}{x + \rho} \biggr)  + \Oo ( t^3)
 \;.
\end{eqnarray}
Introducing as before the spectral decomposition $\rho = \sum_k p_k \ketbra{k}{k}$ and using known integrals, one finds
\begin{eqnarray}  \label{eq-derivation_Hellinger_metric1}
\nn
1- F_\alpha^{\rm (n)} ( \rho + t \D \rho || \rho )^\alpha 
& = & \tr [   B_\alpha (t) \rho^{1-\alpha} ] \\   
& = &    - t \alpha \sum_{k=1}^n  \bra{k} \D \rho \ket{k} 
+ t^2 \sum_{k,l=1}^n  \frac{p_k^{1-\alpha} ( p_k^{\alpha} - p_l^{\alpha} ) }{(p_k-p_l)^{2}}  \bigl| \bra{k} \D \rho \ket{l} \bigr|^2 + \Oo (t^3)\;.
\end{eqnarray} 
Because $\tr ( \D \rho ) =0$, the linear term in $t$ vanishes as it should be. 
Plugging (\ref{eq-derivation_Hellinger_metric1}) into (\ref{eq-infinitesimal_normal_alpha_ent}) one gets
\begin{equation} \label{eq-metrics_assocoiated_to_alpha_ent}
( g_\alpha)_\rho  ( A , A ) = \sum_{k,l=1}^n c_\alpha (p_k,p_l)   |\bra{k} A \ket{l} |^2 \quad , \quad 
c_\alpha ( p, q ) = \frac{(p^{1-\alpha} - q^{1-\alpha} ) ( p^{\alpha} - q^{\alpha} )}{2 (p - q )^2}\;.  
\end{equation}
It is easy to show that $c_\alpha ( p,q ) \leq c_{1/2} (p,q)$ for any $p,q >0$, hence 
\begin{equation}  \label{eq-Hellinger_metric}
\max_{\alpha \in (0,1)} ( g_\alpha)_\rho  ( A , A ) = \bigl( g_{\onehalf} \bigr)_\rho ( A, A ) =  \sum_{k,l=1}^n \frac{ |\bra{k} A \ket{l} |^2}{2( \sqrt{p_k} + \sqrt{p_l} )^2}
\quad , \quad A \in \saobservables \;,
\end{equation}
as claimed above. Furthermore, in view of (\ref{eq-Q_Hellinger_distance}) we deduce  that the quantum Hellinger distance $d_{\rm H}$ is Riemannian and has
a metric $g_{\rm H} = 2 g_{1/2}$.

\subsection{Characterization of the Riemannian contractive distances} \label{sec-charac_contractive_dist}

The complete characterization of Riemannian contractive distances on $\states ( \Hh)$ for finite Hilbert spaces $\Hh$ has been given by Petz~\cite{Petz96}, 
following a work by Morozova and Chentsov~\cite{Morosova90}.  Such distances are induced by metrics $g$ satisfying %
\begin{equation} \label{eq-metric_contractive}
g_{\Mm ( \rho)} \bigl( \Mm (A), \Mm ( A) \bigr) \leq g_\rho ( A , A)
\quad , \quad A \in \saobservables\;,
\end{equation}
for any $\rho \in \states ( \Hh)$ and any \QO $\Mm : \Bb ( \Hh) \rightarrow
\Bb ( \Hh')$.

In the classical setting, it is remarkable that the contractivity
condition leads to a unique  metric (up to a multiplicative constant). Quantum operations correspond classically to Markov mappings 
$\pv \mapsto \Mm^\clas \pv $ on the probability simplex  $\states_{\rm clas} = \{ \pv \in \real_{+}^n ; \sum_{i} p_i = 1\}$, see (\ref{eq-stochastic matrix}),
with stochastic matrices $\Mm^\clas$ having non-negative elements $\Mm^\clas_{ij}$ such that
$\sum_i \Mm^\clas_{ij} =1$ for any $j=1,\ldots , n$. The
contractive distances $d^\clas$ on $\states_\clas$ satisfy $d^\clas ( \Mm^\clas \pv, \Mm^\clas \qv ) \leq d^\clas ( \pv, \qv)$ for any such matrices.
According to a result of Cencov~\cite{Cencov}, a Riemannian distance on $\states_\clas$ with metric $g^\clas$
 is contractive \ifif $g_\pv^\clas ( \av , \av ) = c \sum_k a_k^2/p_k$ for any $\av \in \real^n$ and
some $c>0$, that is, the infinitesimal distance
between a probability vector $\pv$ and a neighboring vector $\pv + \D \pv$  is proportional to
\begin{equation} \label{eq-Fisher_metric}
\D s^2_{\rm Fisher} =  \sum_{k=1}^n \frac{ \D p_k^2}{p_k}\;.
\end{equation}
The associated metric is known as the Fisher metric and plays an important role in statistics. 
It induces the Hellinger distance (\ref{eq-classical_distance}) up to a factor of one fourth.

Let us come back to the quantum case. Although $g_\rho$ is in principle defined on the
real vector space $\saobservables$ (the tangent space of $\states ( \Hh)$), one can extend it as a scalar product on the complex Hilbert space
$\observables$. Without loss of generality, one may require that this scalar product satisfies
\begin{equation} \label{eq-symmetric_g_rho}
g_\rho (A,B) =g_\rho (B^\ast, A^\ast)= \overline{g_\rho(A^\ast,
  B^\ast)} \quad , \quad A,B \in \observables\;.
\end{equation}
(for instance, this is the case for the Hilbert-Schmidt product (\ref{eq-Hilbert_Schmidt_product})).  
We first note that one can associate to $g$ a family 
$\{ \Kk_\rho; \rho \in \states ( \Hh)\}$ of positive operators on the Hilbert space $\observables$ endowed with 
the scalar product (\ref{eq-Hilbert_Schmidt_product}), by
setting
\begin{equation} \label{eq-def_metric_from_K}
g_\rho ( A, B ) = \bigl\langle A , \Kk_\rho^{-1} ( B ) \bigr\rangle \quad , \quad A,B \in \observables \;.
\end{equation}
Let us write $\rho_\Mm = \Mm ( \rho)$.
The monotonicity condition (\ref{eq-metric_contractive}) reads $\Mm^\ast \Kk_{\rho_\Mm}^{-1} \Mm \leq \Kk_\rho^{-1}$, which means that
$\Kk_\rho^{1/2} \Mm^\ast \Kk_{\rho_\Mm}^{-1} \Mm \Kk_\rho^{1/2}$
is a contraction.  This is equivalent to $K_{\rho_\Mm}^{-1/2} \Mm \Kk_\rho \Mm^\ast \Kk_{\rho_\Mm}^{-1/2}$ being a contraction.
Therefore $g$ is contractive \ifif
\begin{equation} \label{eq-condition_monotonous_metric}
\Mm \Kk_{\rho} \Mm^\ast \leq \Kk_{\Mm ( \rho)}
\end{equation}
for any $\rho$ and $\Mm$.

\vspace{2mm}

\begin{lemma} {\rm ~\cite{Petz96}} The contractivity condition (\ref{eq-condition_monotonous_metric}) is fulfilled by the
positive operators
\begin{equation} \label{eq-Kk_rho}
\Kk_\rho = \Rr^\onehalf_\rho f ( \Delta_\rho ) \Rr^\onehalf_\rho\;,
\end{equation}
where $\Rr_\rho$ stands for the right multiplication by $\rho$ (see (\ref{eq-left_and_right_multilication})), 
$\Delta_{\rho}= \Delta_{\rho  | \rho}$ is the modular operator defined in (\ref{eq-def_modular_op}), and
$f : \real_+ \rightarrow \real$ is an operator monotone-increasing function with values in $\real_+$.  
\end{lemma}

\vspace{1mm}

\proof Let us recall that the modular operators
$\Delta_\rho$ and $\Delta_{\rho_\Mm}$ on $\observables$ are (self-adjoint and) positive. In analogy with the proof of Theorem~\ref{theo-Petz}, we
introduce the contraction $\Cc_\Mm$ defined by (\ref{eq-def_VV_m}).
It has been observed in this proof that $\Cc_\Mm^\ast \Delta_\rho \Cc_\Mm \leq \Delta_{\rho_\Mm}$. 
Since asking that a continuous function $f : \real_+ \rightarrow \real$ be operator monotone-increasing and non-negative
is the same as asking that $f$ be operator concave (see Appendix~\ref{app_operator_convex_functions} and \cite{Bhatia}, Theorem~V.2.5),
it follows from  the Jensen-type inequality (\ref{eq-Jensen_for_operator_convex}) and the
monotonicity of $f$ that
\begin{equation} \label{eq-inequality_for_operator_monotone}
\Cc_\Mm^\ast f (\Delta_\rho ) \Cc_\Mm \leq f (\Delta_{\rho_\Mm} )\;.
\end{equation}
Multiplying both sides by $B' \rho^\onehalf_\Mm$ and taking the scalar product by the same vector, this is equivalent to
\begin{equation} \label{eq-inequality_for_operator_monotonebis}
\bigl\langle B' \, , \, \Mm \Rr^\onehalf_\rho f ( \Delta_\rho ) \Rr^\onehalf_\rho \Mm^\ast ( B' ) \bigr\rangle 
\leq 
\bigl\langle B' \, , \,\Rr_{\rho_\Mm}^\onehalf f ( \Delta_{\rho_\Mm} ) \Rr_{\rho_\Mm}^\onehalf ( B' ) \bigr\rangle
\end{equation}
for any $B' \in \Bb (\Hh')$.  Thus the operator $\Kk_\rho$ defined in (\ref{eq-Kk_rho}) satisfies the contractivity condition
(\ref{eq-condition_monotonous_metric}).
\finpro

\vspace{2mm}

Formulas (\ref{eq-def_metric_from_K}) and (\ref{eq-Kk_rho}) yield a family of monotonous metrics, in one-to-one correspondence with
non-negative operator monotone functions $f$.  These metrics are given by 
$g_\rho ( A , B ) = \langle A \rho^{-\onehalf} \, , \, f ( \Delta_\rho )^{-1} (B \rho^{-\onehalf}) \rangle$ for any $A,B \in \observables$.  
More explicitly, for any
$\rho$ with spectral decomposition $\rho = \sum_k p_k \ketbra{k}{k}$ one finds
\begin{equation} \label{eq-characterization_monotone_metrics}
g_\rho ( A , A ) = \sum_{k, l=1}^n c(p_k,p_l) | \bra{k} A \ket{l} |^2 \quad , \quad A \in \saobservables\;,
\end{equation}
where $c(p,q)$ is given by
\begin{equation} \label{eq-def_c}
c ( p, q)= \frac{p f (q/p) + q f( p/q)}{2 pq f (p/q) f (q/p)}
\end{equation}
and satisfies $c ( t p, t q )= t^{-1} c ( p, q)$ for any $t \in \real$, $t \not= 0$, and $c(p,p)= f(1)^{-1} p^{-1}$. By using 
$\Delta_\rho ( B^\ast) = (\Delta^{-1}_\rho ( B ) )^\ast$, it is easy to see that the
 condition (\ref{eq-symmetric_g_rho}) is satisfied \ifif $f(x)=x f(x^{-1})$.  In particular, by choosing the following operator
monotone functions (see Appendix~\ref{app_operator_convex_functions})
:
\begin{equation} \label{eq-monotone_functions_for_contractive_dist}
f_{\rm Harm} (x)= \frac{2 x}{x+1} \;\; \leq \;\;
f_{\rm KM} (x) = \frac{x-1}{\ln x}  \;\;\leq \;\; 
f_{\rm H} = \frac{( 1 + \sqrt{x})^2}{4}  \;\;\leq\;\;
f_{\rm B} (x) = \frac{x+1}{2}
\end{equation}
one is led to
\begin{equation}
c_{\rm Harm} (p,q)= \frac{p+q}{2 p q} \;\;\geq \;\; 
c_{\rm KM} (p,q) = \frac{\ln p - \ln q}{p-q} \;\;\geq \;\; 
c_{\rm H} (p,q) = \frac{4}{(\sqrt{p} + \sqrt{q} )^2} \;\;\geq \;\; 
c_{\rm B} (p,q) = \frac{2}{p+q}\;.
\end{equation}
In view of  (\ref{eq-Bures_metric}) and (\ref{eq-Hellinger_metric}), the last
choice $f_{\rm B}$ gives the Bures metrics and $f_{\rm H}$ gives the Hellinger metric up to a factor of one fourth.
The second choice corresponds to the so-called Kubo-Mori (or Bogoliubov) metric, which is associated to the relative von Neumann entropy.
Actually, by substituting (\ref{eq-metrics_assocoiated_to_alpha_ent}) into (\ref{eq-infinitesimal_normal_alpha_ent}) and taking $\alpha \rightarrow 1$ one obtains
\begin{equation}
S ( \rho + \D \rho || \rho ) = \frac{1}{2} \sum_{k,l=1}^n c_{\rm KM} (p_k,p_l) \bigl| \bra{k} \D \rho \ket{l} \bigr|^2 = \onehalf g_{\rm KM} (\D \rho , \D \rho) \;.
\end{equation}
According to the formula $S(\rho + t \D \rho) = S ( \rho) - t \tr ( \D \rho \ln \rho ) - S ( \rho + t \D \rho ||  \rho)$,
one also gets
\begin{equation}
g_{\rm KM} (\D \rho , \D \rho) = - \frac{\D^2 S( \rho + t \D \rho)}{\D t^2} \biggr|_{t=0}\;,
\end{equation}
$S$ being the von Neumann entropy (since $S$ is concave, the second derivative in the \RHS is non-positive and defines a scalar product
on $\observables$).  As stressed by Balian, Alhassid and
Reinhardt~\cite{Balian86}, this makes the Kubo-Mori metric  quite natural from a physical viewpoint.

\vspace{1mm}

A result due to Kubo and Ando~\cite{Kubo80} states that there is a one-to-one correspondence between operator monotone functions $f$ and
operator means, that is, maps $m : (R,L) \in \observables_+\times \observables_+  \mapsto m( R, L ) \in \observables$ satisfying
\begin{itemize}
\item[(a)] if $0 \leq R \leq T$ and $0 \leq L \leq N$ then $m(R , L) \leq m( T,N)$ (monotonicity);
\item[(b)] $C^\ast m ( R , L ) C \leq m ( C^\ast R C , C^\ast L C)$.
\end{itemize}
This correspondence is given by the formula
\begin{equation}
m_f ( R , L) = R^\onehalf f ( R^{-\onehalf} L R^{-\onehalf} )
R^\onehalf \;.
\end{equation}
By taking $f_{\rm Harm}$ and $f_{\rm B}$ as in (\ref{eq-monotone_functions_for_contractive_dist}) one obtains the
harmonic mean $m_{\rm Harm} (R , L) = (R/2)^{-1} + (L/2)^{-1}$ and the arithmetic mean $m_{\rm B} ( R , L ) = (R+L)/2$, respectively, and for
$f(x) = \sqrt{x}$ one gets the so-called geometric mean (for more detail see e.g.~\cite{Carlen}). 
The positive operators (\ref{eq-Kk_rho}) can be written as
\begin{equation}
\Kk_\rho = m_f ( \Rr_\rho, \Ll_\rho)\;.
\end{equation}
The theory of Kubo and Ando shows that the harmonic mean $m_{\rm Harm}$ and arithmetic mean $m_{\rm B}$ are respectively the smallest and largest
symmetric operator means.  Thus the Bures metric $g_{\rm B}$ is the smallest monotone metric among the family of metrics given by 
(\ref{eq-def_metric_from_K}) and (\ref{eq-Kk_rho})  with the normalization 
$g_\rho ( 1 , 1 ) = \tr ( \rho^{-1})$.  It turns out that
this family contains all contractive metrics, that is, all such metrics have the form (\ref{eq-characterization_monotone_metrics}).

\vspace{2mm}

\begin{theorem} {\rm (Petz~\cite{Petz96})} \label{theo-characterization_cont-metrics}
The distances with metrics $g$ given by
(\ref{eq-characterization_monotone_metrics}) are contractive for any non-negative operator monotone-increasing function $f(x)$ satisfying 
$f(x)=x  f(x^{-1})$. Conversely, any continuous metric $g: \rho \mapsto g_\rho$  on  $\states ( \Hh)$ may be obtained from
(\ref{eq-characterization_monotone_metrics}) by a choice of a suitable function $f$ with these properties.  In particular, there
is a one-to-one correspondence between continuous contractive  metrics satisfying $g_\rho ( 1, 1) = \tr ( \rho^{-1})$ and operator 
means. The Bures distance is the smallest of all contractive Riemannian distances with metrics 
satisfying this normalization condition.
\end{theorem}  

\vspace{1mm}

This theorem is of fundamental importance in geometrical approaches to
quantum information.

\vspace{1mm}

\proof The first statement has been proven above. Conversely, let   $g$ be a
 continuous contractive metric on $\states ( \Hh)$ and let us show that there exists an operator monotone function 
$f: \real_+ \rightarrow \real_+$ such that for any $\rho \in \states ( \Hh)$, $g_\rho $ is given by
 (\ref{eq-def_metric_from_K}) and (\ref{eq-Kk_rho}) or, equivalently, by (\ref{eq-characterization_monotone_metrics}) and 
(\ref{eq-def_c}). We first note that $g$ being contractive it is in particular unitary invariant, 
\ie $g_{U^\ast \rho U} ( U^\ast A U, U^\ast B   U) = g_\rho ( A, B )$ for any unitary $U$ 
(see Sec.\ref{sec-contractive_distances}). More generally, if the \QOs $\Mm$ and $\Ttt$ are 
 such that $\rho$, $A$, and $B$ are invariant under $\Ttt \circ \Mm$, then 
$g_{\Mm ( \rho)} ( \Mm (A), \Mm(B) ) = g_\rho ( A, B )$. 
 The main idea of the proof is to combine this invariance property with
the uniqueness of the contractive  classical distance.  
Denoting by $(g_\rho)_{ij,kl}=g_\rho ( \ketbra{i}{j}, \ketbra{k}{l} ) $
the matrix elements of the scalar product $g_\rho$ in an orthonormal eigenbasis $\{ \ket{k}\}$ of $\rho$, we  
need to prove that
\begin{equation} \label{eq-matrix_elements_metric1}
(g_\rho)_{ij,kl} = \delta_{ik} \delta_{jl}\, c ( p_i, p_j ) 
\end{equation}
where $\delta_{ik}$ is the Kronecker symbol.
To show that the matrix elements of $g_\rho$ vanish for $i\not= j$ and $(k,l) \not= (i,j)$,
 it suffices to establish that
\begin{equation} \label{eq-proof_characterization_contractive_dist1}
g_\rho \bigl( \ketbra{i}{j} + s \ketbra{k}{l} , \ketbra{i}{j} + s
\ketbra{k}{l} \bigr) = g_\rho \bigl( \ketbra{i}{j} -s \ketbra{k}{l} ,
\ketbra{i}{j} - s \ketbra{k}{l} \bigr)
\end{equation}
for $s=1$ and $s=\I$ (the result then follows by polarization).  If one of the indices $i$, $j$, $k$, and $l$ is
different from the three others, say $i \notin \{ j,k,l\}$, this comes from the invariance of $g$ under the 
unitary $U^{(i)} = \sum_k u_k^{(i)} \ketbra{k}{k}$ with $u_k^{(i)} = -1$ if $k=i$ and $1$ otherwise.
Hence $(g_\rho)_{ij,kl}=0$ when $i \not= j$ and $(i,j) \not= (k,l)$, $(l,k)$. Similarly, by choosing 
$u_k^{(i)}= \I$ if $i=k$ and $1$ otherwise, this is also true for $i \not= j$
and $(i,j) = (l,k)$.
The only non-vanishing matrix elements of $g_\rho$ are thus $(g_\rho)_{ii,kk}$ and $(g_\rho)_{ij,ij}$ 
for $i \not= j$.  

To determine $(g_\rho)_{ii,kk}$ we observe that the restriction of
$g_\rho$ to the space of matrices commuting with $\rho$ induces a contractive metric on the probability
simplex $\states_\clas$,
defined by $g^\clas_\pv (\av,\bv ) = g_\rho (\sum_k a_k \ketbra{k}{k}, \sum_k b_k \ketbra{k}{k})$ for any 
$\av, \bv \in \states_\clas$. Indeed,  one can
associate a \QO $\Mm$ to  a stochastic matrix $\Mm^\clas$ by defining 
$\Mm ( \ketbra{k}{l} ) = \delta_{kl} \sum_{j} \Mm^\clas_{jk} \ketbra{j}{j}$
($\Mm$  has the Kraus form (\ref{eq-Kraus_decomp}) as $\Mm^\clas_{jk} \geq 0$ and $\sum_j \Mm^\clas_{jk}=1$ for any $k$). 
Then $\Mm ( \rho)= \sum_j (\Mm^\clas \pv )_j \ketbra{j}{j}$ where $\pv$ is the vector of eigenvalues of $\rho$, 
and (\ref{eq-metric_contractive})
implies that $g^\clas$ is contractive under $\Mm^\clas$. According to the uniqueness of the contractive classical  metrics,
one has
\begin{equation} \label{eq-matrix_elements_metric2}
(g_\rho)_{ii,kk} = g^\clas_\pv ( {\boldsymbol\delta}_i, {\boldsymbol\delta}_k ) = c
  \,\frac{\delta_{ik}}{p_k}\;,
\end{equation}
with $c> 0$ and ${\boldsymbol\delta}_i = (\delta_{il})_{l=1}^n$.

We now turn to the matrix elements $(g_\rho)_{ij,ij}$ for $i \not= j$. By unitary invariance, it is enough to determine
$(g_\rho)_{12,12}$. To this end, we consider the \QOs $\Mm$ from the space $\observables$ of $n \times n$ matrices to the 
space $\Bb (\complex^3)$ of $3 \times 3$ matrices and $\Ttt : \Bb (\complex^3) \rightarrow\observables$
with Kraus operators $\{ A_i\}_{i=2}^n$ and $\{ B_i\}_{i=2}^n$, respectively, given by
\begin{equation}
A_2 = B_2 = \ketbra{1}{1} + \ketbra{2}{2} \quad , \quad A_i =
\ketbra{3}{i} \quad , \quad B_i = \frac{\sqrt{p_i}}{\sqrt{1 - p_1 -
    p_2}} \ketbra{i}{3} \quad ,\quad i = 3, \ldots, n\;.
\end{equation}
A simple calculation yields $\Ttt \circ \Mm ( \rho ) = \rho$. As stressed above, one can deduce from the contractivity of
$g_\rho$ that $(g_\rho )_{12,12}=(g_{\Mm (\rho)} )_{12,12}$, thereby showing that this matrix element depends on
$p_1$ and $p_2$ only.  By unitary invariance, $(g_\rho )_{ij,ij}$ only
depends on $p_i$ and $p_j$ and one can set
$(g_\rho )_{ij,ij} = c ( p_i,p_j)$  for $i \not= j$, $c(p,q)$ being independent of $\rho$.
This complete the proof of (\ref{eq-matrix_elements_metric1}), excepted that it remains to justify that 
$c(p,p)=c/p$.
 
We proceed by showing that $c(q,p)$ is given by (\ref{eq-def_c}) with $f$ having the desired properties.
Thanks to (\ref{eq-symmetric_g_rho}), we know that $c(p,q)$ is real and symmetric.  One verifies that       
$c(p,p)= c/p$ by the following argument. Let us assume that $\rho$ has a degenerate eigenvalue, say
$p_1=p_2$. Then $\rho= U \rho U^\ast$ for any unitary $U$ acting trivially
on $\Span \{ \ket{3}, \ldots, \ket{n}\}$. By unitary invariance, 
$g_\rho ( \ketbra{\psi}{\psi} ,\ketbra{\psi}{\psi})= (g_\rho)_{11,11}=c/p_1$ for any 
$\ket{\psi} \in \Span \{ \ket{1}, \ket{2}\}$. Taking e.g.  $\ket{\psi} = (\ket{1}+\ket{2})/\sqrt{2}$
and using (\ref{eq-matrix_elements_metric1}), we get $(g_\rho)_{12,12} = c(p_1,p_1)=c/p_1$. 
In order to establish
that $c(p,q)$ is homogeneous we consider the \QOs 
$\Mm : \observables \rightarrow \Bb ( \Hh \otimes \Hh_\EE)$ and 
$\Ttt : \Bb ( \Hh \otimes \Hh_\EE) \rightarrow \observables$
defined by $\Mm ( \rho) = \rho \otimes 1 /n_\EE$ and  
$\Ttt (\widehat{\rho}) = \tr_\EE ( \widehat{\rho} )$
(here $n_\EE$ is the dimension of $\Hh_\EE$). Clearly, 
$\Ttt \circ \Mm = 1$, thus by similar arguments as above and by taking advantage of 
(\ref{eq-matrix_elements_metric1}), one finds
\begin{equation}
c (p_i,p_j) = (g_\rho)_{ij,ij} = g_{\Mm ( \rho)} ( \Mm(\ketbra{i}{j}), \Mm(\ketbra{i}{j}) ) 
= n_\EE^{-1} c \Bigl( \frac{p_i}{n_\EE}, \frac{p_j}{n_\EE} \Bigr)\;.
\end{equation}
As this is true for any positive integer $n_\EE$ and any state $\rho$, one concludes that $c (t p,t q) = t^{-1} c (p, q)$
for all $p,q \in [0,1]$ and all rationals $t$ with $tp$, $t q \in [0,1]$.  This is the point where we 
need the continuity of the metric to make sure that $c(p,q)$ is continuous. Then the equality
 holds for all real $t$.  Setting $f ( x) = 1/c ( x, 1)$ and using the symmetry of $c(p,q)$, 
one easily derives the identities (\ref{eq-def_c}) and $f(x^{-1})= x^{-1} f(x)$.  Furthermore, $f(1)^{-1}= c(1,1)=c$.

To complete the proof, we have to show that $f$ is operator concave. 
With this aim, let us
consider the inequality (\ref{eq-condition_monotonous_metric}) which
is equivalent to $g_\rho$ being contractive.  We choose $\Mm$ in this inequality to be the partial trace operation 
$\Ttt : \widehat{\rho} \mapsto \tr_{\complex^2} ( \widehat{\rho} ) \otimes 1/2$ on $\Bb( \Hh \otimes \complex^2)$
and $\widehat{\rho}= (\rho_0 \otimes \ketbra{0}{0} + \rho_1 \otimes \ketbra{1}{1})/2$.  
From (\ref{eq-condition_monotonous_metric}) we find that for any $A \in \observables$,
\begin{equation} \label{eq-proof_classification_cont_metrics}
\bigl\langle \Ttt^\ast ( A \otimes 1 ) \, , \, \Kk_{\widehat{\rho}} \Ttt^\ast ( A \otimes 1) \bigr\rangle 
\leq 
 \bigl\langle A \otimes 1 \, , \, \Kk_{\Ttt (\widehat{\rho})} ( A \otimes 1) \bigr\rangle\;.
\end{equation}
But $\Kk_{\widehat{\rho}} ( A \otimes 1) = (\Kk_{\rho_0} (A) \otimes \ketbra{0}{0} + \Kk_{\rho_1} (A) \otimes \ketbra{1}{1})/2$.
Accordingly, (\ref{eq-proof_classification_cont_metrics}) reduces to
\begin{equation}
\onehalf \bigl\langle A \, , \, ( \Kk_{\rho_0} + \Kk_{\rho_1} ) A \bigr\rangle
\leq \bigl\langle A \, , \, \Kk_{(\rho_0+\rho_1)/2} A \bigr\rangle\;,
\end{equation}
thereby showing that the map
\begin{equation}
\rho \mapsto \Kk_\rho = f ( \Ll_\rho \Rr_\rho^{-1} ) \Rr_\rho
\end{equation}
is  mid-point  concave. By  a  standard  argument  based on  a  dyadic decomposition,  it  follows that  this  map 
is  concave~\cite{Carlen}. Using the  $\ast$-isomorphism between the $C^\ast$-algebras $\Bb (  \Bb ( \Hh))$  and 
$\Bb ( \Hh\otimes  \Hh)$ (Sec.\ref{sec-states_and_observables}),  this is equivalent to say that the map
\begin{equation}
A \mapsto f \bigl( A \otimes (A^T)^{-1} \bigr)\, 1 \otimes A^T
\end{equation}
is concave. One easily deduces from this that the map $(A,B) \mapsto f ( A \otimes (B^T)^{-1} \bigr)\, 1 \otimes B^T$ is jointly concave.
In particular, $A \mapsto f ( A)$ is concave. This shows that $f$ is operator concave. 
\finpro

\newpage
\section{State discrimination and parameter estimation in large systems} \label{sec-QHT_and_parameter_est}

In this section we examine two problems related to the state discrimination talk discussed in Sec.~\ref{sec_qsd}, namely,
the quantum hypothesis testing and parameter estimation. In the first problem, one wants to 
determine asymptotically the probability of error in discriminating two states when one has $N$ independent copies of those states, for
$N \rightarrow \infty$. In the second problem, the goal is to estimate as precisely as possible a real parameter 
from measurements performed on a large number of particles in a state depending smoothly on this parameter.

\subsection{Quantum hypothesis testing: discriminating two states from many identical copies}
\label{eq-Q_hyp_testing}

An important issue in classical information theory is to discriminate two probability measures $\pv_1$ and $\pv_2$ on a measurable space $(\Omega,\Ff)$,
given the  outcomes
of $N$ independent identically distributed (i.i.d.) random variables, whose law is either $\pv_1$ or $\pv_2$. 
 Since one has to decide among two hypothesis -- the first (second) one being that the observed data is distributed according to $\pv_1$
 ($\pv_2$) -- this discrimination task bears the name of ``hypothesis testing''. 
For a given test function, \ie a random variable  $M_\clas$ with values in  $[0,1]$,
the probability of error is
$P_{{\rm err},N} = \eta_1 \pv_{1}^{(N)} ( M_\clas ) + \eta_2 \pv_{2}^{(N)} (1-M_\clas )$, where $\pv_{i}^{(N)} = \pv_i^{\otimes N}$ is the $N$-fold product measure and 
 $\eta_i$ the  prior probability attached to $\pv_i$.
 It is easy to convince oneself that the minimal error is achieved for the 
maximum likelihood  test function defined by\footnote{
Here $1_A$ stands for 
the indicator function on $A \subset \Omega$, \ie  $1_A (\omega) = 1$ if $\omega \in A$ and $0$ otherwise.
}
%
\begin{equation}
M^\opt_\clas  = 1_{\{\eta_2 \rho_{2}^{(N)} -\eta_1 \rho_{1}^{(N)} \geq 0\} }\;,
\end{equation}
 $\rho_{i}^{(N)} = \D \pv_{i}^{(N)}/\D \boldsymbol\mu^{(N)}$ being the density of $\pv_i^{(N)}$ with respect to the measure 
$\boldsymbol\mu^{(N)}=\pv_{1}^{(N)} + \pv_{2}^{(N)} = \boldsymbol\mu^{\otimes N}$. The corresponding error is
\begin{eqnarray} \label{eq-error_proba_class_hyp_testing}
\nn
P_{{\rm err},N}^{\,\opt}  ( \{ \pv_i^{(N)}, \eta_i\} ) 
& =  & 
\min_{0\leq M_\clas \leq 1} \biggl\{ \int_{\Omega^N} \D \boldsymbol\mu^{(N)}   \,\bigl( \eta_1 \rho_{1}^{(N)}  M_\clas  
+ \eta_2 \rho_{2}^{(N)} (1-M_\clas ) \bigr) \biggr\}
\\
& = & \int_{\Omega^N} \D \boldsymbol\mu^{(N)}   \min \bigl\{ \eta_1 \rho_{1}^{(N)} , \eta_2 \rho_{2}^{(N)}  \bigr\}\;.
\end{eqnarray}
One is typically interested in the limit  of a large number of tests, \ie $N\rightarrow \infty$. 
One can show that
the error probability decays  exponentially like $P_{{\rm err},N}^{\,\opt}  \sim \E^{-N \xi ( \pv_1, \pv_2)}$, 
with an  exponent given by the Chernoff bound~\cite{Chernoff52}
\begin{equation} \label{eq-Chernoff_bound}
\xi ( \pv_1,\pv_2)  = - \lim_{N \rightarrow \infty} \frac{1}{N} \ln P_{{\rm err},N}^{\,\opt}  ( \{ \pv_i^{(N)}, \eta_i\} ) 
=  - \inf_{\alpha \in (0,1)} \biggl\{ \ln \biggl( \int_\Omega \D \boldsymbol\mu \, \rho_1^{\alpha} \rho_2^{1-\alpha} \biggr) \biggr\} \;,
\end{equation} 
where we have set $\rho_i = \rho_{i}^{(1)}$. One recognizes in the infimum 
in the \RHS the classical R\'enyi divergence (\ref{eq-Reyni_divergence}) multiplied by $(\alpha-1)$.

In quantum mechanics,  the hypothesis testing  
can be rephrased as the discrimination of two $N$-fold tensor product states $\rho_1^{\otimes N}$ and $\rho_2^{\otimes N}$.
The corresponding  minimal error probability is given by the Helstrom formula (\ref{eq-opt_success_proba}), 
\begin{equation} \label{eq-Helstrom_formula_hypthesis_testing}
 P_{{\rm err},N}^{\,\opt}  ( \{ \rho_i^{\otimes N}, \eta_i\} ) = \onehalf \bigl( 1 - \tr | \Lambda_N | \bigr)
\quad , \quad \Lambda_N = \eta_1  \rho_1^{\otimes N} -  \eta_2  \rho_2^{\otimes N} \;,
\end{equation}
and the optimal measurement consists of the orthogonal projectors $M_\pm^\opt$ on the supports of the positive and negative parts of $\Lambda_N$.
Note that if  $\rho_1$ and $\rho_2$ commute then $M_-^\opt$ can be identified with the maximum likelihood  test function and one recovers
 the classical formula (\ref{eq-error_proba_class_hyp_testing}) from (\ref{eq-Helstrom_formula_hypthesis_testing}).
Surprisingly, the generalization of the Chernoff bound (\ref{eq-Chernoff_bound}) to the quantum setting has been settled out only recently. 
It has been highlighted  in Sec.~\ref{sec-cond_Renyi_entropy} that the R\'enyi divergences appearing in this bound have several natural
quantum extensions, according to the choice of operator ordering. It was proven  by 
Audenaert {\it et al.}~\cite{Audenaert07} and by Nussbaum and Szkola~\cite{Nussbaum09}  that the right extension
is the normal-ordered 
relative R\'enyi entropy $S_\alpha^{\rm (n)} ( \rho||\sigma)$ defined in (\ref{eq-relative_Renyi_entropy_normal}).

\vspace{2mm}

\begin{proposition} {\rm (Quantum Chernoff bound~\cite{Audenaert07,Nussbaum09})}  \label{prop-Q_Chernoff_bound}
One has
\begin{equation} \label{eq-Q_Chernoff_bound}
- \lim_{N \rightarrow \infty} \frac{1}{N} \ln P_{{\rm err},N}^{\,\opt}  ( \{ \rho_i^{\otimes N}, \eta_i\} ) 
 = - \inf_{\alpha \in (0,1)} \Bigl\{ \ln \bigl( \tr [ \rho^\alpha_1 \rho^{1-\alpha}_2 ] \bigr) \Bigr\} 
=  \sup_{\alpha \in (0,1)}\Bigl\{  ( 1 - \alpha) S_\alpha^{\rm (n)} ( \rho_1 || \rho_2 ) \Bigr\} \;.
\end{equation}
This limit defines a  jointly convex function $\xi_Q (\rho_1,\rho_2)$ with values in $\real_+ \cup \{+ \infty\}$, which is
contractive under quantum operations. Moreover, $\xi_Q$ induces the quantum Hellinger metric up to a factor of one half, that is,
if  $\rho$ and $\rho+\D \rho$ are infinitesimally close then $\xi_Q(\rho+ \D \rho,\rho ) = g_{\rm H} ( \D \rho, \D \rho)/2$ is given by (\ref{eq-Hellinger_metric}).
\end{proposition}

\vspace{1mm}

The infimum in
(\ref{eq-Q_Chernoff_bound}) is attained for a unique $\alpha  \in (0,1)$ satisfying
$\tr ( \rho^\alpha_1 \rho^{1-\alpha}_2 (\ln \rho_1 - \ln \rho_2 ))=0$~\cite{Audenaert07}.
Actually, for any fixed $\rho$ and $\sigma$,
the function $\alpha \mapsto F^{\rm (n)}_\alpha ( \rho || \sigma )^\alpha = \tr [ \rho^\alpha \sigma^{1-\alpha} ]$ is convex
(this is a simple consequence of the convexity of $\alpha\mapsto p^\alpha q^{1-\alpha}$ for $p,q>0$) and
 $ F^{\rm (n)}_\alpha ( \rho || \sigma ) \leq F^{\rm (n)}_{0,1} ( \rho || \sigma )  =1$ by
the H\"older inequality~(\ref{eq-main_property_L_P_norm}).
Before entering into the proof, let us also  mention that $\xi_Q ( \rho, \sigma) < \infty$ whenever $\rho$ and $\sigma$ do not have orthogonal supports. 
If 
$\rho = \ketbra{\psi}{\psi}$  is pure, the quantum Chernoff bound is related to the fidelity by  
$\xi_Q ( \rho, \sigma)= - \ln F (\rho,\sigma)=  - \ln \bra{\psi} \sigma \ket{\psi} $
(in fact, then  $F^{\rm (n)}_\alpha ( \rho || \sigma )^\alpha= \bra{\psi} \sigma^{1-\alpha} \ket{\psi}$ is minimum for $\alpha=0$).

\vspace{1mm}

\proof 
To shorten notation we write $P_{{\rm eff},N}^{\,\opt}$ when referring to 
$P_{{\rm err},N}^{\,\opt}  ( \{ \rho^{\otimes N}, \eta, \sigma^{\otimes N} ,  1-\eta \} )$. 
The fact that
\begin{equation} \label{eq-proof_Q_Chernoff_bound0}
\limsup_{N \rightarrow \infty} \frac{1}{N} \ln P_{{\rm eff},N}^{\,\opt} \leq -  \xi_Q ( \rho,\sigma) =  \inf_{\alpha \in (0,1)} \bigl\{ \ln ( \tr [ \rho^\alpha \sigma^{1-\alpha} ] ) \bigr\}
\end{equation}
follows from (\ref{eq-Helstrom_formula_hypthesis_testing}) and the trace inequality
\begin{equation} 
\onehalf \Bigl( \tr (A) + \tr (B) - \tr | A - B| \Bigr) \leq \tr ( A^{\alpha} B^{1-\alpha} ) \;,
\end{equation} 
where $A$ and $B$ are non-negative operators and $\alpha \in [0,1]$. This inequality has been first established in~\cite{Audenaert07}.
 A simple proof  
 due to N. Ozawa is reported in Appendix~\ref{app-norm_inequalities}.  The reverse inequality to (\ref{eq-proof_Q_Chernoff_bound0}) 
is a consequence of the classical Chernoff bound. This can be justified as follows~\cite{Nussbaum09}.
Let us observe that
the optimal \meas is a von Neumann measurement $\{\Pi^\opt , 1 - \Pi^\opt \}$ with $\Pi^\opt$ a projector, so that
\begin{eqnarray}
\nn
 P_{{\rm err},N}^{\,\opt}  =  1 - P_{{\rm S},N}^{\,\opt} 
& = & \eta \tr \bigl( (1-\Pi^\opt)  \rho^{\otimes N} \bigr) + (1-\eta ) \tr \bigl( \Pi^\opt  \sigma^{\otimes N} \bigr)
\\
& = &   
\sum_{\underline{k},\underline{l}} 
\Bigl( 
 \eta p_{\underline{k}} \bigl| \bra{\Phi_{\underline{l}} } (1- \Pi^\opt) \ket{\Psi_{\underline{k}}} \bigr|^2   
+ (1-\eta) q_{\underline{l}}  \bigl| \bra{\Psi_{\underline{k}} } \Pi^\opt  \ket{\Phi_{\underline{l}} } \bigr|^2 
\Bigr)
\;,
\end{eqnarray}
where $\{ \ket{\Psi_{\underline{k}} } \}$ and  $\{ \ket{\Phi_{\underline{l}} } \}$ are orthonormal eigenbases of 
$\rho^{\otimes N}$ and $\sigma^{\otimes N}$, respectively, and
$p_{\underline{k}}$ and $q_{\underline{l}}$ are the corresponding eigenvalues. We may without loss of generality  assume that $\eta \leq 1/2$. 
By using  the inequality $|a|^2 + |b|^2 \geq |a+b|^2/2$ one gets
\begin{equation}
\label{eq-proof_Q_Chernoff_bound}
P_{{\rm err},N}^{\,\opt}    \geq 
\eta  \sum_{\underline{k},\underline{l}}  \onehalf \min\{  p_{\underline{k}}  , q_{\underline{l}}  \} 
\bigl| \braket{\Phi_{\underline{l}} }{\Psi_{\underline{k}} } \bigr|^2  \;.
\end{equation}
But  $\rho^{\otimes N}$ corresponds to $N$ independent copies of the state $\rho=\sum_k p_k \ketbra{\psi_k}{\psi_k}$, hence its
 eigenvalues $p_{\underline{k}}$ and eigenvectors $\ket{\Psi_{\underline{k}} }$ are products of $N$ eigenvalues 
$p_{k}$ and $N$ eigenvectors $\ket{\psi_{k}}$ of $\rho$, respectively, and similarly for $\sigma^{\otimes N}$ with the eigenvalues $q_l$ and eigenvectors
$\ket{\phi_l}$ of $\sigma$.
This means that $p_{\underline{k}} | \braket{\Phi_{\underline{l}} }{\Psi_{\underline{k}} } |^2$ can be viewed as the $N$-fold product of 
the probability ${\boldsymbol\pi}_1$ on $\{ 1, \ldots , n \}^2$ defined by $(\boldsymbol\pi_1)_{kl} = p_{k} | \braket{\phi_{l}}{\psi_{k}} |^2$. Analogously, 
$q_{\underline{l}} | \braket{\Phi_{\underline{l}} }{\Psi_{\underline{k}} } |^2$ is the  $N$-fold product of 
 ${\boldsymbol\pi}_2$ with $(\boldsymbol\pi_2)_{kl}= q_{l} | \braket{\phi_{l}}{\psi_{k}} |^2$. Consequently, 
the sum  in (\ref{eq-proof_Q_Chernoff_bound}) is
 the minimal error probability $P_{{\rm err},N}^{\,\opt}  ( \{ {\boldsymbol\pi}_i^{(N)}, 1/2\})$ for discriminating $\boldsymbol\pi_1$ and 
$\boldsymbol\pi_2$ with equal prior probabilities
(see (\ref{eq-error_proba_class_hyp_testing})). One then deduces from the classical Chernoff bound (\ref{eq-Chernoff_bound}) that
\begin{equation}  \label{eq-proof_Q_Chernoff_bound2}
\liminf_{N \rightarrow \infty} \frac{1}{N} \ln P_{\rm err}^{\,\opt} 
 \geq  
 \inf_{\alpha \in (0,1)} \biggl\{ \ln \biggl( \sum_{k,l=1}^n (\boldsymbol\pi_1)_{kl}^\alpha (\boldsymbol\pi_2)_{kl}^{1-\alpha} \biggr) \biggr\}
= - \xi_Q ( \rho ,\sigma )\;.
\end{equation}
Together  with (\ref{eq-proof_Q_Chernoff_bound0}) this proves the quantum Chernoff bound. 

It is nevertheless instructive to show (\ref{eq-proof_Q_Chernoff_bound2}) directly from (\ref{eq-proof_Q_Chernoff_bound}),
without relying on the classical result, by using the theory of large deviations  for sums of i.i.d. random variables and
the relative modular operator $\Delta_{\sigma|\rho}$ (see Sec.~\ref{sec-states}), which appears here quite naturally~\cite{Jaksic12}. 
Indeed, let us set $\xi = \rho^\onehalf$ and note that for any real function $f : (0,\infty) \rightarrow \real$, according to (\ref{eq-def_modular_op})
and by the functional calculus, it holds
\begin{equation}  \label{eq-proof_Q_Chernoff_bound4}
\langle \xi\, ,\,f ( \Delta_{\sigma|\rho} ) \,\xi \rangle = \sum_{k,l=1}^n p_k f \Bigl( \frac{q_l}{p_k} \Bigr) | \braket{\phi_{l}}{\psi_{k}} |^2  \;.
\end{equation}
In particular,  $\langle \xi\, ,\,\ln ( \Delta_{\sigma|\rho} ) \, \xi \rangle = \tr [ \rho ( \ln \sigma - \ln \rho )] = - S ( \rho|| \sigma)$, as already
observed in Sec.~\ref{sec-relative_entropy}.  Let  $\mv_{\sigma|\rho}$ be the spectral measure of $-\ln \Delta_{\sigma|\rho}$ with respect to the vector 
$\xi$. This is a probability measure ($\xi$ is normalized), which is related to the relative entropy by $S(\rho||\sigma) = \int \D \mv_{\sigma | \rho} (t)\, t$. 
Taking $f(x) = \min \{ x, 1\}= g ( -\ln x )$ with $g(t )=\min\{ e^{-t},1\}$ in (\ref{eq-proof_Q_Chernoff_bound4}), one finds 
\begin{equation}  \label{eq-proof_Q_Chernoff_bound3}
\sum_{k,l=1}^n  \min\{  p_{k}  , q_{l}  \}  | \braket{\phi_{l} }{\psi_{k} } |^2 
 = \bigl\langle \xi \, , \, g ( - \ln  \Delta_{\sigma|\rho}  ) \,\xi \bigr\rangle =
\int_\real \D \mv_{\sigma| \rho} (t ) \,g ( t) 
\; \geq \;  \mv_{\sigma| \rho} ( \real_- )\;. 
\end{equation}
A similar inequality holds for the sum in the \RHS of (\ref{eq-proof_Q_Chernoff_bound}): it suffices to substitute $\Delta_{\sigma|\rho}$ by
$\Delta_{\sigma^{\otimes N}| \rho^{\otimes N}}=\Delta_{\sigma|\rho}^{\otimes N}$. The spectral measure  of 
$-\ln \Delta_{\sigma|\rho}^{\otimes N}$ is a product measure $\mv_{\sigma | \rho}^{(N)}$ and thus 
$-\ln \Delta_{\sigma|\rho}^{\otimes N}$ can be interpreted as a sum of i.i.d. random variables $ -\ln \Delta_{\sigma|\rho}^{(\nu)}$ 
 with law $\mv_{\sigma|\rho}$. The large deviation
principle  ensures that if  $e_{\sigma | \rho } '(0) < \theta <  e_{\sigma | \rho }' (1)$ then~\cite{Durrett}
\begin{equation} \label{eq-large_deviation}
\lim_{N \rightarrow \infty} \frac{1}{N} \ln  
\biggl(  \mv_{\sigma| \rho}^{(N)} 
  \biggl( - \sum_{\nu=1}^N \ln \Delta_{\sigma | \rho}^{(\nu)}  \leq -  \theta N  \biggr) 
\biggr) 
= -  \sup_{\alpha \in [ 0,1 ] } 
\bigl\{ \alpha \theta - e_{\sigma | \rho } ( \alpha) \bigr\}
\end{equation}
is up to a minus sign the Legendre transform of
\begin{equation} 
e_{\sigma | \rho }( \alpha) = \ln \biggl( \int_\real \D \mv_{\sigma | \rho } (t) e^{-t \alpha} \biggr) 
= \ln \bigl( \langle \xi \, ,\,\Delta_{\sigma | \rho}^\alpha\, \xi \rangle \bigr) 
=  \ln \bigl( \tr [ \rho^{1-\alpha} \sigma^\alpha ] \bigr) \;.
\end{equation}
If  $\rho \not = \sigma$ then  $ e_{\sigma | \rho }'(0)= - S ( \rho || \sigma) <0$ and
 $e_{\sigma | \rho }'(1) =  S ( \sigma || \rho ) > 0$ (the second identity follows from the first one by symmetry 
$e_{\sigma | \rho } ( 1 - \alpha) = e_{\rho | \sigma } (\alpha)$). Thus the large deviation bound (\ref{eq-large_deviation}) holds for $\theta=0$. 
Taking advantage of (\ref{eq-proof_Q_Chernoff_bound}) and (\ref{eq-proof_Q_Chernoff_bound3}) one is led to
\begin{eqnarray}
\nn
\liminf_{N \rightarrow \infty} \frac{1}{N} \ln P_{{\rm err},N}^{\,\opt} 
& \geq &  
\liminf_{N \rightarrow \infty} \frac{1}{N} \ln 
\biggl( 
 \sum_{\underline{k},\underline{l}}  \min\{  p_{\underline{k}}  , q_{\underline{l}}  \}  \bigl| \braket{\Phi_{\underline{l}} }{\Psi_{\underline{k}} } \bigr|^2 
\biggr)
\\
& \geq &
\lim_{N \rightarrow \infty} \frac{1}{N}   \ln  \biggl(  \mv_{\sigma | \rho}^{(N)} \biggl( - \sum_{\nu=1}^N \ln \Delta_{\sigma|\rho}^{(\nu)}  \leq 0  \biggr) \biggr) =
 \inf_{\alpha \in [ 0,1]} \bigl\{ e_{\sigma | \rho } ( \alpha) \bigr\} = - \xi_Q ( \rho, \sigma) \;,
\end{eqnarray}
in agreement with (\ref{eq-proof_Q_Chernoff_bound2}).
Note that these arguments justify in particular that the second member  in the classical Chernoff bound
(\ref{eq-Chernoff_bound}) is bounded from above by the third one, as a consequence of the large deviation principle. 
Applying  (\ref{eq-proof_Q_Chernoff_bound0}) for commuting matrices $\rho$ and 
$\sigma$, this gives a full proof of this classical bound.

The joint convexity of $\xi_Q ( \rho, \sigma)$ mentioned in the proposition results from the joint convexity of the relative entropies
$S_\alpha^{\rm (n)} (\rho||\sigma)$ for $\alpha \in (0,1)$, which follows from the Lieb concavity theorem, see
Sec.~\ref{sec-cond_Renyi_entropy}. One then gets the contractivity of $\xi_Q$ with respect to \QOs 
from Proposition~\ref{prop-convexity_implies_monotonicity}.
This concludes the proof.
\finpro

\vspace{2mm}

\begin{exercice}
The quantum Chernoff bound (\ref{eq-Q_Chernoff_bound}) can be generalized to the case where the two states $\rho_{i,N} \in \states (\Hh^{\otimes N})$ to discriminate 
are not product states (\ie for dependent copies). 

\vspace{1mm}

\noindent {\rm Actually, the large deviation principle used in the proof
is not restricted to sums of i.i.d. random variables. It must be assumed that the limit 
$e(\alpha)= \lim_{N \rightarrow \infty}N^{-1} \ln \tr [ \rho_{1,N}^\alpha\rho_{2,N}^{1-\alpha}]$ exists, is continuous in $\alpha$ on $[0,1]$ 
and differentiable on $(0,1)$, and
its right derivative $e'(0)$ is smaller than its left derivative $e'(1)$ (see~\cite{Jaksic12}).}
\end{exercice}

\vspace{1mm}

\begin{exercice}
In asymmetric hypothesis testing one is interested by the minimal error probability 
of identifying the second state under the constraint that the error on the  identification of the first state  is smaller than $\varepsilon$,
\begin{equation}
P_{{\rm err},N, \varepsilon}^{{\rm asym}} = \min_{0 \leq M \leq 1} \bigl\{  \tr [ M \rho_2^{\otimes N} ] \,;\, \tr [ (1-M) \rho_1^{\otimes N} ] \leq \varepsilon \bigr\}\;.
\end{equation}
The quantum Stein's lemma~\cite{Petz91,Ogawa00} shows that this probability decays exponentially with a rate given by the relative von
 Neumann entropy, \ie
\begin{equation}
- \lim_{N \rightarrow \infty} \frac{1}{N} \ln P_{{\rm err},N, \varepsilon}^{{\rm asym}} = S ( \rho_1 || \rho_2 )\;.
\end{equation}
The limit one gets by replacing the fixed parameter $\varepsilon >0$  by
$e^{-r N}$ (that is, asking for an exponentially decaying error on the identification of $\rho_1$) is, in turn, given by the
Hoeffding bound (see e.g.~\cite{Jaksic12} for more detail).
\end{exercice}

\vspace{1mm}

An interesting link between the quantum hypothesis testing  and fluctuation theorems in quantum statistical physics has been found
by Jak\u{s}i\'c {\it et al.} ~\cite{Jaksic12}. They have shown that the quantum Chernoff bound for discriminating the forward and backward time-evolved 
states $\rho_{\pm T/2}$ as $T \rightarrow \infty$ appears in the large deviation principle for the full counting statistics of measurements of the 
energy/entropy flow over the time interval $[0,T]$.

\subsection{Parameter estimation in quantum metrology} \label{sec-interferometry}

The parameter estimation problem is a kind of continuous version of quantum state discrimination, in which
the system state  $\rho (\theta)$ depends on a continuous parameter $\theta$.
One aims at estimating this unknown parameter with the highest
possible precision $\Delta \theta$ by performing  measurements on $\rho (\theta)$. This precision is limited by our ability to distinguish 
the states $\rho (\theta)$ for values of $\theta$ differing by $\Delta \theta$.

\subsubsection{Phase estimation in Mach-Zehnder interferometers} \label{eq-Mach_Zehnder}

An important example is phase estimation in the Mach-Zehnder interferometer represented in Fig.~\ref{fig-2}. 
An input photon passes through a beam splitter~\cite{Bouwmeester} which transforms its state
into a superposition of two modes propagating along different paths. These two modes 
 acquire distinct phases $\theta_1$ and $\theta_2$ during the propagation and are finally recombined in a second beam splitter
to read out interference fringes, from which
the phase difference $\theta=\theta_1-\theta_2$ is inferred.
The interferometric sequence can be described by means of rotation
matrices acting on the two-mode photon state.  We shall assume at this point that the reader is familiar with second quantization\footnote{
A good mathematical introduction to this formalism can be found in~\cite{Bratteli}.
}.
The generators of the aforementioned rotations  are the angular momentum operators
 $J_x$, $J_y$, and $J_z$ related to the bosonic annihilation and creation operators
$b_j$ and $b_j^\ast$ of a photon in mode $j=1,2$ by $J_x=( b^\ast_1 b_2+  b^\ast_2 b_1)/2$,
$J_y=-\I ( b^\ast_1 b_2- b^\ast_2 b_1)/2$, and $J_z = ( b^\ast_1 b_1-  b^\ast_2 b_2)/2$ (Schwinger representation).
These operators act on the bosonic Fock space $\Ff_{\rm b} ( \complex^2)$ associated to  the single photon space $\Hh \simeq \complex^2$.
The output state of the interferometer is given in terms of the input state $ {\rho}_{\inp}$ by~\cite{Yurke86}
\begin{equation}
{\rho}_{\out}(\theta) = e^{-\I \theta  J_{\nv}} {\rho}_{\inp} e^{\I \theta {J}_{\nv}}\;,
\end{equation}
where $\theta$ is the phase to be estimated and ${J}_{\nv}= n_x {J}_x + n_y {J}_y + n_z {J}_z$ the angular momentum 
in the direction specified by the unit vector $\nv \in \real^3$.

One can also realize a Mach-Zehnder  interferometer with ultracold atoms forming a  Bose-Einstein condensate in an optical trap, instead of photons. 
Then the two modes correspond to two distinct atomic energy levels and  
the total number of atoms $N_{\rm p}=N_1+N_2$ in these modes is fixed. In such a case the Hilbert space of the system has finite dimension
$N_{\rm p}+1$ (one deals here with indistinguishable particles).  Atom interferometry in Bose-Einstein condensates is  very promising due to the tunable 
interactions between atoms, which make 
it possible to generate dynamically  entangled states involving a large number of particles\footnote{
In contrast,  because of the absence of direct interactions between photons it is difficult to generate large numbers of 
photons having multipartite entanglement.
}.
We will see below that using such entangled states as inputs 
leads to smaller errors $\Delta \theta$ in the phase estimation than for separable inputs. For independent (\ie separable) particles the precision
is  of the order of $(\Delta \theta )_{\rm SN}\approx 1/\sqrt{N_{\rm p}}$ (shot noise limit).
Higher precisions than  $(\Delta \theta )_{\rm SN}$ 
have been reported experimentally~\cite{Gross10,Riedel10}. Important potential applications of these ultra-precise interferometers include
 atomic clocks and magnetic sensors with  enhanced sensitivities~\cite{Wasilewski10,Schleier10}.

\begin{figure}
\begin{center}
\includegraphics[width=8cm]{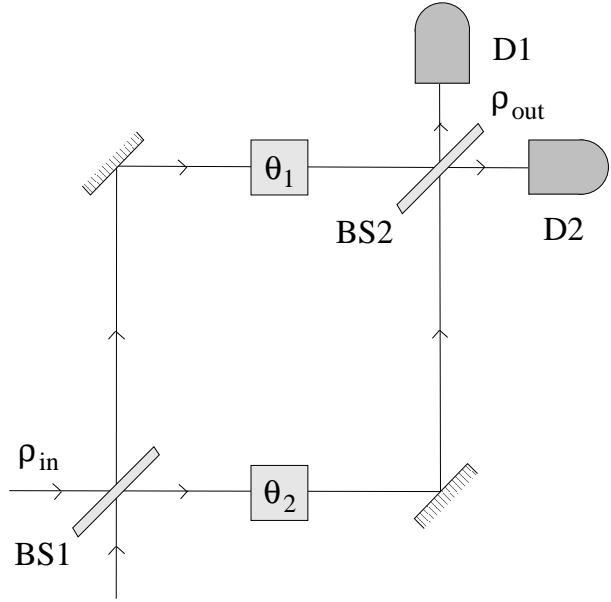}
\end{center}
\caption{In a Mach-Zehnder interferometer, the light entering in one of the two input modes
is split into two beams by a beam splitter (represented by the rectangle BS1 inclined by $45^\circ$).
The photons in the first and second beams acquire some phase shifts $\theta_1$ and $\theta_2$,
respectively. They then go through a second beam splitter (rectangle BS2) and finally into the detectors D1 and D2, which
count the number of photons in the two output modes.
}
\label{fig-2}
\end{figure}

\subsubsection{Quantum Cram\'er-Rao bound}

In the more general setting, the problem of estimating an unknown parameter $\theta$ from a $\theta$-dependent state evolution and  measurements on the output states
can be described as follows. For simplicity we assume that the evolution is given by a self-adjoint operator $H$ (equal to $J_{\nv}$ in the
above Mach-Zehnder interferometer), \ie
\begin{equation} \label{eq-rho(theta)}
\rho ( \theta) = \E^{-\I \theta H } \rho \,\E^{\I \theta H }\;,
\end{equation}
where $\rho= \rho(0)= \rho_{\rm in}$ is the input state. One performs generalized measurements given by a POVM $\{ M_i\}_{i=1}^m$ on the output state 
$\rho ( \theta)= \rho_{\rm out}$.
The probability to get the outcome $i$ is $p_{i| \theta} = \tr [ M_i \rho ( \theta)]$ (Sec.~\ref{sec-generalized_meas}). 
After $N$ independent measurements\footnote{
In practice the experiment is repeated $N$ times, starting  from the same initial state $\rho$ and in similar conditions, so that the quantum evolution
can be considered to be the same at each run.
}
 on copies of $\rho ( \theta)$ yielding the outcomes $i_1, i_2, \ldots, i_N$,
the parameter $\theta$ is estimated by using a statistical estimator  depending on these outcomes, that is,  
 a function $\theta_{\rm est} ( i_1, i_2, \ldots, i_N)$.
The precision of the estimation is defined by the variance
\begin{equation}
\Delta \theta = \biggl\langle \biggl( \Bigl| \frac{\partial \langle \theta_{\rm est} \rangle_\theta}{\partial \theta} \Bigr|^{-1} \theta_{\rm est} - \theta  \biggr)^2 
\biggr\rangle_\theta^\onehalf \;,
\end{equation}
where $\langle\, \cdot \, \rangle_\theta$ denotes the average for the product probability measure 
$\{ p_{i_1|\theta} \ldots  p_{i_N |\theta} \}_{i_1,\ldots, i_N =1}^m$ of the independent outcomes.
The factor $| \partial \langle \theta_{\rm est} \rangle_\theta / \partial \theta |^{-1}$ is put in front of $\theta_{\rm est}$ to remove some possible differences
 in physical units between $\theta$ and its estimator $\theta_{\rm est}$ (see~\cite{Braunstein94}).
We restrict our attention to unbiased estimators satisfying 
$| \partial \langle \theta_{\rm est} \rangle_\theta / \partial \theta |^{-1} \langle  \theta_{\rm est} \rangle_\theta = \theta$.
For a given input state ${\rho}$, one looks for the smallest error $\Delta \theta $  that can be achieved. 
This involves two different optimization steps, associated to the optimization over (i) all possible estimators  $\theta_{\rm est}$ and
(ii) all possible  measurements.
The step (i)  relies on a classical result in statistics known as   the Cram\'er-Rao bound,
\begin{equation}
\label{eq-Cramer_Rao}
\bigl\langle ( \Delta \theta_{\rm est})^2 \bigr\rangle_\theta \geq \frac{1}{N \Ff ( \{ p_{i|\theta} \}) } 
\Bigl( \frac{\partial \langle \theta_{\rm est} \rangle_\theta }{\partial \theta} \Bigr)^2 \;,
\end{equation}
where 
 $\Delta \theta_{\rm est} = \theta_{\rm est} - \langle \theta_{\rm est} \rangle_\theta$ and
\begin{equation}
\Ff ( \{ p_{i|\theta}\})  = \sum_{i=1}^m \frac{1}{p_{i|\theta}} \Bigl( \frac{\partial p_{ i | \theta}}{\partial \theta} \Bigr)^2
\end{equation}
is the Fisher information.
The inequality (\ref{eq-Cramer_Rao}) is saturated asymptotically for $N \rightarrow \infty$ by the maximum-likelihood estimator.
The second optimization step (ii) has been solved in Ref.~\cite{Braunstein94}, leading to the following important statement.
Recall that the  quantum Fisher information is defined as (see Sec.~\ref{sec-Bures_metric})
\begin{equation}
\Ff_Q ( \rho  , H )  = 4 (g_{\rm B})_\rho ( -\I [ H, \rho],  -\I [ H, \rho]) = 4 d_{\rm B} ( \rho, \rho + \D \rho)^2\;,
\end{equation}
where $g_{\rm B}$ is the Bures metric and $\D \rho = (\partial \rho/\partial \theta) \D \theta = - \I [ H, \rho] \D \theta $.

\vspace{2mm}

\begin{proposition} {\rm (Braunstein and Caves~\cite{Braunstein94})} \label{prop_Braunstein_Caves}
The smallest error $\Delta \theta$ that can be achieved in the parameter estimation is
\begin{equation}
\label{eq-quantum_Cramer_Rao}
(\Delta \theta)_{\rm{best}}
= \frac{1}{\sqrt{N} \sqrt{\Ff_Q ( \rho , H ) }}\;,
\end{equation}
where $N$ is the number of measurements and $\Ff_Q ( \rho  , H ) $ is the quantum Fisher information.
Thus $\Delta \theta \geq (\Delta \theta)_{\rm{best}}$ and the equality $\Delta \theta = (\Delta \theta)_{\rm{best}}$ 
can be reached asymptotically as $N \rightarrow \infty$.
\end{proposition}

\vspace{1mm}

It is worth noting that (\ref{eq-quantum_Cramer_Rao}) can be interpreted 
as a generalized uncertainty principle~\cite{Braunstein94}. In fact, if  $\rho= \ketbra{\Psi}{\Psi}$ is a pure state,
in view of the relation (\ref{eq-Fisher_info_for_pure_states})  between $\Ff_Q ( \rho   , H ) $ and the square fluctuation 
$\langle (\Delta H )^2\rangle_\Psi$ of $H$,
 the bound  $\Delta \theta \geq (\Delta \theta)_{\rm{best}}$ can be written as 
\be \label{eq-uncertainty_principle}
\Delta \theta \,\langle ( \Delta H )^2  \rangle^{\onehalf}_\Psi   \geq \frac{1}{2 \sqrt{N}}\,.
\ee
In this  uncertainty relation  $H$ plays the role of the variable conjugated to the parameter $\theta$.

\vspace{1mm}

\proof We present here  a direct proof of (\ref{eq-quantum_Cramer_Rao}) based on the results of Sec.~\ref{sec-Bures_distance} 
(see~\cite{Braunstein94} for an independent proof).
Before that, let us explain how  the classical Cram\'er-Rao bound  is derived.
By differentiating with respect to $\theta$ the identity
\begin{equation}
0 = \langle   \Delta \theta_{\rm est} \rangle_\theta = \sum_{i_1,\ldots , i_N} p_{ i_1 | \theta} \ldots p_{i_N|\theta} \Delta \theta_{\rm est} (i_1,\ldots, i_N)
\end{equation}
one  obtains
\begin{equation}
0  
 = 
\sum_{i_1,\ldots , i_N} p_{i_1 | \theta} \ldots p_{i_N|\theta} \sum_{\nu=1}^N \frac{\partial \ln p_{i_\nu | \theta}}{\partial \theta} \Delta \theta_{\rm est} (i_1,\ldots, i_N)
- \frac{\partial \langle \theta_{\rm est} \rangle_\theta}{\partial \theta}
\;.
\end{equation}
Then the  Cram\'er-Rao bound (\ref{eq-Cramer_Rao}) readily follows from the Cauchy-Schwarz inequality.
Of course, the interesting point is that equality can be achieved in the limit $N \rightarrow \infty$, but we will not 
dwell into that.
Going back to the quantum problem, we  rearrange (\ref{eq-Cramer_Rao}) as
\begin{equation} \label{eq-now_is_time_to_sleep0}
\frac{(\D \theta)^2 }{N} \leq  (\Delta \theta)^2 \sum_{i=1}^m \frac{(\tr [ M_i \D \rho (\theta)]  )^2 }{\tr [ M_i \rho (\theta )]}  
\end{equation}
with $\D \rho (\theta) = ( \partial \rho/\partial \theta ) \D \theta$.
Now, by using Proposition~\ref{prop-link_classical_fidelity} and performing an expansion  up to the second order in $\D \rho$, one finds
\begin{equation} \label{eq-now_is_time_to_sleep1}
\Ff_Q ( \rho(\theta), H )   (\D \theta)^2  
=  \sup_{\{ M_i\}} \biggl\{ \sum_{i=1}^m  \frac{(\tr [ M_i \D \rho (\theta )] )^2 }{\tr [ M_i \rho (\theta )]} \biggr\}\;.
\end{equation}
Here, the supremum is over all POVMs $\{ M_i\}$ and we have used 
$\sum_i \tr [ M_i \D \rho(\theta) ] = \tr [ \D \rho(\theta)] = 0$.
But $\Ff_Q ( \rho (\theta), H ) = \Ff_Q ( \rho, H)$ as a consequence of (\ref{eq-Fisher_info}), since 
$\rho(\theta)$ and $\rho$ are related by a unitary evolution  generated by $H$.
Comparing (\ref{eq-now_is_time_to_sleep0}) and (\ref{eq-now_is_time_to_sleep1}), we conclude that
$\inf_{\{ M_i\}}  \Delta \theta  \geq  (\Delta \theta)_{\text{best}}$, with equality as $N \rightarrow \infty$ for the 
maximum likelihood estimator, as stated in the proposition. 
\finpro

\vspace{2mm}

Before proceeding to derive upper bounds on $\Ff_Q ( \rho, H )$, let us observe that  
the monotonicity of the Bures metric $g_{\rm B}$ implies~\cite{Fujiwara01}:

\vspace{2mm}

\begin{corollary} \label{cor_convexity_Fisher}
The quantum Fisher information $ \Ff_Q ( \rho, H)$ is convex in $\rho$.
\end{corollary}


\proof
Given two states  $\rho_0$ and $\rho_1$ on $\Hh$ and $\eta_0,\eta_1 \geq 0$, $\eta_0 +\eta_1=1$,
we introduce the state $\widehat{\rho} = \eta_0 \rho_0 \otimes \ketbra{0}{0} + \eta_1 \rho_1 \otimes \ketbra{1}{1}$ on $\Hh \otimes \complex^2$
as in the proof of Theorem~\ref{theo-characterization_cont-metrics}.
From the expression of $\Ff_Q$ in the \RHS of (\ref{eq-Fisher_info}) one deduces  that
\begin{equation} \label{eq-proof_convexity_Fisher1}
\Ff_Q ( \widehat{\rho}, H \otimes 1 ) = \eta_0  \Ff_Q ( \rho_0 , H ) + \eta_1  \Ff_Q ( \rho_1 , H )\;.
\end{equation}
Let $\Ttt : \widehat{\sigma} \mapsto \tr_{\complex^2} ( \widehat{\sigma})$ denote the partial trace on $\complex^2$. Then
$\Ttt ( \widehat{\rho}) = \rho= \eta_0 \rho_0 + \eta_1 \rho_1$ and $\Ttt ([ H \otimes 1,  \widehat{\rho} ]  ) = [ H, \rho]$.
As $\Ttt$ is a quantum operation, it results from the contractivity of the Bures metric that
\begin{equation} \label{eq-proof_convexity_Fisher2} 
( g_{\rm B})_{\widehat{\rho}} \bigl( -\I [ H \otimes 1 , \widehat{\rho}], -\I [ H \otimes 1 ,\widehat{\rho}] \bigr)
\geq  (g_{\rm B})_{\rho}  \bigl(  -\I [ H,  {\rho} ], -\I [ H , {\rho}] \bigr)\;.
\end{equation}
Collecting together (\ref{eq-proof_convexity_Fisher1}) and (\ref{eq-proof_convexity_Fisher2}) yields
$\Ff_Q ( \rho , H ) \leq \eta_0 \Ff_Q ( \rho_0 , H ) + \eta_1  \Ff_Q ( \rho_1 , H )$.
\finpro
   
\subsubsection{Interferometer precision and  inter-particle entanglement}

We now show by relying on Proposition~\ref{prop_Braunstein_Caves} that if the input state has $N_{\rm p}$ particles in a maximally entangled state,
 the precision $(\Delta \theta)_{\rm best}$ is smaller by a factor $1/\sqrt{N_{\rm p}}$
with respect to the precision obtained with separable input states.
The Hilbert space of the particles is  $\Hh^{(N_{\rm p})} = \Hh_1 \otimes \cdots \otimes \Hh_{ N_{\rm p} }$, 
$\Hh_\nu$ being the Hilbert space of the $\nu$th particle. Assuming that the particles do not interact between themselves, the Hamiltonian reads
\begin{equation} \label{eq-Hamiltonian_two_body}
H = \sum_{\nu=1}^{N_{\rm p}} 1 \otimes \cdots \otimes H_\nu \otimes \cdots \otimes 1 \;,
\end{equation}
where $ H_\nu$ acts on $\Hh_\nu$.
To simplify the discussion we suppose that the single particle Hamiltonians $H_\nu$ have the same highest eigenvalue $\lambda_\mmax$ 
and the same lowest eigenvalue $\lambda_\mmin$.
This is the case for instance if $H$ is the angular momentum $J_{\nv}$ in the interferometer of Sec.~\ref{eq-Mach_Zehnder} (then 
$H_\nu = (n_x \sigma_{x \nu} + n_y \sigma_{y \nu } + n_z \sigma_{z \nu})/2$ with  $| \nv | = 1$ and $\sigma_{x\nu}$, $\sigma_{y \nu}$, and  $\sigma_{z\nu}$ 
the three Pauli matrices acting on $\Hh_\nu \simeq \complex^2$, so that $\lambda_\mmax = - \lambda_\mmin=1/2$).
Let us recall  that the quantum Fisher information $\Ff_Q ( \ket{\Psi} , H )$ of a pure state $\ket{\Psi}$ is given by the 
square fluctuation $\langle ( \Delta H )^2 \rangle_\Psi =  \bra{\Psi} H^2 \ket{\Psi} -  \bra{\Psi} H \ket{\Psi}^2$ up to a factor of four 
(see Sec.~\ref{sec-Bures_metric}).  
We first observe that the maximum of $\langle ( \Delta H_\nu )^2 \rangle_{\psi_\nu}$ 
over all pure states $\ket{\psi_\nu} \in \Hh_\nu$ is equal
to $(\Delta h)^2 = (\lambda_\mmax - \lambda_\mmin )^2/4$, the maximum being attained when $\ket{\psi_\nu} = (\ket{\phi_{\nu ,\mmax}} + \ket{\phi_{\nu ,\mmin}})/\sqrt{2}$, 
where $\ket{\phi_{\nu ,\mmax}}$ and $\ket{\phi_{\nu ,\mmin}}$ are the eigenvectors of $H_\nu$ with eigenvalues  $\lambda_\mmax$ and $\lambda_\mmin$, respectively. 
Let the $N_{\rm p}$ particles be in a separable state $\rho_{\rm sep}$  and let
$\{ \ket{\Psi_i},\eta_i \}$ be a  decomposition  of $\rho_{\rm sep}$  into pure product
states $\ket{\Psi_i} = \ket{\psi_{i1}}\otimes \cdots \otimes \ket{\psi_{i N_{\rm p} } } \in \Hh^{(N_{\rm p})}$. A simple calculation  gives~\cite{Giovannetti06}
\begin{equation}
\Ff_Q ( \ket{\Psi_i}, H ) = 4 \bigl\langle   ( \Delta H )^2 \bigr\rangle_{\Psi_i} = 4 \sum_{\nu=1}^{N_{\rm p}} \langle ( \Delta H_\nu )^2 \rangle_{\psi_{i \nu}}
\;\leq \; 4 (\Delta h)^2 N_{\rm p}\;.
\end{equation}
By applying  Corollary~\ref{cor_convexity_Fisher} we get
\begin{equation} \label{eq-Fisher_as_sufficient_for_ent}
\rho_{\rm sep} \; \text{ separable } \;\Rightarrow \;\; \Ff_Q ( \rho_{\rm sep},  H ) \leq 4 ( \Delta h)^2  N_{\rm p} \;.
\end{equation}
According to Proposition~\ref{prop_Braunstein_Caves} the phase precision of the interferometer satisfies
for separable inputs
\begin{equation}
 \Delta \theta \geq ( \Delta \theta)_{\rm SN} = \frac{1}{2 \Delta h  \sqrt{N N_{\rm p}}}\;.
\end{equation}
This means that separable input states 
cannot do better than $N_{\rm p}$  independent particles sent one-by-one through the interferometer,  
henceforth producing an error of the order of $1/\sqrt{N_{\rm p}}$. Note that (\ref{eq-Fisher_as_sufficient_for_ent}) provides
a  sufficient condition $\Ff_Q ( \rho ,  H ) > 4 (\Delta h)^2  N_{\rm p}$ for entanglement of $\rho$~\cite{Pezze09}.  
There are, however, entangled states which do not satisfy this criterion~\cite{Pezze09}. Such entangled states are not useful for interferometry,
in the sense that they produce phase errors larger than the shot noise value $( \Delta \theta)_{\rm SN}$.

We now argue that much higher Fisher informations, of the order of $N_{\rm p}^2$, can be achieved for entangled states.
By the same observation as above, $\langle (\Delta H )^2 \rangle_\Psi$ has a maximum given by the square of the half difference of the
maximal and minimal eigenvalues of $H$.  
For the  Hamiltonian (\ref{eq-Hamiltonian_two_body}), one immediately finds
\begin{equation} \label{eq-Heisenberg_bound}
\Ff_Q ( \ket{\Psi}, H )  \leq 4 ( \Delta h )^2  N_{\rm p}^2\;.
\end{equation}
This upper bound is often called the {\it Heisenberg bound} in the literature. It is saturated  for the entangled states~\cite{Giovannetti06}
\begin{equation}
 \ket{\Psi_{\rm ent}^\pm} 
 = \frac{1}{\sqrt{2}} 
   \Bigl( \ket{\phi_{1,\mmax}} \ket{\phi_{2,\mmax}}  \ldots \ket{\phi_{N_{\rm p} ,\mmax} } \pm \ket{\phi_{1,\mmin} } \ket{\phi_{2,\mmin}  } \ldots \ket{\phi_{N_{\rm p} ,\mmin}} \Bigr)\;.
\end{equation}
For large $N_{\rm p}$ such states deserve the name of {\it macroscopic superpositions}, as they are formed by a superposition of two
 macroscopically distinct states in which each particle is in the highest energy eigenstate of the single particle Hamiltonian
(for the first component of the superposition) or in the lowest energy eigenstate (for the second component).
If one uses these superpositions as input states of the interferometer, an error of 
$\Delta \theta = 1/(2 \Delta h \sqrt{N} N_{\rm p}) =  ( \Delta \theta)_{\rm SN}/\sqrt{N_{\rm p}}$ 
can be achieved asymptotically for $N \rightarrow \infty$
on the unknown phase. According to (\ref{eq-quantum_Cramer_Rao}) and (\ref{eq-Heisenberg_bound}), this is the best possible
precision.

\newpage
\section{Measures of entanglement in bipartite systems} \label{sec-entanglement}

Even if it would be better for many computational and communication tasks to work with maximally entangled pure states, 
in practice the coupling of the system with its environment transforms such states into non-maximally entangled mixed states
because of the induced decoherence processes~\cite{Breuer,Giulini,Haroche}. It is thus important to 
quantify the amount of entanglement in an arbitrary quantum state. Unfortunately, this amount of entanglement is not a directly measurable quantity. 
It is quantified by an entanglement measure, which vanishes \ifif the state is separable and
cannot increase under local operations on each subsystems and classical communication  (entanglement monotonicity).
All measures satisfying these two requirements are not equivalent, \ie  a state $\rho$ can be more entangled than a state  $\sigma$ 
for one measure and less entangled  for the other.
In this section, we investigate the properties of entanglement measures, give their general form for pure states,
 and study more especially two of the most popular ones, the entanglement of formation and the concurrence.
We restrict our attention to bipartite entanglement (see \cite{Toth_review,Horedecki_review} for generalizations to entanglement in systems with more 
than two parties).    

\subsection{Entanglement as correlations between local measurements}  \label{sec-concurrence}

Let  $\ket{\Psi}$ be a pure state of a bipartite system $\AB$. 
In view of the discussion in Sec.~\ref{sec-def_entanglement}, it seems natural 
physically to characterize the entanglement in $\ket{\Psi}$
by maximizing the correlator $G_{AB}( \ket{\Psi})$ in (\ref{eq-correlation_function})  over all local observables 
$A \in \Bb ( \Hh_\AAA)_{\rm s.a.}$ and $B \in \Bb ( \Hh_\BB)_{\rm s.a.}$ and to define
\begin{equation} \label{eq-my_def_correlators}
G ( \ket{\Psi} ) 
  =   \max_{A=A^\ast, \| \Delta A \|_{\infty,\Psi}  \leq 1} \;\;\max_{B= B^\ast, \| \Delta B \|_{\infty,\Psi}  \leq 1}
\bigl\{  \bigl| G_{A B} (\ket{\Psi}) \bigr| \bigr\} \; .
\end{equation}
One must face with some arbitrariness on the choice of  the norm used to bound 
$\Delta A= A- \langle A \otimes 1 \rangle_\Psi$ and $\Delta B = B- \langle 1 \otimes B\rangle_\Psi$. In order to obtain an entanglement  measure with the required properties, 
we take the $\Psi$-dependent norm  $\| \Delta A \|_{\infty,\Psi} =  \max_{i,j} | \bra{\alpha_i} \Delta A \ket{\alpha_j} |$, where
$\{ \ket{\alpha_i} \}$ is an orthonormal eigenbasis of the reduced state
$[ \rho_\Psi]_\AAA$, and similarly for $\| \Delta B \|_{\infty,\Psi}$ with the eigenbasis $\{ \ket{\beta_k} \}$  of $[ \rho_\Psi]_\BB$.
These norms correspond to the infinity norms of the vectors in $\Hh_\Atwice$ and 
$\Hh_\Btwice$  associated to $\Delta A$ and $\Delta B$ via the isometry (\ref{eq-ismoemtry_operators_vectors}).
By using the  Schmidt decomposition (\ref{eq-Schmidt_decomposition}) and setting $A_{ij} = \bra{\alpha_i} A \ket{\alpha_j}$ and 
 $B_{ij} = \bra{\beta_i} B \ket{\beta_j}$,  one finds
%
\begin{equation} 
G_{AB} ( \ket{\Psi} )  
 =  \langle \Delta A \otimes \Delta B \rangle_\Psi 
= \sum_{i=1}^n \mu_i ( \Delta A)_{ii} ( \Delta B)_{ii} + \sum_{i \not= j}^n \sqrt{\mu_i \mu_j} A_{ij} B_{ij}  \;.
\end{equation}
The Cauchy-Schwarz inequality immediately yields
\begin{equation}
\label{eq-max_correlator}
G ( \ket{\Psi} )  
 = 
\max_{\| \Delta \av   \|_\infty \leq 1} \bigl\{ \,\overline{(\Delta \av)^2}\, \bigr\}  + C ( \ket{\Psi} ) \;,
\end{equation}
where the overline stands for the average with respect to the Schmidt coefficients $\mu_i$
(e.g. $\overline{\av} = \sum_i \mu_i a_i$), 
$\Delta \av = \av - \overline{\av}$  with $\av = ( A_{11}, \ldots, A_{nn})$, $\| \Delta \av \|_\infty = \max_i | (\Delta \av )_i|$, and
\begin{equation}  \label{eq-my_def_concurrence}
 C ( \ket{\Psi} )  = 
 \sum_{i \not= j}^n \sqrt{\mu_i \mu_j} = \Bigl( \tr \bigl( \sqrt{[\rho_\Psi]_\AAA} \bigr) \Bigr)^2 - 1 \;.
\end{equation}
Thus $C ( \ket{\Psi} )=0$ (similarly, $G ( \ket{\Psi} )  =0$) is equivalent to $\mu_i=0$ save for one index $i$, that is, to $\ket{\Psi}$ being separable.
Furthermore, $ C ( \ket{\Psi} ) \leq  n-1$ with equality \ifif $\mu_i=1/n$ for all $i$, that is, \ifif $\ket{\Psi}$ is 
maximally entangled  (Sec.~\ref{sec-def_entanglement})\footnote{
This last property is not true if one uses  the operator norm instead of $\| \cdot \|_{\infty,\Psi}$ in 
(\ref{eq-my_def_correlators}), except in the two-qubit case $n=2$.
}.
Finally, we note that $G$ and $C$ are invariant under local unitaries, 
\ie $G( U_A \otimes U_B \ket{\Psi}) = G ( \ket{\Psi})$ for any unitaries $U_A$ and $U_B$ on $\Hh_\AAA$ and $\Hh_\BB$. 
For two qubits one obtains
\begin{equation} \label{eq-my_def_concurrence_2_qubits}
G ( \ket{\Psi} ) 
 = \mu_\mmax^{-1} - 1 + C ( \ket{\Psi}) \quad , \quad  C( \ket{\Psi}) = 2 \sqrt{\mu_0 \mu_1}
\end{equation}
with $\mu_\mmax = \max \{ \mu_0, \mu_1\}$.
It is easy to show that
$C( \ket{\Psi}) = | \bra{\Psi} \sigma_y \otimes \sigma_y J \ket{\Psi} |$, where 
$\sigma_y = \I ( \ketbra{0}{1} - \ketbra{1}{0} )$ is the $y$-Pauli matrix and
$J$ the complex conjugation in the canonical basis. 
This quantity has been first introduced by Wootters~\cite{Wootters98} and is known as the {\it concurrence}.

One may wonder how the correlator $G_{AB}$ could be generalized for mixed states.
The first guess would be to replace the expectation value $\langle \cdot \rangle_{\Psi}$ by $\langle \cdot \rangle_\rho = \tr ( \rho \,\cdot\, )$, but one easily sees
that then $G ( \rho)$ can be non-zero even for separable mixed states, because this correlator contains
both the quantum and classical (\ie statistical) correlations in the density matrix $\rho$.
Noting that 
\begin{equation}
G_{AB} ( \ket{\Psi}) = \frac{1}{2}  \bigl\langle \bigl(\Delta ( A \otimes 1 + 1 \otimes B)\bigr)^2 \bigr\rangle_\Psi 
 -  \frac{1}{2} \bigl\langle \bigl( \Delta ( A \otimes 1 ) \bigr)^2 \bigr\rangle_\Psi   -  \frac{1}{2} \bigl\langle \bigl( \Delta ( 1 \otimes B ) \bigr)^2 \bigr\rangle_\Psi \;,
\end{equation}
it is tempting to define a correlator for $\rho$ in terms of the quantum Fisher information (\ref{eq-Fisher_info}), \ie of the Bures metric $g_{\rm B}$,
\begin{eqnarray}
\nn
G_{AB} ( \rho) & = & \frac{1}{8} \Bigl( \Ff_Q ( \rho,  A \otimes 1 + 1 \otimes B ) -  \Ff_Q ( \rho,  A \otimes 1  )   -  \Ff_Q ( \rho,  1 \otimes B  )  \Bigr)
\\ \nn
& = & \re \bigl\{ (g_{\rm B} )_\rho  \bigl( - \I [  A \otimes 1 , \rho ] ,  - \I [  1 \otimes B , \rho ] \bigr) \bigr\}\;.
\end{eqnarray}
By inspection on~(\ref{eq-Fisher_info_for_pure_states}), $G_{AB}(\rho)$ reduces for pure states to the previous correlator. However, 
the maximum of $|G_{AB} ( \rho)|$ over all $A$ and $B$ does not fulfill the axioms of an entanglement measure. 
We will see in Sec.~\ref{sec-EoF} another way to define the concurrence $C$ for mixed states, by using on a convex roof construction.

\subsection{LOCC operations} \label{sec_LOCC}

The main physical postulate on entanglement measures is that they must be  monotonous with respect to certain state transformations. 
Such transformations that cannot increase
entanglement are called {\it Local Operations and Classical Communication} (LOCC) and can
be described as follows~\cite{Bennett96,Horedecki_review}. 
Let us consider an entangled state $\rho$ shared by two observers Alice and Bob. 
Alice and Bob can perform any quantum  operations $\Mm_\AAA : \Bb ( \Hh_\AAA) \rightarrow \Bb ( \Hh_\AAA')$ and
 $\Mm_\BB : \Bb ( \Hh_\BB ) \rightarrow \Bb ( \Hh_\BB ')$ on their respective subsystems $\AAA$ and $\BB$. Here, 
the final spaces  $\Hh_\AAA'$ and $\Hh_\BB '$ may include local ancillae, or may be some subspaces of $\Hh_\AAA$ and $\Hh_\BB$, respectively.
The corresponding transformations on the system $\AB$ are called {\it local quantum operations}.
They are of the form $\Mm_{\rm loc} = \Mm_\AAA \otimes \Mm_\BB$ and are given by families 
$\{ A_i \otimes B_j\} $ of Kraus operators, where $A_i$ and $B_j$ are local observables on $\AAA$ and $\BB$. 
Local operations are performed physically by coupling each subsystem to a local ancilla and by 
making joint unitary evolutions and von Neumann measurements on the subsystem and its ancilla (see Sec.~\ref{sec-QO}).
Such processes can clearly not increase the amount of entanglement between $\AAA$ and $\BB$.
In addition to performing local generalized measurements, Alice and Bob can communicate  their measurement outcomes to each 
other via a classical communication channel (two-way communication). No transfer of quantum systems 
between them is allowed. 
Thanks to classical communication, the observers can increase the classical correlations  between $\AAA$ and $\BB$, but not the $\AB$-entanglement. 
A {\it LOCC operation} is a \QO  on $\Bb ( \Hh_\AB)$ obtained through a succession of the aforementioned actions of Alice and Bob, taken
in arbitrary order. 
For example, if Alice performs a \meas on $\AAA$ and Bob  a \meas on $\BB$ depending on Alice's outcome $i$ (one way communication),  
the post-\meas state in the absence of readout is
\begin{equation}
\Mm_{1-\rm way}  ( \rho) = \sum_i 1 \otimes \Mm_\BB^{(i)} ( A_i \otimes 1 \rho A_i^\ast \otimes 1 )\;.
\end{equation}
This defines a LOCC operation with Kraus operators $A_i \otimes B_j^{(i)}$,
where $\sum_i A_i^\ast A_i = \sum_j (B_j^{(i)})^\ast  B_j^{(i)} = 1$.

Any LOCC operation can be obtained by composing  local operations $\Mm_{\rm loc}$ with the maps
\begin{equation} \label{eq-maps_def_general_LOCC}
\Mm_{\rm LOCC}^{\AAA} ( \rho) = \sum_i \bigl( A_i \otimes 1\,\rho \,A_i^\ast \otimes 1 \bigr) \otimes \ketbra{\kappa_i}{\kappa_i}
\quad , \quad 
\Mm_{\rm LOCC}^\BB ( \rho) = \sum_j\bigl(  1 \otimes B_j \, \rho \, 1 \otimes B_j^\ast  \bigr) \otimes \ketbra{\epsilon_j}{\epsilon_j}
\;,
\end{equation}
where $\sum_i A_i^\ast A_i = \sum_j B_j^\ast B_j =1$ and 
$\{ \ket{\kappa_i} \}$ (respectively $\{ \ket{\epsilon_j} \}$) is an \ONB for Bob's ancilla (respectively Alice's ancilla)~\cite{Horedecki_review}.  
A strictly larger
but much simpler class of transformations, known as the {\it separable quantum operations}~\cite{Vedral98}, 
is the set of all operations with Kraus operators $A_i \otimes B_i$, \ie
\begin{equation}
\Mm_{\rm sep} ( \rho) = \sum_i   A_i \otimes B_i \rho  A_i^\ast \otimes B_i^\ast
\end{equation}
with $A_i \in \Bb (\Hh_\AAA, \Hh_\AAA ')$, $B_i \in \Bb( \Hh_\BB, \Hh_\BB ')$, and $\sum_i A_i^\ast A_i \otimes B_i^\ast B_i  = 1$.  
The local operations and maps (\ref{eq-maps_def_general_LOCC}) being separable, any 
LOCC operation is separable. A result from Ref.~\cite{Bennett99} shows, however, that certain separable operations are not LOCCs.

It is clear that the set $\Ss_\AB$ of separable states  is invariant under separable operations. 
It is also true that every separable state can be converted into any other separable state by a 
separable operation. Actually, any separable state can be obtained from the classical state
$\rho_\clas = \sum_{jk} p_{jk} \ketbra{j}{j} \otimes \ketbra{k}{k}$ by such an operation (take $A_{ijk} = \sqrt{\eta_i} \ketbra{\psi_i}{j}$
and $B_{ijk} = \ketbra{\phi_i}{k}$ with $\eta_i$, $\ket{\psi_i}$, and $\ket{\phi_i}$ as in  (\ref{eq-def_separable_states})).
Furthermore, an arbitrary  state $\rho$ can be transformed into a classical state $\rho_\clas$ by a measurement
in the product basis $\{ \ket{j} \ket{k}\}$, which is a local operation.

When one restricts LOCC transformations to pure states, a great simplification comes from the following observation. 
If the space dimensions of $\AAA$ and
$\BB$ are such that $n_\AAA \geq n_\BB$, 
any \meas by Bob can be simulated by a \meas by Alice followed by a unitary transformation 
by Bob  conditioned to Alice's outcome (such a conditioning is allowed as Alice and Bob can communicate classically). 
In fact,
let $\{ \ket{\alpha_i} \}_{i=1}^{n_\AAA}$ and
$\{ \ket{\beta_i} \}_{i=1}^{n_\BB}$ be orthonormal eigenbasis of the reduced states $[\rho_\Psi ]_\AAA$ and $[\rho_\Psi]_\BB$, and 
let $B_i$ be the Kraus operators  describing Bob's measurement. Consider 
the \meas done by Alice with Kraus operators $A_i= \sum_{j,l} (B_i)_{lj} \ketbra{\alpha_l}{\alpha_j}$, where 
$ (B_i)_{lj}= \bra{\beta_l} B_i \ket{\beta_j}$.
The  unnormalized post-\meas states 
\begin{equation}
\ket{\widetilde{\Phi}_i} = 1 \otimes B_i \ket{\Psi} = \sum_{j,l} \sqrt{\mu_j} (B_i)_{lj} \ket{\alpha_j} \ket{\beta_l}
\quad , \quad 
\ket{\widetilde{\Phi}'_i} = A_i \otimes 1 \ket{\Psi} = \sum_{j,l} \sqrt{\mu_j} (B_i)_{lj} \ket{\alpha_l} \ket{\beta_j}
\end{equation}
have the same Schmidt coefficients because 
$\tr_\BB ( \ketbra{\widetilde{\Phi}_i}{\widetilde{\Phi}_i} )$ and $\tr_\AAA ( \ketbra{\widetilde{\Phi}_i'}{\widetilde{\Phi}_i'} )$ are related
by an isometry $\Hh_\AAA \rightarrow \Hh_\BB$. Thus
 $\ket{\widetilde{\Phi}'_i}=  U_{i} \otimes V_{i} \ket{\widetilde{\Phi}_i}$ for some local unitaries $ U_{i}$ on 
$\Hh_\AAA$ and  $V_{i}$ on $\Hh_\BB$. Consequently, Bob performing the \meas $\{ B_i\}$ is equivalent to Alice performing the \meas $\{ U_{i}^\ast A_i\}$ and Bob
performing the unitary transformation $V_i^\ast$ when Alice gets the outcome $i$.
Applying this result to all Bob's  measurements, we conclude that
a LOCC acting on  a pure state $\ket{\Psi}$ may always be
simulated by a one-way communication protocol involving only three steps: (1) Alice first performs a generalized \meas on subsystem $\AAA$; (2) 
she sends her \meas result to Bob; (3) Bob performs a unitary evolution on $\BB$ conditional to Alice's result.

Based on this observation, we say that a  pure state $\ket{\Psi} \in \Hh_\AB$ can be transformed  by a LOCC into the pure 
state $\ket{\Phi} \in \Hh_\AB$ if there are families of Kraus operators $\{ A_i \}$ on $\Hh_\AAA$ and unitaries 
$\{ V_i\}$ on $\Hh_\BB$ such that all unnormalized conditional states
$A_i \otimes V_i \ket{\Psi}$ are proportional to $\ket{\Phi}$, irrespective of the \meas outcome $i$.
Note that this is equivalent to $\Mm_{\rm LOCC} ( \ketbra{\Psi}{\Psi})$ being equal to $\ketbra{\Phi}{\Phi}$, with  $\Mm_{\rm LOCC}$ the
LOCC operation with Kraus family $\{ A_i \otimes V_i\}$. 
One defines in this way an order relation on the set of pure states.
Nielsen~\cite{Nielsen99} discovered a nice relation between this order and
the theory of majorization for $n$-dimensional vectors~\cite{Bhatia}.
Let  $\xv =(x_1,\ldots, x_n)$ and $\yv =(y_1,\ldots, y_n)$ be two vectors in $\real^n$. We denote by $\xv^\downarrow$ the vector formed by the components of $\xv$
in decreasing order, 
and similarly for $\yv^\downarrow$. One says that $\xv$ is majorized by $\yv$ and write  $\xv \prec \yv$ if 
$\sum_{i=1}^k x^\downarrow_i \leq \sum_{i=1}^k y^\downarrow_i$ for any $k=1,\ldots, n$, with equality instead of inequality for $k=n$.

\vspace{2mm}

\begin{proposition} {\rm (Nielsen~\cite{Nielsen99})} \label{prop_Nielsen}
A pure state $\ket{\Psi}$ of the bipartite system $\AB$ can be transformed into another pure state $\ket{\Phi}$ of $\AB$
by a LOCC if and only if ${\boldsymbol\mu}_\Psi \prec {\boldsymbol\mu}_\Phi$, where ${\boldsymbol\mu}_\Psi$ and ${\boldsymbol\mu}_\Phi$ are the vectors 
formed by the Schmidt coefficients of $\ket{\Psi}$ and $\ket{\Phi}$, respectively.
\end{proposition}

\vspace{1mm}

A detailed proof of this result can be found in~\cite{Nielsen} (Sect. 12.5), so we omit it here.
This proof relies on the following theorem: if $\boldsymbol\lambda_H$ and $\boldsymbol\lambda_K$ are vectors formed by the eigenvalues of two 
Hermitian matrices $H$ and $K$, respectively,
then $\boldsymbol\lambda_H \prec \boldsymbol\lambda_K$ \ifif $H = \sum_i \eta_i U_i K U_i^\ast$ with $\{\eta_i\}$ a set of probabilities and $U_i$ 
some unitary matrices.

\vspace{1mm}

\begin{exercice}
Even if $\ket{\Psi}$ cannot be transformed into $\ket{\Phi}$ by a LOCC, it may still happen that $\ket{\Psi} \otimes \ket{\kappa}$ can 
be transformed into $\ket{\Phi} \otimes \ket{\kappa}$ by a LOCC (here the state of the ancilla does not change 
during the transformation, \ie it acts as catalysts in chemical reactions)~\cite{Jonathan99}.
\end{exercice}

\subsection{Axioms on entanglement measures} \label{sec-axiom_ent_meas}

We are now in position to formulate the physical postulates on entanglement measures~\cite{Bennett96,Vedral98,Vidal00}.

\vspace{2mm}

\begin{definition} \label{def_axioms_ent_meas}
An entanglement measure of a bipartite system $\AB$ is a function $E: \states (\Hh_\AB) \rightarrow \real$ such that
\begin{itemize}
\item[(i)] $E (\rho)=0$ if and only if $\rho$ is separable;
\item[(ii)] $E$ is convex;
\item[(iii)] $E$ cannot increase under LOCCs, \ie if $\Mm_{\rm{LOCC}}$ is a  LOCC operation then
$E ( \Mm_{\rm{LOCC}} (\rho) ) \leq E (\rho)$.
\end{itemize}
\end{definition}

As any two separable states can be transformed one into each other by means of a LOCC operation, the monotonicity (iii) implies that
$E$ is constant on the set of separable states $\Ss_\AB$.
Taking this constant equal to zero yields $\rho \in \Ss_\AB \Rightarrow$ $E (\rho)=0$, so that only the reverse implication  is needed  in (i).
Furthermore, any state $\rho$ can be converted into 
a separable state by a LOCC, thus $E (\rho)$ is minimum for separable states and  $E ( \rho) \geq 0$.
The convexity condition (ii) is motivated by the following observation~\cite{Vidal00}. 
Assume that Alice and Bob share $m$ pairs of particles in the states $\rho_1,\ldots, \rho_m$. 
By classical communication, they can agree to keep the $i$th pair with probability $\eta_i$, thus preparing the ensemble
$\{ \rho_i, \eta_i \}_{i=1}^m$. 
By erasing the information about which state $\rho_i$ was kept, the state becomes $\rho = \sum \eta_i \rho_i$  (see Sec.~\ref{sec-purification}). 
The inequality $E ( \rho ) \leq \sum_i \eta_i E ( \rho_i)$  means that this local loss of information does not increase the average entanglement.

It results from the monotonicity (iii) that entanglement measures are invariant under 
conjugations by local unitaries, \ie $E ( U_\AAA \otimes U_\BB\, \rho \,U_\AAA^\ast \otimes U_\BB^\ast) = E ( \rho)$.
For pure states $\ket{\Psi}$, this implies that $E(\ket{\Psi})$ only depends on the Schmidt coefficients $\mu_i$ of  $\ket{\Psi}$.
Consequently, $E (\ket{\Psi}) = f ( [\rho_\Psi]_\AAA)$ is a  unitary-invariant function of  the reduced state
$[\rho_\Psi]_\AAA= \tr_\BB ( \ketbra{\Psi}{\Psi})$ (or, equivalently, of $[ \rho_\Psi]_\BB= \tr_\AAA ( \ketbra{\Psi}{\Psi})$). 
Given that a pure state is separable \ifif  it has a single non-vanishing Schmidt 
coefficient, one deduces from axiom (i) that $f(\rho_\AAA)$ vanishes \ifif $\rho_\AAA$ is of rank one.
The result below due to Vidal~\cite{Vidal00} characterizes all entanglement measures on pure states
satisfying a slightly stronger  condition than (iii). 
This shows in particular that there are many measures of entanglement
fulfilling the three physical requirements (i-iii) of Definition~\ref{def_axioms_ent_meas}, given by  
concave functions $f$.

\vspace{2mm}

\begin{proposition} {\rm (Vidal~\cite{Vidal00})} \label{prop-entanglement_meas_pure_states}
Let $f : \states ( \Hh_\AAA) \rightarrow \real$ be concave, unitary invariant, and such that $f ( \rho_\AAA)=0$ \ifif $\rho_\AAA$ is a pure state. Then 
\begin{equation} \label{eq-def-E_f}
E_f ( \ket{\Psi} ) =  f ( [ \rho_\Psi]_\AAA ) 
\end{equation}
defines  an entanglement measure on the set of pure states of $\AB$, which satisfies 
the monotonicity condition
\begin{itemize}
\item[(iii')] $\sum_i p_i E (  \ket{\Phi_i} )  \leq E ( \ket{\Psi} )$, where $p_i = \| A_i \otimes B_i \ket{\Psi}\|^2$ and
$\ket{\Phi_i} = p_i^{-1/2} A_i \otimes B_i \ket{\Psi}$  and the  probabilities and conditional 
states of a separable  \meas with Kraus operators $A_i \otimes B_i$. 
\end{itemize}
Conversely, any entanglement measure  on pure states fulfilling (iii')  is given by (\ref{eq-def-E_f}) for some function $f$ satisfying 
the above assumptions. 
\end{proposition}

\vspace{1mm}

It should be noted that asking $E (  \ket{\Phi_i} )  \leq E ( \ket{\Psi} )$ for all outcomes $i$ would put a too strong condition
on $E$. Indeed, local measurements can in principle create entanglement on some conditional states, 
but not on average  (see below).  

\vspace{1mm}

\proof 
Let $f$ be like in the proposition.
We have already argued above that $E_f$ fulfills axiom (i), and (ii) is empty because of the restriction to pure states.  
Recall that for such states any \meas on $\BB$ can be simulated by a \meas on $\AAA$ followed by a 
unitary operation on $\BB$ conditioned
to the \meas result. Hence it  suffices to show the monotonicity (iii') for $B_i = V_i$ unitary. 
Let us set $\rho_{\BB |i} = \tr_\AAA ( \ketbra{\Phi_i}{\Phi_i} )$. Then $\{ V_i^\ast  \rho_{\BB |i} V_i, p_i \}$ is a pure state decomposition of
$[\rho_\Psi ]_\BB$, \ie $\sum_i p_i V_i^\ast \rho_{\BB |i} V_i = [\rho_\Psi ]_\BB$.
This can be interpreted by saying that a local \meas on $\AAA$ does not modify the state of 
$\BB$ when $\BB$ has no information on the \meas outcomes\footnote{
If this would not be true, information
could be sent faster than light in contradiction with Einstein's principle of relativity~\cite{Peres}.
}.
The concavity and unitary invariance of $f$ imply
\begin{equation}
 \sum_i p_i E_f ( \ket{\Phi_i} ) = \sum_i p_i f ( V_i^\ast \rho_{\BB |i} V_i )  \leq  f ( [\rho_\Psi ]_\BB ) = E_f ( \ket{\Psi} )\;.
\end{equation}
This shows (iii'). Thus $E_f$ is an entanglement measure.

Reciprocally, let $E$ be an entanglement measure on pure states satisfying (iii'). From the discussion before the proposition
we  know that 
$E ( \ket{\Psi} ) = f ( [\rho_\Psi]_\AAA )=  f ( [\rho_\Psi]_\BB )$ for some unitary-invariant function $f$ vanishing on pure states only. 
It remains to show that $f$ is concave.
We may assume that the space dimensions of $\AAA$ and
$\BB$ are such that $n_\AAA \leq n_\BB$ (otherwise one can exchange the role of $\AAA$ and $\BB$ in the arguments below).  
Let $\rho_\AAA$ be an arbitrary  state of $\AAA$ and
$\sigma_\AAA^{(1)}$, $\sigma_\AAA^{(2)}$ be such that $\rho_\AAA =p_1 \sigma_\AAA^{(1)} +p_2 \sigma_\AAA^{(2)}$
with $p_1 + p_2 = 1$. As $n_\AAA \leq n_\BB$, one may find a purification $\ket{\Psi}$ of $\rho_\AAA$ on $\Hh_\AB$ (Sec.~\ref{sec-purification}). 
If one can exhibit a \meas on $\BB$ with  outcome probabilities 
$p_i$ and conditional states $\ket{\Phi_i}$ having marginals $\tr_\BB (\ketbra{\Phi_i}{\Phi_i} ) = \sigma_\AAA^{(i)}$ for $i=1,2$, then 
the concavity of $f$ can be deduced from  (iii') thanks to the bound
\begin{equation}
f ( \rho_\AAA ) = E ( \ket{\Psi} ) \geq p_1 E ( \ket{\Phi_1} ) +p_2 E ( \ket{\Phi_2} )
 = p_1 f ( \sigma_\AAA^{(1)} ) + p_2 f ( \sigma_\AAA^{(2)} )\;.
\end{equation} 
The \meas we are looking for is just the square root \meas associated to $\{ \sigma_\AAA^{(i)}, p_i \}$
(Sec.~\ref{sec-least_square_meas}). Indeed, let $\{ \ket{\alpha_j} \}_{j=1}^{n_\AAA}$ and
$\{ \ket{\beta_k} \}_{k=1}^{n_\BB}$ be eigenbases of $[\rho_\Psi]_\AAA$ and $[\rho_\Psi]_\BB$ and
 $M_i^\lsm$, $i=1,2$, be the operators on $\Hh_\BB$ with 
matrix elements given by
(compare with (\ref{eq-definition_LSM}))
\begin{equation}
\bra{\beta_j} M_i^\lsm \ket{\beta_l} 
 = 
\begin{cases}
p_i \bra{\alpha_l}  \rho_\AAA^{-\onehalf} \sigma^{(i)}_\AAA  \rho_\AAA^{-\onehalf}  \ket{\alpha_j} & \text{if $j,l=1,\ldots, n_\AAA$}
\\
0 & \text{otherwise.}
\end{cases}
\end{equation} 
If $n_\BB > n_\AAA$ we add a third \meas operator, equal to the projector onto  
$\Span \{ \ket{\beta_k} ; n_\AAA < k \leq n_\BB \}$. Then $M_1^\lsm + M_2^\lsm + M_3^\lsm = 1$. With the help of the Schmidt decomposition
(\ref{eq-Schmidt_decomposition}) one finds that $\bra{\Psi} 1 \otimes M_i^\lsm \ket{\Psi}$ equals $p_i$ for $i=1,2$ and 
zero for $i=3$, and the conditional state $\ket{\Phi_i} = p_i^{-1/2} 1 \otimes \sqrt{M_i^\lsm} \ket{\Psi}$ 
has marginal $\tr_\BB (\ketbra{\Phi_i}{\Phi_i} )= \sigma_\AAA^{(i)}$ for $i=1,2$. This concludes the proof.
\finpro

\vspace{3mm}

Proposition \ref{prop-entanglement_meas_pure_states} can be partially justified with the help of
Proposition~\ref{prop_Nielsen}.
More precisely, the latter implies that $E_f ( \ket{\Psi} ) \geq E_f ( \ket{\Phi} )$ if $\ketbra{\Phi}{\Phi} = \Mm_{\rm LOCC} ( \ketbra{\Psi}{\Psi})$, that is, if
there exists a LOCC measurement on $\ket{\Psi}$ with all conditional states $\ket{\Phi_i}$  equal to $\ket{\Phi}$.
This comes from the fact that,
by unitary invariance, $f ( [ \rho_\Psi]_\AAA )$ is a symmetric function of
the eigenvalues $(\mu_\Psi)_1, \ldots, (\mu_\Psi)_n$ of $[ \rho_\Psi]_\AAA $. 
But concave symmetric functions $f: \real^n \rightarrow \real$ are Schur-concave, 
\ie $\xv \prec \yv \Rightarrow f (\xv) \geq f(\yv)$  
(see~\cite{Bhatia}, Theorem~II.3.3).

Many entanglement measures satisfying the axioms (i-iii) of Definition~\ref{def_axioms_ent_meas}
have been defined  in the literature. Their restrictions to pure states are all given by (\ref{eq-def-E_f}) for specific
concave functions $f$.  We present in the next subsection a few
of these measures, namely, the entanglement of formation, the concurrence, and the Schmidt number.
An integer-valued entanglement measure has been introduced in~\cite{Sawicki11} 
by using a symplectic geometry approach, but this goes beyond the scope of this article.

\subsection{Entanglement of formation}  \label{sec-EoF}
\subsubsection{Entanglement of formation for pure states}

A natural choice for the function $f$ is the von Neumann entropy. We set
\begin{equation} \label{eq-def-EoF_pure_state}
\EoF ( \ket{\Psi} ) = S \bigl( [ \rho_\Psi ]_\AAA \bigr) = S \bigl( [ \rho_\Psi ]_\BB \bigr) = - \sum_i \mu_i \ln \mu_i\;.
\end{equation}
Then $\EoF ( \ket{\Psi} )=0$ \ifif $\ket{\Psi}$ is separable and $\EoF ( \ket{\Psi} )$ is maximum (and equal to $\ln n$
 with $n = \min \{ n_\AAA, n_\BB\}$) 
\ifif $\ket{\Psi}$ is maximally entangled.
Since the von Neumann entropy is concave, Proposition~\ref{prop-entanglement_meas_pure_states} ensures that $\EoF$ is an entanglement measure on pure states.

An important result due to Bennett {\it et al.}~\cite{Bennett96a} relates $\EoF ( \ket{\Psi} )$ 
to entanglement distillation and entanglement cost, which consist in the following problems.
The EPR two-qubit state $\ket{\Phi_{+}}= (\ket{0} \ket{0} + \ket{1}\ket{1})/\sqrt{2} \in \complex^4$ corresponds to an {\it{e}}-bit of information shared by Alice and Bob.
One such {\it{e}}-bit is required, for  instance, if Alice wants to teleport an unknown quantum state to Bob~\cite{Nielsen}.
Entanglement distillation is the  transformation 
 of $N$ copies of $\ket{\Psi}$ onto $M<N$ copies of $\ket{\Phi_{+}}$.
It was demonstrated by Bennett {\it et al.} that in the large $N$ limit, $\EoF ( \ket{\Psi} )$ is equal to  
the maximal rate of distillation  $M/N$, 
the maximum being over all LOCC operations. Stated differently, $\EoF ( \ket{\Psi} )$ is the highest number of  {\it{e}}-bits
 per input copy of $\ket{\Psi}$ that can be distilled from  $\ket{\Psi}$ via LOCCs. 
Conversely, $\EoF ( \ket{\Psi} )$ is the smallest number of  {\it{e}}-bits per unit copy  of $\ket{\Psi}$ from which $\ket{\Psi}$ may be 
obtained via LOCCs. The precise mathematical statement is given in the  proposition below.

\vspace{2mm}

\begin{proposition} {\rm (Bennett {\it et al.}~\cite{Bennett96a})} \label{prop_Bennett_distillation_rate}
\begin{eqnarray} \label{eq-EoF_and_distillation1}
\frac{\EoF ( \ket{\Psi} )}{\ln 2}
&  = & 
\sup \Bigl\{ r \; ;\; \lim_{N \rightarrow \infty} 
\Bigl( 
\inf_{{\rm LOCC}} 
 \bigl\| \Mm_{\rm LOCC}^{(N)} ( \ketbra{\Psi^{\otimes N}}{\Psi^{\otimes N}} ) - \ketbra{\Phi_{+}^{\otimes r N}}{\Phi_{+}^{\otimes r N}} \bigr\|_1 
\Bigr) = 0 \; \Bigr\}
\\ \label{eq-EoF_and_distillation2}
& = & 
\inf \Bigl\{ r \; ;\; \lim_{N \rightarrow \infty} 
\Bigl( 
\inf_{{\rm LOCC}}  \bigl\|  \ketbra{\Psi^{\otimes N}}{\Psi^{\otimes N}} - \Mm_{\rm LOCC}^{(N)}  ( \ketbra{\Phi_{+}^{\otimes r N}}{\Phi_{+}^{\otimes r N}} )  \bigr\|_1 
\Bigr)
= 0 \; \Bigr\}
\;.
\end{eqnarray}
\end{proposition}

\vspace{1mm}

Let us stress that these identities  are no longer valid for mixed states: then the right-hand sides of (\ref{eq-EoF_and_distillation1}) and
(\ref{eq-EoF_and_distillation2}) are, in general, not equal. They define two measures
of entanglement called the distillable entanglement and the entanglement cost (see~\cite{Horedecki_review} and references therein). 
The fact that these quantities coincide with $\EoF ( \ket{\Psi} )$ for pure states basically indicates that, among all
the possible entanglement measures, only one (namely $\EoF ( \ket{\Psi} )$)  becomes relevant asymptotically when dealing with 
many copies  of $ \ket{\Psi}$. 

\vspace{1mm}

\proof A simple and illuminating proof due to Nielsen~\cite{Nielsen99} is based on Proposition~\ref{prop_Nielsen}  and 
 the Shannon equipartition theorem. It runs as follows.
Let $\mu_i$ be the Schmidt coefficients of $\ket{\Psi}$. Consider $N$ i.i.d. random 
variables with distribution $\{ \mu_i\}$ and values in $I = \{ 1, \ldots, n \}$. 
The joint probabilities of these random variables are $p (\underline{i} )= \mu_{i_1} \ldots \mu_{i_N}$ with $\underline{i} = (i_1, \ldots, i_N)\in I^N$.
Given $\varepsilon>0$, the ``most likely set''  $\Aa_{N,\varepsilon} \subset I^N$ is by definition the set of all 
$\underline{i} \in I^{ N}$ such that  $2^{-N ( H + \varepsilon)} \leq p ({\underline{i}}) \leq 2^{-N ( H - \varepsilon)}$, $H$ being the Shannon
 entropy of $\{ \mu_i\}$, which is defined here by using the binary logarithm  (in our case, $H = \EoF ( \ket{\Psi})/\ln 2$).
The Shannon equipartition theorem~\cite{Shannon48}  tells us that $\Aa_{N,\varepsilon}$ has probability 
$P_{N,\varepsilon}  > 1-\varepsilon$ and cardinality $| \Aa_{N,\varepsilon}|$ satisfying
$(1-\varepsilon ) 2^{N ( H - \varepsilon)} \leq | \Aa_{N,\varepsilon}| \leq  2^{N ( H + \varepsilon)}$ for 
sufficiently large $N$. The idea of Nielsen's proof is to approximate
\begin{eqnarray}
\nn
\ket{\Psi^{\otimes N}} & = & \sum_{\underline{i} \in I^{N}} \sqrt{p ( \underline{i}) } \,
\ket{\alpha_{i_1}} \ldots \ket{\alpha_{i_N}} \otimes \ket{\beta_{i_1}} \ldots \ket{\beta_{i_N}}
\\
& \simeq & 
\ket{\Phi_{N,\varepsilon}} =\sum_{\underline{i} \in \Aa_{N,\varepsilon}} \sqrt{q ( {\underline{i}}) } \,
\ket{\alpha_{i_1}} \ldots \ket{\alpha_{i_N}} \otimes \ket{\beta_{i_1}} \ldots \ket{\beta_{i_N}}  
\end{eqnarray}
with $q ({\underline{i}}) =p ( {\underline{i}}) /P_{N,\varepsilon}$ and $\ket{\alpha_i}$, $\ket{\beta_i}$ as in Theorem~\ref{theo_Schmidt_dec}. 
Observe that the fidelity $| \braket{\Psi^{\otimes N}}{\Phi_{N,\varepsilon}}|^2 = P_{N,\varepsilon}$ is almost one  for small $\varepsilon$.
For any $\Aa \subset | \Aa_{N,\varepsilon}|$, one has
\begin{equation} \label{eq-proof_Bennet_distillation_rate}
\frac{(1-\varepsilon ) | \Aa | \, 2^{-2 N \varepsilon}}{|\Aa_{N,\varepsilon}|} \leq  \sum_{\underline{i} \in \Aa} q ( {\underline{i}}) \leq  
\frac{|\Aa |\, 2^{2 N \varepsilon}}{(1- \varepsilon)  | \Aa_{N,\varepsilon}|}  \;.
\end{equation}
The second inequality  implies that $\qv =(  q ({\underline{i}}) )_{i \in \Aa_{N,\varepsilon}} \prec ( 2^{-M}, \ldots, 2^{-M}, 0 , \ldots ,0 )$
with
\begin{equation}
M = \ln_2 (| \Aa_{N,\varepsilon}| (1 - \varepsilon) ) -  2 N \varepsilon\;.
\end{equation}
By  Proposition~\ref{prop_Nielsen}, this means  that $\ket{\Phi_{N,\varepsilon}}$ can be transformed by a LOCC 
into the $M$-qubit state
\begin{equation}
\ket{\Phi_+^{\otimes M}} = \sum_{\underline{j} \in \{ 0, 1\}^{M}} 2^{-\frac{M}{2}} \ket{j_1} \ldots  \ket{j_{M}} \otimes 
\ket{j_1} \ldots \ket{j_{M}}  \;.
\end{equation}
We conclude
that for $N$ sufficiently large  there exists a LOCC operation $\Mm_{\rm LOCC}^{(N,\varepsilon)}$ from $\Bb ( \Hh_\AB^{\otimes N})$ into $\Bb ( \complex^{\otimes 2M} )$
such that
\begin{eqnarray}
\nn
\bigl\| \Mm_{\rm LOCC}^{(N,\varepsilon )} ( \ketbra{\Psi^{\otimes N}}{\Psi^{\otimes N}}) - \ketbra{\Phi_+^{\otimes M}}{\Phi_+^{\otimes M}} \bigr\|_1
& \leq &
\bigl\| \ketbra{\Psi^{\otimes N}}{\Psi^{\otimes N}} - \ketbra{\Phi_{N,\varepsilon}}{\Phi_{N,\varepsilon}} \bigr\|_1
\\
& \leq &
2 \bigl( 1 - \bigl| \braket{\Psi^{\otimes N}}{\Phi_{N,\varepsilon}} \bigr|^2 \bigr)^\onehalf \; \leq  \; 2 \sqrt{\varepsilon}
\end{eqnarray}
(we have used Propositions~\ref{prop-trace_distance} and~\ref{prop_bounds_between_d_B_and_d_1} to get the first and second inequalities, respectively).
In addition, the distillation rate $M/N$ is bounded from below by $H - 3 \varepsilon + 2 N^{-1} \ln ( 1 - \varepsilon) $. Taking e.g. $\varepsilon = 1/\sqrt{N}$,
this proves that $\EoF ( \ket{\Psi}) \leq E_D ( \ket{\Psi})$, where $E_D ( \ket{\Psi})$ denotes the \RHS of (\ref{eq-EoF_and_distillation1}). 

Similarly, the first inequality in (\ref{eq-proof_Bennet_distillation_rate}) implies that $\ket{\Phi_{N,\varepsilon}}$ can be obtained asymptotically
by transforming $M'$ copies of $\ket{\Phi_{+}}$ with LOCCs,  more precisely it shows the existence of a LOCC operation $\Mm_{\rm LOCC}^{(N,\varepsilon )\,'}$ such that
\begin{equation}
\bigl\| \ketbra{\Psi^{\otimes N}}{\Psi^{\otimes N}} - \Mm_{\rm LOCC}^{(N,\varepsilon )\,'} ( \ketbra{\Phi_+^{\otimes M'}}{\Phi_+^{\otimes M'}}\bigr\|_1 
\leq 2 \sqrt{\varepsilon}
\end{equation}
for $N$ large enough, with 
\begin{equation}
M' = \ln_2 (| \Aa_{N,\varepsilon}|/(1 - \varepsilon)  ) +  2 N \varepsilon\;.
\end{equation}
The production rate $M'/N$ is bounded from above by $H + 3 \varepsilon - N^{-1} \ln ( 1 - \varepsilon) $.
This establishes that $\EoF ( \ket{\Psi}) \geq E_C ( \ket{\Psi})$, where $E_C ( \ket{\Psi})$ denotes the \RHS of (\ref{eq-EoF_and_distillation2}). 
But $E_D ( \ket{\Psi}) \leq  E_C ( \ket{\Psi})$, as otherwise one could transform asymptotically  by a LOCC $r' N$ {\it{e}}-bits  into $r N$ {\it{e}}-bits with
$r'<r$, which is impossible.
Hence $\EoF ( \ket{\Psi} = E_D ( \ket{\Psi}) =  E_C ( \ket{\Psi})$.
\finpro

\subsubsection{Convex roof constructions}  \label{sec-convex_roof}

The extension of $\EoF$ to mixed states  is done via a convex roof construction~\cite{Bennett96}.

\vspace{2mm}

\begin{definition}  \label{def_EoF_mixed_states}
The entanglement of formation of a mixed state $\rho \in \states (\Hh_\AB)$ is 
\begin{equation} \label{eq-def_EoF}
\EoF ( \rho ) = \min_{ \{ \ket{\Psi_i}, \eta_i \}} \biggl\{ \sum_i \eta_i \EoF ( \ket{\Psi_i} ) \biggr\}\;,
\end{equation}
where the minimum is over all pure state decompositions $\rho = \sum_i \eta_i \ketbra{\Psi_i}{\Psi_i}$ of $\rho$.
\end{definition}

\vspace{1mm}

\begin{proposition} {\rm (Vidal~\cite{Vidal00})} \label{prop_EoF_is_entanglement_meas}
$\EoF ( \rho)$ is an entanglement measure with values in the interval $[0, \ln n]$. It
satisfies the  monotonicity condition (which is stronger than (iii)) 
\begin{itemize}
\item[(iii'')] {\it $\sum_i p_i \EoF ( p_i^{-1} \Mm_{{\rm loc}}^{(i)} (\rho) ) \leq \EoF ( \rho)$ with 
 $p_i = \tr [ \Mm_{ {\rm loc}}^{(i)} (\rho) ]$,  for any family of
CP local maps $\Mm_{{\rm loc}}^{(i)}$ with Kraus operators $\{  A_{ij} \otimes B_{ik} \}_{j,k}$ such that
$\sum_{i,j,k} A_{ij}^\ast A_{ij} \otimes B_{ik}^\ast B_{ik} = 1$.}  
\end{itemize}
\end{proposition}

\vspace{1mm}

Note that the maps $\Mm_{ {\rm loc}}^{(i)}$ are not required to be trace preserving (but $\tr [ \Mm_{ {\rm loc}}^{(i)} ( \rho)] \leq 1$). 
Modulo a state  normalization, they
describe wavepacket reduction processes, see (\ref{eq-cond_state_gen_meas}).
 
\vspace{1mm}

\proof 
One has clearly $0 \leq \EoF ( \rho ) \leq \ln n$.
We now argue  that $\EoF$ satisfies all the axioms (i-iii) of an entanglement measure.
In fact, $\EoF$ is convex by construction.
Moreover, it follows from 
the aforementioned properties of $\EoF ( \ket{\Psi} )$ and the definition of mixed state entanglement
(Sec.~\ref{sec-def_entanglement}) that $\EoF ( \rho ) =0$ \ifif $\rho \in \Ss_\AB$. 
Finally, the monotonicity with respect to LOCC operations is a consequence of the convexity and can be shown as follows. Let
$\rho=\sum_i \eta_i \ketbra{\Psi_i}{\Psi_i}$ be the pure state decomposition minimizing the average entanglement in the \RHS of (\ref{eq-def_EoF}). Let
$\Mm$ be a separable operation with Kraus operators $A_j \otimes B_j$. 
We denote by $\eta_{j|i} =\| A_j   \otimes  B_j \ket{\Psi_i}\|^2$ the probability of outcome $j$
 given that the state is $\ket{\Psi_i}$.
From the convexity of $\EoF$ and its monotonicity (iii') for pure states (which holds by Proposition~\ref{prop-entanglement_meas_pure_states})
one finds
\begin{eqnarray}
\nn
\EoF \bigl( \Mm (\rho) \bigr) 
& \leq & 
\sum_{i} \eta_i \EoF \bigl( \Mm ( \ketbra{\Psi_i}{\Psi_i}) \bigr) 
\leq \sum_{ij} \eta_i \eta_{j|i} \EoF \bigl( \eta_{j|i}^{-\onehalf} A_j \otimes B_j  \ket{\Psi_i} \bigr)
\\
& \leq & 
\sum_{i} \eta_i \EoF ( \ketbra{\Psi_i}{\Psi_i}  ) = \EoF ( \rho)\;.
\end{eqnarray}
Thus $\EoF$ is  an entanglement measure. A similar reasoning shows that $\EoF$ satisfies (iii'').
\finpro

\vspace{2mm}

More generally, one can construct entanglement measures by extending to mixed states any entanglement measure on pure states
via a convex roof construction analog to (\ref{eq-def_EoF}). One gets in this way
a family of measures $E_f$ depending on the choice of the function $f$ in Proposition~\ref{prop-entanglement_meas_pure_states}. 
Conversely,  any entanglement measure $E$ satisfying the axiom (iii'') above coincides with
$E_f$ on pure states for some function $f$ fulfilling the assumptions of Proposition~\ref{prop-entanglement_meas_pure_states}~\cite{Vidal00}.
In particular, this suggests to define the concurrence for mixed states as
\begin{equation} \label{eq-my_def_concurrence2}
C ( \rho ) =  \min_{ \{ \ket{\Psi_i}, \eta_i \}} \biggl\{ \sum_i \eta_i C ( \ket{\Psi_i} ) \biggr\} \;,
\end{equation}
where $C ( \ket{\Psi_i} )$ is given by (\ref{eq-my_def_concurrence}).
It is known that $\rho_\AAA \mapsto \| \rho_\AAA\|_{1/2} = (\tr [ \rho_\AAA^{1/2} ] )^2$ is concave 
(see (\ref{eq-Minkowski_inequality}) in Appendix~\ref{app-norm_inequalities}), whence
$C(\rho)$ is an entanglement measure.  
Another measure of entanglement of common use for pure states is the Schmidt number
obtained by choosing $f(\rho_\AAA) = 1/\tr ( \rho_\AAA^2 )$ in  Proposition~\ref{prop-entanglement_meas_pure_states}.

\vspace{1mm}

As stated above, (iii'') means that separable measurements 
cannot  increase the average entanglement, but  entanglement can increase if one considers conditional expectations over subgroups of outcomes, 
\ie   one may have
$ \EoF ( p_i^{-1} \Mm_{{\rm loc}}^{(i)} (\rho) ) \geq \EoF ( \rho)$ for some $i$.
An example is given by the qutrit-qutrit system in the state
\begin{equation}
\rho = \onehalf  \ketbra{\Phi_+}{\Phi_+} + \onehalf \ketbra{2}{2} \otimes \ketbra{2}{2}
\quad , \quad \ket{\Phi_+} = \frac{1}{\sqrt{2}} \bigl(\ket{0} \ket{0} + \ket{1}\ket{1} \bigr)\;.
\end{equation}
Assume that Alice and Bob perform each a von Neumann measurement  with projectors $\Pi_{1}$ onto $\Span \{ \ket{0}, \ket{1} \}$ and 
$\Pi_2$ onto $\complex \ket{2}$. 
The conditional states $\rho_{\AB | 11} = \ketbra{\Phi_+}{\Phi_+}$ and $\rho_{\AB | 22}  =  \ketbra{2}{2} \otimes \ketbra{2}{2}$ have 
entanglement of formations $\ln 2$ and $0$, respectively. The first value is larger than $\EoF (\rho )$, which is
equal to $\ln 2/ 2$ according to the following result. 

\vspace{2mm}

\begin{corollary}
Let $\rho_1$ and $\rho_2$ be two states on $\Hh_\AB$ with bi-orthogonal supports $\range \rho_i \subset \Vv^\AAA_i \otimes \Vv_i^\BB$,  
where $ \Vv_i^\AAA \subset \Hh_\AAA$ and $\Vv_i^\BB  \subset \Hh_\BB$ are such that
$\Vv_2^\AAA = (\Vv_1^\AAA)^\bot$ and $\Vv_2^\BB = (\Vv_1^\BB )^\bot$. Let 
$\rho =  \eta_1 \rho_1 + \eta_2 \rho_2 $ with $\eta_i \geq 0$, $\eta_1 + \eta_2 = 1$. Then
$\EoF (\rho) = \eta_1 \EoF ( \rho_1 ) + \eta_2 \EoF ( \rho_2 )$.
\end{corollary}

\vspace{1mm}

\proof The inequality $\EoF (  \rho) \leq \eta_1 \EoF ( \rho_1 ) + \eta_2 \EoF ( \rho_2)$ follows from convexity.
The reverse inequality is a consequence of the monotonicity property (iii'') applied to the maps
\begin{equation}
\Mm_{\rm loc}^{(i)}(\rho) =\pi_i^\AAA  \otimes \pi_i^\BB \,\rho \,\pi_i^\AAA  \otimes \pi_i^\BB
\; , \; i=1,2 \quad  , \quad  
\Mm_{\rm loc}^{(3)}(\rho) = \pi_1^\AAA \otimes \pi_2^\BB \,\rho \,\pi_1^\AAA \otimes \pi_2^\BB  + \pi_2^\AAA \otimes \pi_1^\BB \,\rho \,\pi_2^\AAA \otimes \pi_1^\BB 
\;,
\end{equation}
where $\pi_i^\AAA$ and $\pi_i^\BB$ are the projectors onto $\Vv_i^\AAA$ and $\Vv_i^\BB$, respectively.
\finpro 
    
\vspace{2mm}

It is worth realizing the link between $\EoF (\rho)$ and the classical mutual information $I_{X:Y}$, where $X=\{ \eta_i\}$ is  associated to 
a pure state decomposition $\{ \ket{\Psi_i}, \eta_i\}$ of $\rho$ and $Y$ to the outcomes  of a local  \meas on $\AAA$ (Sec.~\ref{sec_Holevo_bound}).
Indeed, the maximum of $I_{X:Y}$ over all pure state decompositions and all POVMs on $\AAA$ is bounded by
\begin{equation}
\max_{ \{ \ket{\Psi_i}, \eta_i\}, \{ M_i^\AAA \} }  \bigl\{ I_{X:Y} \bigr\} \leq S( \rho_\AAA ) - \EoF ( \rho)\;.
\end{equation}
This inequality is a direct consequence of the Holevo bound (\ref{eq-Holevo_bound}) and the definition~(\ref{eq-def_EoF}) of $\EoF ( \rho)$. 

\subsubsection{The Wootters formula for two qubits} \label{sec-Wootters_formula}

The main problem with the  convex-roof construction (\ref{eq-def_EoF}) is that
finding the pure state decomposition minimizing the average entanglement is a non-trivial task.
Nevertheless, an astonishing formula enabling to evaluate $\EoF ( \rho )$ explicitly for two qubits was found by Wootters~\cite{Wootters98}.
It reads
\begin{equation} \label{eq-Wootters_formula}
\EoF ( \rho ) = h ( C ( \rho)) 
\end{equation}
where  $C ( \rho)$ is given by (\ref{eq-my_def_concurrence2}) and $h : [0,1] \rightarrow [0,\ln n]$ is the convex increasing function
\begin{equation} 
h ( C) = - \frac{1 + \sqrt{1-C^2}}{2} \ln \Bigl( \frac{1 + \sqrt{1-C^2}}{2}  \Bigr) 
- \frac{1 - \sqrt{1-C^2}}{2} \ln \Bigl( \frac{1 - \sqrt{1-C^2}}{2}  \Bigr)\;.
\end{equation}
The main point is that $C ( \rho)$ can be calculated explicitly as follows.
Let  $\lambda_1 \geq \lambda_2 \geq \lambda_3 \geq \lambda_4$ be the square roots of the eigenvalues of
$\rho \sigma_y \otimes \sigma_y \,\overline{\rho} \,\sigma_y \otimes \sigma_y$ (here $\sigma_y$ is the $y$-Pauli matrix and 
$\overline{\rho}= J \rho J$ the complex conjugate of 
$\rho$ in the canonical basis). Then
\begin{equation}  \label{eq-concurrence_2_qbits}
C ( \rho) = \max \{ 0, \lambda_1 - \lambda_2 - \lambda_3 - \lambda_4 \}\;.
\end{equation}
For pure states this yields $C(\ket{\Psi}) = | \bra{\Psi} \sigma_y \otimes \sigma_y J \ket{\Psi} |^2$, in agreement
with the result of Sec.~\ref{sec-concurrence}.
The proof of (\ref{eq-Wootters_formula}) is somehow tricky but relies on simple linear algebra arguments (see~\cite{Wootters98}).

\subsection{Maximally entangled states} 

One may expect intuitively that the most entangled states  are extremal states in $\states (\Hh_\AB)$, that is,
they are the pure maximally entangled states described in Sec.~\ref{sec-def_entanglement}. If one uses as a criterion 
for being mostly entangled  the property of having the highest 
entanglement of formation, this is indeed correct when the dimensions of $\Hh_\AAA$ and $\Hh_\BB$ are such that 
$n_\AAA/2 < n_\BB < 2 n_\AAA$. When $n_\BB \geq 2 n_\AAA$, convex combinations of pure maximally entangled states
with reduced $\BB$-states living on   orthogonal subspaces of $\Hh_\BB$ are also maximally entangled
(a similar statement holds of course  by exchanging $\AAA$ and $\BB$).

\vspace{2mm}

\begin{proposition} \label{prop_max_ent_states}
Assume that $n=n_\AAA \leq n_\BB$ and let  $r=1,2,\ldots$ be such that  $r n_\AAA \leq n_\BB < (r+1) n_\AAA$. Then
the states $\rho \in \states (\Hh_\AB)$ having a maximal entanglement of formation $\EoF ( \rho) = \ln n$ 
are convex combinations of the $r$ orthogonal maximally entangled states 
\begin{equation}  \label{eq-almost_finish_until_p_66}
\ket{k}= n^{-\onehalf} \sum_{i=1}^{n} \ket{\alpha_i^{(k)}} \otimes \ket{\beta_i^{(k)}}
\quad , \quad k=1,\ldots, r\;,
\end{equation}
with $\braket{\alpha_i^{(k)}}{\alpha_j^{(k)}}=\delta_{ij}$ and 
$\braket{\beta_i^{(k)}}{\beta_j^{(l)}}=\delta_{k l} \delta_{ij}$. 
\end{proposition}

\vspace{1mm}

\proof 
Let $\rho$ be a state with $\EoF ( \rho) = \ln n$. According to Definition~\ref{def_EoF_mixed_states} and given that
$\EoF ( \ket{\Psi} ) \leq \ln n$ with equality \ifif $\ket{\Psi}$ is maximally entangled, this means that
any pure state decomposition of $\rho$ is made of maximally entangled states. This is the case in particular for the
spectral decomposition $\rho = \sum_k p_k \ketbra{k}{k}$, from which one can obtain all other pure state decompositions  $\{ \ket{\Psi_i}, \eta_i\}$ 
by the formula  $\sqrt{\eta_i} \ket{\Psi_i} = \sum_k u_{ik} \sqrt{p_k} \ket{k}$ with
$\eta_i = \sum_k | u_{ik} |^2 p_k$ (see (\ref{eq-link_between_pure_state_dec})).  Let us set $D_{kl} = \tr_\BB ( \ketbra{k}{l} ) $. We would like to show that  $D_{kl}= n^{-1} \delta_{kl}$  if $p_{k} p_{l} \not= 0$.
We already know that $D_{kk} = 1/n$ if $p_k\not= 0$, since $\ket{k}$ is maximally entangled. 
By plugging the above expression of $\sqrt{\eta_i} \ket{\Psi_i}$ into $\tr_\BB ( \ketbra{\Psi_i}{\Psi_i} ) = 1/n$, one is led to 
\begin{equation}
\sum_{k,l, k \not= l} \sqrt{p_k p_l} u_{ik} \overline{u}_{il} D_{kl} = 0\;.
\end{equation}
This equality holds for any $i$ and any unitary matrix $(u_{ik})$, hence $\sqrt{p_k p_l}  D_{kl}=0$ if $k \not= l$ and the above claim is true.
One deduces from $D_{kk}=1/n$ that the eigenvectors $\ket{k}$ with eigenvalues $p_k>0$ have Schmidt decompositions given by (\ref{eq-almost_finish_until_p_66}). 
For $k \not = l$, $D_{kl}=0$ is then equivalent to $\Vv_\BB^{(k)} \bot \Vv^{(l)}_\BB$ with $\Vv_\BB^{(k)} = \Span \{ \ket{\beta_i^{(k)}}  \}_{i=1}^{n} \subset \Hh_\BB$.
If $n_\BB < (r+1) n$ then at most $r$ subspaces $\Vv_\BB^{(k)} $ may
be pairwise orthogonal.  Thus at most $r$ eigenvalues $p_k$ are non-zero. 
\finpro

\newpage
\section{The quantum discord} \label{sec-discord}

 The  quantum discord was introduced by Ollivier and Zurek~\cite{Ollivier01} and Henderson and Vedral~\cite{Henderson01}
as an indicator of the ``degree of quantumness'' of mixed states.
For pure states it coincides with  the entanglement of formation. 
Certain separable mixed states have, however, a non-zero discord.
These states are obtained by preparing  locally mixtures  
of non-orthogonal states, which cannot be perfectly discriminated by local measurements. 
Such separable states cannot be classified as ``classical'' and actually contain quantum correlations that are not captured by the entanglement 
measures reviewed in Sec.~\ref{sec-entanglement}.
Apart from this observation, a motivation for the quantum discord came out
in the last decade from the claim that it could play the role of a resource in certain quantum
algorithms and quantum communication protocols~\cite{Datta08,Lanyon08,Passante11,Madhok11,Gu2012,Dakic12}.  
In particular, it has been suggested~\cite{Datta08,Lanyon08,Passante11} that the discord might capture the quantum correlations at the origin of the 
quantum speedup in the deterministic quantum computation with one qubit  (DQC1)  of  Knill and Laflamme~\cite{Laflamme98}.
The DQC1 algorithm computes the trace of a $2^N \times 2^N$ unitary matrix exponentially faster than all known classical algorithms. 
The entanglement produced during the computation with $(N+1)$ qubits
is bounded independently of $N$, for any bipartition of the $(N+1)$ qubits~\cite{Datta05}. 
This means that the total amount of bipartite entanglement is a negligible 
fraction of the maximal entanglement possible. However, a non-vanishing quantum discord  between the control
qubit and the $N$ target qubits appears during the computation~\cite{Datta08}, save for particular unitaries~\cite{Dakic10}.  
The DCQ1 algorithm is singled out by the fact that it uses mixed states, the $N$ target qubits being initially in a Gibbs state
at infinite temperature.
In contrast, for quantum computations using pure states, Jozsa and Linden~\cite{Jozsa03} have shown that
in order to offer an exponential speedup over classical computers,
the computation must produce  entanglement which is not 
restricted to qubit blocks of fixed size as the problem size increases.

The definition of the quantum discord is given in Sec.~\ref{sec-def_discord}. We then characterize the states with vanishing
discord in Sec.~\ref{sec_A_clas_states} and exhibit some important properties of the discord in Sec.~\ref{sec-prop_didcord}.
The so-called monogamy relation linking the discord and the entanglement of formation in tripartite systems is stated and
proven in Sec.~\ref{sec-monogamy_relation}.

\subsection{Definition of the quantum discord} 
\label{sec-def_discord}

Let us first consider some classical discrete random variables $A$ and $B$ with joint probabilities $p_{ij}$
and marginals $p_A (i) = \sum_j p_{ij}$ and $p_B (j) = \sum_i p_{ij}$.
The correlations between $A$ and $B$ are measured by the mutual information 
$I_{A:B} = H (A) + H(B ) - H(A,B)$.
We recall from Sec.~\ref{sec_Holevo_bound} that
\begin{equation} \label{eq-identity_Shannon}
I_{A:B} = H(B) - H ( B |A )\;,
\end{equation}
where $H( B | A )=\sum_i p_{A} (i) H ( B|i)$ is the conditional entropy, see (\ref{eq-class_cond_entropy}). 
This  conditional entropy describes the amount of information on $B$ left after the value $A=i$ has been measured, averaged over all possible outcomes $i$. 

In the quantum setting, the analog of the  random variables $A$ and $B$ is 
a bipartite quantum system $\AB$ in a state $\rho$.
The marginals are the reduced states  $\rho_\AAA = \tr_{\BB} (\rho)$ and $\rho_\BB=\tr_\AAA(\rho)$.
The generalization of the mutual information reads
\begin{equation} \label{eq-mutual_info}
I_{\AAA:\BB} (\rho) = S (\rho_\AAA) + S (\rho_\BB)  - S(\rho) \;,
\end{equation}
where $S (\cdot)$ is the von Neumann entropy (\ref{eq-def_S}).
Similarly to the classical case, one has $I_{\AAA:\BB} (\rho) \geq 0$ and $I_{\AAA:\BB} (\rho) =0$ \ifif
$\rho$ is a product state, \ie $\rho=\rho_\AAA\otimes \rho_\BB$ (this is nothing but the subadditivity property of $S$,
see Sec.~\ref{sec-von_Neumann_wentropy}). It is easy to verify that $I_{\AAA:\BB} (\rho)$ is related to the relative entropy (\ref{eq-def_relative_S}) by
\begin{equation} \label{eq-mutual_info_as_relative_entropy}
I_{\AAA:\BB} (\rho) = S ( \rho || \rho_\AAA \otimes \rho_\BB )\;.
\end{equation}
By the monotonicity of the relative entropy (Theorem~\ref{theo-Petz}), 
$I_{\AAA : \BB}( \Mm_{\rm loc} (\rho)) \leq I_{\AAA : \BB}(\rho)$ for any local operation $\Mm_{\rm loc} = \Mm_\AAA \otimes \Mm_\BB$, where 
the operations $\Mm_\AAA : \Bb ( \Hh_\AAA) \rightarrow \Bb ( \Hh_\AAA')$  and $\Mm_\BB : \Bb( \Hh_\BB) \rightarrow \Bb ( \Hh_\BB')$ may have 
different initial and final  spaces (for instance, $\Mm_\AAA$  can be the partial trace over a part of $\AAA$).

However, there is no quantum analog of the identity (\ref{eq-identity_Shannon}).
Let us define a conditional entropy of $\BB$ given a von Neumann measurement $\{ \pi_i^\AAA \}$
on  $\AAA$ by $S_{\BB | \AAA} (\rho | \{\pi_i^\AAA\} ) = \sum_i \eta_i S ( \rho_{\BB|i} )$, where
\begin{equation} \label{eq-def_cond_state_discord}
\rho_{\BB|i} = \eta_i^{-1} \tr_\AAA (  \pi_i^\AAA \otimes 1 \,\rho )
\quad , \quad \eta_i = \tr (  \pi_i^\AAA \otimes 1 \,\rho )\;.
\end{equation}
Here $\eta_i$ is the probability of the \meas outcome $i$ and
$\rho_{\BB|i} = \tr_\AAA (\rho_{\AB|i})$ is the corresponding conditional state of $\BB$
(see Sec.~\ref{sec-measurements}). The ensemble $\{ \rho_{\BB|i} , \eta_i \}$ defines a convex decomposition of $\rho_\BB$ 
(\ie $\rho_\BB = \sum_i \eta_ i \rho_{\BB|i}$) describing a state preparation of the subsystem $\BB$ realized by the \meas on $\AAA$.
The quantum version of the right-hand side of (\ref{eq-identity_Shannon}) is
the maximal reduction of entropy of $\BB$ due to a von Neumann measurement on $\AAA$,
\begin{equation} \label{eq-class_correl}
J_{\BB|\AAA}^\vN(\rho)  = S (\rho_\BB) - \min_{ \{ \pi_i^\AAA \}} \biggl\{  \sum_i \eta_i S ( \rho_{\BB|i} ) \biggr\} \;,
\end{equation}
the minimum being over all orthonormal families of projectors on $\Hh_\AAA$.  
This quantity represents the classical correlations between $\AAA$ and $\BB$ (see the discussion after Proposition~\ref{prop-discord1} below). 
Note that $J_{\BB|\AAA}^\vN(\rho) $ places an upper bound on the classical mutual information  between the ensemble $\{ \rho_{\BB | i}, \eta_i\}$ and 
the outcome probabilities when performing measurements on $\BB$ to discriminate the states $\rho_{\BB| i}$ (Sec.~\ref{sec_Holevo_bound}). 
Actually,  $J_{\BB|\AAA}^\vN(\rho) $ coincides with the corresponding Holevo quantity (\ref{eq-Holevo_bound}). 
By concavity of the von Neumann entropy, one has $ J_{\BB|\AAA}^\vN (\rho) \geq 0$. Furthermore, (\ref{eq_entropy_as_min_over_pure_state_dec2_gen}) entails
$J_{\BB|\AAA}^\vN (\rho) \leq \max_{ \{ \pi_i^\AAA \}} H( \{ \eta_i\})$.

It also follows from the concavity of $S$ that the minimum in (\ref{eq-class_correl})   is achieved for
rank-one projectors. In fact, by decomposing each projector $\pi_i^\AAA$ of rank $r_i$ as a sum of $r_i$
rank-one projectors $\pi_{ik}^\AAA$, one finds that
$ \rho_{\BB |i} = \sum_k (\eta_{ik}/\eta_i) \rho_{\BB |ik}$ is a convex combination of the states 
$\rho_{\BB |ik} = \eta_{ik}^{-1}\tr_\AAA ( \pi_{ik}^\AAA \otimes 1 \,\rho)$
 if $\eta_i =\sum_k \eta_{ik} >0$. Thereby $\sum_i \eta_i S(\rho_{\BB
   |i}) \geq \sum_{ik} \eta_{ik} S ( \rho_{\BB |ik})$.

Ollivier and Zurek~\cite{Ollivier01} and Henderson and Vedral~\cite{Henderson01} proposed 
in two independent works published in 2001 to characterize the amount of non-classicality in the state $\rho$ by forming the difference
between the total correlations given by $I_{\AAA:\BB} (\rho)$ and the classical correlations given by 
$J_{\BB|\AAA}^\vN (\rho)$.

\vspace{2mm}

\begin{definition} \label{def-Q_discord}
The quantum discord of the bipartite system $\AB$ in state $\rho$ is 
\begin{equation} \label{eq-def_QD}
\delta_\AAA^\vN (\rho) 
= I_{\AAA:\BB} (\rho) - J_{\BB|\AAA}^\vN (\rho) = S ( \rho_\AAA) - S( \rho) + \min_{ \{ \pi_i^\AAA \}} \biggl\{ \sum_i \eta_i S ( \rho_{\BB|i} ) \biggr\}
\;.
\end{equation}
\end{definition}

\vspace{1mm}

In~\cite{Henderson01}, the minimization is done over generalized measurements
given by POVMs $\{ M_i^\AAA\}$ on $\Hh_\AAA$, instead of von Neumann measurements. 
The conditional states and  outcome probabilities are then (Sec.~\ref{sec-measurements})
\begin{equation} \label{eq-def_cond_state_discord2}
\rho_{\BB|i} = \eta_i^{-1} \tr_\AAA (  M_i^\AAA \otimes 1 \,\rho )
\quad , \quad \eta_i = \tr ( M_i^\AAA \otimes 1 \,\rho )\;.
\end{equation}
 We denote the corresponding discord by $\delta_\AAA (\rho)$.
As in the case of von Neumann measurements, the minimum is achieved for rank-one \meas operators $M_i^\AAA$. 
In general, the inequality   $\delta_\AAA ( \rho ) < \delta_\AAA^\vN ( \rho)$ is strict\footnote{
See e.g.~\cite{Zaraket04,Galve11} for a comparison of the von Neumann and POVM  discords for two qubits.  
}.
Nevertheless,  by the Neumark extension theorem, $\delta_\AAA$ coincides with $\delta_\AAA^\vN$ up to an enlargement 
of the space $\Hh_\AAA$. More precisely, by plugging $M_i^\AAA = \bra{\epsilon_0} \Pi^{\AAE} \ket{\epsilon_0}$   (see Remark~\ref{exo-Peres_Found_Phys})
into  (\ref{eq-def_cond_state_discord2}) and using the additivity of $S$ under tensor products, a simple calculation gives
\begin{equation} \label{eq-link_between_von_Neumann_and_POVM_discords}
\delta_\AAA ( \rho) = \delta_\AAE^\vN ( \rho \otimes \ketbra{\epsilon_0}{\epsilon_0} )\;,
\end{equation}
the \RHS being independent of the ancilla state $\ket{\epsilon_0} \in \Hh_\EE$.

The discords $\delta_\AAA^\vN (\rho) $ and $\delta_\AAA (\rho) $ thus measure the amount of total correlations between $\AAA$ and $\BB$ which cannot be accessed  
by local measurements on the subsystem $\AAA$. 
Note that they are asymmetric under the exchange $\AAA \leftrightarrow \BB$. One can define similarly the discords $\delta_\BB^\vN (\rho) $ and
$\delta_\BB (\rho)$ by performing the measurements  on  the subsystem $\BB$. 

For pure states $\rho_\Psi = \ketbra{\Psi}{\Psi}$, the mutual information $I_{\AAA : \BB} ( \rho_\Psi)$ is equal to $2 S ( [\rho_\Psi]_\BB)$, see 
(\ref{eq-equality_entropy_reduce_states}),  and 
 the \meas minimizing the conditional entropy of $\BB$ is the \meas in the eigenbasis $\{ \ket{\alpha_i} \}$ of the reduced state $[\rho_\Psi]_\AAA$.
In fact, according to (\ref{eq-Schmidt_decomposition}) the corresponding post-measurement states 
$\rho_{\BB | i} = \ketbra{\beta_i}{\beta_i}$ are pure and thus have zero entropy. Then (\ref{eq-class_correl}) yields
$J_{\BB | \AAA} ( \rho_\Psi ) =  S ( [\rho_\Psi]_\BB)$. As a result, the discords coincide for pure states with the entanglement of formation,
\begin{equation} 
\delta_\AAA ( \ket{\Psi} ) = \delta_\AAA^{\rm v.N.} ( \ket{\Psi} )  = \delta_\BB ( \ket{\Psi} ) = \delta_\BB^{\rm v.N.} ( \ket{\Psi} ) = \EoF ( \ket{\Psi} ) \;.
\end{equation}

For mixed states, it was pointed out in~\cite{Ollivier01} that if the \meas operators $M_i^\AAA$ are of rank one then
\begin{equation} \label{eq-Ollivier_Zurek_identity}
 \sum_i \eta_i S ( \rho_{\BB |i} ) = S \bigl( \Mm_\AAA \otimes 1 (\rho) \bigr) - S  \bigl( [\Mm_\AAA \otimes 1 (\rho)]_\AAA \bigr)
= -I_{\AAA:\BB} \bigl( \Mm_\AAA \otimes 1 (\rho) \bigr) + S ( \rho_\BB )\;,
\end{equation}
where $\Mm_\AAA$ is the quantum operation on $\AAA$ associated to the measurement.
Actually, consider the family of Kraus operators for $\Mm_\AAA$ given by $\{ A_i = \ketbra{i}{\widetilde{\mu_i}} \}$, where 
$\ket{\widetilde{\mu}_i}$ are unnormalized vectors  such that $M_i^\AAA = \ketbra{\widetilde{\mu_i}}{\widetilde{\mu_i}}$
and $\{ \ket{i} \}$ is an \ONB of a pointer space $\Hh_\PP$.
Then $\Mm_\AAA \otimes 1 (\rho)  =  \sum_{i} \eta_i \ketbra{i}{i} \otimes \rho_{\BB |i}$
and the reduced state $[ \Mm_\AAA \otimes 1 ( \rho) ]_\AAA = \sum_i \eta_i  \ketbra{i}{i}$ has entropy $-\sum_i \eta_i \ln \eta_i$.    
A simple calculation yields the first equality in (\ref{eq-Ollivier_Zurek_identity}).
The second equality is clear once one notices that $[ \Mm_\AAA \otimes 1 ( \rho) ]_\BB = \rho_\BB$.

Therefore, by combining (\ref{eq-class_correl}), (\ref{eq-def_QD}), and (\ref{eq-mutual_info_as_relative_entropy}) one
obtains the following result. 

\vspace{2mm}

\begin{proposition} \label{prop-discord1} {\rm \cite{Luo_Fu10}}
The discord $\delta ( \rho) = I_{\AAA:\BB} (\rho) - J_{\BB | \AAA} ( \rho)$
is the minimal difference of mutual information of 
$\AB$ before and after a  measurement on $\AAA$, \ie
\begin{equation} \label{eq-classical_correl_as_max_mutual_info}
J_{\BB | \AAA} ( \rho) = \max_{ \{ M_i^\AAA \} } \bigl\{  I_{\AAA:\BB} ( \Mm_\AAA  \otimes 1  (\rho) ) \bigr\}\;,
\end{equation}
where the maximum is over all POVMs on $\AAA$ with rank-one operators $M_i^\AAA$ and $\Mm_\AAA $ is the associated \QO on $\Bb ( \Hh_\AAA)$.
As a result, 
\begin{equation} \label{eq-other_def_Q_discord}
\delta_\AAA  (\rho) 
= \min_{ \{ M_i^\AAA \} } \Bigl\{ S \bigl( \rho ||  \rho_\AAA \otimes \rho_\BB \bigr) - 
 S \bigl( \Mm_\AAA \otimes 1 (\rho) || \Mm_\AAA(\rho_\AAA )\otimes \rho_\BB ) \bigr) \Bigr\}
\;.
\end{equation}
Similarly, $J_{\BB | \AAA}^\vN  (\rho)$ is  given by maximizing $ I_{\AAA:\BB} ( \Mm_{\pi^\AAA}  \otimes 1  (\rho) )$
over all von Neumann measurements $\Mm_{\pi^\AAA}$ on $\AAA$ of the form (\ref{eq-von_Neumann_measurement})
with rank-one projectors $\pi_i^\AAA$.
\end{proposition}

\vspace{1mm}

Observing that a \meas on $\AAA$ with no readout 
 removes the quantum correlations between $\AAA$ and $\BB$, the \RHS of (\ref{eq-classical_correl_as_max_mutual_info})
can be interpreted as the amount of classical correlations between the two subsystems. These subsystems are not correlated classically, 
\ie $J_{\BB | \AAA} ( \rho) =0$, \ifif $\rho = \rho_\AAA \otimes \rho_\BB$ is a product state. This result holds for $J_{\BB | \AAA}^{\rm v.N.} ( \rho)$ as well.  
Actually, by (\ref{eq-classical_correl_as_max_mutual_info}),
$J_{\BB | \AAA} ( \rho) =0$ is equivalent to
  $ \Mm_\AAA  \otimes 1  (\rho)$ being a product state for any collection of operators 
$M_i^\AAA= \ketbra{\widetilde{\mu}_i}{\widetilde{\mu}_i}$ forming a POVM. This implies 
$\eta_i \rho_{\BB | i} = \bra{\widetilde{\mu_i}} \rho \ket{\widetilde{\mu}_i} =\eta_i  \rho_\BB$ for all $i$ 
(see the discussion before Proposition~\ref{prop-discord1}). Choosing the $\ket{\widetilde{\mu_i}}$ to be the eigenvectors of the observable $A$,
one obtains that $\langle A \otimes B \rangle_\rho =  \langle A \otimes 1 \rangle_\rho \langle 1 \otimes B \rangle_\rho$
for any $A \in \Bb ( \Hh_\AAA)_{\rm s.a.}$ and $B \in \Bb ( \Hh_\BB)_{\rm s.a.}$, with $\langle \cdot \rangle_\rho = \tr ( \cdot \rho)$.

Let us emphasize that finding the optimal \meas which maximizes the post-\meas mutual information,
 and hence calculating  the discords $\delta_\AAA^\vN (\rho)$ and $\delta_\AAA (\rho)$, is a difficult problem  in general.  
Even for two qubits, this problem has been solved so far for a restricted family of states only, namely, the states $\rho$ with maximally 
mixed marginals $\rho_\AAA = \rho_\BB = 1/2$~\cite{Luo08}. In other cases\footnote{
An incorrect work~\cite{Ali10} claiming to extend the result of Ref.~\cite{Luo08} to the larger family of the so-called  $X$-states has generated a
profusion of articles.  Comparing with numerical evaluations, the result of~\cite{Ali10} apparently 
gives good approximations of the discord for randomly chosen $X$-states  (see the discussion in~\cite{Modi_review}). 
}
the discords must be evaluated numerically
(however, $\delta_\AAA (\rho)$ can be determined analytically for low-rank density matrices with the help of the
monogamy relation, see Sec.~\ref{sec-monogamy_relation} and~\cite{Modi_review}).

\subsection{The $\AAA$-classical states} \label{sec_A_clas_states}

 The monotonicity property of the relative entropy and formula (\ref{eq-other_def_Q_discord}) imply that $\delta_\AAA (\rho)$ is non-negative.
The states with vanishing discord can be determined with the help of Theorem~\ref{theo-Petz}, leading to the following result\footnote{
In Ref.~\cite{Ollivier01}, the authors argue that the non-negativity of $\delta_\AAA^\vN ( \rho)$ is a direct consequence of 
(\ref{eq-Ollivier_Zurek_identity}) and the concavity of
$S(\rho) - S ( \rho_\AAA)$ with respect to $\rho$. I do not see how such a claim could be justified and believe that the simplest proof of  
Proposition~\ref{prop-classical_states} is to rely on Theorem~\ref{theo-Petz}. Alternatively, the non-negativity of the discord can be 
justified with the help of the strong subadditivity of the von Neumann entropy (which is closely related to Theorem~\ref{theo-Petz}, 
see Sec.~\ref{sec-relative_entropy}), as shown in Ref.~\cite{Madhok13}.
}.

\vspace{2mm}

\begin{proposition} \label{prop-classical_states}
The quantum discord is non-negative and $\delta_\AAA (\sigma )=0$ \ifif
\be \label{eq-A-classical_state}
\sigma 
=
\sum_{i=1}^{n_\AAA}  q_i \ketbra{\varphi_i}{\varphi_i} \otimes \sigma_{\BB|i} \;,
\ee
where $\{ \ket{\varphi_i} \}_{i=1}^{n_\AAA}$ is an orthonormal basis of $\Hh_\AAA$, 
$\sigma_{\BB|i}$ are some (arbitrary)
states of $\BB$ depending on the index $i$, and $q_i \geq 0$ are some probabilities, $\sum_i q_i = 1$.
\end{proposition}

\vspace{1mm}

The non-negativity of $\delta_\AAA (\rho)$ means that one cannot gain more information on a bipartite system $\AB$ by performing a \meas on the 
subsystem $\AAA$ than the entropy of $\AAA$, namely, $S ( \rho_\AB ) - \sum \eta_i S ( \rho_{\AB |i}) \leq S ( \rho_\AAA)$ for any 
$\rho_\AB \in \states (\Hh_\AB)$. The important point is that if $\rho_\AB$ is not of the form  (\ref{eq-A-classical_state}), then any 
measurement on $\AAA$ gives {\it less} information on $\AB$ than $S(\rho_\AAA)$. Stated differently, one can not retrieve
all the information on $\AAA$ by a local measurement, because of the presence of quantum correlations between $\AAA$ and $\BB$.

\vspace{2mm}

\Proof
It remains to show the second affirmation. It is easy to convince oneself that the states (\ref{eq-A-classical_state}) have a vanishing discord. 
In fact, one finds  
$I_{\AAA:\BB}(\sigma  )=S (\sigma_\BB ) - \sum_i q_i S (\sigma_{\BB |i}) \leq J_{\BB| \AAA }^\vN (\sigma )$ (the inequality follows by noting that
$\sigma_{\BB | i}$ and $q_i$ are the conditional state and outcome probability for a \meas on $\AAA$ in the basis $\{\ket{\varphi_i}\}$). Hence  
 $\delta_\AAA ( \sigma) = \delta_\AAA^\vN ( \sigma)=0$ as a consequence of the non-negativity of $\delta_\AAA$. Reciprocally, let 
$\sigma \in \states (\Hh_\AB)$ be such that $\delta_\AAA^\vN ( \sigma)=0$.
As we shall see below it is enough to work with the 
von Neumann discord, the result for $\delta_\AAA$ will then follow from (\ref{eq-link_between_von_Neumann_and_POVM_discords}).
 According to (\ref{eq-other_def_Q_discord}) and
Theorem~\ref{theo-Petz}, $\delta_\AAA^\vN (\sigma )=0$ \ifif there exists a von Neumann measurement $\Mm_\AAA$ on $\AAA$
with rank-one projectors $\pi_i^\AAA = \ketbra{\varphi_i}{\varphi_i}$ such that
$\sigma = {\Rr}_{\AAA} \Mm_\AAA \otimes 1 (\sigma)$, where $\Rr_\AAA = {\Rr}_{\Mm_\AAA \otimes 1 , \sigma_0}$ is the
transpose operation of $\Mm_\AAA \otimes 1$ for the state  $\sigma_0 = \sigma_\AAA \otimes \sigma_\BB$.
Without loss of generality we may assume $\eta_i = \bra{\varphi_i} \sigma_\AAA \ket{\varphi_i} >0$ for all $i$.  
Thanks to  (\ref{eq-transpose_operation}) and to the identity
$\Mm_\AAA \otimes 1 ( \sigma_0 ) = \sum_i \eta_i \ketbra{\varphi_i}{\varphi_i} \otimes \sigma_\BB$, the transpose operation $\Rr_\AAA$ has Kraus operators
$R_i =  \eta_{i}^{-1/2} \sqrt{\sigma_\AAA} \ketbra{\varphi_i}{\varphi_i} \otimes 1$.
We now argue that this implies that $\sigma = \widehat{\Mm}_\AAA \otimes 1 ( \sigma)$ with 
$\widehat{\Mm}_\AAA$ the  von Neumann \meas with projectors $\widehat{\pi}_k^\AAA$ onto 
the subspaces $\Span \{ \ket{\varphi_i} ; i \in I_k \}$, where $\{ I_1, \ldots, I_d \}$ 
is a partition of $\{ 1, \ldots , n_\AAA \}$. Actually, the condition $\sigma = \Rr_\AAA \Mm_\AAA \otimes 1 (\sigma)$ reads
\begin{equation} \label{eq-condition_hatM_M_rho=rho}
\bra{\varphi_i} \sigma \ket{\varphi_j}  
 = 
  \sum_{l=1}^{n_\AAA} \eta_l^{-1}  (\sqrt{\sigma_\AAA})_{il} (\sqrt{\sigma_\AAA} )_{lj}
    \bra{\varphi_l} \sigma \ket{\varphi_l} 
\quad , \quad i,j=1, \ldots ,n_\AAA
\end{equation}
with $(\sqrt{\sigma_\AAA})_{ij} = \bra{\varphi_i} \sqrt{\sigma_\AAA} \ket{\varphi_j} \in \real$.
Let us set $\sigma_{\BB | i} = \eta_i^{-1} \bra{\varphi_i} \sigma \ket{\varphi_i}$ and $\eta_{l|i} = | (\sqrt{\sigma_\AAA})_{il}|^2/\eta_i$. 
This defines respectively a state on $\Hh_\BB$ and a probability distribution  for any fixed $i$. With this notation, (\ref{eq-condition_hatM_M_rho=rho}) can be rewritten 
for $i=j$ as
\begin{equation} \label{eq-proof_prop_vanishing_discord}
\sigma_{\BB | i} = \sum_{l=1}^{n_\AAA} \eta_{l|i} \sigma_{\BB | l}
\quad , \quad i=1,\ldots ,n_\AAA \;.
\end{equation} 
Let $I_i = \{ j  ; \sigma_{\BB | j } = \sigma_{\BB |i} \} \subset \{1, \ldots, n_\AAA\} $.
Clearly, the sets $I_i$ are either equal or disjoint. Hence one can extract from them a partition
$\{ I_{i_1} , I_{i_2} , \ldots , I_{i_d} \}$ of $\{1, \ldots, n_\AAA\}$. We claim that (\ref{eq-proof_prop_vanishing_discord}) 
implies $\eta_{l|i}=0$ for $l \notin I_i$.  This is a consequence of the following lemma.

\vspace{2mm}

\begin{lemma} \label{lemma_proof_vanishing_discord}
Let $\xv= (x_1,\ldots, x_d) $ be a vector of $\Xx^d$ with distinct components $x_k$, where $\Xx$ is a real vector space, and  
$\{ \xi_{k|m}\}_{k=1}^d$ be some probability distributions such that $\xi_{k|m} =0 \Leftrightarrow \xi_{m|k} =0$ 
and the components of $\xv$ have convex decompositions
\begin{equation}
x_m= \sum_{k=1}^d \xi_{k|m} x_k \quad \forall \;\; m=1,\ldots, d\;.
\end{equation}
Then $\xi_{k|m} = \delta_{km}$ for any $k,m =1,\ldots , d$.
\end{lemma}

\vspace{1mm}

We postpone the proof of this result to the next paragraph. By rewriting (\ref{eq-proof_prop_vanishing_discord}) as
\begin{equation}
\sigma_{\BB | i_m} = \sum_{k=1}^{d} \xi_{k|m} \sigma_{\BB | i_k}
\quad \text{ with } \quad \xi_{k|m}= | I_{i_m}|^{-1} \sum_{(l,i) \in I_{i_k}  \times I_{i_m}} \eta_{l|i}\;,
\end{equation}
one concludes from 
Lemma~\ref{lemma_proof_vanishing_discord} that $\xi_{k|m} = 0$ for $k \not=m$, \ie $\eta_{l|i} = ( \sqrt{\sigma_\AAA} )_{il}  = 0$ for any $(i,l)$ 
such that $l \notin I_i$. One then obtains from (\ref{eq-condition_hatM_M_rho=rho}) 
\begin{equation}
\sigma = \sum_{i,j=1}^{n_\AAA} \sum_{ l \in I_i \cap I_j} ( \sqrt{\sigma_\AAA})_{il}  ( \sqrt{\sigma_\AAA} )_{lj} \ketbra{\varphi_i}{\varphi_j} \otimes \sigma_{\BB | l}
= \sum_{k=1}^d \sum_{i,j \in I_{i_k}} (\sigma_\AAA)_{ij} \ketbra{\varphi_i}{\varphi_j} \otimes \sigma_{\BB | i_k}
\;.
\end{equation} 
This gives
\begin{equation} \label{eq-local_meas_leave_state_unchanged}
\sigma = \sum_{k=1}^d \widehat{\pi}_k^\AAA \sigma_\AAA \widehat{\pi}_k^\AAA \otimes \sigma_{\BB |i_k} 
\quad , \quad  \widehat{\pi}_k^\AAA = \sum_{i \in I_{i_k}} \ketbra{\varphi_i}{\varphi_i}\;.
\end{equation}
The last expression is of the form (\ref{eq-A-classical_state}) (note that the vectors $\ket{\varphi_i}$ in 
the latter formula are the eigenvectors of $\widehat{\pi}_k^\AAA \sigma_\AAA \widehat{\pi}_k^\AAA$, so that they are in general linear combinations 
of the vectors $\ket{\varphi_i}$ defined above). 
To get the result for the discord $\delta_\AAA$ we take advantage of (\ref{eq-link_between_von_Neumann_and_POVM_discords}). From the foregoing result,
 $\delta_\AAA ( \sigma)= 0$ 
is equivalent to $ \sigma  \otimes \ketbra{\epsilon_0}{\epsilon_0}$ being of the form (\ref{eq-A-classical_state}) for some
\ONB $\{ \ket{\varphi_i^\AAE }\}$ of $\Hh_\AAE$. This straightforwardly  implies $\ket{\varphi_i^\AAE} = \ket{\varphi_i} \ket{\epsilon_0}$ with 
$\{ \ket{\varphi_i }\}$ an \ONB of $\Hh_\AAA$.
\finpro

\vspace{3mm}

\Proofof{Lemma~\ref{lemma_proof_vanishing_discord}}
One proceeds by induction on $d$. The result is trivial for $d=2$. Let us assume that it holds true for $d\geq 2$ and that 
one can find a vector $\xv \in \Xx^{d+1}$ and some probabilities $\{ \xi_{k|m} \}_{k=1}^{d+1}$ like in the lemma such that
$\xi_{k_0|k_0} < 1$ for some $k_0 \in \{ 1, \ldots, d+1\}$. We are going to show that this leads to a contradiction.
By plugging $x_{k_0} = (1-\xi_{k_0|k_0}  )^{-1}\sum_{k \not= k_0} \xi_{k|k_0} x_k$ into
the $p$ other convex decompositions, one gets $x_m = \sum_{k \not= k_0} \zeta_{k|m} x_k$ for $k \not= k_0$, with 
$\zeta_{k|m} = \xi_{k|m} + (1-\xi_{k_0|k_0}  )^{-1} \xi_{k_0|m} \xi_{k|k_0}$. As $\{ \zeta_{k|m} \}_{k\not= k_0}$ is a probability distribution 
satisfying $\zeta_{k|m} =0 \Leftrightarrow \zeta_{m|k} =0$, by the induction hypothesis one has  $\zeta_{k|m} =\delta_{km}$ for 
any $k,m \in \{ 1, \ldots, d+1\} \setminus\{  k_0\}$. Now  $\xi_{m_0|k_0} >0$ for some index $m_0 \not= k_0$ (because $\xi_{k_0|k_0} < 1$). 
One deduces  from the above identities and the hypothesis on $\xi_{k|m}$ that the only non-vanishing probabilities are 
 $\xi_{k_0|m_0}$,   $\xi_{m_0|k_0}$, and $\xi_{k|k}$, $k=1,\ldots, p+1$. The problem then reduces to the case $p=2$. Thus $\xi_{k_0|k_0}=\xi_{m_0|m_0}=1$, 
in contradiction with our assumption.
\finpro

\vspace{2mm}

\begin{definition} \label{def-AAA_classical_states}
The  zero-discord states of the form (\ref{eq-A-classical_state}) are   called the {\it $\AAA$-classical states}.
We denote by $\Cc_\AAA$ the set of all $\AAA$-classical states.
Similarly,  $\Cc_\BB$ is the set of all $\BB$-classical states, namely, the states with vanishing $\BB$-discord.
A classical state is a state which is both $\AAA$- and $\BB$-classical. We write $\Cc_\AB = \Cc_\AAA \cap \Cc_\BB$.
\end{definition}

\vspace{2mm}

Our terminology can be justified by noting that if $\AB$ is in a state  of the form (\ref{eq-A-classical_state}) then 
the subsystem $\AAA$ is in one of the orthogonal states $\ket{\varphi_i}$ with probability $q_i$, 
whence $\AAA$ behaves as a classical system being in state $i$ with probability $q_i$. 
Alternatively, a state $\sigma$ is $\AAA$-classical \ifif there exists a von Neumann \meas on $\AAA$ with rank-one projectors $\pi_i^\AAA= \ketbra{\varphi_i}{\varphi_i}$
which does not perturb it in the absence of readout, \ie $\sigma = \Mm_{\{ \pi_i^\AAA\} } \otimes 1 ( \sigma)$.
The unfortunate name  ``classical-quantum states'' has become popular in the literature to refer to the $\AAA$-classical states, 
the $\BB$-classical states being called ``quantum-classical''.  Using the spectral decompositions of the $\sigma_{\BB|i}$'s, any $\AAA$-classical
state $\sigma_{\Aclass} \in \Cc_\AAA$ can be decomposed as
\be \label{eq-A-classical_state_bis}
\sigma_{\Aclass} 
=
\sum_{i=1}^{n_\AAA} \sum_{j=1}^{n_\BB} q_{ij} \ketbra{\varphi_i}{\varphi_i} \otimes \ketbra{\chi_{j|i}}{\chi_{j|i}}\;,
\ee
where  $q_{ij} \geq 0$, $\sum_{i,j} q_{ij} = 1$ and, for any $i$, $\{ \ket{\chi_{j|i}} \}_{j=1}^{n_\BB}$ is an orthonormal basis of
$\Hh_\BB$ (note that the 
$\ket{\chi_{j|i}}$ need not be orthogonal for distinct $i$'s). 
A classical  state  $\sigma_\clas \in \Cc_\AAA \,\cap\, \Cc_\BB$ possesses an
eigenbasis  $\{ \ket{\varphi_i} \otimes \ket{\chi_j}\}_{i=1,j=1}^{n_\AAA,n_\BB}$ of product vectors. It
is fully classical, in the sense that any quantum system in this state can be
``simulated'' by a classical apparatus being in the state  $(i,j)$ with probability $q_{ij}$.

Let us point out that $\Cc_\AAA$, $\Cc_\BB$, and $\Cc_\AB$ are not convex. Their convex hull
is the set $\Ss_\AB$ of separable states. It is also important to realize that for pure states, $\AAA$-classical, 
$\BB$-classical, classical, and separable  states all coincide. Actually, according to  (\ref{eq-A-classical_state_bis}) the 
pure $\AAA$-classical (and, similarly, the pure $\BB$-classical) states are product states.  
In contrast, one can find mixed separable states which are not $\AAA$-classical.
An example for two qubits is
\begin{equation}
\rho = \frac{1}{4} \bigl( \ketbra{+}{+} \otimes \ketbra{0}{0} + \ketbra{-}{-} \otimes \ketbra{1}{1} + \ketbra{0}{0} \otimes \ketbra{-}{-} 
+ \ketbra{1}{1} \otimes \ketbra{+}{+} \bigr)
\end{equation} 
 with $\ket{\pm} = (\ket{0} \pm \ket{1} )/\sqrt{2}$.
It is clear that $\rho \in \Ss_\AB$, but $\rho$ is neither $\AAA$-classical nor $\BB$-classical. 
A schematic picture of the  sets $\Ss_\AB$, $\Cc_\AAA$, $\Cc_\BB$, and $\Cc_{\AB}$ for a general bipartite system $\AB$ is displayed in Fig.~\ref{fig3}.

\subsection{Properties of the quantum discord} \label{sec-prop_didcord}
\subsubsection{Invariance and monotonicity properties}

Unlike entanglement measures, the quantum discord is {\it not} monotonous with respect to LOCCs. In particular, local operations on
the measured subsystem $\AAA$ can create discord. For instance, consider the classical state
\begin{equation}
\sigma = \onehalf \Bigl( \ketbra{0}{0} \otimes \ketbra{0}{0} + \ketbra{1}{1} \otimes \ketbra{1}{1} \Bigr)
\end{equation}
of two qubits. One can transform this state by a local operation $\Mm_\AAA$ on $\AAA$
into 
\begin{equation}
\rho = \Mm_\AAA \otimes 1 ( \sigma) = \onehalf \Bigl( \ketbra{0}{0} \otimes \ketbra{0}{0} + \ketbra{+}{+} \otimes \ketbra{1}{1} \Bigr)\;,
\end{equation}
where $\Mm_\AAA$ has Kraus operators $A_0=\ketbra{0}{0}$ and $A_1 = \ketbra{+}{1}$. 
The final state $\rho$ has less total correlations than $\sigma$,
its mutual information $I_{\AAA : \BB} ( \rho ) = - p \ln p - (1-p) \ln (1-p)$ 
being smaller than  $I_{\AAA : \BB} ( \sigma )= \ln 2$  (here $p=1/2+\sqrt{2}/4$). However, 
it has a positive discord $\delta_\AAA ( \rho )> \delta_\AAA ( \sigma)=0$. This  means that 
the loss of classical correlations $J_{\BB | \AAA} (\sigma) -  J_{\BB | \AAA} (\rho)$ is larger than the loss of total correlations 
$I_{\AAA : \BB} ( \sigma )- I_{\AAA : \BB} ( \rho )$.

In contrast, as far as local operations on $\BB$ are concerned everything goes as expected, as shown by the following result. 

\vspace{2mm}

\begin{proposition} \label{prop-monotonicity_Q_discord}
The quantum discord $\delta_\AAA$ and classical correlations $J_{\BB | \AAA} ( \rho)$ 
are invariant with respect to unitary conjugations $\Uu_\AAA : \rho_\AAA \mapsto U_\AAA \rho_\AAA  U_\AAA^\ast$ on $\AAA$ and monotonous
 with respect to \QOs  $\Mm_\BB$ on $\BB$, namely,
\begin{equation} \label{eq-monotonicity_discord}
\begin{array}{ccccccc}
\delta_\AAA (  \Uu_\AAA \otimes 1 ( \rho ) ) & = & \delta_\AAA ( \rho)  &  , & 
\delta_\AAA ( 1 \otimes \Mm_\BB ( \rho ) ) & \leq &  \delta_\AAA ( \rho ) \\
J_{\BB | \AAA}  (  \Uu_\AAA \otimes 1 ( \rho ) ) & = & J_{\BB | \AAA}  ( \rho) & , & 
J_{\BB | \AAA}  ( 1 \otimes \Mm_\BB ( \rho ) ) & \leq  & J_{\BB | \AAA}  ( \rho )
\end{array} 
\end{equation}
and similarly for $\delta_\AAA^\vN$ and $J_{\BB | \AAA}^\vN $.
\end{proposition}


\proof The unitary invariance is trivial. The monotonicity of $J_{\BB | \AAA } ( \rho) $ with respect to operations on $\BB$ comes from the 
monotonicity of the relative entropy and the formula
\begin{equation}
J_{\BB | \AAA } ( \rho) = \max_{\{ M_i^\AAA \} }  \Bigl\{  \sum_i\eta_i S ( \rho_{\BB |i} || \rho_\BB ) \Bigr\} \;,
\end{equation}
which is a consequence of the definition (\ref{eq-class_correl}) and of $\rho_\BB = \sum_i \eta_i \rho_{\BB | i}$.
A simple justification of the monotonicity of $\delta_\AAA$ with respect to operations on $\BB$
 uses the following reasoning~\cite{Piani12}. Let us consider a generalized \meas $\{ M_i^\AAA\}$ on $\AAA$ with associated \QO $\Mm_\AAA$.
By invoking the Stinespring theorem, one can represent $\Mm_\AAA$ as $\Mm_\AAA \otimes 1 ( \rho) = \tr_\EE ( \sigma_\ABE )$ with
$ \sigma_\ABE = U_\AAE \rho \otimes \ketbra{\epsilon_0}{\epsilon_0} U_\AAE^\ast$ pertaining to an enlarged space $\Hh_{\AB\EE}$ and 
$U_\AAE$ a unitary  on $\Hh_\AAE$. Thanks to the additivity and unitary invariance of the von Neumann entropy and to the 
relation $\tr_\AAE ( \sigma_\ABE ) = \rho_\BB$, one finds
\begin{equation}
I_{\AAA : \BB} ( \rho ) =  I_{\AAE : \BB} ( \sigma_\ABE )
\quad , \quad
I_{\AAA : \BB} ( \Mm_\AAA \otimes 1 (\rho) ) = I_{\AAA : \BB} ( \sigma_\AB )\;.
\end{equation}
Plugging these expressions into (\ref{eq-other_def_Q_discord}) gives 
 the following  expression of $\delta_\AAA ( \rho)$ in terms of the conditional mutual informations
\begin{equation}
\label{eq-other_def_Q_discord2}
\delta_\AAA  (\rho) 
 =  \min_{ \{ M_i^\AAA \} } \bigl\{ I_{\AAE:\BB} (\sigma_\ABE ) -   I_{\AAA:\BB} (\sigma_\AB)  \bigr\}   
 =  \min_{ \{ M_i^\AAA \} } \bigl\{ I_{\AB:\EE} (\sigma_\ABE ) -   I_{\AAA:\EE} (\sigma_\AAE)  \bigr\}
\;.
\end{equation}
The monotonicity of $\delta_\AAA$ then  
follows from the monotonicity of the mutual information with respect to local operations (Sec.~\ref{sec-def_discord}).
\finpro

\subsubsection{States with the highest discord} \label{sec-state_with_highest_discord}

As stated at the beginning of this section, the quantum discord $\delta_\AAA(\rho)$ is an indicator of the degree of quantumness of $\rho$.
It is thus natural to ask whether the `` most quantum'' states having the highest discord are  the maximally entangled states characterized 
in  Proposition~\ref{prop_max_ent_states}. The answer is affirmative when $n_\AAA \leq n_\BB$.

\vspace{2mm}

\begin{proposition} \label{prop-maximal_discord}
For any state $\rho$ of the bipartite system $\AB$, one has
\begin{equation} \label{eq-bound_on _Q_discord}
\delta_\AAA ( \rho ) \leq \delta_\AAA^\vN ( \rho ) \leq S ( \rho_\AAA ) \leq \ln n_\AAA\;.
\end{equation}
If $n_\AAA \leq n_\BB$ then the maximal value of $\delta_\AAA ( \rho )$ over all states $\rho \in \states (\Hh_\AB)$ is equal to $\ln n_\AAA$ and 
$\delta_\AAA(\rho) = \ln n_\AAA$ 
\ifif $\rho$ has highest entanglement of formation. Thus, the states $\rho_{\rm ent}$ with highest discord are the maximally entangled states  
given by Proposition~\ref{prop_max_ent_states}, which satisfy 
\begin{equation}
\delta_\AAA ( \rho_{\rm ent} ) = \delta_\AAA^\vN ( \rho_{\rm ent} ) = \EoF ( \rho_{\rm ent} ) = \ln n_\AAA \; .
\end{equation}
\end{proposition}

\vspace{1mm}

The statements in this proposition are probably well known in the literature, 
although I have not found an explicit reference.

\vspace{1mm}

\proof Let $\rho= \sum_k p_k \ketbra{k}{k}$ be the spectral decomposition of $\rho$ and $r = \rank (\rho)$. 
As mentioned earlier, the von Neumann \meas minimizing the conditional entropy $\sum_i \eta_i S ( \rho_{\BB | i})$ consists of rank-one projectors
$\pi_i^\AAA = \ketbra{\varphi_i}{\varphi_i}$. The conditional states (\ref{eq-def_cond_state_discord})
take the form
\begin{equation}
\rho_{\BB | i} = \sum_{k=1}^{r} p_{k|i} \ketbra{\phi_{ki}}{\phi_{ki}} \quad \text{ with } \quad p_{k|i} = \frac{p_k \eta_{i|k}}{\eta_i} \quad \text{ and } \quad  
  \sqrt{\eta_{i|k}} \ket{\phi_{ki}} = \braket{\varphi_i }{k} \in \Hh_\BB\;, 
\end{equation}
where $\eta_{i|k}= \| \braket{\varphi_i}{k} \|^2$ is the probability of outcome $i$ given the state $\ket{k}$
and $ p_{k|i}$ is the ``a posteriori'' probability that the state is $\ket{k}$ given the \meas outcome $i$ (Bayes rules).
Since $\{ \ket{\phi_{ki}}, p_{k |i} \}$ is a pure state decomposition of $\rho_{\BB | i}$, the formula (\ref{eq_entropy_as_min_over_pure_state_dec}) yields
\begin{equation} \label{eq-proof_upper_bound_on_discord1}
\sum_{i} \eta_i S ( \rho_{\BB | i} ) \leq \sum_{i} \eta_i H ( \{ p_{k|i} \} )\;.
\end{equation}
The \RHS is the classical  conditional entropy given the \meas outcomes, see (\ref{eq-class_cond_entropy}).
By the non-negativity of the classical  mutual information, it is bounded from above by the Shannon entropy 
$H ( \{ p_k\} ) = - \sum_k p_k \ln p_k = S ( \rho)$. 
Hence  $\delta_\AAA^\vN ( \rho) \leq S ( \rho_\AAA)$ by (\ref{eq-def_QD}).
But $S( \rho_\AAA) \leq \ln n_\AAA$, thus we have proven (\ref{eq-bound_on _Q_discord}). 

 Let us assume that $\delta_\AAA ( \rho) = S( \rho_\AAA)$. We know from Sec.~\ref{sec-von_Neumann_wentropy} that a necessary and sufficient condition for 
(\ref{eq-proof_upper_bound_on_discord1}) to be an equality is that $\{ \ket{\phi_{ki}}, p_{k |i} \}$ be a spectral decomposition of $\rho_{\BB | i}$,
for any $i$. Setting $D_{kl} = \tr_\BB ( \ketbra{k}{l} )$ as in the proof of Proposition~\ref{prop_max_ent_states}, one gets
$\sqrt{\eta_{i|l} \eta_{i|k}} \braket{\phi_{li}}{\phi_{ki}} = \bra{\varphi_i} D_{kl} \ket{\varphi_i} = 0$ if $k \not= l$ and
$p_{k|i} p_{l|i} > 0$. Since $\delta_\AAA ( \rho) = S( \rho_\AAA)$, (\ref{eq-proof_upper_bound_on_discord1}) holds with equality 
for any \ONB $\{ \ket{\varphi_i}\}$ and thus $D_{kl} = 0$ for such $k$ and $l$. 
In addition, the conditional entropy in the \RHS of (\ref{eq-proof_upper_bound_on_discord1}) is equal to its upper bound $H ( \{ p_k\} )=S(\rho)$.
This can happen only if $p_{k|i}=p_k$, \ie   
$\eta_{i|k} = \bra{\varphi_i} D_{kk} \ket{\varphi_i} = \eta_i$, for all $i$ and $k$ (indeed, the mutual information vanishes for independent random variables only).
Hence $\delta_\AAA ( \rho) = S ( \rho_\AAA)$ \ifif $D_{kk}$ is independent of $k$ and $D_{kl} = 0$ when $k\not= l$ and $p_k p_l >0$.
Suppose now that $\delta_\AAA ( \rho ) = \ln n_\AAA$. Then $\delta_\AAA ( \rho) = S ( \rho_\AAA) = \ln n_\AAA$ and the foregoing conditions on $D_{kl}$ are fulfilled.
In addition, $\rho_\AAA = \sum p_k D_{kk} = 1/n_\AAA$, whence $D_{kk} = 1/n_\AAA$ for all $k$ with $p_k>0$. 
One concludes that the eigenvectors $\ket{k}$ are as in  Proposition~\ref{prop_max_ent_states} by following the same steps 
as in the proof of this proposition.   
\finpro

\vspace{2mm}

Note that when $n_\AAA > n_\BB$,  $\delta_\AAA (\rho)$ is strictly smaller than $\ln n_\AAA$ for any $\rho \in \states (\Hh_\AB)$. In fact, in that case
$\rank ( D_{kk} ) \leq n_\BB < n_\AAA$ by the Schmidt decomposition (\ref{eq-Schmidt_decomposition}), and
the necessary condition $D_{kk} = 1/n_\AAA$ for having $\delta_\AAA (\rho) = \ln n_\AAA$ cannot be fulfilled.

\subsubsection{Monotonicity when disregarding a part of the measured subsystem}

We close this review of the properties of the discord by a simple remark concerning  tripartite systems  $\ABC$. If such a 
system is in the state $\rho_\ABC$, it is easy to show that 
\begin{equation} \label{eq-inequality_class_correl}
J_{\BB | \AAA \CC } (  \rho_\ABC ) \geq J_{\BB | \AAA } (  \rho_\AB )\;.
\end{equation}
This means that if $\BB$ is coupled to both $\AAA$ and $\CC$, the gain of information on $\BB$ from joint measurements on $\AAA$ and 
$\CC$ is larger than the gain of information by measuring $\AAA$ only and ignoring $\CC$, as this sounds reasonable. 
A similar bound exists for the total correlations: by (\ref{eq-mutual_info_as_relative_entropy}) and the monotonicity of the relative entropy 
(or, equivalently, the strong subadditivity of $S$),
\begin{equation} \label{eq-monotonicity_mutual_info}
I_{\AAA \CC : \BB } (  \rho_{\ABC} ) \geq I_{\AAA : \BB } (  \rho_{\AB} ) \, .
 \end{equation}

\begin{exercice} \label{exo-Holevo_bound}
The Holevo bound (\ref{eq-Holevo_bound}) can be derived by using the 
monotonicity of the quantum  mutual information  under operations acting on one subsystem (Sec.~\ref{sec-def_discord})
and the property (\ref{eq-monotonicity_mutual_info}).

\vspace{1mm}

\noindent Sketch of the proof {\rm ~\cite{Nielsen}. 
Given  an ensemble  $\{ \rho_i, \eta_i\}_{i=1}^m$ of states on $\Hh_\AAA$ and  a family $\{ A_j\}_{j=1}^p$ of Kraus operators describing the measurement on $\AAA$,
consider the state 
$\rho_\ARP = \sum_i \eta_i \rho_i  \otimes \ketbra{\nu_i}{\nu_i} \otimes \ketbra{0}{0}$ on $\Hh_\ARP$, where  $\RR$ and $\PP$ are
 auxiliary systems with \ONBs  $\{ \ket{\nu_i} \}_{i=1}^{m}$ and  $\{ \ket{j} \}_{j=0}^{p-1}$. These systems represent a register of the state preparation 
and a pointer for the
measurement, respectively.
Let $\Mm_\AP$ be the \QO on $\Bb ( \Hh_\AP)$ with Kraus operators $A_j \otimes U_j$, $U_j$ being the unitary on $\Hh_\PP$ defined by
$U_j \ket{l} = \ket{l+j}$ for any $l=0,\ldots, p -1$ (the addition is modulo $p$). 
It is an easy exercise to show that the Holevo bound (\ref{eq-Holevo_bound}) is
equivalent to $I_{\RR : \PP} ( [\Mm_\AP \otimes 1 ( \rho_\ARP)]_\RP ) \leq I_{\AP  : \RR } (\rho_\ARP )$.
}    
\end{exercice}

\begin{figure}
\begin{center}
\includegraphics[width=10cm]{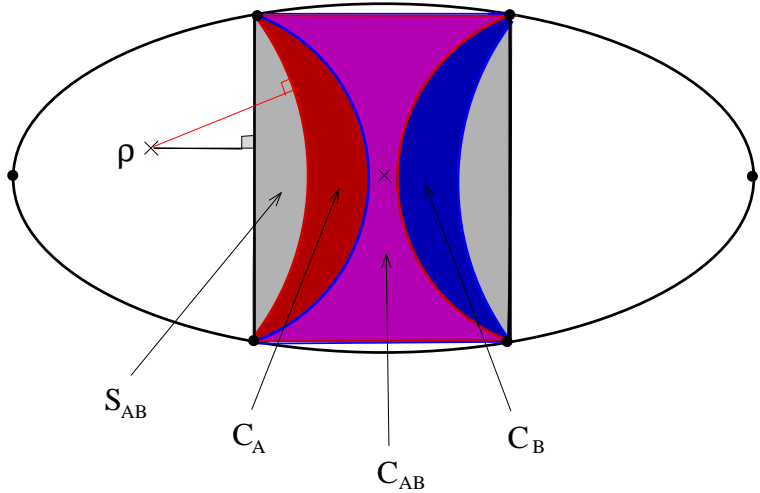}
\end{center}
\caption{
Schematic view of the set of quantum states $\states_\AB= \states (\Hh_\AB)$ of a bipartite system $\AB$. The subset $\Cc_\AB$  of classical states (in magenta) 
is the intersection of the subsets $\Cc_\AAA$ and $\Cc_\BB$ of $\AAA$- and $\BB$-classical states  (in red and blue). The convex hull of 
$\Cc_\AAA$ (or $\Cc_\BB$) is the subset $\Ss_\AB$ of separable states  (gray square). All  these subsets intersect the border of 
$\states_\AB$ (pure states of $\AB$) at the pure product states, represented by the four vertices of the square.
The maximally mixed state $\rho_\AB = 1/(n_\AAA n_\BB)$ lies at the center (cross).  
The two points at the left and right extremities of the ellipse represent the maximally entangled  pure states, which are the most distant states 
from $\Ss_\AB$ (and also from $\Cc_\AAA$, $\Cc_\BB$, and $\Cc_\AB$).
The closest distances of a state $\rho$ to $\Ss_\AB$ (black line) and  of $\rho$ to $\Cc_\AAA$ (red line)  
define the square roots of the geometric measure of entanglement $E_{\rm Bu} ( \rho)$  and of the geometric discord $D_{\AAA} ( \rho)$, respectively.
Note that this picture is for illustrative purposes and does not reflect all geometrical aspects
  (in particular, $\Cc_\AAA$, $\Cc_\BB$, and $\Cc_\AB$ typically have a lower dimensionality than 
$\states_\AB$ and $\Ss_\AB$).
}
\label{fig3}
\end{figure}

\subsection{Monogamy relation} \label{sec-monogamy_relation}

Consider a tripartite system $\ABC$ in a pure state $\ket{\Psi_\ABC}$.
If $\BB$ and $\CC$ are entangled,  is there a limit on the
amount of entanglement $\BB$ can have with $\AAA$? 
In other words, can entanglement be freely shared between different subsystems?
A negative answer to the last question has been highlighted in~\cite{Coffman00}, where it is shown that  when  $\AAA$, $\BB$, and $\CC$ are qubits, 
the sum $C ( \rho_\AB)^2 + C ( \rho_\BC)^2$ of the square
concurrences  is smaller or equal to $4 \det (\rho_\BB )$. 
It is instructive to consider the limiting case where $\BB$ and $\CC$ are maximally entangled.
Then, if one also assumes that  $r n_\BB \leq n_\CC < (r+1) n_\BB$  with $1 \leq  r\leq n_\AAA$, 
$\AAA$ and $\BB$ cannot be entangled and even have vanishing discords $\delta_\AAA ( \rho_\AB) =\delta_\BB ( \rho_\AB)=0$.
In fact, the state of $\BC$ being maximally entangled, one has $\rho_\BC = \sum_k p_k \ketbra{k}{k}$ for some orthogonal maximally entangled states $\ket{k}$ satisfying
$D_{kl}= \tr_\CC ( \ketbra{k}{l} ) = n_\BB^{-1} \delta_{kl}$ (see Proposition~\ref{prop_max_ent_states}).
Hence the pure state of $\ABC$ is 
 $\ket{\Psi_\ABC} = \sum_k \sqrt{p_k} \ket{\alpha_k} \ket{k} $  with $\{ \ket{\alpha_k} \}$ an orthonormal family of $\Hh_\AAA$ 
(Sec.~\ref{sec-purification}). Consequently, $\rho_\AB =(  \sum_k p_k \ketbra{\alpha_k}{\alpha_k}) \otimes ( 1 / n_\BB)$ is a product state and 
thus a classical state.

The proposition below exhibits an astonishing bound, called the monogamy relation,  
between the entanglement of formation of $\rho_\BC$ and the POVM-discord of $\rho_\AB$ measuring $\AAA$. 

\vspace{2mm}

\begin{proposition} {\rm (Koashi and Winter~\cite{Koashi04})}  Let $\ABC$ be a tripartite system in the
state $\rho_\ABC$. Let $\rho_\AB = \tr_\CC ( \rho_\ABC )$ and 
$\rho_{\BC} = \tr_\AAA ( \rho_\ABC)$ denote the reduced states of the bipartite systems $\AB$ and 
$\BC$, respectively. Then 
\begin{equation} \label{eq-monogamy_inequality}
\EoF ( \rho_\BC ) \leq S(\rho_\BB )  - J_{\BB |\AAA} ( \rho_{\AB}) = \delta_\AAA ( \rho_{\AB}) + S(\rho_{\AB} ) - S (\rho_\AAA )\;.
\end{equation} 
Moreover, the inequality is an equality 
if $\rho_{\ABC}$ is a pure state.
\end{proposition}

\vspace{1mm}

The inequality (\ref{eq-monogamy_inequality}) tells us that the more
classically correlated are  $\AAA$ and $\BB$, the less  $\BB$ can be entangled to a third system $\CC$. If  $n_\BB \leq n_\CC$ and
$\BB$ and $\CC$ are maximally entangled, \ie  $\EoF ( \rho_\BC) = \ln ( n_\BB)$, then this inequality entails $J_{\BB |\AAA} ( \rho_{\AB})=0$ 
(since $S(\rho_\BB) \leq \ln (n_\BB)$). Thus
$\AAA$ and $\BB$ are not correlated  classically, in agreement with the above statement that $\rho_\AB$ is a product state.

The entropy difference $S_{\BB | \AAA} (\rho_\AB ) = S(\rho_{\AB} ) - S (\rho_\AAA )$ in the \RHS of (\ref{eq-monogamy_inequality}) 
is called the {\it conditional von Neumann entropy}. It is known that 
$S_{\BB | \AAA} (\rho_\AB ) \geq 0$ if $\rho_{\AB}$ is separable~\cite{Horodecki96,Cerf97}. 
Thanks to the subadditivity of $S$ 
one has $- S( \rho_\BB ) \leq S_{\BB | \AAA} (\rho_\AB) \leq S( \rho_\BB )$ (the first inequality is obtained by considering a purification of $\rho_\AB$
on $\Hh_\ABC$ and using the subadditivity for $\rho_\BC$ together with the identities $S ( \rho_\BC ) = S ( \rho_\AAA)$ and $S(\rho_\CC ) = S ( \rho_\AB)$).
The quantity $- S_{\BB | \AAA} (\rho_\AB )$ is the coherent
information introduced by Schumacher and Nielsen in the context of the quantum channel capacity~\cite{Schumacher_Nielsen96}.

Two consequences of the claim that (\ref{eq-monogamy_inequality}) is an equality for tripartite systems $\ABC$ in pure states 
deserve further comments. First,  one easily deduces from this claim and the  identity (\ref{eq-equality_entropy_reduce_states}) that~\cite{Fanchini11} 
\begin{equation}
\EoF ( \rho_\AB ) + \EoF ( \rho_\BC ) = \delta_\AAA ( \rho_\AB) + \delta_\CC ( \rho_\BC) \;.
\end{equation}
Hence the sum of all entanglement of formations describing the
bipartite entanglement shared by $\BB$ is equal to the sum of the corresponding quantum discords 
with measurements on the other subsystems.
Second, if $\BB$ is a qubit and $\rho_\AB$ is of rank two, then $\rho_\AB$ admits a purification $\ket{\Psi_\ABC}$ on $\Hh_\AB \otimes \complex^2$ 
(see (\ref{eq-example_purification}))
and  the entanglement of formation of the two-qubit  state $\rho_\BC$ can be computed with the help of the Wootters formula (\ref{eq-Wootters_formula}).
One may in this way determine $\delta_\AAA ( \rho_\AB)$ via (\ref{eq-monogamy_inequality}).

\vspace{3mm}

\Proof
We first assume that $\ABC$ is in a pure state $\ket{\Psi_\ABC}$. Let $\{ {M}_{\AAA, i}^\opt  \}$ be an optimal measurement on $\AAA$ 
maximizing the gain of information on $\BB$, that is, such that $J_{\BB | \AAA} (\rho_\AB ) = S ( \rho_\BB ) - \sum_i \eta_i^\opt S ( \rho_{\BB | i}^\opt)$, where
$\eta_i^\opt$ and $\rho_{\BB | i}^\opt$ are the outcome probabilities and conditional states of $\BB$ for this measurement.  
Without loss of generality one may assume that $M_{\AAA, i}^\opt= \ketbra{\widetilde{\mu}_i^\opt}{\widetilde{\mu}_i^\opt}$ are of rank one
(see the discussion after (\ref{eq-class_correl})). Since $\rho_\AB = \tr_\CC ( \ketbra{\Psi_\ABC}{\Psi_\ABC} )$, one has
$\eta_i^\opt = \tr ( \rho_\AB M_{\AAA, i}^\opt \otimes 1 )  = \| \braket{\widetilde{\mu}_i^\opt}{\Psi_\ABC} \|^2$.
Moreover, the post-measurement conditional state of $\BC$ is the pure state
\begin{equation}
\ket{\Psi_{\BC | i}} = (\eta_i^\opt )^{-\onehalf} \braket{\widetilde{\mu}_i^\opt }{\Psi_\ABC}
\end{equation}
and the conditional state of $\BB$ is  $\rho_{\BB | i}^\opt = \tr_\CC ( \ketbra{\Psi_{\BC|i}}{\Psi_{\BC|i}})$.
The ensemble $\{  \ket{\Psi_{\BC |i}}, \eta_i^\opt \}$ gives a pure state decomposition of $\rho_\BC$.
Actually, let us consider the post-\meas state  of $\ABC$ in the absence of readout,  $\rho_\ABC' = \Mm_{\AAA}^\opt\otimes 1 ( \ketbra{\Psi_\ABC}{\Psi_\ABC})$.
The measurement being performed on $\AAA$, it does not change the state of  $\BC$, \ie
\begin{equation} \label{eq-proof_manogamy_rel}
\rho_\BC = \rho_\BC' =  \sum_i \eta_i^\opt \ketbra{\Psi_{\BC | i}}{\Psi_{\BC | i}}\;.
\end{equation}
From the definition (\ref{eq-def_EoF})
of the entanglement of formation one has 
\begin{equation} \label{eq-proof_monogamy_rel2}
\EoF ( \rho_\BC ) \leq \sum_i \eta_i^\opt S (  \rho_{\BB | i}^\opt ) =  S ( \rho_\BB )-  J_{\BB | \AAA} (\rho_\AB ) \;.
\end{equation}
Conversely, let $\{  \ket{\Psi_{\BC, i} } , \eta_i \}$ be a pure state decomposition of $\rho_\BC$ which
achieves the minimum in the definition 
of the entanglement of formation.
Let us show that there exists a generalized measurement $\{ {M}_i^\AAA  \}$ on $\AAA$ such that 
 $\eta_i$ is the probability of outcome $i$ and $\ket{\Psi_{\BC, i}}$ the corresponding conditional state of $\BC$, \ie  
\begin{equation} \label{eq-proof_monogamy_rel3}
\tr_\AAA ( M_i^\AAA \otimes 1 \ketbra{\Psi_\ABC}{\Psi_\ABC} ) = \eta_i \ketbra{\Psi_{\BC, i} }{\Psi_{\BC, i} } \;.
\end{equation}
In fact, let us observe that
$\ket{\Psi_\ABCE ' } = \sum_i \sqrt{\eta_i}  \ket{\Psi_{\BC, i}} \ket{\phi_i}$ is a purification of $\rho_\BC$ on $\Hh_\ABCE$ for some
 ancilla $\EE$, where  $\{ \ket{\phi_i} \}$  is  an orthonormal family of $\Hh_\AAE$. Given an arbitrary  state 
$\ket{\epsilon_0}\in \Hh_\EE$, $\ket{\Psi_\ABC} \ket{\epsilon_0}$ is also a purification of $\rho_\BC$ on the same space.
As a result, there is a unitary $U_\AAE$ on $\Hh_\AAE$ such that $\ket{\Psi_\ABCE ' } = 1  \otimes U_\AAE \ket{\Psi_\ABC} \ket{\epsilon_0}$ (see Sec.~\ref{sec-purification}).
Define
\begin{equation} 
M_i^\AAA = \bra{\epsilon_0} U_\AAE^\ast \ketbra{\phi_i}{\phi_i} U_\AAE \ket{\epsilon_0}
\end{equation}
(note the analogy with (\ref{eq-Neumark_extension_bis})).
Then (\ref{eq-proof_monogamy_rel3}) is satisfied.
Let $\rho_{\BB | i} = \tr_\CC (\ketbra{\Psi_{\BC, i} }{\Psi_{\BC, i} })$ be the post-measurement states of $\BB$, so that 
$\EoF ( \ket{\Psi_{\BC, i}}) = S ( \rho_{\BB | i})$. Since by assumption $\EoF ( \rho_{\BC}) = \sum_i \eta_i  \EoF ( \ket{\Psi_{\BC, i}})$,
one infers from  the definition (\ref{eq-class_correl}) of the classical correlations that 
\begin{equation}
J_{\BB| \AAA} (\rho_\AB ) \geq S(\rho_\BB) - \sum_i \eta_i S (  \rho_{\BB | i} )
=  
S(\rho_\BB) - \EoF ( \rho_\BC )\;.
\end{equation}
Together with (\ref{eq-proof_monogamy_rel2}) this proves that
\begin{equation} \label{eq-monogamy_equality}
\EoF ( \rho_\BC )   =
S(\rho_\BB) - J_{\BB| \AAA} (\rho_\AB ) \;.
\end{equation}
Let us now turn to the case of a tripartite system $\ABC$ in a mixed state $\rho_\ABC$. Consider a purification $\ket{\Psi_{\ABC \EE}}$ of  
$\rho_\ABC$ in the Hilbert space $\Hh_\ABC  \otimes \Hh_\EE$. Thanks to  (\ref{eq-inequality_class_correl}) one
then has $J_{\BB | \AAA} (\rho_\AB) \leq J_{\BB | \AAE} ( \rho_{\AB \EE} )$. The inequality (\ref{eq-monogamy_inequality}) then
follows by applying (\ref{eq-monogamy_equality}) with $\AAA \rightarrow \AAE$. 
\finpro

\newpage
\section{Distance and entropic measures of quantum correlations} \label{sec-geometric_measures}

In this section we study the measures of entanglement and quantum correlations based
on the Bures distance and the relative entropies. First, we introduce in Sec.~\ref{sec-geometric_entanglement}
the geometric measure of entanglement, defined as the minimal square distance between the state $\rho$ and a separable state, 
as well as similar measures obtained by replacing the square distance by relative entropies.
We define analogously in Sec.~\ref{sec-geo_discord} the geometric discord as the minimal square distance between $\rho$ and an $\AAA$-classical state. 
We show there that this discord is related to a quantum state discrimination task and determine the closest $\AAA$-classical states to $\rho$
in terms of the corresponding optimal measurements.

\subsection{Geometric and  relative-entropy measures of entanglement}\label{sec-geometric_entanglement}
\subsubsection{Definition and main properties}

From a geometrical point of view, it is natural to quantify the amount of entanglement in a state $\rho$ of a bipartite system $\AB$ by
the distance $d( \rho, \Ss_\AB)$ of $\rho$ to the subset $\Ss_\AB \subset \states (\Hh_\AB)$ of separable states (see Fig~\ref{fig3}). 
As it will become clear below, in order to obtain an entanglement  monotone measure the  distance $d$ must be contractive.
Choosing the Bures distance, it is easy to verify that 
\begin{equation} \label{eq-def_geo_entanglement}
E_{\rm Bu} (\rho) = d_{\rm B} ( \rho, \Ss_\AB )^2 = \min_{\sigma_\sep \in \Ss_\AB} \bigl\{ d_B ( \rho,\sigma_\sep )^2 \bigr\}
\end{equation} 
satisfies all the axioms of an entanglement measure in Definition~\ref{def_axioms_ent_meas}.
Actually, the axiom (i) holds because $d_{\rm B}$ is a distance on $\states ( \Hh_\AB)$. The convexity property (ii) is a consequence of the convexity of 
$\Ss_\AB$ and the joint convexity of the square Bures distance\footnote{
This justifies the square in our definition (\ref{eq-def_geo_entanglement}).
}
(Corollary~\ref{prop-d_B_contractive}). Finally, the monotonicity (iii) is shown in the following way. 
Let $\sigma_\rho \in \Ss_\AB$ be a closest separable state to $\rho$, \ie $E_{\rm Bu} (\rho) = d_{\rm B} ( \rho, \sigma_\rho )^2$. 
Let us recall from Sec.~\ref{sec-axiom_ent_meas} 
that any LOCC is a separable \QO and can be written as $ \Mm ( \rho) = \sum_{i} A_i \otimes B_i \rho A_i^\ast \otimes B_i^\ast$. Furthermore, one has
$ \Mm ( \Ss_\AB) \subset \Ss_\AB$. 
One can then use  the contractivity of $d_{\rm B}$ to obtain
\begin{equation}
E_{\rm Bu} (\rho)  \geq  d_{\rm B} ( \Mm ( \rho) ,  \Mm (\sigma_\rho ) )^2 \geq   E_{\rm Bu} ( \Mm (\rho))\;.
\end{equation}
This shows  that $E_{\rm Bu} $ is monotonous with respect to separable operations and, in particular, to LOCCs. The entanglement measure  $E_{\rm Bu} $ has been
first introduced by Vedral and Plenio~\cite{Vedral98}. Another measure was considered in~\cite{Vedral97,Vedral98} 
by replacing the square distance in (\ref{eq-def_geo_entanglement}) by the relative entropy $S(\rho||\sigma_\sep)$. More generally, we can define
\begin{equation} \label{eq-def_entropic_ent}
E_{\alpha} (\rho)  = \min_{\sigma_\sep \in \Ss_\AB} \bigl\{ S_\alpha ( \rho || \sigma_\sep ) \bigr\} \;,
\end{equation} 
where $S_\alpha$ is the quantum relative R\'enyi entropy (Sec.~\ref{sec-cond_Renyi_entropy}). For  $1/2 \leq \alpha \leq 1$, this defines an entanglement measure 
by the same arguments as above, because $S_\alpha$ is jointly convex and contractive (see Theorem~\ref{theo-Renyi_relative_entropy};
the property (i) in this theorem ensures that $E_{\alpha} (\rho) \geq 0$ with equality \ifif $\rho \in \Ss_\AB$). 
One establishes the following result by invoking the fact  that $S_\alpha$ is  non-decreasing in $\alpha$ 
(Proposition~\ref{prop-monotonicity_in_alpha_of_S_alpha}) and by using (\ref{eq-def_Bures_dist_2}) and
the relation (\ref{eq-def_alpha_fidelity}) between $S_{1/2}(\rho || \sigma)$ and the fidelity $F ( \rho, \sigma)$.

\vspace{2mm}

\begin{corollary} \label{eq-relative_entropy_entanglement_meas}
$\{ E_{\alpha} \}_{1/2 \leq \alpha \leq 1}$  constitutes a non-decreasing family of  entanglement measures and
\begin{equation} \label{eq-bound_between_E_B_and_E_1}
E_\onehalf ( \rho ) = - 2 \ln \Bigl( 1 - \frac{E_{\rm Bu} ( \rho)}{2} \Bigr) \; \leq \; E_\alpha ( \rho ) \quad , \quad  \onehalf \leq \alpha \leq 1 \;.
\end{equation}
\end{corollary}

\vspace{1mm}

The measure  $E_1$ associated to the relative entropy (\ref{eq-def_relative_S})  is less 
geometrical than $E_{\rm Bu}$ (it is not associated to a distance) but has the following interesting property.

\vspace{2mm}

\begin{proposition} {\rm (Vedral and Plenio~\cite{Vedral98})}  \label{eq-bound_E__and_EoF}
The entanglement measure $E_1$ coincides with the entanglement of formation $\EoF$ for pure states, and for mixed states $\rho \in \states (\Hh_\AB)$ 
it is bounded from above by $\EoF$,
\begin{equation} \label{eq-inequality_E_1_and_EoF}
E_1 ( \rho ) \leq \EoF ( \rho)\;.
\end{equation}
\end{proposition}

\vspace{1mm}

\proof We refer the reader to~\cite{Vedral98} for a detailed proof of the first statement. It is based on the observation that for a pure state
 with Schmidt decomposition $\ket{\Psi} = \sum_i \sqrt{\mu_i} \ket{\alpha_i} \ket{\beta_i}$, the minimum in (\ref{eq-def_entropic_ent}) 
is achieved when $\sigma_\sep$ is the classical state
\begin{equation} \label{eq-closest_state_relative_entropy_pure_states}
\sigma_\ast = \sum_{i=1}^n \mu_i \ketbra{\alpha_i}{\alpha_i} \otimes \ketbra{\beta_i}{\beta_i}\;.
\end{equation}
Since $S ( \rho_\Psi || \sigma_\ast ) = - \bra{\Psi} \ln \sigma_\ast  \ket{\Psi} = - \sum_i \mu_i \ln \mu_i$, the equality $E_1 ( \ket{\Psi} ) = \EoF ( \ket{\Psi} )$
follows once one has proven that 
$S ( \rho_\Psi || \sigma_{\rm sep} ) \geq  S ( \rho_\Psi || \sigma_\ast ) $ for all $\sigma_{\rm sep} \in \Ss_\AB$. This is done in Ref.~\cite{Vedral98} by showing that 
for any $\sigma_{\rm sep} \in \Ss_\AB$,
\begin{equation} \label{eq-proof_E_1_entnaglement_measure}
\frac{\D f_\Psi ( t, \sigma_{\rm sep} )}{\D t } \Bigr|_{t=0} = 1 - \int_0^\infty \D t \, \tr \bigl( (\sigma_\ast + t)^{-1} \rho_\Psi (\sigma_\ast + t)^{-1} \sigma_{\rm sep} \bigr)
\;\geq \; 0 
\end{equation}
 with  $f_\Psi (t, \sigma) = S ( \rho_\Psi || (1-t) \sigma_\ast + t \sigma)$. 
Indeed, assume that $S( \rho_\Psi || \sigma_{\rm sep} ) < S( \rho_\Psi || \sigma_\ast)$ for some $\sigma_{\rm sep} \in \Ss_\AB$. By taking advantage of the right convexity 
of the relative entropy, one then finds for any $t \in (0,1]$ 
\begin{equation}
\frac{ f_\Psi ( t , \sigma_{\rm sep} )- f_\Psi ( 0, \sigma_{\rm sep} )}{t} \leq - S ( \rho_\Psi || \sigma_\ast ) +  S( \rho_\Psi || \sigma_{\rm sep} ) < 0\; ,
\end{equation}
in contradiction with (\ref{eq-proof_E_1_entnaglement_measure}). Note that it suffices to prove the non-negativity in (\ref{eq-proof_E_1_entnaglement_measure}) 
for the pure product states  $\sigma_{\rm sep} = \ketbra{\phi \otimes \chi}{\phi \otimes \chi}$, because of
the linearity in $\sigma_{\rm sep}$ of the trace in the right-hand side.

The second statement in the proposition is a consequence of the first one and of the convexity  of
$E_1$. Actually, if $\{  \ket{\Psi_i} , \eta_i \}$ is a pure state decomposition of $\rho$ minimizing the average entanglement, then
\begin{equation}
 \EoF ( \rho) = \sum_i \eta_i \EoF ( \ket{\Psi_i} ) = \sum_i \eta_i E_1 ( \ket{\Psi_i} ) \geq E_1 \Bigl( \sum_i \eta_i \ketbra{\Psi_i}{\Psi_i} \Bigr) 
= E_1 ( \rho) \;.
\end{equation}
\finpro

\vspace{1mm}

Note that the  inequality (\ref{eq-inequality_E_1_and_EoF}) can be strict. Examples of two-qubit states $\rho$ for which $E_1 ( \rho ) < \EoF ( \rho)$
are given in~\cite{Vedral97}. 
Thanks to (\ref{eq-bound_between_E_B_and_E_1}) and (\ref{eq-inequality_E_1_and_EoF}), one can place an upper bound on 
$E_{\rm Bu} (\rho) $ by a function of the entanglement of formation $\EoF$. Such a bound does not seem to be known in the literature, but it 
is not optimal  for pure states as a consequence of the next proposition.

\vspace{1mm}

\begin{exercice}
As shown in~\cite{Vedral98},  $E_1$ fulfills the stronger monotonicity condition\footnote{
The property (iii'') is established in~\cite{Vedral98} for local maps of the form 
$\Mm_{{\rm loc}}^{(i)} ( \rho ) = A_{i} \otimes B_{i} \rho  A_{i}^\ast \otimes B_{i}^\ast$, but the proof in this reference can be extended  
without difficulty to  CP maps with Kraus operators $\{ A_{ik} \otimes B_{ik} \}$.
}
 (iii'') of Sec.~\ref{sec-convex_roof}. 
\end{exercice}

\subsubsection{Relation between the geometric measure of entanglement and convex roof constructions}
\label{sec_geometric_entnaglement_and_convex_roof}

Let $F (\rho,\Ss_\AB) $ denote the maximal fidelity between $\rho$ and a separable state, 
\begin{equation}
F (\rho,\Ss_\AB) = \max_{\sigma_\sep \in \Ss_\AB} \bigl\{ F ( \rho, \sigma_\sep ) \bigr\} \;.
\end{equation}
%

\vspace{2mm}

\begin{proposition} {\rm (Streltsov, Kampermann, and Bru\ss~\cite{Streltsov10})}  \label{prop-fidelity_convex_roof}
The geometric measure of entanglement is given for pure states by 
\begin{equation} \label{eq-geo_ent_for_pure_states}
E_{\rm Bu} (\ket{\Psi} )  =  2 - 2 \sqrt{F (\ket{\Psi}, \Ss_\AB )}
 =  2 ( 1 - \sqrt{\mu_{\rm max}} )\;,
\end{equation}
where $\mu_{\rm max} = \max \{ \mu_i \}$ is the largest Schmidt coefficient  of  $\ket{\Psi}$. For mixed states,
$F ( \rho, \Ss_\AB )$
is obtained via a maximization  over the pure state decompositions of $\rho$, 
\begin{equation} \label{eq-fidelity_as_convex_roof}
F (\rho,\Ss_\AB) = \max_{ \{ \ket{\Psi_i}, \eta_i \}} \biggl\{ \sum_i \eta_i F ( \ket{\Psi_i} ,\Ss_\AB ) \biggr\} \;.
\end{equation}
\end{proposition}

The nice relation (\ref{eq-fidelity_as_convex_roof}) is intimately related to Uhlmann's theorem (Sec.~\ref{sec-def-Bures_distance}) and to the convexity of $\Ss_\AB$.
Note that  the relative-entropy measure $E_1$ does not fulfill a similar property  
(compare with Proposition~\ref{eq-bound_E__and_EoF}).
Even though $E_{\rm Bu}$ is not a convex roof, it is a simple function of another entanglement 
measure $E_{\rm G}$ defined via a convex-roof construction  like in (\ref{eq-def_EoF}) and from its expression for pure states~\cite{Shimony95,Wei03} 
\begin{equation} \label{eq-def_Grover_measure}
E_{\rm G} ( \ket{\Psi}) = 1 - \max_{\ket{\Phi} \in \Ss_\AB} \bigl\{ | \braket{\Phi}{\Psi} |^2 \bigr\} \;.
\end{equation}
Actually, we will see that a pure state always admits a pure product state as closest separable state, hence the maximum in (\ref{eq-def_Grover_measure}) 
coincides with $F ( \ket{\Psi}, \Ss_\AB )$ and $E_{\rm G} (\rho) = 1 - F ( \rho,\Ss_\AB )$
 by the proposition above.  According to (\ref{eq-geo_ent_for_pure_states}), $E_{\rm G} ( \ket{\Psi}) = 1 - \mu_\mmax$ is of the form
(\ref{eq-def-E_f}) with $f_{\rm G} ( \rho_\AAA ) = 1 - \| \rho_\AAA \|$ satisfying all hypothesis of Proposition~\ref{prop-entanglement_meas_pure_states}.
Therefore, by a similar reasoning as in the proof of Proposition~\ref{prop_EoF_is_entanglement_meas}, 
$E_{\rm G}$ is an entanglement measure which fulfills the strong  monotonicity property (iii''). In contrast,
 $E_{\rm Bu} ( \ket{\Psi}) = f_{\rm Bu}( [\rho_\Psi]_\AAA) = 2 ( 1- \sqrt{\| [\rho_\Psi]_\AAA\|})$ but $f_{\rm Bu}$ is not concave, whence
Proposition~\ref{prop-entanglement_meas_pure_states} indicates that $E_{\rm Bu}$ does not fulfill (iii').
We should not be bothered too much about that, the two measures $E_{\rm Bu}$ and $E_{\rm G} $ being equivalent 
(that is, they define the same order of entanglement) and simply related to each other.

\vspace{1mm}

\proof
For a pure state $\rho_\Psi = \ketbra{\Psi}{\Psi}$, the fidelity reads $F (\rho_\Psi, \sigma_\sep ) = \bra{\Psi} \sigma_\sep \ket{\Psi}$. 
Writing the decomposition  of separable states into pure product states, 
$\sigma_\sep = \sum_i \xi_i \ketbra{\varphi_i \otimes \chi_i }{\varphi_i \otimes \chi_i}$, we get
\be \label{eq-max_fidelity_pure_state}
F ( \rho_{\Psi} , \Ss_\AB ) 
 =  \max_{ \{ \ket{\varphi_i} , \ket{\chi_i} , \xi_i \} }
 \Bigl\{ \sum_{i} \xi_{i} | \braket{\varphi_i \otimes \chi_i}{\Psi} |^2 \Bigr\}
 =  
\max_{\| \varphi\| = \| \chi\|=1}  \bigl\{ | \braket{\varphi \otimes \chi}{\Psi} |^2 \bigr\}  \;,
\ee
where we have used $\sum_{i} \xi_{i} =1$. For any normalized vectors
 $\ket{\varphi}\in \Hh_A$ and $\ket{\chi}\in \Hh_B$, one derives from the Schmidt decomposition (\ref{eq-Schmidt_decomposition}) and the 
Cauchy-Schwarz inequality that 
\begin{eqnarray} \label{eq-bound_by_mu_max}
\nn
|\braket{\varphi \otimes \chi}{\Psi} | 
& \leq &  
\sum_{j=1}^n \sqrt{\mu_j} \bigl| \braket{\varphi}{\alpha_j} \braket{\chi}{\beta_j} \bigr|
\leq 
\sqrt{\mu_{\rm max}}  \sum_{j=1}^n  \bigl| \braket{\varphi}{\alpha_j} \braket{\chi}{\beta_j} \bigr| 
 \\ \label{eq-bound_by_mu_max2}
& \leq & 
\sqrt{\mu_{\rm max}} \biggl( \sum_{j=1}^n | \braket{\varphi}{\alpha_j} |^2 \biggr)^{1/2} 
   \biggl( \sum_{j=1}^n | \braket{\chi}{\beta_j} |^2 \biggr)^{1/2}
    \leq \sqrt{\mu_{\rm max}} \;.
\end{eqnarray}
All bounds are saturated for $\ket{\varphi} = \ket{\alpha_{j_\mmax} }$ and $\ket{\chi} = \ket{\beta_{j_\mmax}}$, where $j_\mmax$ is the index for which
$\mu_j$ is maximum.
Thus $F( \rho_\Psi , \Ss_\AB )=\mu_{j_\mmax} = \mu_\mmax$ and the formula (\ref{eq-geo_ent_for_pure_states}) is proven. 
It is of interest to note that  the pure product state $\ket{\alpha_{j_\mmax}} \ket{\beta_{j_\mmax} } $ is a closest separable state
to  $\ket{\Psi}$ (a characterization of all these closest separable states will be given in Proposition~\ref{prop_closest_classical_state_pure} below). 

We now proceed to show (\ref{eq-fidelity_as_convex_roof}).
Consider a  fixed separable state $\sigma_\sep = \sum_{i=1}^p \xi_i \ketbra{\Phi_i}{\Phi_i}$ with $\ket{\Phi_i} \in \Ss_\AB$ and $\xi_i \geq 0$.
Without loss of generality one may assume $p=(n_\AAA n_\BB)^2 +1$ (see the discussion after Definition~\ref{eq-def_entangled_state}).
Let  $\{ \ket{f_i}\}_{i=1}^p$ be an \ONB of an ancilla space $\Kk$  and
$\ket{\Phi}= \sum_i \sqrt{\xi_i} \ket{\Phi_i} \ket{f_i}$ be a purification of  $\sigma_\sep$ on $\Hh \otimes \Kk$.
Thanks to Theorem~\ref{theo-Uhlmann}, $F ( \rho, \sigma_\sep ) $ is the maximum over all purifications $\ket{\Psi}$ of $\rho$ on $\Hh \otimes \Kk$
of the transition probability $| \braket{\Psi}{\Phi} |^2$. Writing $\ket{\Psi}$ in the form (\ref{eq-time_to_go_to_bed})
and using  the one-to-one correspondence  between pure state decompositions and purifications 
(see Sec.~\ref{sec-purification}), one can equivalently maximize over all pure state decompositions  
$\{ \ket{\Psi_i} , \eta_i \}$ of $\rho$.  Moreover, the maximization of $F ( \rho, \sigma_\sep ) $  over the separable states $\sigma_\sep$ leads to 
a maximization over the pure state ensembles $\{ \ket{\Phi_i}, \xi_i \}$ in $\Ss_\AB$. This yields
\begin{equation} \label{eq-stop_go_to_bed1}
F ( \rho, \Ss_\AB) 
 = \max_{ \{ \ket{\Phi_i}, \xi_i \}} \max_{ \{ \ket{\Psi_i} , \eta_i \}} \biggl\{ \biggl| \sum_{i=1}^p \sqrt{\eta_i \xi_i } \braket{\Psi_i}{\Phi_i} \biggr|^2\biggr\}
\;.
\end{equation}
But, using once more the Cauchy-Schwarz inequality and $\sum_i \xi_i = 1$, one has 
\begin{equation} \label{eq-stop_go_to_bed2}
 \max_{ \{ \ket{\Phi_i}, \xi_i \}} \biggl\{ \biggl| \sum_{i=1}^p \sqrt{\eta_i \xi_i} \braket{\Psi_i}{\Phi_i} \biggr|^2 \biggr\}
   = \sum_{i=1}^p \eta_i \max_{\ket{\Phi} \in \Ss_\AB} \bigl\{ | \braket{\Psi_i}{\Phi} |^2 \bigr\}\;.
\end{equation}
 It has been argued above that the maximal fidelity between $\ket{\Psi_i}$ and a separable state  is attained for pure product states, thus 
$F ( \ket{\Psi_i} , \Ss_\AB ) =  \max_{\ket{\Phi} \in \Ss_\AB} | \braket{\Psi_i}{\Phi} |^2$. 
Substituting this expression into (\ref{eq-stop_go_to_bed2}) and  (\ref{eq-stop_go_to_bed1}), we arrive at the required relation 
(\ref{eq-fidelity_as_convex_roof}).
\finpro

\vspace{2mm}

According to (\ref{eq-geo_ent_for_pure_states}), $E_{\rm Bu}  (  \ket{\Psi} )=0$ \ifif 
$\ket{\Psi}$ is a product state, in agreement with the fact that  by definition  separable 
pure states are product states.  
Another consequence of  (\ref{eq-geo_ent_for_pure_states}) and of 
the bound $\mu_{\rm max} \geq 1/n$ (which follows from $\sum_i \mu_i = 1$) is
$F (\ket{\Psi}, \Ss_\AB )  \geq 1/n$, with  $n=\min\{ n_\AAA,n_\BB\}$. Furthermore, $F (\ket{\Psi}, \Ss_\AB )  = 1/n$ \ifif $\ket{\Psi}$ is maximally entangled (Sec.~\ref{sec-def_entanglement}).
One deduces from  (\ref{eq-fidelity_as_convex_roof}) that
\begin{equation} \label{eq-maximal_geo_ent}
E_{\rm Bu} ( \rho) \leq 2-\frac{2}{\sqrt{n}}\;.
\end{equation}
By the same arguments as in the proof of Proposition~\ref{prop_max_ent_states}, this bound is saturated 
\ifif $\rho$ has maximal entanglement of formation $\EoF ( \rho)= \ln n$.
This means that $E_{\rm Bu}$ and $\EoF$ capture the same maximally entangled states.

\subsubsection{Geometric measure of entanglement for two qubits}

In the case of two qubits, a closed formula for $E_{\rm Bu} ( \rho)$ can be obtained with the help of 
Proposition~\ref{prop-fidelity_convex_roof} and of Wootters's result on the concurrence (Sec.~\ref{sec-Wootters_formula}).
It reads \cite{Streltsov10}
\begin{equation} \label{eq-geo_ent_for_2_qubits}
E_{\rm Bu}(\rho)=  2 - \sqrt{2} \bigl(1 +  \sqrt{1-C(\rho)^2} \bigr)^{\onehalf}
\end{equation}
with $C(\rho)$ given by (\ref{eq-concurrence_2_qbits}). Actually,  for pure states one finds 
by comparing $C ( \ket{\Psi} ) = 2 \sqrt{\mu_0 \mu_1}$ and (\ref{eq-geo_ent_for_pure_states})
that $F( \ket{\Psi}, \Ss_\AB ) = g ( C ( \ket{\Psi} ) )$ with $g(C) = (1 + \sqrt{1-C^2})/2$. As $g$ is decreasing and concave,
(\ref{eq-my_def_concurrence2}) and (\ref{eq-fidelity_as_convex_roof}) yield $F( \rho, \Ss_\AB) \leq g ( C (\rho))$. 
 But it is shown in~\cite{Wootters98} that there is an optimal pure state decomposition $\{  \ket{\Psi_i} , \eta_i \}$ of $\rho$ such that
$C ( \rho) = C ( \ket{\Psi_i})$ for any $i$.  Thus 
\begin{equation}
g \bigl(  C ( \rho) \bigr) \geq F  ( \rho, \Ss_\AB) \geq \sum_i \eta_i F ( \ket{\Psi_i} , \Ss_\AB ) = \sum_i \eta_i g \bigl( C ( \ket{\Psi_i} ) \bigr)
= g \bigl(  C ( \rho) \bigr)
\;,
\end{equation}
which justifies (\ref{eq-geo_ent_for_2_qubits}).

\subsection{Geometric quantum discord} \label{sec-geo_discord}
\subsubsection{Discord-like measures of quantum correlations} 

In the same spirit as for the geometric measure of entanglement, one  defines the geometric quantum discord as    
\begin{equation} \label{eq-max_fidelity} 
D_\AAA ( \rho) = d_{\rm B} (\rho, \Cc_\AAA )^2 = 2  ( 1 - \sqrt{ F (\rho ,\Cc_\AAA)} ) 
\quad , \quad  F (\rho, \Cc_\AAA )= \max_{\sigma_\Aclass \in \Cc_\AAA} \bigl\{ F (\rho, \sigma_\Aclass) \bigr\} \;,
\end{equation}
where $\Cc_\AAA$ is the (non-convex) set of $\AAA$-classical states (see Definition~\ref{def-AAA_classical_states}).
One can introduce similarly the relative-entropy discords
\begin{equation} \label{eq-relative_ent_discord} 
D^{(\alpha)}_\AAA ( \rho) = \min_{\sigma_\Aclass \in \Cc_\AAA} \bigl\{ S_\alpha  (\rho || \sigma_\Aclass) \bigr\} \;.
\end{equation}
As in Corollary~\ref{eq-relative_entropy_entanglement_meas}
one has $D_\AAA^{(1/2)} ( \rho ) = -2 \ln ( 1 - D_\AAA ( \rho)/2 ) \leq D_\AAA^{(\alpha)} ( \rho)$  for any $\alpha \in [1/2,1]$.

An analog of the geometric discord $D_\AAA$  based on the Hilbert-Schmidt distance $d_2$ has been  first introduced 
by Daki\'c, Vedral, and Brukner~\cite{Dakic10}.
We hope to have convinced the reader in Sec.~\ref{sec-Bures_distance} that  the Bures distance is a more natural choice in quantum information.
We will see that the discord (\ref{eq-max_fidelity})  shares many of the properties of the quantum discord $\delta_\AAA$ of Sec.~\ref{sec-discord},
while  its analog with the $d_2$-distance has unpleasant features. 
In particular,  like $\delta_\AAA$ the Bures geometric discord is
 invariant under conjugations by local unitaries 
and contractive with respect to \QOs $\Mm_\BB$ on $\BB$. 
For indeed,  the set of $\AAA$-classical states is invariant under such transformations (see  (\ref{eq-A-classical_state})), 
whence
\begin{equation} \label{eq-monotonicity_geo_discord}
D_\AAA ( U_\AAA \otimes U_\BB\, \rho \,U_\AAA^\ast \otimes U_\BB^\ast ) = D_\AAA ( \rho) \quad , \quad D_\AAA ( 1 \otimes \Mm_\BB ( \rho) ) \leq  D_\AAA (\rho)
\end{equation}
 by  unitary invariance and contractivity of $d_{\rm B}$. These properties also hold for $D_\AAA^{(\alpha)}$, $1/2 \leq \alpha \leq 1$,
because the relative R\'enyi entropy is also contractive (Theorem~\ref{theo-Renyi_relative_entropy}). 
This should be contrasted with the non-monotonicity with respect to operations on $\BB$ of the Hilbert-Schmidt geometric discord, 
which is due to the lack of monotonicity of $d_2$ (Sec.~\ref{sec-contractive_distances}). An explicit counter-example is given in~\cite{Piani12}.
We now precise the axioms on discord-like correlation measures.

\vspace{2mm}

\begin{definition} \label{def-measures_QCs}
A measure of quantum correlations of a bipartite system $\AB$ with respect to subsystem 
$\AAA$ is a function $D_\AAA : \states ( \Hh_\AB) \rightarrow [0, \infty)$ satisfying
\begin{itemize}
\item[(i)] $D_\AAA ( \rho) = 0$ \ifif $\rho$ is $\AAA$-classical;
\item[(ii)] $D_\AAA$ is invariant under local unitary transformations and contractive under \QOs on $\BB$, 
that is, (\ref{eq-monotonicity_geo_discord}) holds true;
\item[(iii)] $D_\AAA$ coincides with an entanglement measure for pure states.
\end{itemize}
\end{definition}

\vspace{1mm}

This definition is at the time of writing of this article believed to capture all relevant physical requirements
for quantifying the amount of  quantum correlations in $\AB$ given that one can access to subsystem $\AAA$ only~\cite{Roga14}. 
The axioms (i-iii) are in particular satisfied by the quantum discord $\delta_\AAA$
(Propositions~\ref{prop-classical_states}  and~\ref{prop-monotonicity_Q_discord}). This is also
true for the geometric discord $D_\AAA$. Actually,
we have just shown above that $D_\AAA$ satisfies (ii), and (i) is trivial.
Since the closest separable state to a pure state is a pure product state, which is
$\AAA$-classical, $D_\AAA$ coincides with the geometric measure of entanglement $E_{\rm Bu}$ for pure states 
(see~(\ref{eq-equality_distances}) below). Hence $D_\AAA$ is a measure of quantum correlations. Similarly, the relative-entropy based discord $D_\AAA^{(1)}$ 
is a measure of quantum correlations. The property (iii) follows in this case from the fact that  if $\rho_\Psi$ is a pure
state then a separable state $\sigma_\sep$ minimizing $S ( \rho_\Psi || \sigma_\sep )$ is the classical state given by 
(\ref{eq-closest_state_relative_entropy_pure_states})
 (see the proof of Proposition~\ref{eq-bound_E__and_EoF}), so that $D_\AAA^{(1)} ( \rho_\Psi)$ coincides with the entanglement measure $E_1 (\rho_\Psi)$
defined in (\ref{eq-def_entropic_ent}). It is an open problem to show that $D_\AAA^{(\alpha)}$ satisfies (iii) when $\alpha \not= 1/2, 1$.

The $\BB$-discords $D_\BB$ and $D^{(\alpha)}_\BB$ are defined by exchanging $\AAA$ and $\BB$
in (\ref{eq-max_fidelity}) and (\ref{eq-relative_ent_discord}). As for the 
quantum discord of Sec.~\ref{sec-discord},  in general $D_\AAA \not= D_\BB$.
Symmetric measures of quantum correlations are obtained by considering the square distance
 to the set of classical states $\Cc_\AB = \Cc_\AAA \cap \Cc_\BB$,
\begin{equation}
D_\AB ( \rho) = 2 \Bigl( 1 - \max_{\sigma_\clas \in \Cc_\AB} \bigl\{ \sqrt{ F ( \rho, \sigma_\clas )} \bigr\} \Bigr)
\quad , \quad 
D_\AB^{(\alpha)} (\rho) = \min_{\sigma_\clas \in \Cc_\AB} \bigl\{ S ( \rho || \sigma_\clas ) \bigr\}\;.
\end{equation}
Let us mention that a similar symmetric information-based discord can be defined by modifying the maximization procedure 
in (\ref{eq-other_def_Q_discord}) so as to involve projectors $\pi_i^\AAA \otimes \pi_i^\BB$ (or generalized measurement operators 
$M_i^\AAA \otimes M_i^\BB$), instead of $M_i^\AAA \otimes 1$. It is  called the {\it measurement-induced  disturbance}~\cite{Luo08a}.  
The  relative-entropy symmetric discord $D_\AB^{(1)}$ has been studied in~\cite{Modi10}, together with other quantities characterizing 
quantum and classical correlations.
We will not elaborate further here on the numerous discord-like  measures defined in the literature  and their operational interpretations
(see e.g.~\cite{Modi_review}).

We emphasize that since 
$\Cc_\AB \subset \Cc_\AAA \subset \Ss_\AB$ (see Fig.~\ref{fig3}), the geometric measures are ordered as 
\begin{equation} \label{eq-order_geometrical_meas}
E_{\rm Bu} (\rho) \leq D_\AAA (\rho)  \leq D_\AB (\rho)\;.
\end{equation} 
This ordering  is a nice feature of the geometrical approach. It also holds for the relative-entropy measures. In contrast, depending on $\rho$
the 
entanglement of formation $\EoF (\rho)$ 
can be larger or smaller than the quantum discord $\delta_\AAA (\rho)$.

Before going on to general results, let us say few words about explicit calculations of the  discords.
In the special case of two-qubit states $\rho$ with maximally mixed marginals $\rho_\AAA = \rho_\BB = 1/2$, the relative-entropy 
measure $D_\AB^{(1)}  ( \rho)$ coincides with the usual discord
$\delta_\AAA^\vN ( \rho)$~\cite{Modi10,Mazzola10}. 
For the same  states, a closed formula for $D_\AAA ( \rho)$ has been found in~\cite{Aaronson13,moi_JPA} and the 
closest $\AAA$-classical states to $\rho$ have been determined explicitly\footnote{
This is done in~\cite{moi_JPA} with the help of Corollary~\ref{prop-geo_discord_qubit} below.
}.
The Hilbert-Schmidt  geometric discord is much easier to calculate. A simple formula for arbitrary 2-qubit states
is derived in~\cite{Dakic10} and has been later on extended to higher dimensions.
The geometric discord defined with the trace distance $d_1$ 
has been determined recently for certain families of two-qubit states 
(the so-called $X$-states, containing in particular the states with maximally mixed marginals, 
and the $\BB$-classical states)~\cite{Ciccarello14,Nakano13}. Note that since
$d_1$ is contractive, this geometric discord fulfills the axiom (ii) of Definition~\ref{def-measures_QCs}.

\subsubsection{Geometric discord for pure states} \label{sec_pure_states}

We now proceed to determine the geometric discord $D_\AAA$ for pure states. It has been seen in the proof of Proposition~\ref{prop-fidelity_convex_roof}
that the family of closest separable states of a pure state $\ket{\Psi}$ contains a pure product state, which is a
classical state. By inspection of (\ref{eq-geo_ent_for_pure_states}) and (\ref{eq-order_geometrical_meas}), one gets
\be \label{eq-equality_distances}
D_\AAA ( \ket{\Psi} ) = D_\BB (\ket{\Psi} ) = D_\AB (\ket{\Psi} )  = E_{\rm Bu} ( \ket{\Psi} )  
 =  2 ( 1 - \sqrt{\mu_{\rm max}} )\;.
\ee
One deduces from the bound $\mu_{\rm max} \geq 1/n$ (which follows from $\sum_{i=1}^n \mu_i = 1$) that
\begin{equation}
D_\AAA  (\ket{\Psi} ) \leq 2  \Bigl(  1 - \frac{1}{\sqrt{n}} \Bigr)\
\quad , \quad n = \min \{ n_\AAA, n_\BB\} \;.
\;
\end{equation}
This bound is saturated when 
$\mu_i = 1/n$ for any $i$, that is, for the maximally entangled states. We will see below that this statement is still true
for mixed states provided that $n_\AAA \leq n_\BB$.

The identities (\ref{eq-equality_distances}) are analogous to the equality between the entanglement of formation $E_{\rm EoF}$
and the discord $\delta_\AAA$ for pure states (Sec.~\ref{sec-def_discord}).
As said before, they reflect the existence of a pure product state 
which is closer or at the same distance from $\ket{\Psi}$ than any other separable state. 
It is of interest to find all the closest $\AAA$-classical states to $\ket{\Psi}$. 
This is done in the next proposition.

\vspace{2mm}

\begin{proposition} {\rm (Spehner and Orszag~\cite{moi_NJP})} \label{prop_closest_classical_state_pure}
Let $\rho_\Psi = \ketbra{\Psi}{\Psi}$ be a pure state of $\AB$ with largest Schmidt coefficient   $\mu_{\rm max}$.
If $\mu_{\rm max}$ is non-degenerate, then
the closest $\AAA$-classical (respectively classical, separable) state to $\rho_{\Psi}$ for the Bures distance is unique. 
It is given by the pure product 
state $\ket{\alpha_{{\rm max}}}\ket{\beta_{{\rm max}}}$, where $\ket{\alpha_{{\rm max}}}$ and $\ket{ \beta_{{\rm max}}}$ are
eigenvectors  with eigenvalue $\mu_{\rm max}$ of $[ \rho_\Psi ]_\AAA$ and $[ \rho_\Psi ]_\BB$, respectively.
If $\mu_{\rm max}$ is $r$-fold degenerate, say 
$\mu_{\rm max}=\mu_1=\ldots = \mu_r> \mu_{r+1}, \ldots , \mu_{n}$, then infinitely many
$\AAA$-classical (respectively classical, separable) states $\sigma$ minimize $d_{\rm B}(\rho_\Psi,\sigma )$.
These closest states  are convex combinations of the pure product states 
$\ket{\varphi_{l}}\ket{\chi_{l}}$
with 
\begin{equation} \label{eq-Schmidt_vectors}
\ket{\varphi_{l}} = \sum_{i=1}^r u_{il} \ket{\alpha_i}
\quad , \quad 
\ket{\chi_{l} }= \sum_{i=1}^r \overline{u_{il}} \ket{\beta_i} 
\quad , \quad l=1,\ldots, r\;,
\end{equation}
where  $\{ \ket{\alpha_i}\}_{i=1}^r$ and $\{ \ket{\beta_i}\}_{i=1}^r $ are orthonormal  families 
of Schmidt vectors associated to $\mu_{\rm max}$ in the Schmidt decomposition (\ref{eq-Schmidt_decomposition}), and
$(u_{il})_{i,l=1}^r$ is an arbitrary $r\times r$ unitary matrix.

\end{proposition}

\vspace{1mm}

It should be noticed that when $\mu_{\rm max}$ is  degenerate,  the vectors (\ref{eq-Schmidt_vectors}) provide together with
$\ket{\alpha_i}$, $\ket{\beta_i}$, $i=r+1,\ldots, n$,  a Schmidt decomposition of $\ket{\Psi}$ (in that case this decomposition is not unique, 
see Sec.~\ref{eq-Schmist_dec}). 
Conversely, disregarding the degeneracies of the other eigenvalues $\mu_i < \mu_\mmax$,
all Schmidt decompositions of $\ket{\Psi}$ are of this form for some unitary matrix $(u_{il})_{i,l=1}^r$.
Thus, the existence of an infinite family of closest $\AAA$-classical states to $\ket{\Psi}$ is related to
the non-uniqueness of the Schmidt vectors associated to $\mu_\mmax$, and this family contains
the products $\ket{\varphi_{l}}\ket{\chi_{l}}$ of these vectors and convex combinations thereof. 
This  shows in particular that the maximally entangled pure states are the pure states with the largest  family of closest states\footnote{
This family forms a  $(n^2+n-2)$ real-parameter sub-manifold of $\states ( \Hh_\AB)$.
}.

\vspace{1mm}

\proof 
An arbitrary $\AAA$-classical state $\sigma$ can be decomposed  as
$\sigma  =\sum_{ij} q_{ij} \ketbra{\varphi_i}{\varphi_i} \otimes \ketbra{\chi_{j|i}}{\chi_{j|i}}$. 
In much the same way as in the proof 
of Proposition~\ref{prop-fidelity_convex_roof}, $F ( \ket{\Psi}, \Cc_\AAA ) = \mu_\mmax$  and
the closest $\AAA$-classical states to $\rho$ fulfill 
\begin{equation} \label{eq-proof_all_closest_clas_states}
| \braket{\varphi_i \otimes \chi_{j|i}}{\Psi} |^2  = \max_{\| \varphi\| = \| \chi\|=1}  \bigl\{ | \braket{\varphi \otimes \chi}{\Psi} |^2 \bigr\} = \mu_{\rm max}
\quad \text{ when\; $q_{ij}>0$.}
\end{equation}
We have thus  to determine all $\ket{\varphi} \in \Hh_\AAA$ and $\ket{\chi} \in \Hh_\BB$ such that
$|\braket{\varphi \otimes \chi}{\Psi} |^2 = \mu_{\rm max}$. This occurs if all inequalities in (\ref{eq-bound_by_mu_max}) are equalities.
Let us first assume that $\mu_1=\mu_{\rm max}>\mu_2, \ldots , \mu_n$. 
After a close look to  (\ref{eq-bound_by_mu_max}) one immediately finds  that $|\braket{\varphi \otimes \chi}{\Psi} |^2 = \mu_{\rm max}$ if and only if 
$\ket{\varphi} = \ket{\alpha_1}$ and $\ket{\chi} = \ket{\beta_1}$ up to irrelevant phase factors. Hence 
(\ref{eq-proof_all_closest_clas_states}) is satisfied for a single pair $(i,j)$. 
Therefore, all the $q_{ij}$ vanish except one and 
the closest $A$-classical state to $\ket{\Psi}$ is the pure product state $\ket{\alpha_{1}}\ket{\beta_{1} }$.

We now proceed  to the degenerate case  
$\mu_1=\ldots = \mu_r=\mu_{\rm max}> \mu_{r+1}, \ldots , \mu_{n}$. 
Let us establish the necessary and sufficient conditions for the inequalities in (\ref{eq-bound_by_mu_max}) to  be equalities.
For the first inequality, the condition is  $\arg ( \braket{\varphi}{\alpha_j} \braket{\chi}{\beta_j}) =\theta$ 
with $\theta$ independent of $j$. For the second one, the condition is that
$\ket{\varphi} $ belongs to $V_{\rm max} = \Span \{ \ket{\alpha_j} \}_{j=1}^r$ or $\ket{\chi}$ belongs to $W_{\rm max} = \Span \{ \ket{\beta_j} \}_{j=1}^r$. 
The Cauchy-Schwarz inequality in (\ref{eq-bound_by_mu_max}) is saturated if and only if
$| \braket{\varphi}{\alpha_j} | =  \lambda | \braket{\chi}{\beta_j} |$ for all $j$, with $\lambda\geq 0$. 
Finally, the last inequality holds with equality if and only if
$\ket{\varphi} \in \Span \{ \ket{\alpha_j}\}_{j=1}^n$ and $\ket{\chi} \in \Span \{ \ket{\beta_j}\}_{j=1}^n$. Putting all conditions together, we obtain
$\ket{\varphi} \in V_{\rm max}$, $\ket{\chi} \in W_{\rm max}$, and 
$ \braket{\chi}{\beta_j} = e^{\I \theta} \braket{\alpha_j}{\varphi}$ for $j=1,\ldots,r$.
Therefore,
from any orthonormal family $\{ \ket{\varphi_{l}} \}_{l=1}^r$ of $V_{\rm max}$  one can  construct  
$r$ orthogonal vectors $\ket{\varphi_{l} \otimes \chi_{l}}$ satisfying
$|\braket{\varphi_{l} \otimes \chi_{l}}{\Psi} |^2 = \mu_{\rm max}$ for all $l=1,\ldots, r$, with   
$ \braket{\chi_l}{\beta_j} = \braket{\alpha_j}{\varphi_l}$.  
The probabilities $\{ q_{ij}\}$ 
are then given by $q_{ij}=q_i$  if $i=j \leq r$ and zero otherwise,
$\{ q_l\}_{l=1}^r$ being an arbitrary probability distribution.
The corresponding $\AAA$-classical states $\sigma$ maximizing the fidelity $F(\rho_\Psi,\sigma)$ are the classical states
\begin{equation}
\sigma  =   \sum_{l=1}^r q_l    \ketbra{\alpha_{l} \otimes \beta_{l}}{\alpha_{l} \otimes \beta_{l}}\;.
\end{equation}
\finpro

\subsubsection{Geometric discord for mixed states and quantum state discrimination}

As for all other  measures of entanglement and quantum correlations, determining $D_\AAA (\rho)$ is harder for mixed states  
than for pure states. Interestingly, this problem is related  to an ambiguous quantum state discrimination task.

\vspace{2mm}

\begin{proposition} {\rm (Spehner and Orszag~\cite{moi_NJP})} \label{prop_link_geo_discord_QSD}
For any state $\rho$ of the bipartite system $\AB$,
the maximal fidelity between $\rho$ and an $\AAA$-classical state reads 
\begin{equation} \label{eq-variationnal_formula_bis} 
F (\rho, \Cc_\AAA ) = \max_{\{ \ket{\varphi_i} \} } \Bigl\{ P_{\rm S,a}^{\,\rm{opt\,v.N.}} ( \{ \rho_i,\eta_i \} )\Bigr\}
= \max_{\{ \ket{\varphi_i} \} }  \max_{ \{ \Pi_i \} } \biggl\{ \sum_{i=1}^{n_\AAA} \eta_i \tr ( \Pi_i \rho_i)  \biggr\} \;,
\end{equation}
where the maxima are over all orthonormal bases $\{ \ket{\varphi_i} \}_{i=1}^{n_\AAA}$ 
of $\Hh_\AAA$ and  all von Neumann measurements given by orthonormal families $\{ \Pi_i\}_{i=1}^{n_\AAA}$  of  projectors 
of $\Hh_\AB$ with rank $n_\BB$. Here,  
$P_{\rm S,a}^{\,\rm{opt\,v.N.}} ( \{ \rho_i,\eta_i \})$ is the maximal success probability in discriminating ambiguously by such measurements the 
states $\rho_i$ with probabilities $\eta_i$ defined by 
\begin{equation} \label{eq-state_Q_discrimination}
\eta_i = \bra{\varphi_i}  \rho_\AAA \ket{\varphi_i} \quad , \quad 
\rho_i = \eta_i^{-1} \sqrt{\rho} \ketbra{\varphi_i}{\varphi_i} \otimes 1 \sqrt{\rho}
\quad , \quad i=1 , \ldots , n_\AAA
\end{equation}
(if $\eta_i = 0$ then $\rho_i$ is not defined but does not contribute to the sum in
(\ref{eq-variationnal_formula_bis})).
Furthermore, the closest $\AAA$-classical states to $\rho$ are given by
\begin{equation} \label{eq-again_I_was_stupid}
\sigma_\rho = \frac{1}{F(\rho, \Cc_\AAA)} 
\sum_{i=1}^{n_\AAA} \ketbra{\varphi_i^{\opt}}{\varphi_i^{\opt}} 
\otimes \bra{\varphi_i^{\opt}} \sqrt{\rho}\, \Pi_i^{\opt} \sqrt{\rho} \ket{\varphi_i^{\opt}}  \;,
\end{equation}
where $\{ \ket{\varphi_i^{\opt}} \}$ and $\{ \Pi_i^{\rm{opt}} \}$ are any \ONB of $\Hh_\AAA$ and von Neumann measurement maximizing  
the \RHS of (\ref{eq-variationnal_formula_bis}).
\end{proposition}

\vspace{1mm}

The $\rho_i$ are quantum states if $\eta_i >0$ because $\rho_i \geq 0$  and $\eta_i$ is chosen such that $\tr(\rho_i)=1$.  
Moreover, $\{ \eta_i \}_{i=1}^{n_\AAA}$ is a probability distribution (since $\eta_i \geq 0$ and $\sum_i \eta_i = \tr (\rho)=1$) and 
the ensemble $\{ \rho_i,\eta_i\}_{i=1}^{n_\AAA}$ is a convex decomposition of $\rho$, \ie 
$\rho = \sum_i \eta_i \rho_i$.

\vspace{2mm}

\begin{corollary} \label{cor-case_rho_invertible}
If $\rho$ is invertible then
  one can substitute $P_{\rm S,a}^{\,\rm{opt\,v.N.}} ( \{ \rho_i,\eta_i \})$ in (\ref{eq-variationnal_formula_bis}) by
the maximal success probability $P_{\rm S,a}^{\opt} ( \{ \rho_i,\eta_i \})$  over all POVMs, given by (\ref{eq-max_success_proba_POVM}).
\end{corollary}


\proof This is a simple consequence of Corollary \ref{prop_Eldar_Megretski}. 
Actually, if  $\rho>0$ then the states $\rho_i$ defined in (\ref{eq-state_Q_discrimination}) are linearly independent, thus
the optimal \meas to discriminate them is a von Neumann \meas with projectors of rank $r_i=\rank ( \rho_i )$. 
The linear independence can be justified as follows. 
Let us first notice that  $\rho_i$ has
rank $r_i = n_\BB$ (for indeed, it  has the same rank as 
$\eta_i \rho^{-1/2} \rho_i = \ketbra{\varphi_i}{\varphi_i} \otimes 1 \sqrt{\rho}$). A necessary and sufficient condition for 
$\ket{\xi_{ij}}$ to be an eigenvector of $\rho_i $  with eigenvalue $\lambda_{ij}>0$ is 
$\ket{\xi_{ij}}= (\lambda_{ij} \eta_i )^{-1} \sqrt{\rho} \ket{\varphi_i}\otimes \ket{\zeta_{ij}}$, where
$\ket{\zeta_{ij}} \in \Hh_\BB$ is an eigenvector of $R_i= \bra{\varphi_i} \rho \ket{\varphi_i}$
with eigenvalue $\lambda_{ij} \eta_i >0$. 
For any $i$, the  Hermitian invertible matrix
 $R_i$ admits an orthonormal 
eigenbasis $\{ \ket{\zeta_{ij}} \}_{j=1}^{n_\BB}$. Thanks to the invertibility of $\sqrt{\rho}$,
$\{ \ket{\xi_{ij}} \}_{i=1, \ldots , n_\AAA}^{j=1,\ldots, n_\BB}$ is a basis of $\Hh_\AB$ and thus
the states $\rho_i$ are linearly independent and span $\Hh_\AB$.
\finpro

\vspace{2mm}

Before going into the proof of the proposition, let us discuss the state discrimination problems when
$\rho$  is pure or $\AAA$-classical.
Of course, the values of $D_\AAA ( \rho)$ are already known in these cases, being given by (\ref{eq-equality_distances}) and by 
$D_\AAA ( \rho)=0$, respectively, but it is instructive to recover that from   Proposition~\ref{prop_link_geo_discord_QSD}.
If $\rho=\rho_\Psi$ is pure then all states $\rho_i$ with $\eta_i >0$ are identical and equal 
to $\rho_\Psi$, so that 
$P_{\rm S,a}^{\,\rm{opt\,v.N.}} = \max_{\{ \Pi_i\}} \{ \sum_i \eta_i \bra{\Psi} \Pi_i \ket{\Psi} \} = \eta_{\rm max}$.
One gets $F (\rho_\Psi, \Cc_\AAA )=\mu_{\rm max}$ by optimization over the basis $\{ \ket{\varphi_i} \}$.
If $\rho$ is an  $\AAA$-classical state, \ie if it can be decomposed as in 
(\ref{eq-A-classical_state}), then the optimal basis $\{ \ket{\varphi_i^\opt}\}$ coincides
with the basis appearing in this decomposition. With this choice one obtains $\eta_i = q_i$ and 
$\rho_i = \ketbra{\varphi_i}{\varphi_i} \otimes \sigma_{\BB|i}$ for all $i$ such that $q_i >0$. 
The states $\rho_i$ are orthogonal and can thus be perfectly discriminated by von Neumann measurements. 
This yields
$F(\rho,\Cc_\AAA )=1$ and $D_\AAA ( \rho)=0$ as it should be.
Reciprocally, if $F (\rho, \Cc_\AAA )=1$ then $P_{\rm S,a}^{\,\rm{opt\,\vN}}( \{ \rho_i,\eta_i \})=1$ for some
basis $\{ \ket{\varphi_i} \}$ and  the corresponding $\rho_i$ must be orthogonal (Sec.~\ref{sec_qsd}). Hence one can find  an 
orthonormal family $\{ \Pi_i\}$ of   projectors with rank  $n_\BB$ such that
$\rho_i = \Pi_i \rho_i \Pi_i$ for any $i$ with $\eta_i >0$. It is an easy exercise to show that this implies that
$\Pi_i = \ketbra{\varphi_i}{\varphi_i} \otimes 1$ if $\rho|_{\Pi_i \Hh}$ is invertible. Thus $\rho= \sum_i \eta_i \rho_i $ is 
$\AAA$-classical, in agreement with  
the fact (following directly from the definition) that $D_\AAA (\rho )=0$ if and only if $\rho$ is $\AAA$-classical. 

The above discussion provides a clear  interpretation of the result of Proposition~\ref{prop_link_geo_discord_QSD}:
the  states $\rho$ with non-zero discord are characterized by ensembles    $\{\rho_i, \eta_i\}$ 
of non-orthogonal states, which thereby are not perfectly distinguishable, for any orthonormal basis
$\{ \ket{\varphi_i} \}$ of $\Hh_\AAA$. The less distinguishable are the $\rho_i$'s, the most distant is $\rho$ from the
set of zero-discord states.

We will establish Proposition \ref{prop_link_geo_discord_QSD} by relying on the slightly  more general statement summarized in the following lemma.

\vspace{2mm}

\begin{lemma} \label{lemma_link_geo_discord_QSD}
For a fixed family $\{ \sigma_{\AAA | i} \}_{i=1}^n$  of states  $\sigma_{\AAA |i} \in \states ( \Hh_\AAA)$ 
having orthogonal supports and spanning $\Hh_\AAA$, with $1 \leq n \leq n_\AAA$,  
let us define
\begin{equation}
\Cc_\AAA ( \{ \sigma_{\AAA | i} \} ) = \Bigl\{ \sigma = \sum_{i=1}^n q_i \sigma_{\AAA | i} \otimes \sigma_{\BB | i} \; ;\; \{ q_i, \sigma_{\BB | i} \}_{i=1}^n 
\;\text{ is a state ensemble on $\Hh_\BB$} \;\Bigr\}\;.
\end{equation}
Then
\begin{equation} \label{eq-general_link_geo_discord_QSD}
F \bigl( \rho, \Cc_\AAA ( \{ \sigma_{\AAA | i} \} ) \bigr) 
 = \max_{ \sigma \in \Cc_\AAA ( \{ \sigma_{\AAA | i} \} )} \bigl\{ F ( \rho, \sigma ) \bigr\}
= \max_{U} \biggl\{  \sum_{i=1}^n \| W_i ( U) \|_2^2 \biggr\}\;,
\end{equation}
where the last maximum is over all unitaries $U$ on $\Hh_\AB$ and
\begin{equation} 
W_i ( U ) = \tr_\AAA \bigl( \sqrt{\sigma_{\AAA | i}} \otimes 1\,\sqrt{\rho}\, U \bigr)\;.
\end{equation}
Moreover, there exists  a unitary $U_\opt$ achieving the maximum in (\ref{eq-general_link_geo_discord_QSD}) which is such that $W_i ( U_\opt ) \geq 0$.
The states $\sigma_\opt$ satisfying $F(\rho, \sigma_\opt ) =  F ( \rho, \Cc_\AAA ( \{ \sigma_{\AAA | i} \} ))$ are given in terms of this unitary  by
\begin{equation} \label{eq-general_closest_A_class_state}
\sigma_\opt =  \frac{1}{F ( \rho, \Cc_\AAA ( \{ \sigma_{\AAA | i} \} ))} \sum_{i=1}^n  \sigma_{\AAA | i} \otimes W_i ( U_\opt)^2\;.
\end{equation}
\end{lemma}

\vspace{1mm}

\proof
Using the spectral decompositions of the states $\sigma_{\BB |i}$, any $\sigma \in \Cc_\AAA ( \{ \sigma_{\AAA | i} \} )$ can be written as
\begin{equation} \label{eq-proof_geo_discord_and_QSD0}
\sigma = \sum_{i=1}^n \sum_{j=1}^{n_\BB} q_{ij} \sigma_{\AAA | i} \otimes \ketbra{\chi_{j|i}}{\chi_{j|i}} \quad \text{ with }
\quad q_{ij} \geq 0\;,\;\sum_{ij} q_{ij} = 1\;,
\end{equation}
where $\{ \ket{\chi_{j|i}} \}_{j=1}^{n_\BB}$ is an \ONB of $\Hh_\BB$ for any $i$ (compare with (\ref{eq-A-classical_state_bis})).
By assumption,  if $i \not= i'$ then $\range \sigma_{\AAA | i}  \,\bot\,\range \sigma_{\AAA | i'}$, so that 
$\sqrt{\sigma} = \sum_{i,j} \sqrt{q_{ij}} \sqrt{\sigma_{\AAA | i}} \otimes \ketbra{\chi_{j|i}}{\chi_{j|i}} $.
We start
by evaluating the trace norm in the definition (\ref{eq-fidelity}) of the fidelity by means of the formula
$\| O \|_1 = \max_{U} | \tr ( U O ) |$ to obtain
\begin{eqnarray} \label{eq-proof_geo_discord_and_QSD1}
\nn
F \bigl( \rho, \Cc_\AAA ( \{ \sigma_{\AAA | i} \} ) \bigr) 
& = & 
\max_{\sigma \in  \Cc_\AAA ( \{ \sigma_{\AAA | i} \} )} \max_U \Bigl\{ \bigl| \tr ( U^\ast \sqrt{\rho} \sqrt{\sigma} ) \bigr|^2 \Bigr\}
\\
& = & \max_U \biggl\{ \max_{ \{ q_{ij} \}, \{ \ket{\chi_{j|i}} \} } \biggl|  \sum_{i,j} \sqrt{q_{ij}} \bra{\chi_{j|i}} W_i (U)^\ast \ket{\chi_{j|i}} \biggr|^2 \biggr\}\;.
\end{eqnarray}
The square modulus can be bounded by invoking twice the Cauchy-Schwarz inequality and $\sum_{ij} q_{ij} =1$,
\begin{eqnarray} \label{eq-proof_geo_discord_and_QSD2}
\nn
\biggl|  \sum_{i,j}  \sqrt{q_{ij}} \bra{\chi_{j|i}} W_i (U)^\ast \ket{\chi_{j|i}} \biggr|^2
& \leq & \sum_{i,j} \bigl|  \bra{\chi_{j|i}} W_i (U)^\ast \ket{\chi_{j|i}} \bigr|^2 \\
&\leq & \sum_{i,j}  \bigl\| W_i (U) \ket{\chi_{j|i}} \bigr\|^2 = \sum_i \| W_i (U ) \|_2^2\;.
\end{eqnarray}
The foregoing inequalities are equalities if the following conditions are satisfied:
\begin{itemize}
\item[(1)] $W_i ( U )= W_i (U)^\ast  \geq 0$;
\item[(2)] $q_{ij} =\bra{\chi_{j|i}}  W_i ( U) \ket{\chi_{j|i}}^2 / ( \sum_{i,j}  \bra{\chi_{j|i}}  W_i ( U) \ket{\chi_{j|i}}^2 )$;
\item[(3)] $\{ \ket{\chi_{j|i}} \}_{j=1}^{n_\BB}$ is an eigenbasis of $W_i ( U)$ for any $i$.
\end{itemize}
Therefore, (\ref{eq-general_link_geo_discord_QSD}) holds true 
provided that there is a unitary $U$ on $\Hh_\AB$ satisfying (1). For a given $U$, let us define $U_\opt = U \sum_{i} \pi_i^\AAA \otimes V_i^\ast$,
where $\pi_i^\AAA$ is the projector onto $\range \sigma_{\AAA | i}$ and $V_i$ a unitary on $\Hh_\BB$ such that
$W_i ( U ) = | W_i(U)^\ast |  V_i$ (polar decomposition). 
Then $U_\opt$ is unitary since by hypothesis $\pi_i^\AAA \pi_{i'}^\AAA = \delta_{i i'} \pi_i^\AAA$ and $\sum_i \pi_i^\AAA = 1$, and
one readily shows that $W_i ( U_\opt ) = W_i ( U ) V_i^\ast \geq 0$. 
As $\sum_i \| W_i ( U ) \|_2^2 = \sum_i \| W_i ( U_\opt ) \|_2^2$, the identity (\ref{eq-general_link_geo_discord_QSD}) follows from 
(\ref{eq-proof_geo_discord_and_QSD1}) and (\ref{eq-proof_geo_discord_and_QSD2}). 
From condition (3) one has $W_i ( U_\opt ) \ket{\chi_{j|i}^\opt} = w_{ji} \ket{\chi_{j|i}^\opt}$ with 
$\sum_{i,j} w_{ji}^2 = F ( \rho, \Cc_\AAA ( \{ \sigma_{\AAA | i} \} ) )$, see (\ref{eq-proof_geo_discord_and_QSD2}). Condition (2) entails
\begin{equation}
\sigma_{\BB | i}^\opt = \sum_j q_{ij}^\opt \ketbra{\chi_{j|i}^\opt}{\chi_{j|i}^\opt}  = \frac{W_i ( U_\opt )^2 }{F ( \rho, \Cc_\AAA ( \{ \sigma_{\AAA | i} \} ) )} 
\;,
\end{equation}
which together with (\ref{eq-proof_geo_discord_and_QSD0}) leads to (\ref{eq-general_closest_A_class_state}).
\finpro

\vspace{3mm}

\Proofof{Proposition~\ref{prop_link_geo_discord_QSD}}
Let  $\{ \ket{\varphi_i} \}_{i=1}^{n_\AAA}$ be an \ONB of $\Hh_\AAA$.
Applying Lemma~\ref{lemma_link_geo_discord_QSD} with $\sigma_{\AAA | i} = \ketbra{\varphi_i}{\varphi_i}$ one gets 
\begin{eqnarray} \label{eq-proof_link_geo_discord_QSD1}
\nn
F \bigl( \rho , \Cc_\AAA ( \{ \ket{\varphi_i} \} ) \bigr)
& = & \max_U \left\{ \sum_{i=1}^{n_\AAA}  \tr \bigl[ U \ketbra{\varphi_i}{\varphi_i} \otimes 1 \, U^\ast \sqrt{\rho} \,\ketbra{\varphi_i}{\varphi_i} \otimes 1\, \sqrt{\rho} \bigr] \right\}\;,
\\
& = &
\max_{ \{ \Pi_i\}} \left\{ \sum_{i=1}^{n_\AAA} 
\tr \bigl[ \Pi_i \sqrt{\rho} \ketbra{\varphi_i}{\varphi_i} \otimes 1 \, \sqrt{\rho} \bigr] \right\} 
= P_{\rm S,a}^{\,\rm{opt\,v.N.}} ( \{ \rho_i,\eta_i \} )\;.
\end{eqnarray}
The last maximum is over all orthonormal families $\{ \Pi_i\}_{i=1}^{n_\AAA}$ of projectors of rank $n_\BB$ and
$P_{\rm S,a}^{\,\rm{opt\,v.N.}} ( \{ \rho_i,\eta_i \} )$ is given by (\ref{eq-max_success_proba_von_Neumann}).
Since the fidelity $F ( \rho, \Cc_\AAA )$ is the maximum of $F ( \rho , \Cc_\AAA ( \{ \ket{\varphi_i} \} ) )$ over all bases $ \{ \ket{\varphi_i } \}$,
this leads to (\ref{eq-variationnal_formula_bis}) and (\ref{eq-again_I_was_stupid}).
\finpro

\subsubsection{The qubit case}

It has been emphasized in Sec.~\ref{sec_qsd} that the optimal success probability and measurement for discriminating
 ambiguously more than two states are not known  explicitly in general.  
Nonetheless, if the subsystem $\AAA$ is a qubit, the ensemble $\{ \rho_i, \eta_i \}$ in Proposition~\ref{prop_link_geo_discord_QSD}
contains only $n_\AAA=2$ states and the optimal probability and measurement are easily determined.
Following the steps yielding to  (\ref{eq-opt_success_proba}) we find 
\begin{equation}
\label{eq-opt_success_proba_vN}
P_{\rm S,a}^{\,\rm opt\,\vN}( \{ \rho_i,\eta_i\})
= \frac{1}{2}\bigl(  1- \tr  \Lambda \bigr) + 
\sum_{l=1}^{n_\BB} \lambda_l \;,
\end{equation}
where $\lambda_1  \geq \cdots \geq \lambda_{n_\BB}$ are the $n_\BB$ largest eigenvalues of $\Lambda = \eta_0 \rho_0 - \eta_1 \rho_1$.
The optimal von Neumann \meas  is formed by the spectral projector $\Pi_0^\opt$ of $\Lambda$ for 
these $n_\BB$ eigenvalues and its complement $\Pi_1^\opt = 1  -\Pi_0^\opt$.
For the states $\rho_i$ associated to the \ONB $\{ \ket{\varphi_i}\}_{i=0}^1$ of $\complex^2$ 
via formula (\ref{eq-state_Q_discrimination}),  one has 
$\Lambda=\sqrt{\rho} \,( \ketbra{\varphi_0}{\varphi_0} - \ketbra{\varphi_1}{\varphi_1} ) \otimes 1\, \sqrt{\rho}$. 
The operator inside the parenthesis in the last identity
is equal to  $\sigma_{\uv} \equiv \sum_{m=1}^3 u_m \sigma_m$ for some unit vector $\uv \in \real^3$ depending on $\{ \ket{\varphi_i}\}$
(here $\sigma_1$, $\sigma_2$, and $\sigma_3$  are the Pauli matrices).
Conversely,  one can associate to  any unit vector $\uv \in \real^3$ the 
eigenbasis $\{ \ket{\varphi_i}\}_{i=0}^1$ of ${\sigma}_{\uv}$. According to Proposition~\ref{prop_link_geo_discord_QSD}, $F (\rho, \Cc_\AAA)$ is  obtained
by maximizing the \RHS of (\ref{eq-opt_success_proba_vN}) over all Hermitian matrices 
\begin{equation} \label{eq-Lambda}
 \Lambda (\uv )  =  \sqrt{\rho} \, {\sigma}_{\uv} \otimes 1 \, \sqrt{\rho} 
\end{equation}
with $\uv \in \real^3$, $| \uv | = 1$.
The following corollary of  Proposition~\ref{prop_link_geo_discord_QSD} is a refinement of a result in~\cite{moi_JPA}.

\vspace{2mm}
 
\begin{corollary}  \label{prop-geo_discord_qubit}
Let $\AAA$ be a qubit, \ie $n_\AAA = 2$. The fidelity between $\rho$ and the set of $\AAA$-classical states is given by 
\begin{equation} \label{eq-fidelity_as_min_success_discrimination_qubit}
F (\rho, \Cc_\AAA) =  \frac{1}{2} \max_{\| \uv \|=1 } \bigl\{  1  +  \| \Lambda (\uv) \|_1  \bigr\}\;,
\end{equation}
where
$\Lambda (\uv)$ is the $2n_B \times 2 n_B$ matrix  (\ref{eq-Lambda}).
The closest $\AAA$-classical states to $\rho$ are given by (\ref{eq-again_I_was_stupid}) where  $\Pi_0^\opt$ is the spectral projector 
associated to the $n_\BB$ largest eigenvalues of $\Lambda ( \uv^\opt )$ and $\uv^\opt \in \real^{3}$ is a unit vector 
achieving the maximum in (\ref{eq-fidelity_as_min_success_discrimination_qubit}).
\end{corollary}

\vspace{1mm}

\proof Let $\lambda_l ( \uv)$ be the eigenvalues of $\Lambda ( \uv)$ in non-increasing order. We claim that
\begin{equation} \label{eq-identity_with_Helstrom_formula}
 -\onehalf  \tr ( \Lambda (\uv)) +  \sum_{l=1}^{n_\BB} \lambda_l  (\uv) = \onehalf \sum_{l=1}^{n_\BB} \lambda_l  (\uv) - \onehalf \sum_{l=n_\BB +1}^{2 n_\BB} \lambda_l (\uv)
= \onehalf \tr | \Lambda (\uv) | \;.
\end{equation}
To prove this claim it suffices to show  that 
$\Lambda (\uv)$ has at most $n_\BB$ positive eigenvalues $\lambda_l ( \uv) >0$ and at most $n_\BB$ negative eigenvalues $\lambda_l ( \uv) <0$, 
counting multiplicities. 
As $\ker \rho \subset \ker \Lambda (\uv)$ one may without loss of generality restrict
$\Lambda (\uv)$ to the subspace $\Pi \Hh_\AB$, with $\Pi$ the projector onto $\range ( \rho)$. 
A standard linear algebra argument shows that  if $S$ is a finite invertible matrix and $\Sigma$ a self-adjoint matrix, 
then the number of positive (respectively negative) eigenvalues of $\Sigma$ is equal to the number
of positive (respectively negative) eigenvalues of $S^\ast \Sigma S $.
Let $P_\Sigma^\pm$ be  the spectral projectors of $\Sigma= \Pi \, \sigma_{\uv} \otimes 1 \, \Pi$  on $\real_{\pm} \setminus \{ 0 \}$.
Since $\sqrt{\rho} : \Pi \Hh_\AB \rightarrow \Pi \Hh_\AB$ is invertible, 
in order to prove (\ref{eq-identity_with_Helstrom_formula}) it is thus enough   
to verify that $\rank ( P_\Sigma^\pm ) \leq n_\BB$. This is evident if $\rank ( \Pi) \leq n_\BB$. 
If $\rank( \Pi) > n_\BB$, then $\pm \bra{\Psi} \sigma_{\uv} \otimes 1   \ket{\Psi} = \pm \bra{\Psi} \Sigma \ket{\Psi}   > 0$ for 
any $\ket{\Psi} \in P_\Sigma^\pm \Hh_\AB \subset \Pi \Hh_\AB$. This implies that $ \rank ( P_\Sigma^\pm ) \leq \rank ( P_{\sigma_{\uv} \otimes 1}^\pm ) =n_\BB$, as 
otherwise one could find a non-vanishing vector $\ket{\Psi} \in P_\Sigma^\pm  \Hh_\AB$ belonging to the 
$n_\BB$-dimensional eigenspace of  $\sigma_{\uv} \otimes 1$ with eigenvalue $\mp 1$, in contraction with 
the foregoing inequality. This establishes (\ref{eq-identity_with_Helstrom_formula}).
Then (\ref{eq-fidelity_as_min_success_discrimination_qubit}) follows from  (\ref{eq-opt_success_proba_vN}) and Proposition~\ref{prop_link_geo_discord_QSD}. 
\finpro

\subsubsection{States with the highest geometric discord}

The geometric discord $D_\AAA$,  as the quantum discord $\delta_\AAA$, quantifies the degree of quantumness of a state.
Let us recall from Sec.~\ref{sec-state_with_highest_discord} that when the space dimensions of $\AAA$ and $\BB$ are such that
$n_\AAA \leq n_\BB$, the ``most quantum'' states $\rho$ having  the highest discord $\delta_\AAA ( \rho)$  are the maximally entangled states, \ie
the states with the highest entanglement of formation $\EoF ( \rho) = \ln n_\AAA$.
It is comforting that a similar result holds for the geometric discord. 

\vspace{2mm}

\begin{corollary}  \label{prop_maximal_value_geo_discord}
If $n_\AAA \leq n_\BB$, the highest value of $D_\AAA ( \rho)$ on $\states (\Hh_\AB)$ is equal to $2-2/\sqrt{n_\AAA}$. 
The most distant states $\rho$ from the set of $\AAA$-classical states,
which are such that $D_\AAA ( \rho)= 2-2/\sqrt{n_\AAA}$, are the maximally entangled states given by Proposition~\ref{prop_max_ent_states}. 
\end{corollary}

\vspace{1mm}

Comparing with the results of Sec.~\ref{sec_geometric_entnaglement_and_convex_roof}, we see that when $n_\AAA \leq n_\BB$
the  most distant states from $\Cc_\AAA$ are also the most distant from the set of separable states $\Ss_\AB$.
If $n_\AAA \leq n_\BB < 2 n_\AAA$, these most distant states are maximally entangled pure states, as illustrated in Fig.~\ref{fig3}.

\vspace{1mm}

\proof This is again a corollary of Proposition~\ref{prop_link_geo_discord_QSD}.
The success probability  $P_{\rm S,a}^{\,\rm{opt\,\vN}}$ is clearly larger or equal to the highest prior probability\footnote{
A receiver would obtain $P_{\rm S,a} = \eta_{\rm max}$ by simply guessing that his state is
$\rho_{i_{\rm max}}$, with $\eta_{i_{\rm max}}=\eta_{\max}$, whatever the \meas outcomes. A better strategy is of course to perform the 
von Neumann measurement $\{ \Pi_i \}$ such that  $\Pi_{{i}_{\rm max}}$ projects onto 
a $n_\BB$-dimensional subspace containing $\range ( \rho_{i_{\rm max}} )$. This range
has a dimension $\rank ( \rho_{i_{\rm max}} ) \leq n_\BB$  by a similar reasoning as in the proof of
Corollary~\ref{cor-case_rho_invertible}.
}
$\eta_{\max}  = \max_i \{ \eta_i\}$. 
In view of Proposition~\ref{prop_link_geo_discord_QSD} and  $\eta_{\max} \geq 1/n_\AAA$, we get 
\begin{equation} \label{eq-bound_F_A}
F (\rho, \Cc_\AAA ) \geq \frac{1}{n_\AAA} 
\end{equation}
for any state $\rho$. When $n=n_\AAA \leq n_\BB$ this bound  is optimal, the value $1/n$ being achieved for 
the maximally entangled pure states (Sec.~\ref{sec_pure_states}). This proves the first statement.
Let $\rho$ be a state such that $F (\rho, \Cc_\AAA ) = 1/n$.  According to (\ref{eq-variationnal_formula_bis}) and since it has been argued above that
$P_{\rm S,a}^{\,\rm{opt\,v.N.}} \geq \eta_{\rm max} \geq 1/n$, this implies
that $P_{\rm S,a}^{\,\rm{opt\,v.N.}} ( \{ \rho_i,\eta_i \})=1/n$  whatever the orthonormal basis $\{ \ket{\varphi_i} \}$.
It is intuitively clear\footnote{
An explicit proof of this fact can be found in~\cite{moi_NJP}.
}
 that this can  happen only if the receiver gets a collection of identical states $\rho_i$ with equal prior 
probabilities $\eta_i = 1/n$. From (\ref{eq-state_Q_discrimination})
and $\rho= \sum \eta_i \rho_i$ one obtains $\rho_\AAA = 1/n$
and $\rho_i = \rho$ for any $i$ and $\{ \ket{\varphi_i} \}$. Plugging the spectral decomposition 
$\rho = \sum_{k} p_k \ketbra{k}{k}$ into  (\ref{eq-state_Q_discrimination}),
the second equality yields
$D_{kl}= \tr_\BB ( \ketbra{k}{l} )= n^{-1} \delta_{kl}$
for all $k$ and $l$ such that $p_k p_l \not= 0$. 
One concludes that $\rho$ has maximal entanglement of formation by following the same steps 
as in the proof of  Proposition~\ref{prop_max_ent_states}. 
\finpro

\vspace{2mm}

One may wonder if Corollary~\ref{prop_maximal_value_geo_discord} could also hold for $n_\AAA > n_\BB$ 
(modulo  the exchange $n_\AAA \leftrightarrow n_\BB$), as what happens for the
geometric measure of entanglement (see Sec.~\ref{sec_geometric_entnaglement_and_convex_roof}). 
However, unlike $E_{\rm Bu}(\rho)$ the geometric discord is not symmetric under the 
exchange of the two subsystems. The problem of determining its highest value and the corresponding ``most quantum'' states  
 is still open for $n_\AAA > n_\BB$.
For such space dimensions the bound (\ref{eq-bound_F_A}) is still correct but it is not optimal, that is, 
there are no states $\rho$ with fidelity $F (\rho, \Cc_\AAA )= 1/n_\AAA$. 
Indeed, one can show as in the proof above that if $F (\rho, \Cc_\AAA )= 1/n_\AAA$ then
the eigenvectors $\ket{k}$ of $\rho$ with eigenvalues $p_k>0$
have maximally mixed marginals $D_{kk}= (\ketbra{k}{k})_\AAA = 1/n_\AAA$. But this  is impossible 
since $\rank ( D_{kk} ) \leq n_\BB$  by (\ref{eq-Schmidt_decomposition}).

\vspace{1mm}

\begin{exercice}
One can place a lower bound on $F (\rho, \Cc_\AAA )$ for $n_\AAA > n_\BB$ by invoking the inequality~\cite{moi_NJP}
\begin{equation} \label{eq-refined_bound_n_A>n_B}
F (\rho, \Cc_\AAA ) \geq \frac{\| \rho\|}{n_\BB} + \frac{1-\|\rho\|}{n_\AAA} \frac{n_\BB-\delta_\rho}{n_\BB}
\end{equation}
where $\delta_\rho = 0$ if $\rank (\rho) \leq n_\BB$ and $1$ otherwise.
\end{exercice}

Table~\ref{tab1} presents a comparison of the properties of the entanglement of formation, the quantum discord, 
and their geometrical analogs based on the Bures distance.

\begin{table}
\scriptsize
\begin{tabular}{|l||c|c|c|c|}
\hline
 \begin{tabular}{c}  \end{tabular}  &  \begin{tabular}{c} Entanglement of \\  formation \end{tabular}
& Quantum discord & \begin{tabular}{c} Geometric  \\  entanglement \end{tabular}  &  Geometric discord
\\[1mm]
\hline
\hline 
$\AB$ in a pure state &  \multicolumn{2}{|c|}{$\EoF ( \ket{\Psi} ) = \delta_\AAA ( \ket{\Psi} ) = H ( \{ \mu_i\} )$}  
 &  \multicolumn{2}{|c|}{$E_{\rm Bu} ( \ket{\Psi} ) = D_\AAA ( \ket{\Psi} ) = 2 ( 1 - \sqrt{\mu_\mmax})$}   
\\[1mm]
\hline 
$\AB$ in a mixed state  
&  $\begin{array}{c} \EoF ( \rho) =  \min  \\ \bigl\{ \sum_i \eta_i \EoF ( \ket{\Psi_i} ) \bigr\} 
\\[2mm] \text{(convex roof)} \end{array}$
&  $\begin{array}{c} \delta_\AAA ( \rho) 
= I_{\AAA : \BB} ( \rho) - \\ \underbrace{\max \{ I_{\AAA : \BB} ( \Mm_\AAA \otimes 1 ( \rho))\} }_{\text{\scriptsize{classical\,\,correlations}}}  \end{array}$
& $\begin{array}{c} E_{\rm Bu} ( \rho) = 2  ( 1 - \\ \underbrace{\max \{ \sqrt{F ( \rho, \sigma_\sep)} \} }_{=\; \text{\scriptsize{convex roof}}} ) \end{array} $
&  $\begin{array}{c} D_\AAA ( \rho) 
= 2  ( 1 - \\  \underbrace{\max\{ \sqrt{F ( \rho, \sigma_\Aclass)}\}  )}_{\text{\scriptsize \begin{tabular}{l}= max.\,\,success proba. \\ in state discrimination \end{tabular}} } 
 \end{array}$
\\[1mm]
\hline 
Vanishes iff  &  $\rho$ is separable  & $\rho$ is $\AAA$-classical   &  $\rho$ is separable  & $\rho$ is $\AAA$-classical 
\\[1mm]
\hline 
\begin{tabular}{l} Maximal iff \\ with maximal value  \end{tabular}  
& 
\multicolumn{2}{|c|}{ 
$\left. \begin{array}{c} \rho \text{ is max.\,entangled} \\ \ln n \end{array} \right\}
\begin{array}{r} \text{$\EoF$: true $\forall\,\,n_{\AAA,\BB}$}  \\ \text{$\delta_\AAA$: true\,\,if\,\,$n_\AAA \leq n_\BB$ } \end{array}$
} 
& 
\multicolumn{2}{|c|}{ 
$\left. \begin{array}{c} \rho \text{ is max.\,entangled} \\ 2(1-1/\sqrt{n}) \end{array} 
\right\}
\begin{array}{r} \text{$E_{\rm Bu}$: true $\forall\,\,n_{\AAA,\BB}$}  \\ \text{$D_\AAA$: true\,\,if\,\,$n_\AAA \leq n_\BB$} \end{array}
$ 
}
\\[1mm]
\hline 
Local unit.\,invariance  &  \checkmark & \checkmark  &  \checkmark  &  \checkmark
\\[1mm]
\hline 
Monotonicity w.r.t. &  LOCCs  & operations on $\BB$  &  LOCCs  &  operations on $\BB$
\\[1mm]
\hline 
Convexity         &    \checkmark   &  no  &  \checkmark  & no
\\[1mm]
\hline 
Ordering             & \multicolumn{2}{|c|}{no}  & \multicolumn{2}{|c|}{$E_{\rm Bu} ( \rho) \leq D_\AAA ( \rho)$}
\\[1mm]
\hline
$\ABC$ in a pure state   &  \multicolumn{2}{|c|}{$\EoF (\rho_\BC) = \delta_\AAA ( \rho_\AB) + S( \rho_\AB) - S ( \rho_\AAA)   $}  & 
\multicolumn{2}{|c|}{?} 
\\
\hline
\end{tabular}
\caption{\label{tab1} Summary of the definitions and properties of the entanglement of formation (Sec.~\ref{sec-entanglement}), 
quantum discord (Sec.~\ref{sec-discord}), geometric measure of entanglement (Sec.~\ref{sec-geometric_entanglement}), 
and geometric discord (Sec.~\ref{sec-geo_discord}).  Here $n_\AAA$ and $n_\BB$ are the space dimensions of the subsystems $\AAA$ and $\BB$,
$n= \min \{ n_\AAA, n_\BB\}$, and $\mu_i$ are the Schmidt coefficients in~(\ref{eq-Schmidt_decomposition}). 
}
\end{table}

\subsubsection{Geometric discord and least square measurements}

The ensemble $\{ \rho_i, \eta_i\}$ in the discrimination task associated to the geometric discord  in
Proposition~\ref{prop_link_geo_discord_QSD} turns out to be related to the transpose operation of the
von Neumann \meas in the basis $\{ \ket{\varphi_i}\}$. In fact, let us denote by $\Mm_\AAA$ the \meas 
on $\AAA$ with rank-one orthonormal projectors $\pi_i^\AAA = \ketbra{\varphi_i}{\varphi_i}$. Let
\begin{equation} \label{eq-conditional_state_geo_discord}
\eta_i = \bra{\varphi_i} \rho_\AAA \ket{\varphi_i} \quad , \quad 
\rho_{\AB | i } = \eta_i^{-1} \ketbra{\varphi_i}{\varphi_i} \otimes \bra{\varphi_i} \rho \ket{\varphi_i}
\end{equation}
be the corresponding probabilities and post-\meas conditional states when the initial state is $\rho$. 
The transpose operation of $\Mm_\AAA$ for $\rho$ is (see (\ref{eq-def_widehat_Mm}))
\begin{equation}
\Rr_{\Mm_\AAA , \rho } ( \sigma ) 
 = \sum_{i=1}^{n_\AAA} \sqrt{\rho} \ketbra{\varphi_i}{\varphi_i} \otimes 
  \bra{\varphi_i} \rho \ket{\varphi_i}^{-\onehalf}   \bra{\varphi_i} \sigma \ket{\varphi_i} \bra{\varphi_i} \rho \ket{\varphi_i}^{-\onehalf} \sqrt{\rho}
\;.
\end{equation}
We observe that
\begin{equation} \label{eq-misleading_formula_for_rho_i}
\rho_i = \Rr_{\Mm_\AAA , \rho } ( \rho_{\AB | i} ) \quad , \quad i = 1 , \ldots, n_\AAA \;.
\end{equation}
Comparing (\ref{eq-falta_poco}) and
(\ref{eq-misleading_formula_for_rho_i}), one  expects from the discussion in Sec.~\ref{sec-least_square_meas}  
that the least square \meas $\{ M_i^\lsm \}$ for the ensemble $\{ \rho_i, \eta_i\}$ is associated to the transpose operation of
$\Rr_{\Mm_\AAA , \rho }$ for $\Mm_\AAA ( \rho) = \sum_i \eta_i \rho_{\AB |i}$. But this two-fold transpose operation coincides with $\Mm_\AAA$,
hence $\{ M_i^\lsm \}$ is nothing but the von Neumann \meas on $\AAA$ in the basis $\{ \ket{\varphi_i} \}$.
This can be readily checked: since $\{ \rho_i, \eta_i\}$ is a convex decomposition of $\rho$,  (\ref{eq-definition_LSM}) leads to
\begin{equation}
M_i^\lsm  = \eta_i \rho^{-1/2}  \rho_i \rho^{-1/2} = \pi_i^\AAA \otimes 1 \;.
\end{equation}
One can bound $P_{\rm S,a}^{\,\rm{opt\,v.N.}} ( \{ \rho_i,\eta_i \})$ from below  by
the success probability obtained by discriminating the $\rho_i$ with $\{ M_i^\lsm \}$, and from above by the square root of this probability, see (\ref{eq-upper_bound_P_S_in_term_of_P_lsm}). 
By Proposition~\ref{prop_link_geo_discord_QSD}, this yields
\begin{equation} \label{eq-inequality_square_root_meas}
 \max_{\{ \ket{\varphi_i} \} }\biggl\{  \sum_{i=1}^{n_\AAA}  \tr_\BB \bigl[ \bra{\varphi_i} \sqrt{\rho} \ket{\varphi_i}^2 \bigr]\biggr\}
\leq 
F (\rho, \Cc_\AAA ) 
\leq  
\max_{\{ \ket{\varphi_i} \} }  \biggl\{ \sum_{i=1}^{n_\AAA}   \tr_\BB \bigl[ \bra{\varphi_i} \sqrt{\rho} \ket{\varphi_i}^2 \bigr] \biggr\}^\onehalf \;.
\end{equation}
The left- and right-hand sides become nearly equal when $F(\rho,\Cc_\AAA)$ is almost one, that is, if $\rho$ is close to 
$\Cc_\AAA$. Other inequalities on $ F (\rho, \Cc_\AAA ) $ can be obtained in terms of the fidelities $F(\rho_i,\rho_j)$ with the help of 
Proposition~\ref{prop_lower_and_upper_bounds_on_P_S_amb}.

The aforementioned observations are summarized by Fig.~\ref{fig4}.

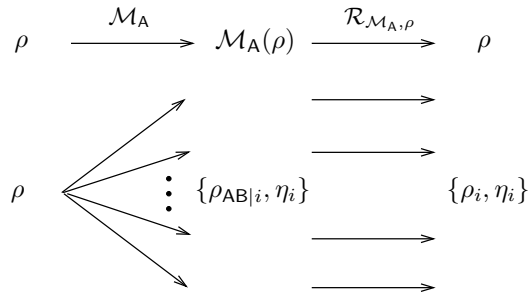
\begin{figure}
\begin{center}
\vspace*{4mm}
\input{fig3_with_latex.pstex_t} 
\end{center}
\caption{State changes under the von Neumann measurement $\Mm_\AAA$ with rank-one projectors $\pi_i^\AAA = \ketbra{\varphi_i}{\varphi_i}$ followed by
its transpose operation $\Rr_{\Mm_\AAA, \rho}$. The upper line corresponds to a \meas without readout and the other lines to the different \meas outcomes.}
\label{fig4}
\end{figure}

\vspace{2mm}

\noindent {\bf Acknowledgments.}
I am grateful to V. Eremeev, G. Ferrini, A. Joye, M. Orszag, and A. Smerzi for interesting discussions and to   
V. Jak\u{s}i\'c for pointing out to me the works of Refs.~\cite{Mueller-Lennert13,Frank13}.
I acknowledge support from the ANR project no. ANR-13-JS01-0005-01.

\vspace{5mm}
\appendix
\renewcommand{\theequation}{\Alph{section}\arabic{equation}}
\setcounter{equation}{0}

\section{Operator monotone and operator convex functions} \label{app_operator_convex_functions}

We recall in this appendix some basic facts about operator monotone and operator convex functions.
We refer the reader to the lecture notes~\cite{Carlen} and the book~\cite{Bhatia} 
for more complete presentations of these notions.

We denote by $\observables_{+}$ the set of non-negative operators on  $\Hh$, with $\dim ( \Hh) = n < \infty$.
A function  $f : \real_+ \rightarrow \real$ is {\it operator convex} if for any $n\times n$
matrices $A, B \in \observables_{+}$ and any $0 \leq \eta \leq 1$,  it holds
$f (  (1-\eta) A +\eta B ) \leq (1-\eta)  f(A) +  \eta f(B)$. It is {\it strictly operator convex} 
if the inequality holds with equality \ifif $\eta\in {0,1}$ or $A=B$. It is {\it operator concave} if $-f$ is operator convex.
It is {\it operator monotone-increasing} if for any $A, B \in \observables_+$, 
$ A \leq B$ $\Rightarrow f(A) \leq f(B)$, and {\it operator monotone-decreasing} if the reverse equality holds.

It is not hard to show (see e.g.~\cite{Carlen}) that $f(x)=x^{-1}$ is operator monotone-decreasing and strictly operator convex.  
Clearly, this is then also true for $f(x) = (x+t)^{-1}$ for any $t \geq 0$. According to the integral representation
\begin{equation} \label{eq-int_repres_A^alpha}
A^\alpha = \frac{\sin (\alpha \pi )}{\pi} \int_0^\infty \D t \,t^\alpha \Bigl( \frac{1}{t} - \frac{1}{t + A} \Bigr)\;,
\end{equation}
it follows that
$f_\alpha (x) = x^\alpha $ is operator monotone-increasing and strictly operator concave for $0 <  \alpha < 1$.
Similarly, one shows that $f_\alpha $ is operator monotone-decreasing and operator convex for  
$\alpha \in [-1,0]$ and operator convex for $\alpha \in [1,2]$. However, for instance 
the square function $f_2$ is not operator monotone and the cube function $f_3$ is not operator convex. 
One can establish that $g(x)=\ln x$ and $f(x)=x \ln x$ are operator concave and operator convex,  respectively, thanks to the identities 
\begin{equation} \label{eq-log_as_limit}
\ln A = \lim_{\alpha \rightarrow 0} \alpha^{-1} (A^\alpha - 1 )
\quad , \quad 
A \ln  A = \lim_{\alpha\rightarrow 1} \frac{A^\alpha - A}{\alpha - 1}\;.
\end{equation}
Another example of monotone-increasing function is $f(x) = (x-1)/\ln x = \int_0^1 \D \alpha \,x^\alpha$.

Operator monotonicity is  much stronger than usual monotonicity of real functions. This is clear 
from L\"owner's theorem, which states that if $f: (-1,1) \rightarrow \real$ is operator monotone
and non-constant, then  $f$ admits the integral representation
\begin{equation} \label{eq-int_rep_monotone_functions}
f(x) = f(0) + f' (0) \int_{-1}^1 \D \mu (t) \frac{x}{1 - x t} \;,
\end{equation}
where $\mu$ is a probability measure  on $[-1,1]$ (see \cite{Bhatia}, Corollary V.4.5).
Furthermore, if $f: \real_+ \rightarrow \real_+$ is continuous, then $f$ is operator monotone
\ifif it is operator concave (\cite{Bhatia}, Theorem~V.2.5). The fact that concavity implies 
monotonicity is easily obtained by
noting that if $ 0 \leq A \leq B$, $C = B - A \geq 0$, and $0 \leq \eta < 1$, then $f ( \eta B ) \geq \eta f ( A) + ( 1 - \eta) f ( \eta ( 1 - \eta)^{-1} C)$
(by concavity). As $f(x) \geq 0$ the second term in the \RHS
is non-negative and thus $f(\eta B ) \geq \eta f (A)$. Letting $\eta \rightarrow 1$ 
we get $f(B) \geq f(A)$.
The converse implication can be shown by similar arguments as those used to establish 
(\ref{eq-Jensen_for_operator_convex}) below and by invoking the fact that 
if (\ref{eq-Jensen_for_operator_convex})  is satisfied for any contraction $C$ then $f$ is operator convex (see~\cite{Bhatia} for more detail).

Another remarkable result valid for continuous functions $f : [0,a) \rightarrow \real$ is that 
$f$ is operator convex and $f(0)\leq 0$ \ifif 
$g(x) = x^{-1} f(x)$ is operator monotone on $(0,a)$ (\cite{Bhatia}, Theorem~V.2.9). Similarly, 
for functions $f : (-1,1) \rightarrow \real$ of class $C^2$, if $f$ is operator convex and $f(0)=0$ then $g(x)$ is operator monotone (\cite{Bhatia}, Corollary V.3.11).
An integral representation  for non-linear operator convex functions
$f$ can be obtained with the help of the last property, by applying (\ref{eq-int_rep_monotone_functions}) to $g(x)$.

If $f: \real_+ \rightarrow \real$ is operator convex and $f(0) \leq 0$, then
\begin{equation} \label{eq-Jensen_for_operator_convex}
f ( C^\ast A C  ) \leq C^\ast f ( A ) C
\end{equation}
for any contraction $C \in \observables$, $\| C \| \leq 1$, and any $A \in \observables_+$.
This inequality can be shown as follows~\cite{Hansen82}. 
Let us consider the matrices
\begin{equation}
\widehat{A} = 
\left( \begin{array}{cc} A  & 0   \\ 0 & 0 \end{array} \right)
\quad , \quad 
\widehat{U}_\pm =  \left( \begin{array}{cc} C  & \pm D   \\ E & \mp C^\ast \end{array} \right)
\end{equation}
with $D= \sqrt{ 1 - C C^\ast}$ and $E = \sqrt{ 1 - C^\ast C }$ (the latter operators are well defined since $\| C  \| \leq 1$). 
An explicit calculation shows that $\widehat{U}_\pm$ is unitary and 
\begin{equation}
\left(  \begin{array}{cc} C^\ast A C & 0 \\ 0 & D A D \end{array} \right) 
 = \onehalf \sum_{\epsilon=\pm} \widehat{U}_\epsilon^\ast  \widehat{A} \,\widehat{U}_\epsilon 
\;.
\end{equation}
If $f$ is operator convex and $f(0) \leq 0$, then
\begin{eqnarray}
\nn
\left( \begin{array}{cc} f ( C^\ast A C ) & 0 \\ 0 & f ( D A D )\end{array} \right)
& = &  
f \left( \begin{array}{cc} C^\ast A C  & 0 \\ 0 & D A D \end{array} \right)
\\ \nn
& \leq &  
\onehalf \sum_{\epsilon=\pm} f ( \widehat{U}_\epsilon^\ast  \widehat{A} \,\widehat{U}_\epsilon )
\\ 
& \leq &  \onehalf \sum_{\epsilon=\pm} \widehat{U}_\epsilon^\ast 
 \left( \begin{array}{cc} f (A ) & 0 \\ 0 & 0 \end{array} \right) \widehat{U}_\epsilon
= \left( \begin{array}{cc} C^\ast f (A ) C & 0 \\ 0 & D f ( A ) D  \end{array} \right)\;.
\end{eqnarray}
This implies in particular the bound (\ref{eq-Jensen_for_operator_convex}).
Conversely, it is  shown in~\cite{Hansen82}  that 
if this bound  is satisfied for any orthogonal projection $C$ and any $A \in \Bb ( \Hh)_+$,
then $f$ is operator convex and $f(0) \leq 0$.

Let $\Mm$ be a \QO on $\observables$ and $f: \real_+ \rightarrow \real$ be operator convex. 
Then the following Jensen-type inequality  holds~\cite{Davis57}:
\begin{equation} \label{eq-Jensen_type_ineq}
f ( \Mm^\ast (A)) \leq \Mm^\ast ( f (A)) \quad , \quad A \in \observables_+\;.
\end{equation}
A simple justification of this inequality is as follows. 
Since $\Mm^\ast (c \,1 ) = c\,1$ for any constant $c \in \real$, one may assume without loss of generality that $f(0)=0$.
Let $A \in \observables_+$.
According to Stinespring's theorem  (Sec.~\ref{sec-measurements}) 
one can find a unitary operator $U$ on an enlarged space $\Hh \otimes \Hh_\EE$ and a vector  $\ket{\epsilon_0} \in \Hh_\EE$
such that $\Mm^\ast ( A ) = \bra{\epsilon_0} U^\ast A \otimes 1 U \ket{\epsilon_0}$.
Let us set $P_0 = \ketbra{\epsilon_0}{\epsilon_0}$. 
Applying (\ref{eq-Jensen_for_operator_convex}) with $C = 1 \otimes P_0$, one gets
\begin{eqnarray} \label{eq-proof_Davis}
\nn
f( \Mm^\ast ( A) ) \otimes P_0
& = & f ( 1 \otimes P_0\, U^\ast A \otimes 1 U \,1\otimes P_0 )
\\
& \leq & 1 \otimes P_0 f (  U^\ast A \otimes 1 U )  1 \otimes P_0
= \Mm^\ast ( f(A) ) \otimes P_0\;.
\end{eqnarray}
%

\setcounter{equation}{0}
\section{Trace inequalities}  
\label{app-norm_inequalities}

In  this appendix  some inequalities
involving the $\|\cdot \|_p$-norms  are stated or derived. 

\begin{enumerate}

\item
Let us first recall the triangle and ``inverse triangle'' inequalities: for any matrices $A$ and $B$ one has
\begin{equation} \label{eq-Minkowski_inequality}
\| A + B \|_p \;\;
\begin{cases}
\;\;
\leq \| A \|_p +  \| B \|_p & \text{ if $p\geq 1$} 
\\
\;\; \geq   \| A \|_p +  \| B \|_p & \text{ if $0 < p < 1$.}
\end{cases}
\end{equation}
This shows that the map $A \mapsto \| A \|_p$  defined by  (\ref{eq-Lp_norm}) is a norm  
for $p \geq 1$, but this is not the case for $p <1$. 
One deduces  the bound
\begin{equation} \label{eq-inequality_convex_function}
\tr [ \sqrt{ |A|^2+|B|^2 }] \leq \tr |A| + \tr |B| 
\end{equation}
by applying (\ref{eq-Minkowski_inequality}) for $p=1$ to the matrices
\begin{equation*}
\widehat{A} = 
\left( \begin{array}{cc} A  & 0 \\ 0 & 0 \end{array} \right)
\quad , \quad 
\widehat{B} = 
\left( \begin{array}{cc} 0 & 0 \\ B & 0 \end{array} \right)
\;.
\end{equation*}

\item 
Another standard result is the Lieb-Thirring inequality~\cite{Lieb76}.
We quote here without proof a generalization of this inequality derived by Araki~\cite{Araki90}. Let $k>0$ and  $A$ and $B$ be non-negative operators. 
If $\alpha \geq 1$ then
\begin{equation} \label{eq-Lieb-Thirring}
\bigl\| B^\onehalf A B^\onehalf \bigr\|_{\alpha k}^{\alpha} \leq \bigl\| B^{\frac{\alpha}{2}} A^\alpha  B^{\frac{\alpha}{2}} \bigr\|_k \;.
\end{equation}
Taking $\alpha \rightarrow \alpha^{-1}$ and $k \rightarrow k/\alpha$, one can deduce that 
the reverse inequality holds true if $0 \leq \alpha \leq 1$. 

\item
Next, let us show that for any square matrices $A$, $B$, $C$, and $D$ of the same size, the following bound generalizing the Cauchy-Schwarz inequality 
$\| A B \|_1 \leq \| A \|_2 \| B \|_2$ holds true~\cite{Montanaro08}
\begin{equation} \label{eq-norm_1_and_2_inequality}
\| A B + C D \|_1^2 \leq \bigl( \| A \|_2^2 + \| D \|_2^2 \bigr) \bigl( \|  B \|_2^2 + \| C \|_2^2 \bigr) \;.
\end{equation}
Actually, let us  form the $2 \times 2$ block matrices
\begin{equation*}
\widehat{E} = 
\left( \begin{array}{cc} A^\ast & 0 \\ C^\ast & 0 \end{array} \right)
\quad , \quad 
\widehat{F} = 
\left( \begin{array}{cc} B & 0 \\ D & 0 \end{array} \right)
\;.
\end{equation*}
Then 
\begin{equation*}
\| A B  + C D  \|_1^2 = \bigl\| \widehat{E}^\ast \widehat{F} \bigr\|^2_1 
 \leq \| \widehat{E} \|_2 ^2 \| \widehat{F} \|_2^2
  = \bigl( \| A \|_2^2 + \| C \|_2^2 \bigr) \bigl( \| B \|_2^2 + \| D \|_2^2 \bigr)\;.
\end{equation*} 
But  $C D = U D^\ast C^\ast U$ with $U$ unitary by the polar decomposition. Applying the above inequality with
$C$ and $D$ replaced by $U D^\ast$ and $C^\ast U$ and using the unitary invariance of $\|\cdot \|_2$, one gets 
the desired result (\ref{eq-norm_1_and_2_inequality}).

\item
Let $B=(B_{ij})_{i,j=1}^m$ be a non-negative $m \times m$ operator-valued matrix, whose entries $B_{ij}$ are given by  $p_i \times p_j$ matrices. 
Denote by $A=\sqrt{B}=(A_{ij})_{i,j=1}^m $ the square root of $B$. Then for any $j=1,\ldots, m$, one has~\cite{Barnum02}
\begin{equation} \label{eq-lower_and_upper_bounds_on_P_S_amb_app}
 \sum_{i,i \not= j} \bigl\| A_{ij} \bigr\|_2^2 \leq  \onehalf \sum_{i,i \not= j} \bigl\| B_{ij} \bigr\|_1\;.
\end{equation}
Let us first establish (\ref{eq-lower_and_upper_bounds_on_P_S_amb_app}) for $m=2$. 
Thanks to the singular value decomposition and the unitary invariance of the $\| \cdot \|_p$-norms, 
we may assume without loss of generality that $A_{12}$ is a diagonal $p_1 \times p_2$ matrix, \ie 
$A_{12} = \sum_{k=1}^{p} \sqrt{\nu_k} \ketbra{k}{k}$ with $p=\min\{ p_1,p_2\}$. By a standard argument, the non-negativity of $A$ implies 
\begin{equation*} \label{eq-keep_zen}
| \bra{\varphi_1} A_{12} \ket{\varphi_2}|^2 \leq \bra{\varphi_1} A_{11} \ket{\varphi_1} \bra{\varphi_2} A_{22} \ket{\varphi_2} 
\end{equation*} 
for any vectors $\ket{\varphi_1} \in \complex^{p_1}$ and $\ket{\varphi_2} \in \complex^{p_2}$. Using this bound 
and the relation $B_{12}= A_{11} A_{12} + A_{12} A_{22}$, we find
\begin{equation*}
\| A_{12} \|_2^2
 =  
 \sum_{k=1}^p \nu_k  \leq \sum_{k=1}^p \sqrt{\nu_k \bra{k} A_{11} \ket{k} \bra{k} A_{22} \ket{k}}
  \leq  \onehalf \sum_{k=1}^p  \sqrt{\nu_k } \bigl( \bra{k} A_{11} \ket{k} + \bra{k} A_{22} \ket{k} \bigr)
 =  \onehalf  \| B_{12} \|_1\;.
\end{equation*}
Consider now the general case $m \geq 2$. The idea is to write $B$ as a  $2 \times 2$ block matrix such that the upper left and
 lower right blocks are the $(m-1)\times (m-1)$ matrix $( B_{ij})_{i,j=1}^{m-1}$ and the single entry $B_{m m}$, respectively, whereas the
upper right (lower left) block forms 
a column (line) vector with entries $B_{i m}$ ($B_{m i}$). A similar block decomposition can be made for $A$. 
Applying the foregoing result for $m=2$, one gets
\begin{equation*}
\sum_{i,i \not= m} \bigl\| A_{im} \bigr\|_2^2 = \left\| \left( \begin{array}{c} A_{1m} \\ \vdots \\ A_{(m-1)m} \end{array} \right) \right\|_2^2 
\leq \onehalf 
 \left\| \left( \begin{array}{c} B_{1m} \\ \vdots \\ B_{(m-1)m} \end{array} \right) \right\|_1 = \onehalf \biggl\| \sqrt{\sum_{i,i \not= m} | B_{im} |^2} 
\biggr\|_1
\leq \onehalf  \sum_{i,i \not= m}  \| B_{im} \|_1\;,
\end{equation*}
where we have used (\ref{eq-inequality_convex_function}) in the last bound.
This proves (\ref{eq-lower_and_upper_bounds_on_P_S_amb_app}) for $j=m$. By an appropriate unitary conjugation, one deduces that the bound holds for any $j$. 

\item
The following trace inequality  plays a central role in the derivation of the quantum Chernoff bound~\cite{Audenaert07}:
for any positive square matrices $A>0$ and $B>0$ and any $0 \leq s \leq 1$,
\begin{equation} \label{eq-ineq_Audenaert}
\onehalf \Bigl( \tr (A) + \tr (B) - \tr | A - B| \Bigr) \leq \tr ( A^{1-s} B^s )\;.
\end{equation} 
This inequality was first shown in~\cite{Audenaert07}, but the proof in this reference is not very transparent.
We present here a much simpler proof  due to Ozawa, which has been first reported in~\cite{Jaksic12}.
Denoting by $O_\pm =( |O| \pm  O )/2 \geq 0$ the positive and negative parts  of $O$, one may express
$\tr | A - B |$ as  $2 \tr ( A-B )_+ - \tr ( A) + \tr (B)$. Thus (\ref{eq-ineq_Audenaert}) is equivalent to
\begin{equation*}
 \tr \bigl( ( A^s - B^s )  A^{1-s} \bigr) \leq \tr ( A-B )_+  \;.
\end{equation*}
Since $f(x)=x^s$ is operator monotone (see Appendix~\ref{app_operator_convex_functions}) and
$ A \leq A + ( A-B)_- = B + (A-B)_+$, one has $A^s \leq ( B + (A-B)_+ )^s$. Hence
\begin{eqnarray*}
  \tr \bigl( ( A^s - B^s )  A^{1-s} \bigr) 
& \leq  & \tr \bigl( \bigl[  ( B + (A-B)_+ )^s  - B^s \bigr] A^{1-s} \bigr)
\\
& \leq &
\tr \bigl( \bigl[ ( B + (A-B)_+ )^s  - B^s \bigr] ( B + (A-B)_+ )^{1-s} \bigr)\;,
\end{eqnarray*}
where the second inequality relies on the similar bound $B^s \leq ( B + (A-B)_+ )^s$.
By rearranging the product in the last trace and using the latter bound with $s \leftrightarrow (1-s)$, one gets
\begin{equation*}
 \tr \bigl( ( A^s - B^s )  A^{1-s} \bigr)
 \leq  
\tr (B) + \tr ( A-B)_+ - \tr \bigl( B^s  ( B + (A-B)_+ )^{1-s}  \bigr) 
\leq \tr ( A-B)_+\;.
\end{equation*}
This concludes the justification of (\ref{eq-ineq_Audenaert}).

\end{enumerate}

\newpage

\end{document}

%% file: fig3_with_latex.pstex_t
\begin{picture}(0,0)%
\includegraphics{fig3_with_latex.pstex}%
\end{picture}%
\setlength{\unitlength}{2072sp}%
\begingroup\makeatletter\ifx\SetFigFont\undefined%
\gdef\SetFigFont#1#2#3#4#5{%
  \reset@font\fontsize{#1}{#2pt}%
  \fontfamily{#3}\fontseries{#4}\fontshape{#5}%
  \selectfont}%
\fi\endgroup%
\begin{picture}(5610,3436)(2596,-4628)
\put(2611,-3481){\makebox(0,0)[lb]{\smash{{\SetFigFont{10}{12.0}{\rmdefault}{\mddefault}{\updefault}{\color[rgb]{0,0,0}$\rho$}%
}}}}
\put(2656,-1681){\makebox(0,0)[lb]{\smash{{\SetFigFont{10}{12.0}{\rmdefault}{\mddefault}{\updefault}{\color[rgb]{0,0,0}$\rho$}%
}}}}
\put(4816,-3481){\makebox(0,0)[lb]{\smash{{\SetFigFont{10}{12.0}{\rmdefault}{\mddefault}{\updefault}{\color[rgb]{0,0,0}$\{ \rho_{{\sf{AB}}|i},\eta_i\}$}%
}}}}
\put(7831,-3481){\makebox(0,0)[lb]{\smash{{\SetFigFont{10}{12.0}{\rmdefault}{\mddefault}{\updefault}{\color[rgb]{0,0,0}$\{ \rho_i , \eta_i \}$}%
}}}}
\put(8191,-1681){\makebox(0,0)[lb]{\smash{{\SetFigFont{10}{12.0}{\rmdefault}{\mddefault}{\updefault}{\color[rgb]{0,0,0}$\rho$}%
}}}}
\put(3781,-1411){\makebox(0,0)[lb]{\smash{{\SetFigFont{9}{10.8}{\rmdefault}{\mddefault}{\updefault}{\color[rgb]{0,0,0}${\mathcal{M}}_{\sf{A}}$}%
}}}}
\put(6601,-1411){\makebox(0,0)[lb]{\smash{{\SetFigFont{9}{10.8}{\rmdefault}{\mddefault}{\updefault}{\color[rgb]{0,0,0}${\mathcal{R}}_{ {\mathcal{M}}_{\sf A},\rho}$}%
}}}}
\put(5056,-1696){\makebox(0,0)[lb]{\smash{{\SetFigFont{10}{12.0}{\rmdefault}{\mddefault}{\updefault}{\color[rgb]{0,0,0}${\mathcal{M}}_{\sf{A}} (\rho)$}%
}}}}
\end{picture}%